\documentclass[times, twocolumn]{aastex631}

\newcommand{\herschel}{\textit{Herschel}}

\usepackage{savesym}
\savesymbol{tablenum}

\usepackage[binary-units]{siunitx}
\restoresymbol{SIX}{tablenum}

\renewcommand\micron{\mbox{\si{\micro\metre}}}

\renewcommand\arcdeg{\mbox{$^\circ$}}%
\renewcommand\arcmin{\mbox{$^\prime$}}%
\renewcommand\arcsec{\mbox{$^{\prime\prime}$}}%
\newcommand\ha{H$\mathrm{\alpha}$}

\newcommand\msun{\mbox{\si{M_\odot}}}
\newcommand\lsun{\mbox{\si{L_\odot}}}
\newcommand\smpy{\mbox{\si{M_\odot. yr^{-1}}}}

\newcommand{\lya}{Ly$\alpha$}
\newcommand{\cii}{[\ion{C}{2}]}

\newcommand{\oiii}{[\ion{O}{3}]}
\newcommand{\nii}{[\ion{N}{2}]}
\newcommand{\sii}{[\ion{S}{2}]}
\newcommand{\feii}{[\ion{Fe}{2}]}

\defcitealias{rawle16}{R16}

\newcommand{\textred}[1]{{#1}}
\newcommand{\textblue}[1]{{#1}}

\graphicspath{{./}{figures/}}

\received{May 1, 2023 }
\revised{October 11, 2023}
\accepted{October 17, 2023}
\submitjournal{ApJ}

\shorttitle{JADES: Stellar Component and Overdense Environment of HDF850.1 at $z=5.18$}
\shortauthors{Sun et al.}

\begin{document}

\title{JADES: Resolving the Stellar Component and Filamentary Overdense Environment of HST-Dark Submillimeter Galaxy HDF850.1 at $z=5.18$}


\correspondingauthor{Fengwu Sun}
\email{fengwusun@arizona.edu}


\author[0000-0002-4622-6617]{Fengwu Sun}
\affiliation{Steward Observatory, University of Arizona, 933 N. Cherry Avenue, Tucson, AZ 85721, USA}

\author[0000-0003-4337-6211]{Jakob M. Helton}
\affiliation{Steward Observatory, University of Arizona, 933 N. Cherry Avenue, Tucson, AZ 85721, USA}

\author[0000-0003-1344-9475]{Eiichi Egami}
\affiliation{Steward Observatory, University of Arizona, 933 N. Cherry Avenue, Tucson, AZ 85721, USA}

\author[0000-0003-4565-8239]{Kevin N. Hainline}
\affiliation{Steward Observatory, University of Arizona, 933 N. Cherry Avenue, Tucson, AZ 85721, USA}

\author[0000-0003-2303-6519]{George H. Rieke}
\affiliation{Steward Observatory, University of Arizona, 933 N. Cherry Avenue, Tucson, AZ 85721, USA}

\author[0000-0001-9262-9997]{Christopher N. A. Willmer}
\affiliation{Steward Observatory, University of Arizona, 933 N. Cherry Avenue, Tucson, AZ 85721, USA}

\author[0000-0002-2929-3121]{Daniel J.\ Eisenstein}
\affiliation{Center for Astrophysics $|$ Harvard \& Smithsonian, 60 Garden St., Cambridge MA 02138 USA}

\author[0000-0002-9280-7594]{Benjamin D.\ Johnson}
\affiliation{Center for Astrophysics $|$ Harvard \& Smithsonian, 60 Garden St., Cambridge MA 02138 USA}

\author[0000-0002-7893-6170]{Marcia J.\ Rieke}
\affiliation{Steward Observatory, University of Arizona, 933 N. Cherry Avenue, Tucson, AZ 85721, USA}

\author[0000-0002-4271-0364]{Brant Robertson}
\affiliation{Department of Astronomy and Astrophysics, University of California, Santa Cruz, 1156 High Street, Santa Cruz, CA 95064, USA}

\author[0000-0002-8224-4505]{Sandro Tacchella}
\affiliation{Kavli Institute for Cosmology, University of Cambridge, Madingley Road, Cambridge, CB3 0HA, UK}
\affiliation{Cavendish Laboratory, University of Cambridge, 19 JJ Thomson Avenue, Cambridge, CB3 0HE, UK}

\author[0000-0002-8909-8782]{Stacey Alberts}
\affiliation{Steward Observatory, University of Arizona, 933 N. Cherry Avenue, Tucson, AZ 85721, USA}

\author[0000-0003-0215-1104]{William M.\ Baker}
\affiliation{Kavli Institute for Cosmology, University of Cambridge, Madingley Road, Cambridge, CB3 0HA, UK}
\affiliation{Cavendish Laboratory, University of Cambridge, 19 JJ Thomson Avenue, Cambridge, CB3 0HE, UK}

\author[0000-0003-0883-2226]{Rachana Bhatawdekar}
\affiliation{European Space Agency (ESA), European Space Astronomy Centre (ESAC), Camino Bajo del Castillo s/n, 28692 Villanueva de la Cañada, Madrid, Spain}
\affiliation{European Space Agency, ESA/ESTEC, Keplerlaan 1, 2201 AZ Noordwijk, NL}

\author[0000-0003-4109-304X]{Kristan Boyett}
\affiliation{School of Physics, University of Melbourne, Parkville 3010, VIC, Australia}
\affiliation{ARC Centre of Excellence for All Sky Astrophysics in 3 Dimensions (ASTRO 3D), Australia}

\author[0000-0002-8651-9879]{Andrew J.\ Bunker}
\affiliation{Department of Physics, University of Oxford, Denys Wilkinson Building, Keble Road, Oxford OX1 3RH, UK}

\author[0000-0003-3458-2275]{Stephane Charlot}
\affiliation{Sorbonne Universit\'e, CNRS, UMR 7095, Institut d'Astrophysique de Paris, 98 bis bd Arago, 75014 Paris, France}

\author[0000-0002-2178-5471]{Zuyi Chen}
\affiliation{Steward Observatory, University of Arizona, 933 N. Cherry Avenue, Tucson, AZ 85721, USA}

\author[0000-0002-7636-0534]{Jacopo Chevallard}
\affiliation{Department of Physics, University of Oxford, Denys Wilkinson Building, Keble Road, Oxford OX1 3RH, UK}

\author[0000-0002-9551-0534]{Emma Curtis-Lake}
\affiliation{Centre for Astrophysics Research, Department of Physics, Astronomy and Mathematics, University of Hertfordshire, Hatfield AL10 9AB, UK}

\author[0000-0002-9708-9958]{A.\ Lola Danhaive}
\affiliation{Kavli Institute for Cosmology, University of Cambridge, Madingley Road, Cambridge, CB3 0HA, UK}
\affiliation{Cavendish Laboratory, University of Cambridge, 19 JJ Thomson Avenue, Cambridge, CB3 0HE, UK}

\author[0000-0002-4781-9078]{Christa DeCoursey}
\affiliation{Steward Observatory, University of Arizona, 933 N. Cherry Avenue, Tucson, AZ 85721, USA}

\author[0000-0001-7673-2257]{Zhiyuan Ji}
\affiliation{Steward Observatory, University of Arizona, 933 N. Cherry Avenue, Tucson, AZ 85721, USA}

\author[0000-0002-6221-1829]{Jianwei Lyu}
\affiliation{Steward Observatory, University of Arizona, 933 N. Cherry Avenue, Tucson, AZ 85721, USA}

\author[0000-0002-4985-3819]{Roberto Maiolino}
\affiliation{Kavli Institute for Cosmology, University of Cambridge, Madingley Road, Cambridge, CB3 0HA, UK}
\affiliation{Cavendish Laboratory - Astrophysics Group, University of Cambridge, 19 JJ Thomson Avenue, Cambridge, CB3 0HE, UK}
\affiliation{Department of Physics and Astronomy, University College London, Gower Street, London WC1E 6BT, UK}

\author[0000-0002-0303-499X]{Wiphu Rujopakarn}
\affiliation{National Astronomical Research Institute of Thailand, Don Kaeo, Mae Rim, Chiang Mai 50180, Thailand}
\affiliation{Department of Physics, Faculty of Science, Chulalongkorn University, 254 Phayathai Road, Pathumwan, Bangkok 10330, Thailand}

\author[0000-0001-9276-7062]{Lester Sandles}
\affiliation{Kavli Institute for Cosmology, University of Cambridge, Madingley Road, Cambridge, CB3 0HA, UK}
\affiliation{Cavendish Laboratory, University of Cambridge, 19 JJ Thomson Avenue, Cambridge, CB3 0HE, UK}

\author[0000-0003-4702-7561]{Irene Shivaei}
\affiliation{Steward Observatory, University of Arizona, 933 N. Cherry Avenue, Tucson, AZ 85721, USA}

\author[0000-0003-4891-0794]{Hannah \"Ubler}
\affiliation{Kavli Institute for Cosmology, University of Cambridge, Madingley Road, Cambridge, CB3 0HA, UK}
\affiliation{Cavendish Laboratory, University of Cambridge, 19 JJ Thomson Avenue, Cambridge, CB3 0HE, UK}

\author[0000-0002-4201-7367]{Chris Willott}
\affiliation{NRC Herzberg, 5071 West Saanich Rd, Victoria, BC V9E 2E7, Canada}

\author[0000-0002-7595-121X]{Joris Witstok}
\affiliation{Kavli Institute for Cosmology, University of Cambridge, Madingley Road, Cambridge, CB3 0HA, UK}
\affiliation{Cavendish Laboratory, University of Cambridge, 19 JJ Thomson Avenue, Cambridge, CB3 0HE, UK}

\begin{abstract}
HDF850.1 is the brightest submillimeter galaxy (SMG) in the Hubble Deep Field. 
It is known as a heavily dust-obscured star-forming galaxy embedded in an overdense environment at $z = 5.18$.
With nine-band NIRCam images at 0.8--5.0\,\micron\ obtained through the JWST Advanced Deep Extragalactic Survey (JADES), we detect and resolve the rest-frame UV-optical counterpart of HDF850.1, which splits into two components because of heavy dust obscuration in the center.
The southern component leaks UV and H$\alpha$ photons, bringing the galaxy $\sim$100\, times above the empirical relation between infrared excess and UV continuum slope (IRX--$\beta_\mathrm{UV}$).
The northern component is higher in dust attenuation and thus fainter in UV and H$\alpha$ surface brightness.
We construct a spatially resolved dust attenuation map from the NIRCam images, well matched with the dust continuum emission obtained through millimeter interferometry.
The whole system hosts a stellar mass of $10^{10.8\pm0.1}\,\mathrm{M}_\odot$ and star-formation rate of $10^{2.8\pm0.2}\,\mathrm{M}_\odot\,\mathrm{yr}^{-1}$, placing the galaxy at the massive end of the star-forming main sequence at this epoch.
We further confirm that HDF850.1 resides in a complex overdense environment at $z=5.17-5.30$, which hosts another luminous SMG at $z=5.30$ (GN10). 
The filamentary structures of the overdensity are characterized by 109 \ha-emitting galaxies confirmed through NIRCam slitless spectroscopy at 3.9--5\,\micron, of which only eight were known before the JWST observations. 
Given the existence of a similar galaxy overdensity in the GOODS-S field, our results suggest that $50\pm20$\% of the cosmic star formation at $z=5.1-5.5$ occur in protocluster environments.
\end{abstract}



\section{Introduction}
\label{sec:01_intro}

In the local Universe, the population of ultra-luminous infrared galaxies (ULIRGs, with IR luminosities $L_\mathrm{IR}>10^{12}$\,\lsun) hosts vigorous dust-obscured star formation.
In the distant Universe at redshift $z \gtrsim 1$, galaxies with similar luminosities have been routinely discovered with submillimeter / millimeter and infrared sky surveys. 
The Spitzer Space Telescope provided detailed information out to redshifts of $z \simeq 2 - 3$ \citep[e.g.,][]{armus20}, but at higher redshifts the infrared excess emission moves into the far infrared and the negative K-correction strongly favors discoveries at submillimeter and millimeter wavelengths. 
Some of these galaxies, with flux densities of $\gtrsim 1$\,mJy at submillimeter wavelengths, are classified as submillimeter galaxies (SMGs; see a recent review by \citealt{hodge20}).

Since the first discoveries around the end of the last century \citep[e.g.,][]{smail97,hughes98,ivison98}, the identification of the rest-frame optical (i.e., stellar) counterparts has remained ambiguous for certain high-redshift SMGs with strong dust obscuration.
This was partially caused by the coarse angular resolution of single-dish submillimeter telescopes, for example, $\theta \sim$15\arcsec\ with the 15-m James Clerk Maxwell Telescope (JCMT) at 850\,\micron.
However, even with (sub)-arcsec resolution dust continuum images obtained with millimeter interferometers including the Atacama Large Millimeter Array (ALMA), Submillimeter Array (SMA), Plateau de Bure Interferometer (PdBI) and its successor Northern Extended Millimeter Array (NOEMA), it is still challenging to detect the counterparts of certain SMGs in optical and near-IR wavelengths. 

Recent ALMA continuum observations have suggested that 15--20\% of SMGs remain undetected in deep ground-based $K$-band images ($K>24.4$, \citealt{simpson14}; $K>25.3$, \citealt{dudze20}, \citealt{smail21}; AB magnitude system is used throughout this paper).
Similar percentages of optical/near-IR-dark SMGs are also seen with the SMG samples observed with HST/WFC3-IR ($H_{160} \gtrsim 27$, e.g., \citealt{chen15}, \citealt{franco18}, \citealt{yamaguchi19}, \citealt{gomez21}, \citealt{casey21}, \citealt{sun22a} and \citealt{fujimoto23}), and these galaxies are often referred to as HST-dark, $H$-dropout (not to confuse with Lyman-break galaxies at $z\gtrsim12$) or $H$-faint galaxies.
These galaxies are thought to be objects with high dust attenuation ($A_V \gtrsim 3$) and at high redshifts ($z\simeq 3 - 6$), which could contribute to $\sim$10\% of the cosmic star-formation rate density at this epoch (e.g., \citealt{alcalde19}, \citealt{wang19}, \citealt{williams19}, \citealt{yamaguchi19}, \citealt{dudze20} and \citealt{sun21b}).

As the brightest SMG discovered in the Hubble Deep Field-North \citep[HDF;][]{hughes98}, HDF850.1 is known as one of the earliest examples of optical/near-IR-dark SMGs.
Early tentative optical/near-IR counterpart identifications \citep[]{hughes98,downes99,dunlop04} have been ruled out with increasing accuracy of dust continuum position through high-resolution SMA \citep{cowie09} and PdBI \citep{walter12,neri14} observations.
The source remains undetected even with the deep HST/WFC3-IR imaging obtained through the CANDELS survey \citep[$J_{125}>28.2$, $H_{160}>27.3$;][]{grogin11,koekemoer11,serjeant14}.
With the detections of multiple CO and \cii\ transitions using the PdBI, \citet{walter12} spectroscopically confirmed that HDF850.1 is an SMG at $z=5.183$ (also \citealt{neri14} and \citealt{riechers20}), which resides in a galaxy overdensity at $z\sim5.2$ in the GOODS-N field \citep[see also][]{ah18,calvi21,calvi23}.

The great imaging sensitivity and angular resolution of the JWST/NIRCam \citep{rieke23} have offered us an unprecedented opportunity to resolve the rest-frame optical morphology of HST-dark dusty star-forming galaxies at $z\simeq3-6$ \citep[e.g.,][]{barrufet22,cheng23,kokorev23,mckinney23,nelson22,pg23,rodighiero23,smail23}, which was previously impossible with HST, Spitzer/IRAC or any ground-based facilities unless with the aid of gravitational lensing (see \citealt{sun21b}).

In this paper, we present nine-band NIRCam imaging observation of HDF850.1 obtained through the JWST Advanced Deep Extragalactic Survey (JADES), a guaranteed-time observation (GTO) program aiming to study the formation and evolution of galaxies from $z\sim15$ to the present Universe.
We detect and resolve the stellar component of HDF850.1, solving the puzzle of its optical counterpart identification which has been unclear since its discovery \citep{hughes98}.
Moreover, combining with public NIRCam wide-field slitless spectroscopy (WFSS) observations at 3.9--5.0\,\micron\ \citep{oesch23}, we also confirm 109 galaxies at $z=5.17-5.30$ through the detections of \ha\ emission, including the eight galaxies previously confirmed through ground-based spectroscopy.
Our study unveils four galaxy groups across a comoving line-of-sight distance of $\sim$60\,Mpc, whose 3D distribution resembles filaments of cosmic webs seen with simulations \citep[e.g.,][]{bond96}.
Through this work, we demonstrate the capability of JWST/NIRCam in resolving the dust-obscured stellar component and large-scale structures of high-redshift galaxies through imaging and spectroscopy.

This paper is arranged as follows: In Section \ref{sec:02_obs}, we describe the observations and corresponding data reduction techniques. 
Section~\ref{sec:03_res} presents the photometric and spectroscopic measurements of HDF850.1 and \ha-emitting galaxies in the GOODS-N field at $z_\mathrm{spec}=5.1-5.5$.
We study the physical properties and overdense environment of HDF850.1 in Section~\ref{sec:04_hdf}.
The summary is in Section~\ref{sec:05_sum}.
Throughout this work, we assume a flat $\Lambda$CDM cosmology with $H_0= 70$\,\si{km.s^{-1}.Mpc^{-1}} and $\Omega_\mathrm{M} = 0.3$. 
We also assume a \citet{chabrier03} initial mass function.
We define the IR luminosity ($L_\mathrm{IR}$) as the integrated luminosity over a rest-frame wavelength range from 8 to 1000\,\micron.

\section{Observation and Data Reduction}
\label{sec:02_obs}

\subsection{Imaging}
\label{ss:02a_img}

We conducted JWST/NIRCam imaging observations in the GOODS-N field through the JADES program (PID: 1181; PI: Eisenstein) in early February, 2023. 
The detailed observation design \textred{is presented in \citet{eisenstein23}}.
Our NIRCam observations consist of three medium-depth pointings with MIRI as parallel instrument (12.2\,hours in total; the NIRCam data were partially presented in \citealt{tacchella23}), and four deeper pointings with NIRSpec as primary instrument (31.5\,hours in total; the NIRSpec data were partially presented in \citealt{bunker23}).
The total NIRCam survey area is $\sim$56\,arcmin$^2$ in the GOODS-N field \textred{by February 8, 2023}.
We used nine photometric filters from 0.8 to 5.0\,\micron, including F090W, F115W, F150W, F200W, F277W, F335M, F356W, F410M and F444W filters.
The total exposure time at the location of HDF850.1 (R.A.: 12:36:51.980, Decl.: $+$62\arcdeg12\arcmin25\farcs7) is 6.30\,hours in the F115W band and 3.15\,hours in each of the other eight bands.

The NIRCam imaging data reduction techniques will be presented in a forthcoming paper from the JADES Collaboration (S.\ Tacchella et al., in preparation), and we also refer the readers to \citet{tacchella23} \textred{and \citet{rieke23b}}.
The data were initially processed with the standard JWST calibration pipeline v1.9.2.
The JWST Calibration Reference Data System (CRDS) context map ``\texttt{jwst\_1039.pmap}'' is used, including the flux calibration for JWST/NIRCam from Cycle 1.
Customized steps were included for the removal of the so-caleld ``1/f'' noise, ``wisp'' and ``snowball'' artifacts \citep[see][]{rigby22}.
Because the current long-wavelength (LW; 2.5--5.0\,\micron) flat-field data in the pipeline could introduce small artifacts in the background, we also used customized sky-flats for the LW filters, which were constructed using the deep imaging data obtained with JADES and other public programs (see \citealt{tacchella23}).
The world coordinate system (WCS) positions of all individual images were registered to a reference catalog, which was constructed from the HST mosaics in the GOODS-N field with astrometry tied to the Gaia-EDR3 catalog (\citealt{gaiaedr3}; private communication from G.\ Brammer). 
Finally, the calibrated images are mosaicked to a common world-coordinate system with a pixel size of 0\farcs03 and drizzle parameter of \texttt{pixfrac}=1.

When applicable, we also used the HST/ACS mosaics in the F435W, F606W, F775W, F814W and F850LP bands that were produced as part of the Hubble Legacy Fields project (HLF v2.5; \citealt{whitaker19}).
Spitzer/IRAC mosaics in Channel 1/2 produced by the GOODS Re-ionization Era wide-Area Treasury from Spitzer project (GREATS; \citealt{stefanon21}) are also used, but only for display purpose.

\subsection{Grism Spectroscopy}

The JWST/NIRCam WFSS observations of the GOODS-N field were obtained on UT February 11-13, 2023 through the First Reionization Epoch Spectroscopic COmplete Survey (FRESCO, \citealt{oesch23}; PI: Oesch, PID: 1895).
FRESCO observed both the GOODS-N and GOODS-S fields using the row-direction grisms on both modules of NIRCam with the F444W filter (3.9--5.0\,\micron). 
The spectral resolution is $R\sim1600$ at 4\,\micron.
The total overlapping area between the JADES and FRESCO footprints is $\sim35$\,arcmin$^2$ in GOODS-N. 
The total spectroscopic observing time for FRESCO per GOODS field is $\sim$16\,hours with a typical on-source time of $\sim$2\,hours.

A detailed description of the JWST/NIRCam grism data reduction can be found in \citet{sun22c} and \citet{helton23}.
Starting from the stage-1 products of the JWST standard pipeline, we assigned WCS information, performed flat-fielding and subtracted out the sigma-clipped median sky background from each individual exposure.
Because we are interested in conducting a targeted emission line search, and we do not expect any of our $z>5$ \ha-emitting galaxies to have a strong continuum that is detectable with grism spectroscopy, we utilized a median-filtering technique to subtract out any remaining continuum or background on a row-by-row basis following the methodology of \citet{kashino22}. 
This produces emission-line maps for each grism exposure without continuum emission. 
\textred{Short-wavelength (SW) NIRCam observations were used for the astrometric calibration of the LW WFSS data taken simultaneously. 
Under the assumption that the internal alignment between NIRCam SW and LW detectors is stable, this effectively provides wavelength calibration of grism spectra (see \citealt{sun22c})}. 
We used the spectral tracing, grism dispersion and flux calibration models that were produced using the JWST/NIRCam commissioning or Cycle-1 calibration data (PID: 1076, 1536, 1537, 1538;  \citealt{sun22b,sun22c})\footnote{\href{https://github.com/fengwusun/nircam_grism}{https://github.com/fengwusun/nircam\_grism}}.

\begin{figure*}[!ht]
\centering
\includegraphics[width=\linewidth]{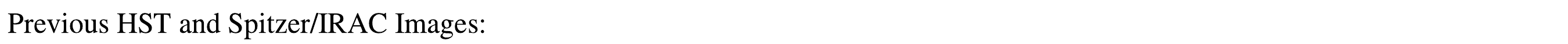}
\includegraphics[width=0.8\linewidth]{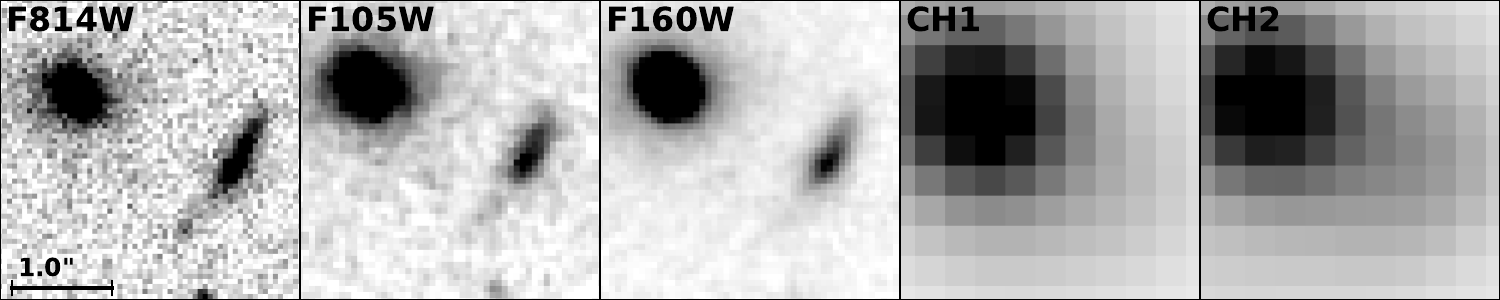}
\includegraphics[width=\linewidth]{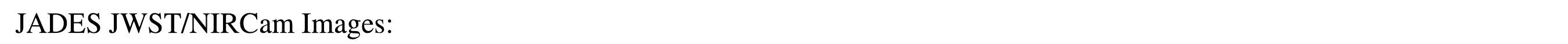}
\includegraphics[width=0.8\linewidth]{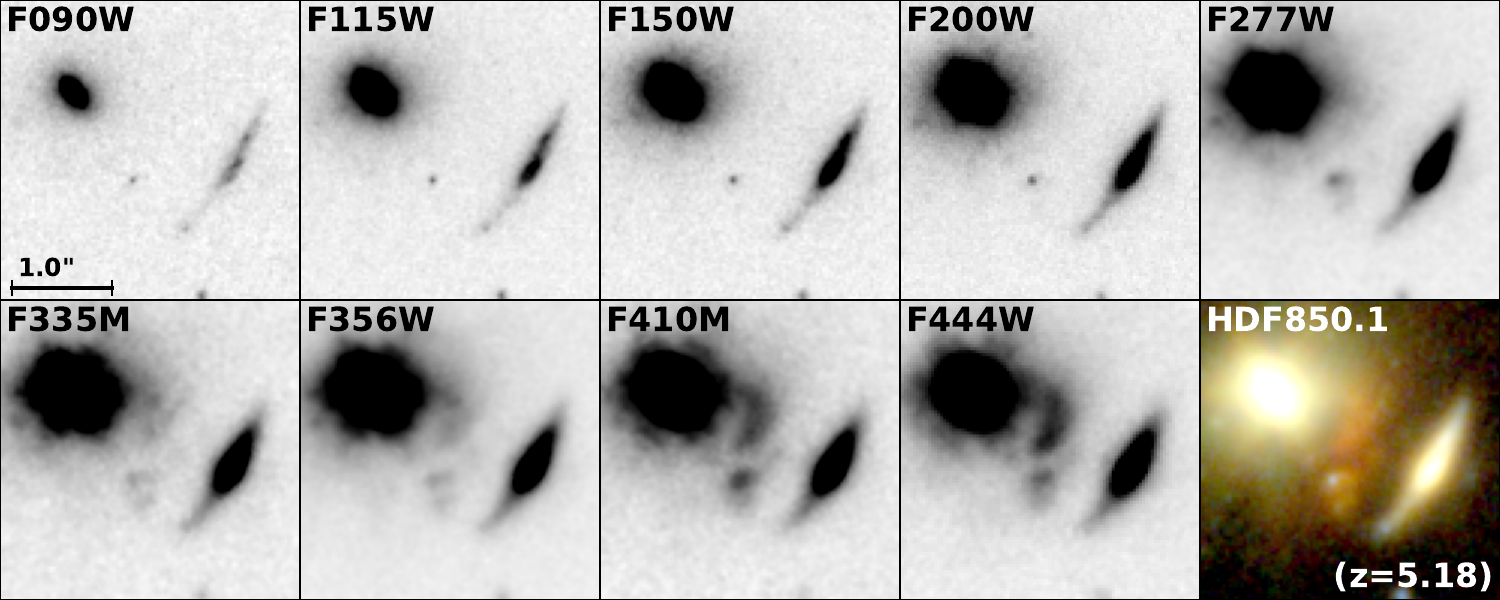}
\caption{\textbf{Top}: HST F814W/F105W/F160W and Spitzer/IRAC Channel 1/2 (3.6/4.5\,\micron) images at the location of HDF850.1. Image sizes are 3\arcsec$\times$3\arcsec\ (north up, east left). HDF850.1 is not detected in HST images and blended with two foreground galaxies in IRAC images. 
\textbf{Bottom}: Same but for JWST/NIRCam cutout images. The last panel shows a false-color image produced using all NIRCam data. The southern component of HDF850.1 is detected in all nine NIRCam bands and the northern component is only detected with the LW filters (F277W--F444W).
}
\label{fig:01_cutout}
\end{figure*}

\section{Results}
\label{sec:03_res}

\subsection{Detection and Photometry of HDF850.1}
\label{ss:03a_smg_phot}

Figure~\ref{fig:01_cutout} presents the HST, Spitzer/IRAC and JWST/NIRCam cutout images of HDF850.1 at $z=5.185$ \citep{neri14}.
HDF850.1 locates between the $z=1.224$ elliptical galaxy in the top-left corner \citep{barger08} and the $z_\mathrm{phot}=2.02_{-0.15}^{+0.08}$ elongated galaxy on the right.
The photometric redshift of this elongated galaxy is estimated from our NIRCam data \citep{hainline23a}, which is slightly higher than the literature value, $z_\mathrm{phot}\sim 1.75$ \citep[e.g.,][]{fs99,re03}.

HDF850.1 is not detected in these HST images, and it is blended with the two foreground galaxies in the Spitzer/IRAC bands. 
Therefore, there was no robust detection of the stellar continuum for HDF850.1 before the launch of the JWST, although we note that \citet{serjeant14} reported the detection of HDF850.1 in deblended IRAC Channel 3/4 (5.8 and 8\,\micron) images. 
Given the coarse resolution of IRAC (the full width of half maximum, FWHM, of the point-spread function, PSF, is $\sim$2\arcsec), HDF850.1 is not resolved separately at these wavelengths.

HDF850.1 is detected and resolved in our nine-band NIRCam images. 
At $z=5.185$, the 4000\,\AA\ break of HDF850.1 is redshifted to 2.5\,\micron, which is similar to the dichroic wavelength between the NIRCam SW and LW channel.
The source is split into a northern and a southern component in the LW filters (F277W--F444W; rest-frame optical), dubbed as HDF850.1-N and HDF850.1-S, respectively.
In the SW filters (F090W--F200W; rest-frame UV), the northern component is not detected, while the southern component is detected as a compact source.

To measure the flux densities of HDF850.1 in a consistent way across all NIRCam wavelengths, we first subtract the two foreground galaxies to the northeast and west of HDF850.1 through morphological modeling with \textsc{galfit} \citep{galfit}. 
The $z=1.224$ elliptical galaxy is fitted with two S\'ersic profiles to account for compact and extended light components, and the $z\sim2$ galaxy is fitted with single S\'ersic profile. 
Source centroids of these two galaxies are fixed in the fitting, while their effective radii ($R_\mathrm{e}$), axis ratios ($b/a$), S\'ersic indices and position angles are allowed to float.
We also include one or two S\'ersic profiles to model the light from HDF850.1.
The best-fit models of two foreground galaxies are then subtracted from our NIRCam images.
\textred{The remaining residual within 0\farcs3 of the northeast elliptical galaxy is about $\sim$1\% of its total brightness, suggesting an overall good quality of subtraction at the present PSF accuracy.
}

In order to study the spatially resolved color information, we homogenize the PSF of NIRCam images in the F090W--F410M filters to that of the F444W filter.
As illustrated in \citet{tacchella23}, we first derive empirical PSF (ePSF) using $\sim10$ stars with the \textsc{Photutils} package \citep{photutils}.
We then construct PSF matching kernels with \textsc{Photutils} assuming a top-hat or cosine bell window function.
The PSF encircled energy functions after kernel convolution agree with that in the F444W band within 1\%, similar to the accuracy reported in \citet{chen23}.

\begin{figure*}[!ht]
\centering
\includegraphics[width=\linewidth]{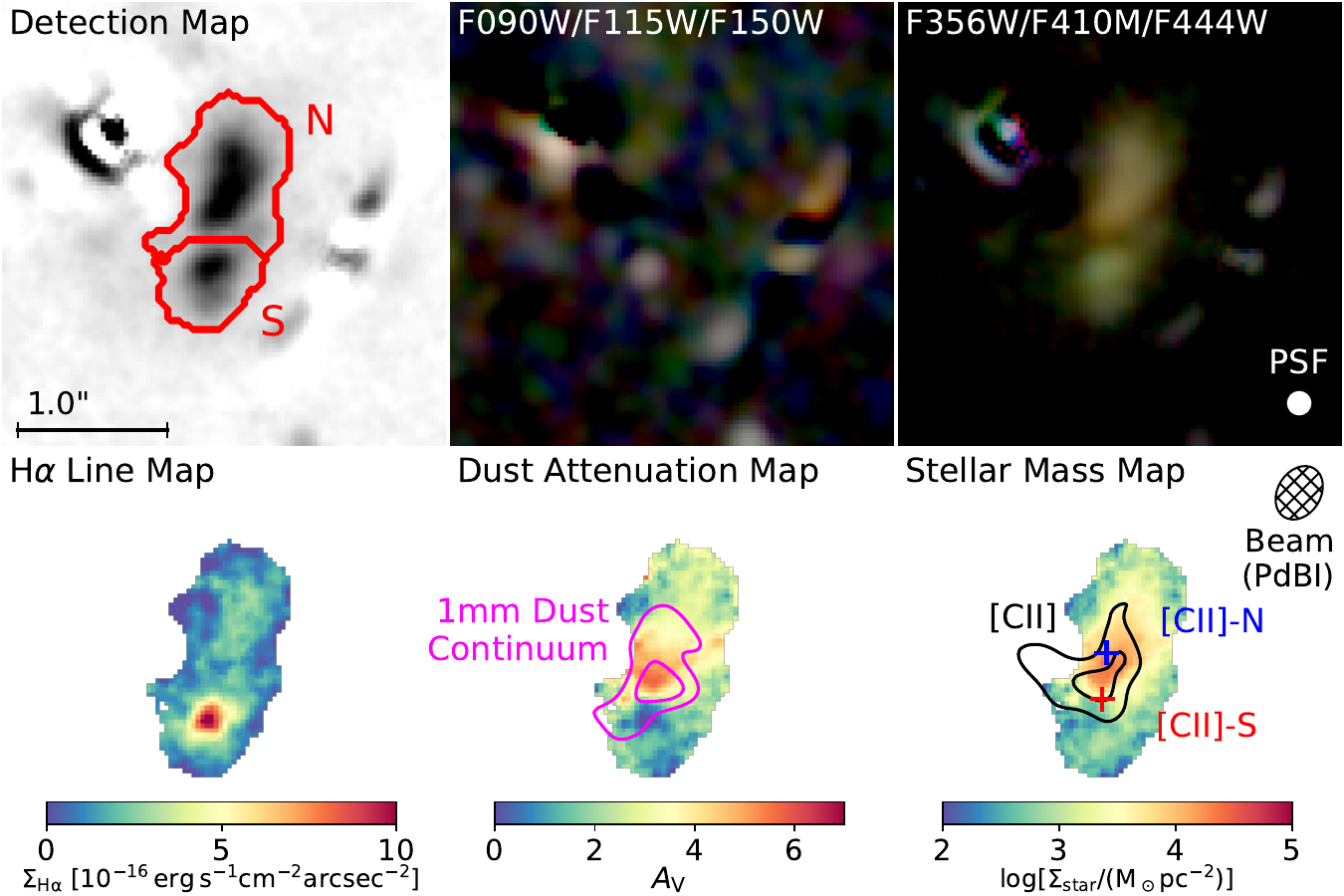}  
\caption{JWST NIRCam images after neighbor subtraction and PSF matching.
Image sizes are 3\arcsec$\times$3\arcsec\ (north up, east left).
The FWHM of the matched NIRCam PSF (F444W) is shown in the top-right panel.
\textbf{Top-left}: Detection map obtained by stacking the F335M, F356W, F410M and F444W images. The deblended segments of the HDF850.1-N and -S components are shown in solid red lines.
\textbf{Top-center}: Rest-frame UV image of HDF850.1 (red: F150W; green: F115W; blue: F090W). The southern component is detected in these images \textred{(4--9$\sigma$)}. 
\textbf{Top-right}: Rest-frame optical image of HDF850.1 (red: F444W; green: F410M; blue: F356W). The \ha\ line  is within the bandwidths of F410M and F444W filters. The northern component is strongly dust-reddened and the southern component exhibits bright \ha\ emission.
\textbf{Bottom-left}: \ha\ line map derived from pixelated SED modeling (Section~\ref{ss:03e_pix}). The \ha\ emission from the southern component is higher in terms of surface brightness.
\textbf{Bottom-center}: Dust-attenuation ($A_V$) map derived from pixelated SED modeling. The center of HDF850.1 is highly dust-obscured ($A_V \gtrsim 5$), and the high-$A_V$ region matches well with the 1\,mm dust continuum emission observed with the PdBI (magenta contours at 4 and $6\sigma$; \citealt{neri14}).
\textbf{Bottom-right}:Stellar mass map derived from the $A_V$ and observed F356W (i.e., rest-frame $V$-band) maps.
The integrated intensity map of the \cii\,158\,\micron\ line is shown as black contours (4 and $6\sigma$; \citealt{neri14}).
The blue and red plus signs mark the centroids of northern and southern \cii\ components in \citet{neri14}, respectively.
The beam size of PdBI observation is indicated in the upper-right corner.
}
\label{fig:02_matched}
\end{figure*}

The neighbor-subtracted PSF-matched NIRCam images of HDF850.1 are presented in Figure~\ref{fig:02_matched}.
We construct the detection map of HDF850.1 by stacking the NIRCam images obtained above 3\,\micron.
We then produce image segmentation with \textsc{Photutils} by finding sources with 40 pixels detected above $5\sigma$ in the detection map. 
Sources are deblended with standard parameters \texttt{nlevels=32} and \texttt{contrast=0.001}.
This allows the deblending of HDF850.1 into two components (-N/S) as shown in solid red lines in Figure~\ref{fig:02_matched}.
We conduct aperture photometry using these segments. 
The northern segment is affected by the residual light from the $z=1.224$ elliptical galaxy in the SW filters, resulting in artificial detections of HDF850.1-N in these bands that are rejected by visual inspection. 
Therefore, in practice we use smaller segments constructed with a higher signal-to-noise ratio (S/N) threshold at 10. 
An aperture correction factor of 1.29 is utilized, which is derived from the median flux ratio between $5\sigma$ and $10\sigma$ segments in the LW filters.
The photometric uncertainty is derived using both the error extension of our image mosaics and the root mean square (RMS) of the residual images. 
Similar to \citet{tacchella23}, we adopt a 5\% error floor for our photometric measurements (Table~\ref{tab:01_hdf}) to account for potential uncertainties from photometric zeropoints and instrument flat fields.

\begin{figure*}[!ht]
\centering
\includegraphics[width=0.45\linewidth]{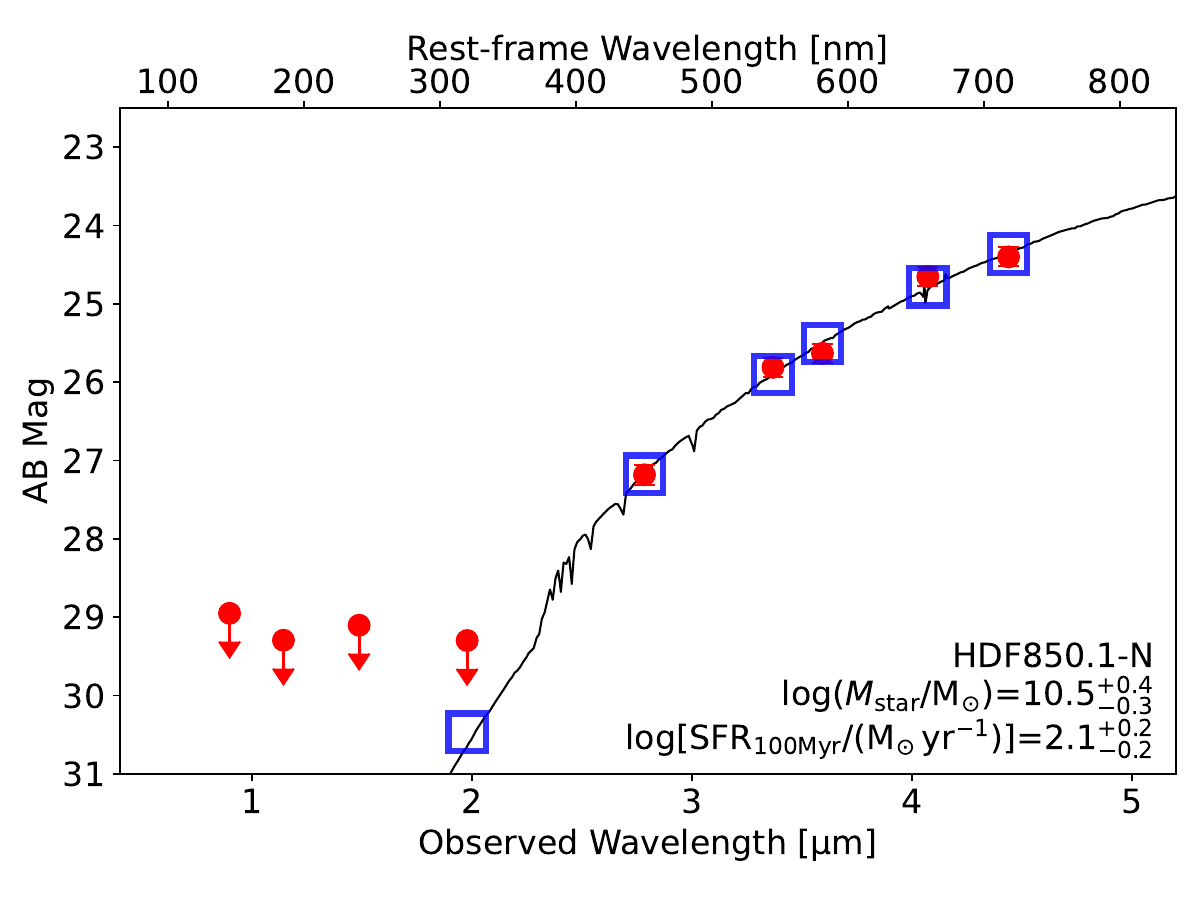}
\includegraphics[width=0.45\linewidth]{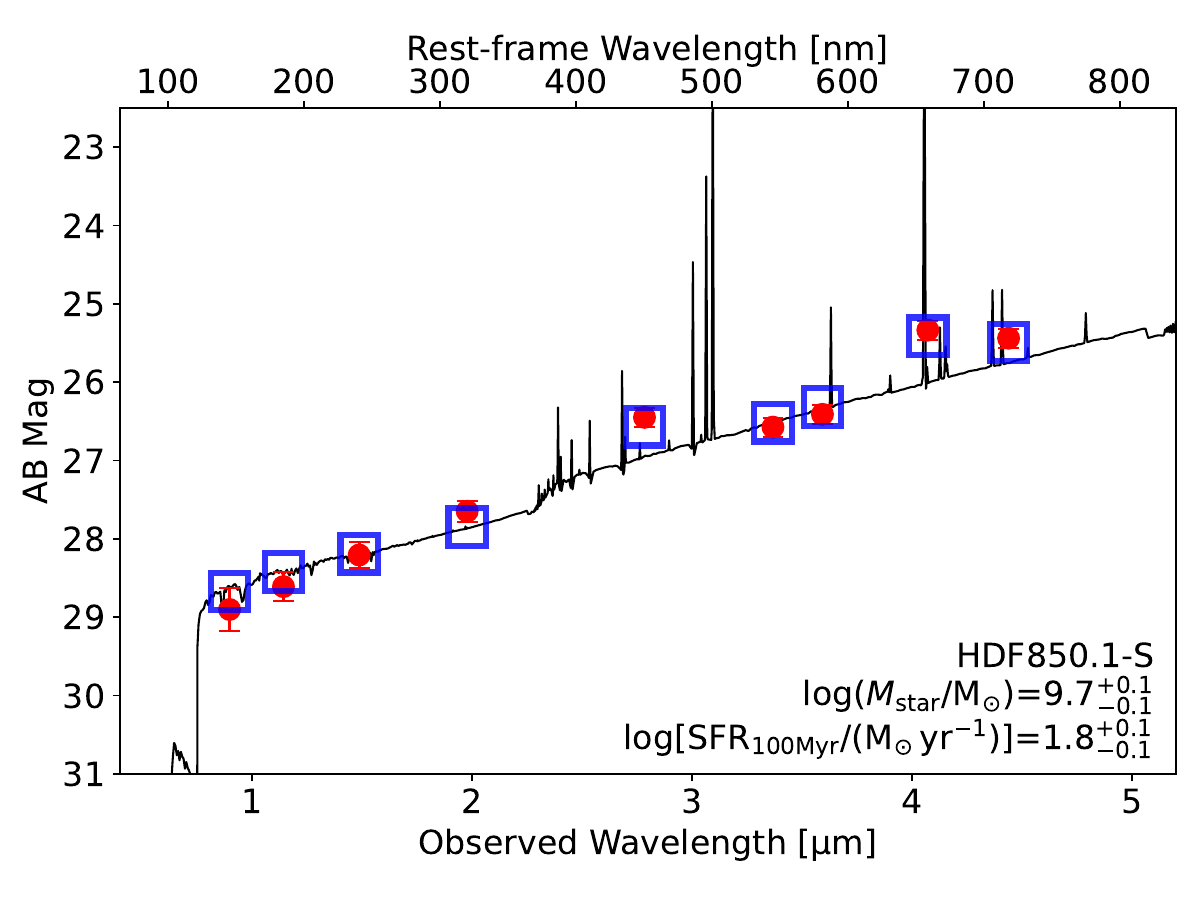}
\includegraphics[width=0.6\linewidth]{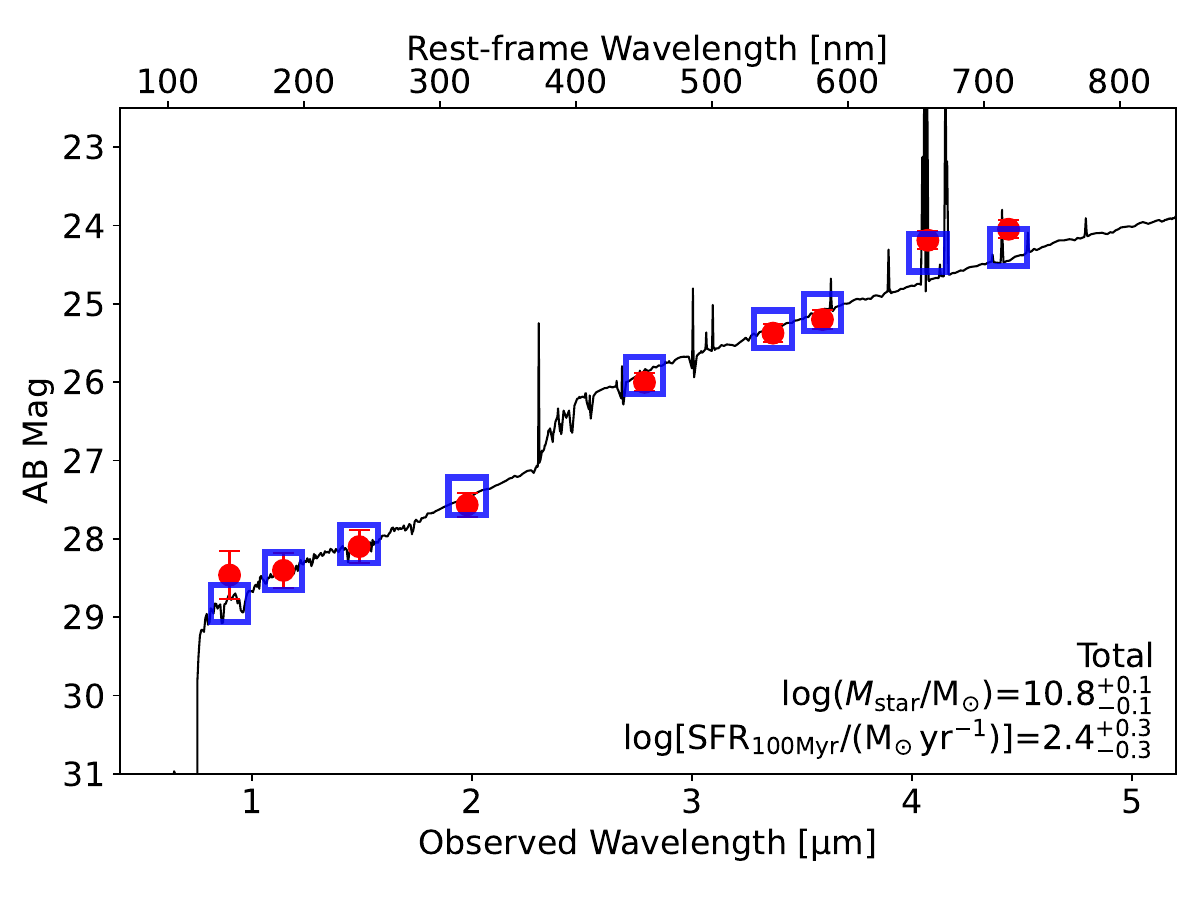}
\caption{Best-fit \textsc{cigale} SED models of HDF850.1-N (top-left), HDF850.1-S (top-right) and the whole system (bottom). 
JWST/NIRCam photometric measurements are shown as red circles, and best-fit source flux densities are shown as open blue squares.
Best-fit source properties have been corrected for lensing magnification (Section~\ref{ss:03b_lens}).
}
\label{fig:07_sed}
\end{figure*}

\subsection{Lens Model}
\label{ss:03b_lens}

HDF850.1 is mildly lensed by the $z=1.224$ elliptical galaxies to the northeast \citep[e.g.,][]{hogg96,hughes98,downes99,dunlop04,cowie09,walter12,neri14}.
With the latest JWST/NIRCam observation, we reconstruct the lens model using software \textsc{lenstool} \citep{lenstool}.
Similar to that in \citet{neri14}, we model the lens as an elliptical singular isothermal sphere with a redshift of $z_l=1.224$ \citep{barger08}.
\textblue{Instead of assuming a velocity dispersion of 150\,\si{km.s^{-1}} used by \citet{dunlop04} and \citet{neri14}, we performed JWST/NIRCam spectral energy distribution modeling of the lens with code \textsc{cigale} \citep{cigale}, and we derived a stellar mass of $M_\mathrm{star} = 10^{10.8\pm0.1}$\,\msun.
With an effective radius of $R_\mathrm{e} = 1.7\pm0.2$\,kpc measured by NIRCam, we estimate a velocity dispersion of $\sigma = \sqrt{GM_\mathrm{star}/5R_\mathrm{e}} \sim 180$\,\si{km.s^{-1}}.
}
The coordinate and ellipticity (0.40) are adopted from the best-fit S\'ersic model of the NIRCam image with \textsc{galfit}.
From this modeling, we derive a mean lensing magnification factor of \textblue{$\mu_\mathrm{S} = 1.9$} for the southern component and \textblue{$\mu_\mathrm{N} = 2.7$} for the northern component.
The mean lensing magnification for HDF850.1 is \textblue{$\bar{\mu} = 2.5$, which is slightly larger those assumed} in \citet{neri14} and \citet{serjeant14} (\textblue{$\bar{\mu} = 1.7$} in the latter case based on the positions of the [CII] emission).
The magnification factor is corrected in the following section when we discuss the intrinsic physical properties of HDF850.1.

\subsection{Physical Model of SED}
\label{ss:03c_sed}

We carry out physical SED modeling of HDF850.1 using two software versions, \textsc{cigale} \citep[][as fiducial results]{noll09,cigale} and \textsc{bagpipes} \citep[][for comparison]{bagpipes}.
Both approaches assume energy balance in dust absorption and emission.
For \textsc{cigale} modeling, we assume a commonly used delayed-$\tau$ star-formation history (SFH), in which $\mathrm{SFR}(t) \propto  t \exp(-t / \tau )$ and $\tau$ is the peak time of SF.
An optional late starburst is allowed in the last 20\,Myr. 
We use \citet{bc03} stellar population synthesis models. 
We also allow a metallicity range of $0.2 Z_\odot - Z_\odot$, and a broad ionization parameter ($\log U$) range of $-1$ to $-3.5$. 
We adopt a modified \citet{calzetti00} attenuation curve, allowing the variation of the power-law slope by $\pm0.3$, and also the dust emission model of \citet{dl07}.
Similar parameters are also assumed for \textsc{bagpipes} modeling. 

We model the SEDs of HDF850.1-N, -S and two components combined.
In rest-frame UV/optical bands, we only use the nine-band NIRCam photometry because the galaxy is undetected with HST (the data are shallower) and heavily blended with foreground sources in the Spitzer/IRAC bands.
To avoid invoking unrealistic dust attenuation and IR luminosity, we also include the \herschel/SPIRE non-detection limits \citep{walter12} in the fitting.
Other (sub)-millimeter flux densities are not used in the fitting because it is challenging to decompose the dust continuum flux densities from the northern and southern components.
The inaccuracy of the energy balance assumption because of patchy dust geometry in contrast to that of the stellar component is also a concern (e.g., see discussions in \citealt{kokorev21} \textred{and evidence in Section~\ref{ss:04c_leak}}).
Far-IR SED modeling of HDF850.1 is presented in Section~\ref{ss:03d_fir}.
However, we confirm that the inclusion of millimeter photometry does not change our best-fit results with \textsc{cigale} presented below, \textred{despite a much larger $\chi^2$ of the best-fit SED model}.

The best-fit \textsc{cigale} SEDs of HDF850.1-N/S and the whole system combined are show in Figure~\ref{fig:07_sed}.
The reduced $\chi^2$ is smaller than 1 for all three fittings, indicating a good fit quality.
The best-fit physical parameters are presented in Table~\ref{tab:01_hdf}.
We find that the northern component is higher in dust attenuation, stellar mass and SFR than the southern component, while the nebular emission in the northern component is much fainter \textred{(see further pixelated SED modeling presented in Section~\ref{ss:03e_pix})}.
The best-fit stellar mass and SFR of the whole system is consistent with the combination of best-fit results for the two components.
We also find  good agreement between the best-fit results of \textsc{cigale} and \textsc{bagpipes}.

The intrinsic $B-V$ of the unattenuated stellar continuum of the best-fit SED for HDF850.1 is 0.04, and the best-fit $A_V$ is $4.6\pm0.7$.
The intrinsic (lensing-corrected) stellar mass of HDF850.1 is \textblue{$\log(M_\mathrm{star}/\mathrm{M}_\mathrm{\odot})=10.8\pm0.1$}. 
This is broadly consistent with the stellar mass estimated by \citet{serjeant14} with deblended IRAC 5.8 and 8.0\,\micron\ flux densities assuming the same IMF.
However, we note that the expected IRAC 5.8 and 8.0\,\micron\ flux densities from our best-fit SED (1.5 and 3.0\,\si{\micro Jy}) are somewhat smaller than those reported in \citet[$2.7\pm0.5$ and $5.9 \pm 0.7$\,\si{\micro Jy}, respectively]{serjeant14}.
The stellar mass that we derived is much larger than the gas mass of \textblue{$1.6\times10^{10}$\,\msun\ found by \citet[updated with new lens model]{neri14}}, implying a gas fraction of $f_\mathrm{gas} = M_\mathrm{gas} / (M_\mathrm{star} + M_\mathrm{gas}) = 0.18\pm0.10$, which is lower than the typical gas mass fraction of $z\sim5$ SMGs reported in \citet[$f_\mathrm{gas}\sim0.55$]{dudze20} \textred{but consistent with certain $z=4-6$ SMGs in literature \citep[e.g.,][]{gomez22,zavala22}}.
We also derived a lensing-corrected \textblue{$\log[\mathrm{SFR} / (\mathrm{M}_\mathrm{\odot}\mathrm{yr}^{-1})]=2.4\pm0.3$.}

\subsection{Mid--Far Infrared Properties}
\label{ss:03d_fir}

\begin{figure}[!t]
\centering
\includegraphics[width=1.07\linewidth]{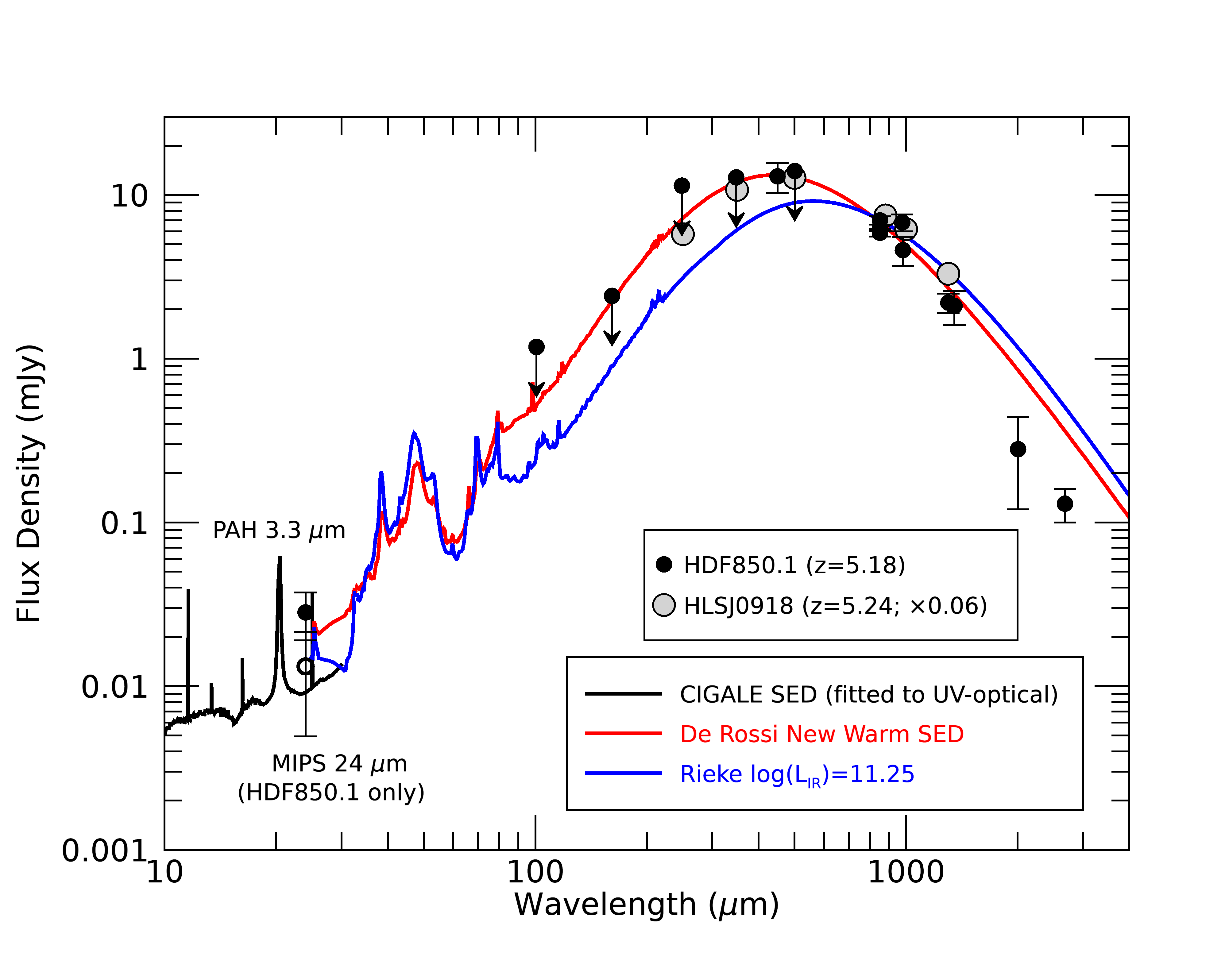}
\caption{Far-IR (rest frame) photometry of HDF850.1. The measurements are from \citet{cowie09} (24\,\micron; the empty circle shows the HDF850.1-only flux density of 13 $\mu$Jy), \citet{walter12} (100--500\,\micron\ and 0.98/2.7 mm), \citet{cowie17} (450 $\mu$m), 
\citet{hughes98} (850\,\micron, 1.35\,mm), \citet{chapin09} (850\,\micron; the measurement at 1.1\,mm is not plotted because of its low signal to noise), \citet{neri14} (980\,\micron), \citet{downes99} (1.3\,mm), and \citet{staguhn14} (2\,mm). 
The SED templates (scaled to the 850 \micron\ points) are from \citet{rieke09} ($\log(L_\mathrm{IR}/L_\odot) = 11.25$; blue line) and the newly derived warm SED template preferred for the redshift of HDF850.1 (red line). The latter is broadly consistent with the measured SED shape of the bright lensed galaxy HLSJ0918 at $z=5.24$ \citep{rawle2014}.  The 24 \micron\ points have an error bar determined by combining the statistical error and expected confusion noise by root sum square;
For the 100, 160, 250, 350, and 500 $\mu$m points, the 2-$\sigma$ upper limits from \citet{walter12} are shown.  The best-fit \textsc{cigale} SED for the whole system is also plotted up to 30 \micron\ (black line).
\textblue{Please refer to \citet{walter12} for a far-IR SED model fit based on a modified blackbody using the \textsc{magphys} program \citep{dacunha08}.}
}

\label{fig:firsed}
\end{figure}

Figure~\ref{fig:firsed} shows the available measurements of the HDF850.1 SED from mid-IR to millimeter, along with two template SEDs. The galaxy is so faint that confusion noise is an issue for the measurements with Spitzer and Herschel \citep{dole03,dole04,nguyen10,berta11}. 
At 24 \micron, we have combined the  statistical error with the 1 $\sigma$ photometric criterion confusion noise by root sum square. In the Herschel bands, significant detections were not achieved, so we show 2-$\sigma$ upper limits from \citet{walter12}. 

\citet{cowie09} measured a Spitzer/MIPS 24\,\micron\ flux density of $28.2 \pm 4.4$\,\si{\micro Jy} for HDF850.1 and treated it as an upper limit because of source confusion.
Based on the HST and JWST photometry and SED modeling, we conclude that the $z = 1.224$ foreground elliptical galaxy is quiescent and therefore its 24-\micron\ flux density is negligible ($\sim1$\,\si{\micro Jy}). 
However, the lensed arc at $z_\mathrm{phot} \sim 2.02$ is star-forming ($\mathrm{SFR}\sim 20$\,\smpy), which could contribute to a 24-\micron\ flux density of $\sim$15\,\si{\micro Jy} according to the LIRG ($L_\mathrm{IR} = 10^{11}$\,\lsun) template in \citet{rieke09}.
Therefore, the 24-\micron\ flux density of HDF850.1 is $\sim 13$\,\si{\micro Jy}, consistent with the predicted value from the best-fit \textsc{cigale} SED in Section~\ref{ss:03c_sed}.

The figure demonstrates that the far-IR measurements only loosely constrain the SED and hence the IR luminosity. 
As a result, the behavior in this spectral range needs to be determined relative to other galaxies from which reasonably well-constrained templates have been constructed. 
These issues are discussed in detail in \citet{derossi18}, where it is shown that the template of local (U)LIRGs from \citet{rieke09} for $\log(L_\mathrm{IR}/L_{\odot}) = 11.25$ is well suited for infrared galaxies at $2 \le z < 4$, and a template based on the behavior of Haro 11, with greatly enhanced output in the 20 $\mu$m range,  is preferred for $5 \le z < 7$. 
The underlying cause for this shift toward a warmer far-IR SED is the compact size and intense heating of dust by star formation in the high-redshift galaxies, properties potentially shared by HDF850.1. 

As shown in Figure~\ref{fig:firsed}, the $\log(L_\mathrm{IR} / L_\odot) = 11.25$ template does a poor job of matching the SED slope of HDF850.1 from 450\,\micron\ through the submillimeter and millimeter-wave range. As discussed in \citet{derossi18}, it is also too cold in the 100--300\,\micron\ (observed) range to match theoretical expectations. The Haro 11 based template is too bright in the 100--300\,\micron\ range, perhaps because HDF850.1 is at the low-redshift end of the $5 < z < 7$ bin where it was derived. We have generated a new template specifically for the $4 < z \le 5$ range by combining the $\log(L_\mathrm{IR}/L_{\odot}) = 12.25$ and Haro 11 templates to fit the measurements in \citet{derossi18}. The best fit is with about 65\% of the Haro 11 template and 35\% of the $\log(L_\mathrm{IR}/L_{\odot}) = 12.25$ one. This template closely resembles the one proposed for this redshift by \citet{schreiber18}, although it provides a somewhat better fit even than that one. The new template is shown in Figure~\ref{fig:firsed} to fit the data for HDF850.1 well, although it is clearly not unique.
\textred{We note that the template overpredicts the 2-3\,mm flux densities, which is caused by the large dust emissivity index ($\beta\sim2.5$) of HDF850.1 as pointed out in \citet{walter12}.}

In Figure~\ref{fig:firsed}, we also show the far-IR SED of the exceptionally bright lensed galaxy HLS~J091828.6+514223 (HLSJ0918) at $z=5.24$ \citep{combes2012,rawle2014}. The intrinsic IR luminosity of this galaxy is $1.8\times10^{13}$ L$_{\sun}$, a factor of \textblue{$\sim$\,3} larger than that of HDF850.1, but because of a lensing magnification of $\sim$9, the peak of the far-IR SED reaches 212 mJy at 500 \micron, producing high S/N detections in all three SPIRE band and providing a tight constraint on the far-IR SED shape around the peak. The figure shows that the measured far-IR SED of HLSJ0918 is broadly consistent with that of HDF850.1 and the newly derived warm SED template.

The indicated IR luminosity of HDF850.1 is $\sim 1.2 \times 10^{13}$\,\si{\mu^{-1}.L_\odot} before lensing correction. 
Corrected for the lensing magnification, the intrinsic \textblue{$ L_\mathrm{IR}$ is $ \sim 5 \times 10^{12}$\,\lsun}. The potential systematic errors dominate the uncertainty in this value, and could be as much as \textblue{$2 \times 10^{12}$\,\lsun}.
In comparison, \citet{walter12} and \citet{serjeant14} estimated values of $(8.7 \pm 1)\times 10^{12}$ and $1 \times 10^{13}$\,\si{\mu^{-1}.L_\odot}, the former value $\sim$70\% of our value uncorrected for lensing. This difference is a direct result of their using templates that are weak in the mid-IR compared with the typical behavior of ULIRGs at similar redshift \citep{derossi18}.

Applying the relation between total IR luminosity and the SFR from \citet{ke12} would lead to a SFR estimate of \textblue{$700\pm300$\,M$_\odot$\,yr$^{-1}$}. This value is substantially larger than the \textblue{$\sim 270 \pm 130$}\,M$_\odot$\,yr$^{-1}$ estimated by modeling the rest wavelength UV-optical spectrum in the preceding section. The difference is not surprising given the very large extinction and the resulting possibility that a significant part of the ongoing star formation is deeply hidden at those wavelengths.

\subsection{Maps of H$\alpha$ Line and Stellar Continuum}
\label{ss:03e_pix}

We also perform a pixel-by-pixel modeling of the SED of HDF850.1 in the LW filters.
We assume a simple power-law continuum model and a delta function at 4.06\,\micron\ for the \ha\ emission line at $z=5.185$.
This also effectively assumes negligible \nii\ $\lambda\lambda$6548, 6583 and \sii\ $\lambda\lambda$6716, 6731 emission ($\lesssim$10\%\ of \ha; see Section~\ref{ss:03g_ha}).
The \ha\ line is within the bandwidths of F410M and F444W filters.

From the best-fit models, we derive the \ha\ emission-line map as shown as the bottom-left panel of Figure~\ref{fig:02_matched}.
The southern component is higher in \ha\ surface brightness, while the \ha\ emission from the northern component is more diffuse and lower in surface brightness.
The total \ha\ line flux integrated from the image segments is $(1.7\pm0.1)\times10^{-17}$\,\si{erg.s^{-1}.cm^{-2}}, where the uncertainty is computed from the error extension of the best-fit \ha\ line map. 
We find that $\sim$55\% of the \ha\ flux is from the southern component.

The best-fit slope of the underlying continuum reflects the rest-frame optical color indices (e.g., $B-V$) of HDF850.1 in a spatially resolved manner. 
Under the assumption that (\romannumeral1) the stellar population within the galaxy is homogeneous with a common intrinsic $B-V$ of $0.04\pm0.20$ (\textred{based on the} physical SED modeling in Section~\ref{ss:03c_sed}), and (\romannumeral2) the dust attenuation follows the \citet{calzetti00} law, we convert the map of rest-frame optical continuum slope into a dust attenuation map ($A_V$; bottom-center of Figure~\ref{fig:02_matched}).
The highest dust attenuation ($A_V > 5$) is observed around the southern tip of HDF850.1-N, which is clearly seen reddened in the F444W-F410M-F356W RGB image.
This highly dust-attenuated region is also spatially consistent with the 1\,mm dust continuum emission obtained with the PdBI \citep{neri14}.
The lowest dust attenuation ($A_V < 1$) is seen in HDF850.1-S, exactly at the location of the compact clump seen in rest-frame UV.
It is worth reminding that rest-frame UV data are not used in this simple SED modeling.

Combining the $A_V$ map and the rest-frame $V$-band image observed with the F356W filter, we are able to reconstruct a rest-frame $V$-band image of HDF850.1 that is free from dust attenuation (Figure~\ref{fig:02_matched}, bottom-right panel).
Under our previous assumption of a homogeneous stellar population, this is effectively a map of the stellar mass distribution.
\textred{The total stellar mass derived from this map is consistent with that derived from integrated physical SED modeling (Section~\ref{ss:03c_sed}) assuming the same mass-to-light ratio.}
We also find that the stellar mass can be well described by one component instead of two components as seen in the LW detection map.
The stellar mass centroid is also on the southern tip of HDF850.1-N, consistent with the region with highest dust attenuation, suggesting that HDF850.1 is split into two components at rest-frame optical wavelength because of heavy dust obscuration in the center.
The stellar mass distribution is also broadly consistent with the \cii\,158\,\micron\ intensity map (i.e., momemt 0) obtained by \citet{neri14}, although the \cii\ emission shows signs of extension toward the $z=1.224$ elliptical galaxy.
This region is subject to the residual of neighbor subtraction in our NIRCam images and no stellar continuum can be robustly detected.

\textred{
We also experiment pixel-by-pixel physical SED modeling of HDF850.1 with the same software and assumption as that in Section~\ref{ss:03c_sed} (similar to those performed in recent JWST studies including \citealt{gimenez23} and \citealt{pg23}).
The patterns of stellar mass and dust attenuation map are similar to those presented in Figure~\ref{fig:02_matched} derived with simple power law and delta function model.
However, we observe a clear degeneracy between the best-fit stellar age and $A_V$ for each pixel, and the vast majority of pixels can be modeled with a mass-weighted stellar age of $\sim$\,100\,Myr except for the UV-bright, \ha-emitting region of HDF850.1-S ($\sim$\,30\,Myr).
This is because our current dataset does not cover the rest-frame $J$ band for HDF850.1, and most of the pixels are not detected in the rest-frame $U$ band.
It is therefore very difficult to obtain spatially resolved diagnostics of stellar population and $A_V$ at the same time for HDF850.1 on a pixel-to-pixel level with the current dataset.
Therefore, we opt not to report the physical parameters of HDF850.1 obtained through this method.
}

\begin{figure}[!t]
\centering
\includegraphics[width=\linewidth]{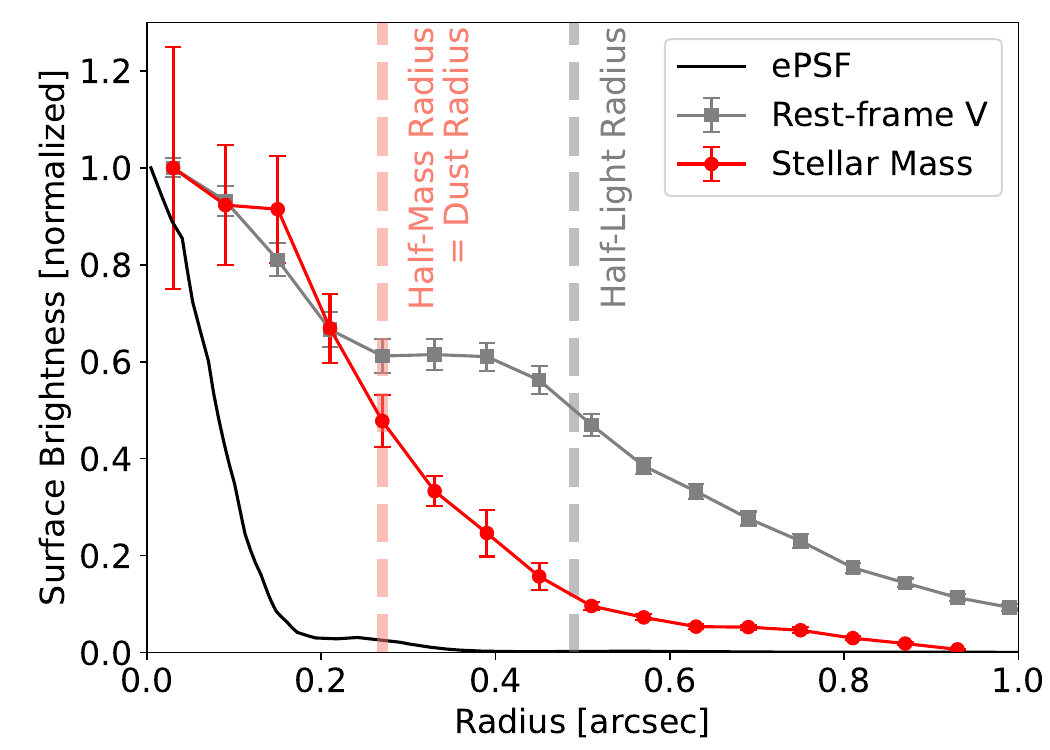}
\caption{1D profile of HDF850.1 along the direction of major axis \textblue{in the image plane}. 
Surface brightness profile in the rest-frame $V$ band (F356W) is shown in grey, and stellar mass profile (Section~\ref{ss:03e_pix}) is shown in red.
Half-light and half-mass radii are highlighted as vertical dashed lines.
\textred{The half-mass radius is consistent with the half-light radius of 1\,mm dust continuum measured by \citet{neri14}.}
Empirical PSF profile in the F444W band is shown in solid black line for comparison.
All profiles are normalized to 1 at the centroids.
}
\label{fig:03_size}
\end{figure}

\subsection{Size of HDF850.1}
\label{ss:03f_size}

We measure the size of HDF850.1 \textblue{in the image plane} at rest-frame $V$-band with and without dust attenuation correction.
Through a simple 2D Gaussian fitting to the stellar mass map derived in Section~\ref{ss:03e_pix}, we obtain a half-mass radius of HDF850.1 along the major axis as $R_\mathrm{e} = 0\farcs27\pm0\farcs02$.
We also model the map with 2D S\'ersic profile using \textsc{galfit} \citep{galfit}, and we find $R_\mathrm{e}=0\farcs21\pm0\farcs02$ with best-fit S\'ersic index $n=1.3\pm0.1$.
Both effective radii have been deconvolved with the PSF.

To account for the irregular shape of HDF850.1 and the uncertainty of the dust attenuation modeling, we also measure the size from the 1D profile of HDF850.1 along its major axis.
We use a set of elliptical annulus apertures from the mass centroid of HDF850.1 with $b/a=0.54$ and constant annulus width of $\Delta r = 0\farcs06$.
We also perform Monte-Carlo simulations of the stellar-mass map, which considers the uncertainty from photometry, intrinsic $B-V$ and dust attenuation.
We then derive the uncertainty of the 1D mass profile from the standard deviation of the profiles measured with these simulations.
The radial mass profile of HDF850.1 is shown as the filled red circles and solid line in Figure~\ref{fig:03_size}.
In comparison, we also show the 1D profile of ePSF and rest-frame $V$-band (F356W) light in the same figure.
The half-mass radius derived from 1D profile is $0\farcs27\pm0\farcs02$ after PSF deconvolution.
This is much smaller than the half-light radius in rest-frame $V$-band ($0\farcs49\pm0\farcs02$).
The smaller half-mass radius in contrast to half-light radius is a natural consequence of higher dust attenuation in the galaxy center, which flattens out the rest-frame optical light profile \citep[e.g.][]{nelson16, mosleh17, tacchella18, lang19, chen22, cheng22, suess22, cheng23, wu23}.

\begin{figure*}
\centering
\includegraphics[width=\linewidth]{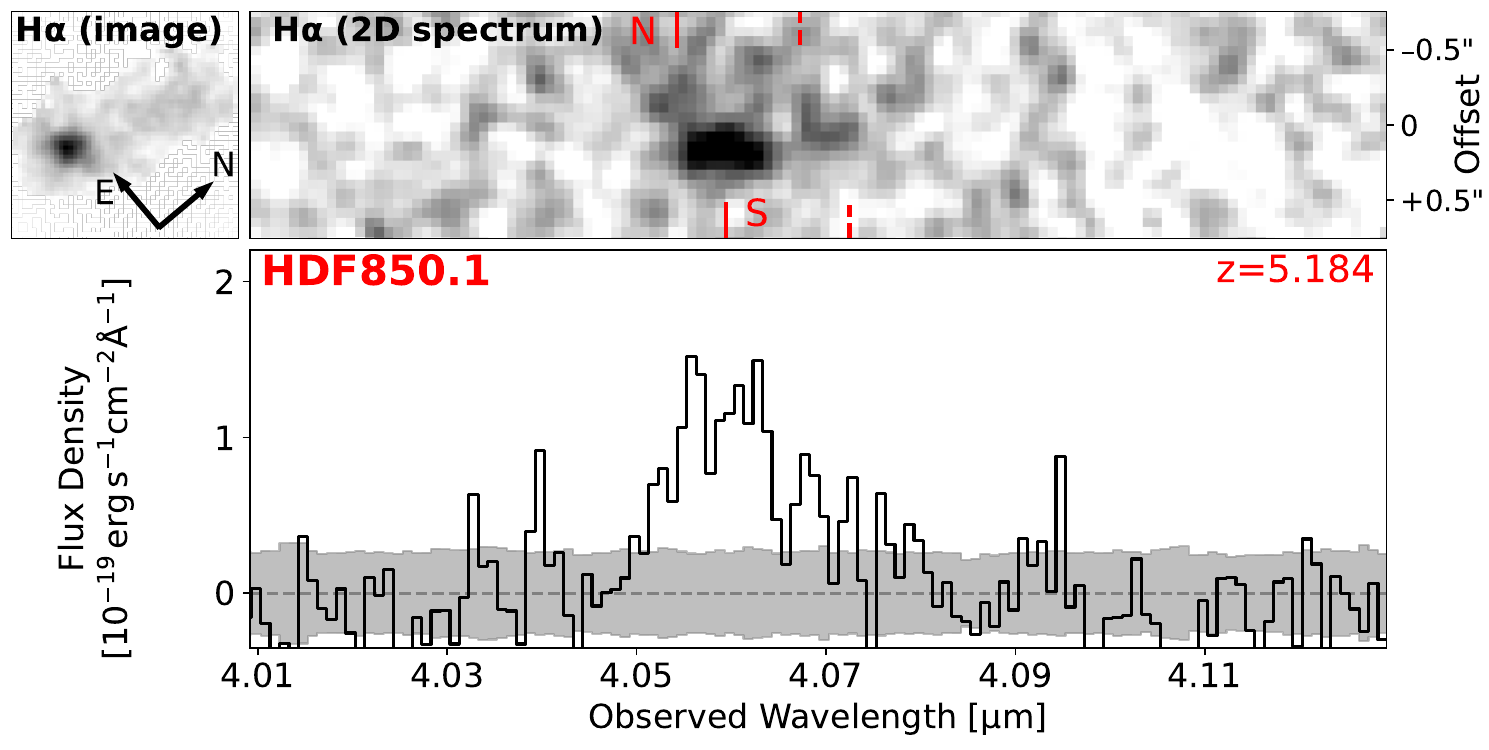}
\caption{NIRCam grism spectrum of HDF850.1 obtained from FRESCO. \textbf{Top-left}: Same \ha\ emission line map as in Figure~\ref{fig:02_matched}, but rotated to align with the dispersion direction.
\textbf{Top-Right}: 2D grism spectrum of HDF850.1. \ha\ line emissions from the northern and southern components are indicated with solid vertical red lines, and expected \nii\,$\lambda$6583 line emissions are indicated by dashed lines.
The relative positions of \ha\ emission from HDF850.1-N and S are different from those in the top-left panel, suggesting different velocities of the two components.
\textbf{Bottom}: 1D spectrum of HDF850.1 obtained by collapsing the 2D spectrum within $\pm$0\farcs5\ spatial offset. The uncertainty spectrum is shown as the filled gray region.
\textred{The full 1D spectrum is displayed in Appendix~\ref{apd:01_spec}, Figure~\ref{fig:full_spec}.}
}
\label{fig:04_spec}
\end{figure*}

\subsection{H$\alpha$ Spectroscopy of HDF850.1}
\label{ss:03g_ha}

We extract the NIRCam grism spectrum of HDF850.1 at 3.9--4.8\,\micron\ obtained through the FRESCO program \citep{oesch23}. 
The spectrum was obtained with Module B of NIRCam and the dispersion direction is 40\arcdeg\ from the north in clockwise direction. The most notable feature in the spectrum is the \ha\ emission (Figure~\ref{fig:04_spec}; \textred{for completeness, the full spectrum of HDF850.1 is displayed in Appendix~\ref{apd:01_spec}, Figure~\ref{fig:full_spec}}).
The 2D grism spectrum of HDF850.1 reveals similar \ha\ morphology as that derived with the NIRCam images (Section~\ref{ss:03e_pix}), including bright and compact \ha\ emission from the southern component and diffuse emission from the northern component.
The \ha\ redshift of the whole system is $z=5.184\pm0.002$ in extracted 1D spectrum, consistent with the \cii\ redshift ($z_\mathrm{[C\,II]} = 5.185$) reported in \citet{neri14}.
The total \ha\ flux measured from the grism spectrum is $(1.8\pm0.1)\times10^{-17}$\,\si{erg.s.^{-1}.cm^{-2}}, consistent with that derived from NIRCam image.
\nii\ lines are not detected, which is likely a combined result of (\romannumeral1) blending with \ha\ emission because the NIRCam WFSS resolution decreases for extended sources, and (\romannumeral2) a low \nii/\ha\ line ratio ($\lesssim10$\%) as seen for $z> 5$ galaxies reported in recent NIRSpec and NIRCam grism studies \citep[e.g.,][]{cameron23, helton23, sanders23, shapley23}.

The NIRCam grism spectrum also resolves the kinematics of HDF850.1.
The best-fit \ha\ redshift for HDF850.1-S is $z=5.192 \pm 0.001$, indicating a velocity offset of $\Delta v = 330\pm70$\,\si{km.s^{-1}} from the galaxy center where the redshift is measured through the \cii\ line.
The \ha\ emission of HDF850.1-N is likely blueshifted ($z=5.179\pm0.005$) when compared with the galaxy center, although the velocity offset ($\Delta v = -310 \pm 240$\,\si{km.s^{-1}}) is much more uncertain because of the low surface brightness.
The kinematic information derived from the NIRCam grism is broadly consistent with that from the \cii\ observations presented in \citet{neri14}.

\section{The Nature of HDF850.1}
\label{sec:04_hdf}




\begin{figure*}[!t]
\centering
\includegraphics[width=0.49\linewidth]{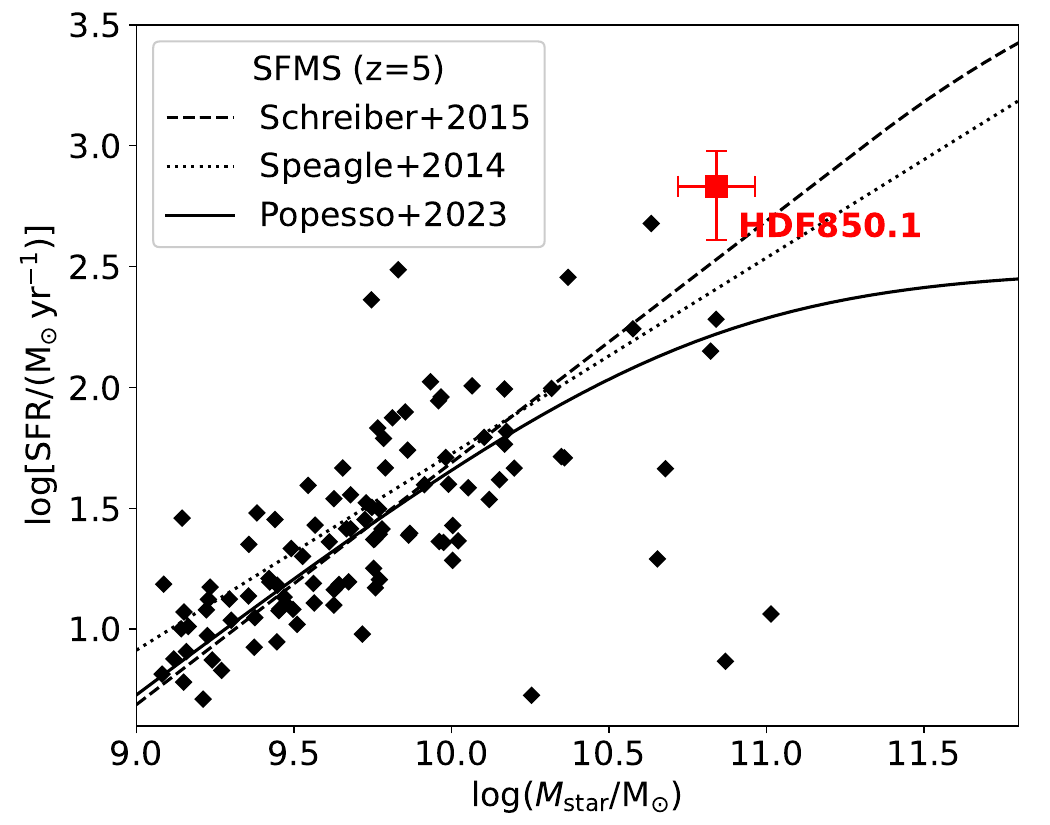}
\includegraphics[width=0.49\linewidth]{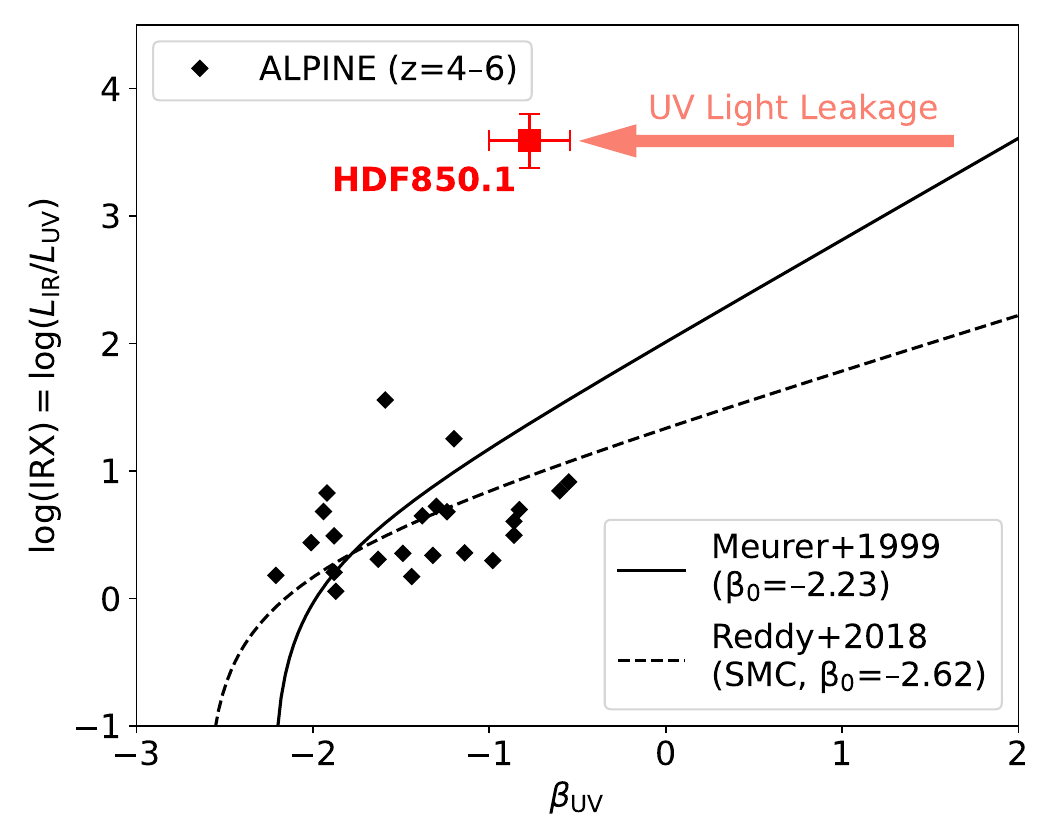}
\caption{\textbf{Left}: Star-formation rate versus stellar mass of HDF850.1 (red square), compared with $z=4-6$ galaxies observed with the ALMA ALPINE survey \citep{bethermin20,lefevre20}. The best-fit star-forming main sequences at $z=5$ in \citet{speagle14}, \citet{schreiber15}, and \citet{popesso23} are also plotted for comparison.
\textbf{Right}: Infrared excess versus UV continuum slope of HDF850.1, compared with $z=4-6$ galaxies in the ALPINE sample whose dust continua are detected with ALMA \citep{fudamoto20}.
IRX-$\beta_\mathrm{UV}$ relations of \citet{meurer99} and \citet{reddy18} are also plotted for comparison.
HDF850.1 is well above (or bluer than) these relations because of inhomogeneous dust attenuation and leakage of UV light in the southern component.
}
\label{fig:08_prop}
\end{figure*}

\subsection{SFR versus Stellar and Gas Mass}
\label{ss:04a_ms}

The left panel of Figure~\ref{fig:08_prop} compares the SFR and stellar mass of HDF850.1 with massive star-forming galaxies at $z=4-6$ observed through the ALMA large program ALPINE \citep{bethermin20, lefevre20}, and also the so-called star-forming main sequence at $z=5$ \citep{speagle14,schreiber15,popesso23}.
Here we adopt the SFR derived from the \textred{mid-to-far-IR SED modeling}. 

As one of the most massive galaxies found in the $z>5$ Universe, the SFR of HDF850.1 is \textred{slightly above} that expected from the star-forming main sequence.
We note that $z>5$ galaxies with similar stellar mass are rare with regard to volume density ($\lesssim10^{-5}$\,\si{cMpc^{-3}.dex^{-1}}; e.g., \citealt{weaver22}), and thus such a mass and redshift parameter space can be under-weighted for the determination of star-forming main sequence across cosmic time. 
The high dust attenuation, high obscured fraction of SFR ($\mathrm{SFR}_\mathrm{UV} / \mathrm{SFR}_\mathrm{total} < 10^{-3}$ for HDF850.1) and high incompleteness in spectroscopic confirmation and (sub)-millimeter SED constraints make the main-sequence SFR of such galaxies even more uncertain before the era of the JWST.
If the star-forming main sequence does exist at $z\gtrsim5$ and $M_\mathrm{star}\sim10^{11}$\,\msun, given the accurate JWST photometry and stellar SED modeling, HDF850.1 could be an example of galaxy that anchors the massive end of such a main sequence at $z \gtrsim 5$ (see also \citealt{serjeant14}).

\textred{With the IR-based SFR and molecular gas mass measured by \citet{neri14}, the molecular gas depletion time is only $t_\mathrm{dep} = M_\mathrm{gas} / \mathrm{SFR} = 25^{+25}_{-15}$\,Myr ($60_{-35}^{+70}$\,Myr if we adopt the SFR from NIRCam SED fitting). 
Both short timescales indicate that HDF850.1 will likely quench at $z\simeq4.7-5.0$ if there is no further molecular gas replenishment (cf.\ the massive quiescent galaxy at $z=4.658$ confirmed with \citealt{carnall23}).}

\subsection{HDF850.1 Is Not Necessarily a Major Merger}
\label{ss:04b_merger}

With 0\farcs3-resolution PdBI observation of the \cii\,158\,\micron\ line of HDF850.1, \citet{neri14} found two components of \cii\ emission with distinct velocity and spatial offsets (330\,\si{km.s^{-1}} and 0\farcs3, respectively). 
A number of physical properties of HDF850.1 are similar to those of major-merger ULIRGs in the local Universe (e.g., Arp\,220), including IR luminosity, CO line luminosities and strong dust obscuration.
Together with other arguments including the irregular shape of the \cii\ moment\,0 map (Figure~\ref{fig:02_matched}, bottom-right panel), \citet{neri14} concluded that HDF850.1 is a galaxy merger. 

With the latest JWST observation of HDF850.1 and other studies of high-redshift ($z\gtrsim5$) star-forming galaxies over the past decade, we argue that HDF850.1 is not necessarily a major merger system, but our observations cannot rule out the potential existence of a minor merger component, especially given the overdense environment at $z\sim5.2$ \textred{(see Section~\ref{sec:05_od})}.
Indeed, one can speculate that the southern component is an infalling galaxy that is lower in dust-attenuation than the center of HDF850.1 (or simply the northern component), but the stellar mass ratio between HDF850.1-S and -N ($\lesssim 1/10$) is lower than the canonical threshold of 1/4 that differentiates minor and major mergers.

Despite the two-component morphology seen in the deblended rest-frame optical image, we found  excellent spatial agreement between the dust attenuation and dust emission map in Section~\ref{ss:03e_pix}, which suggests that the gap between the two components results from a high amount of dust attenuation.
The derived stellar mass map is also well described with a single component model, indicating that it is not necessary to include a secondary galaxy component for the interpretation.

After a decade of ALMA operation at sub-arcsec resolution (see a recent review by \citealt{hodge20}), it is now known that in the high-redshift Universe, luminous SMGs are not necessarily major mergers in the final coalescence phase. 
Minor mergers and secular evolution can be important trigger mechanisms for vigorous dusty starbursts at $z\gtrsim2$ \citep[e.g.,][]{fujimoto17,gomez18,rujopakarn20,sun21a}.
This has been recently confirmed with JWST rest-frame optical observations \citep[e.g.,][]{chen22,cheng22,cheng23,lebail23,rujopakarn23}, in which SMGs are often observed as \textblue{fairly isolated disks (with or without noticeable substructures)}, and some of them exhibit spiral-arm features that could be triggered as a result of minor mergers \citep[e.g.,][]{wu23}.
This is different from the local ULIRGs that are predominantly gas-rich major mergers \citep[e.g.,][]{sanders88}.
Therefore, the similarity between HDF850.1 and local ULIRGS in physical properties (luminosity and dust obscuration) is not a strong argument to justify a major merger interpretation for HDF850.1.

However, the \cii\ morphology and kinematics could favor a major-merger scenario as pointed out by \citet{neri14}. 
Alternatively, the \cii\ emission may be a tracer of gas outflow \citep[e.g.,][]{maiolino15,cicone15,ginolfi20,pizzati20,herrera21,akins22}, which could extend out to a $\sim10$-kpc scale, as in \cii\ halos \citep[e.g.,][]{fujimoto19,fujimoto20,lambert23,pizzati23}.
The outflow direction can differ  from the major axis of galaxy \citep[e.g., HZ4 at $z=5.5$;][]{herrera21}, which is the case for HDF850.1 where the \cii\ extends toward the $z=1.224$ elliptical galaxy.
The \cii\ extension of HDF850.1 is closer to the lensing critical curve and magnified by a higher factor (\textblue{$\mu \sim 5$}), making it easier to detect.

The broad \cii\ and CO line widths (total \cii\ width $\sim940$\,\si{km.s^{-1}}; \citealt{neri14}) do not necessarily require an interpretation as a major merger. 
\citet{walter12} derived a dynamical mass of $M_\mathrm{dyn}\sim 1.3\times10^{11}$\,\msun\ (lensing-uncorrected) through the far-IR line widths and a simple rotating disk assumption with an inclination of 30\arcdeg.
Indeed, the blueshifted and redshifted \cii\ components in \citet{neri14} peak at opposite directions from the newly determined stellar-mass centroid (Figure~\ref{fig:02_matched}), favoring a rotation disk interpretation.
With the half-mass radius of \textblue{$\sim1.0$\,kpc} in the source-plane, an inclination of $\sim55$\arcdeg\ from the axis ratio of stellar mass distribution (assuming a thin circular disk), and also the velocity dispersion from \cii\ and \ha\ lines ($\sigma_V\sim300$\,\si{km.s^{-1}}), we estimate a dynamic mass of $M_\mathrm{dyn}\sim 8 \times 10^{10}$\,\msun\ for the whole HDF850.1 system following the method adopted by \citet{daddi10} and \citet{walter12}.
\textblue{This is actually consistent with the total baryonic mass if we include both our stellar mass estimate and the gas mass measured by \citet{neri14}.}
A major merger interpretation could imply a lower dynamical mass, potentially resulting in tension with the total baryonic mass of the system.
It is also worth noticing that the reconcilability between baryonic and dynamic mass of HDF850.1 does not require the use of very top-heavy IMF (e.g., \citealt{steinhardt22} \textred{and C.\ Woodrum et al.\ submitted}) suggested for the interpretation of luminous $z\gtrsim8$ galaxies. 

Combining all evidence presented above, we conclude that HDF850.1 is not necessarily a major merger system in the coalescence phase. 
However, our observations cannot rule out the potential existence of a minor merger component or a major merger in the previous formation history of HDF850.1.

\subsection{Leakage of UV and H$\alpha$ Photons}
\label{ss:04c_leak}

The southern component of HDF850.1 is detected in the rest-frame UV bands as a compact source, and our spatially resolved dust attenuation analysis also suggests low attenuation ($A_V < 1$) at the location of the UV source (Section~\ref{ss:03e_pix}).
\ha\ emission is also detected at this location.
This is in great contrast to the centroid of HDF850.1 with high dust attenuation ($A_V \gtrsim 5$).

The right panel of Figure~\ref{fig:08_prop} shows the infrared excess ($\mathrm{IRX} = L_\mathrm{UV} / L_\mathrm{IR}$) versus UV continuum slope $\beta_\mathrm{UV}$ of HDF850.1, which is further compared with the ALPINE sample of $z=4-6$ galaxies whose dust continua are detected with ALMA \citep{fudamoto20}.
Here the $\beta_\mathrm{UV}$ of HDF850.1 is derived from a simple power-law fitting of four-band SW photometry (F090W--F200W), and the UV luminosity $L_\mathrm{UV}$ is derived as $\nu L_\nu$ at rest-frame 1500\,\AA.
Assuming an intrinsic UV continuum slope of $\beta_0$, uniform dust screen and energy balance of dust absorption and emission, IRX will increase monotonically with the reddening of $\beta_\mathrm{UV}$ depending on the dust extinction law.
The best-fit IRX--$\beta_\mathrm{UV}$ relation of local starburst galaxies in \citet{meurer99} and the SMC relation with a blue $\beta_0 = -2.62$ \citep{reddy18} are plotted for comparison.

In contrast to galaxies in the ALPINE sample that generally follow the SMC-like IRX--$\beta_\mathrm{UV}$ relation, the IRX of HDF850.1 is $\sim100$ times above the empirical relations at its $\beta_\mathrm{UV}$. 
Previous studies have shown that SMGs can host relatively blue UV continuum slopes at large IRX \citep[e.g.,][]{penner12,oteo13,casey14}.
Theoretical works interpret these results through patchy dust screen models, resulting in a decrease in the far-UV optical depth compared with that at optical wavelengths, and also the turbulence of the dust screen \citep[e.g.,][]{popping17,narayanan18}.
With the high-resolution NIRCam images at rest-frame UV, we confirm that the bluer-than-expected $\beta_\mathrm{UV}$ of HDF850.1 is caused by inhomogeneous dust attenuation and therefore leakage of UV photons.

\begin{figure}[!t]
\centering
\includegraphics[width=\linewidth]{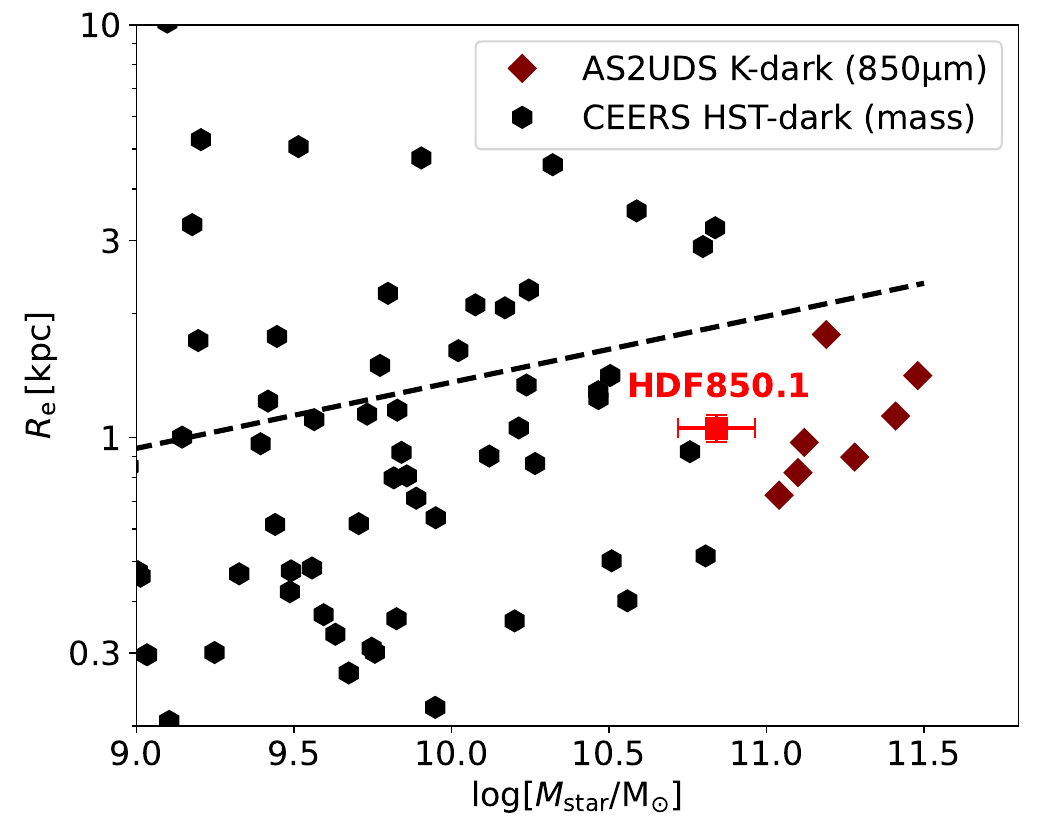}
\caption{Half-mass radius versus stellar mass of HDF850.1 (red square), compared with those of HST-dark galaxies at $z\sim4.4$ selected from the CEERS JWST program (black hexagons, \citealt{pg23}).
The $K$-dark SMGs at $z\sim3.4$ in AS2UDS sample are also shown as maroon diamonds for comparison, with their sizes measured at 850\,\micron\ (i.e., dust continuum; \citealt{gullberg19}, \citealt{smail21}).
The dashed line denotes the relation between half-mass radii and stellar masses of star-forming galaxies at $z=2.0-2.5$ \citep{suess19}, but scaled to $z=5$ assuming the ratio of scale factors of $(1+z)^{-1}$ at the two redshifts.
}
\label{fig:09_size_mass}
\end{figure}

\subsection{The Assembly of the Stellar and Dust Profiles}
\label{ss:04d_profile}

In the source plane, the half-mass radius of HDF850.1 is measured as \textblue{$1.0\pm0.1$\,kpc}.
We note that in our resolved SED modeling, we assume a constant intrinsic $B-V$ for the stellar population at each pixel, which can be an oversimplified assumption as the spatial variation of stellar age and color index has been observed with JWST for galaxies across $z=2-8$ \citep[e.g.,][]{miller22,chen23,duncan22}.
Even though we have assumed an intrinsic scattering of $B-V$ of 0.2\,mag and taken that into account through Monte Carlo simulations, the derived map of stellar mass can still be biased if an intrinsic color gradient exists for such massive star-forming galaxies, especially when most of the dust-obscured star formation occurs around the galaxy centers.
If the center of HDF850.1 is younger than its outskirts and intrinsically bluer, we will then underestimate the dust attenuation in the center, and therefore the underlying stellar mass distribution could be even more compact.

Despite the potential bias in half-mass radius estimate, we find that the half-mass radius of HDF850.1 is broadly consistent with the half-mass radii of massive HST-dark galaxies found with the CEERS sample at similar redshift (\citealt{pg23}; Figure~\ref{fig:09_size_mass}).
The half-mass radius of HDF850.1 is  slightly smaller than the half-light radius of the CEERS HST-dark galaxies at similar mass and redshift measured in the F444W band \citep[$R_\mathrm{e}\sim2$\,kpc]{nelson22}, which is likely a result of increasing dust attenuation in the galaxy center and flattening of the rest-frame optical light profile.
We also extrapolate the relation between half-mass radii and stellar masses of star-forming galaxies at $z=2.0-2.5$ \citep{suess19} to $z=5$ assuming a redshift dependence of $R_\mathrm{e,mass}\propto(1+z)^{-1}$ \citep[e.g.,][]{oesch10,shibuya15}, and we conclude that the size-mass of HDF850.1 is consistent with such an extrapolated relation within a dispersion of 0.2\,dex.

It is interesting that the dust continuum radius of HDF850.1 measured by \citet{neri14} is consistent with the half-mass radius derived in this paper.
\citet{smail21} studied $K$-faint SMGs in the AS2UDS sample \citep{dudze20} with similar stellar mass at $z\sim3.5$, and found a median effective radius of their dust continuum emission of $\sim$1.0\,kpc, which is also consistent with the half-mass radius of HDF850.1.
The similar compactness of the stellar and dust components of HST-dark SMGs at high-redshift is likely the cause of the high dust attenuation seen in this type of galaxy (median $A_V=5.2$ in the sample of \citealt{smail21} and 4.6$\pm$0.7 in this work) as suggested by previous studies \citep[e.g.,][]{smail21,sun21b}.

\section{The Environment of HDF850.1}
\label{sec:05_od}

\begin{figure*}[!t]
\centering
\includegraphics[width=\linewidth]{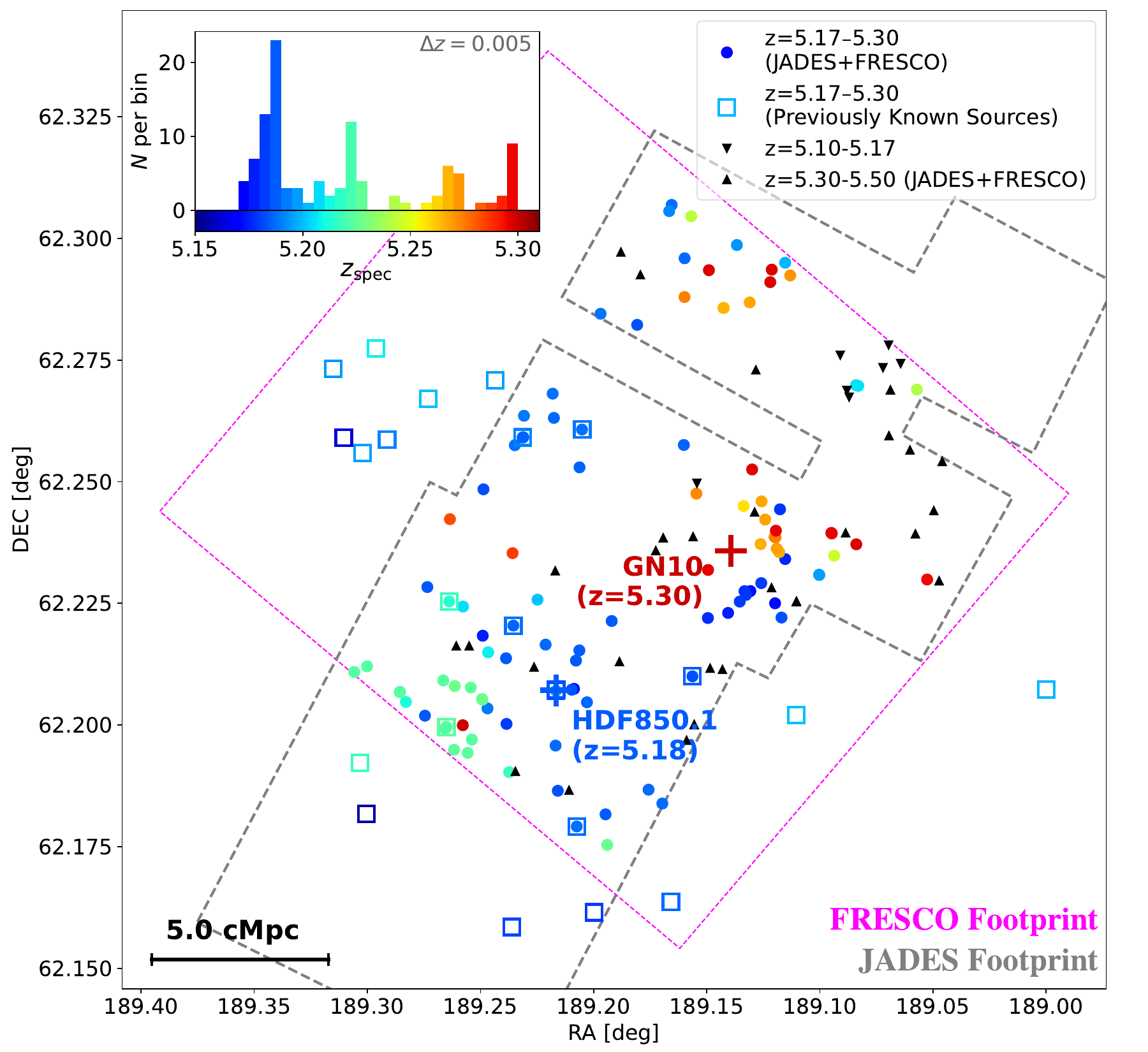}
\caption{On-sky distribution of 146 \ha-emitting galaxies at $z=5.1-5.5$ selected with JADES photometry and confirmed with FRESCO. 
Galaxies at $z=5.17-5.30$ are shown in filled circles color-coded by their redshifts, and the redshift distribution (bin size $\Delta z = 0.005$) is shown as the inset panel in the upper-left corner.
Galaxies at $z=5.10-5.17$ and $z=5.30-5.50$ are denoted by \textred{downward and upward black triangles, respectively}.
Spectroscopically confirmed galaxies in the $z\sim5.2$ overdensity before the JWST \citep{walter12,calvi21} are shown in open squares, color-coded by their redshifts.
HDF850.1 and another luminous SMG, GN10 at $z=5.30$ \citep{wang04,pope05,riechers20}, are shown in blue and red plus signs, respectively.
The footprint of JADES and FRESCO are shown in dashed gray and magenta polygons, respectively.
A scale bar of 5 comoving Mpc at $z=5.2$ is shown in the lower-left corner for comparison.
}
\label{fig:05_sky}
\end{figure*}

\begin{figure*}[!t]
\centering
\includegraphics[width=\linewidth]{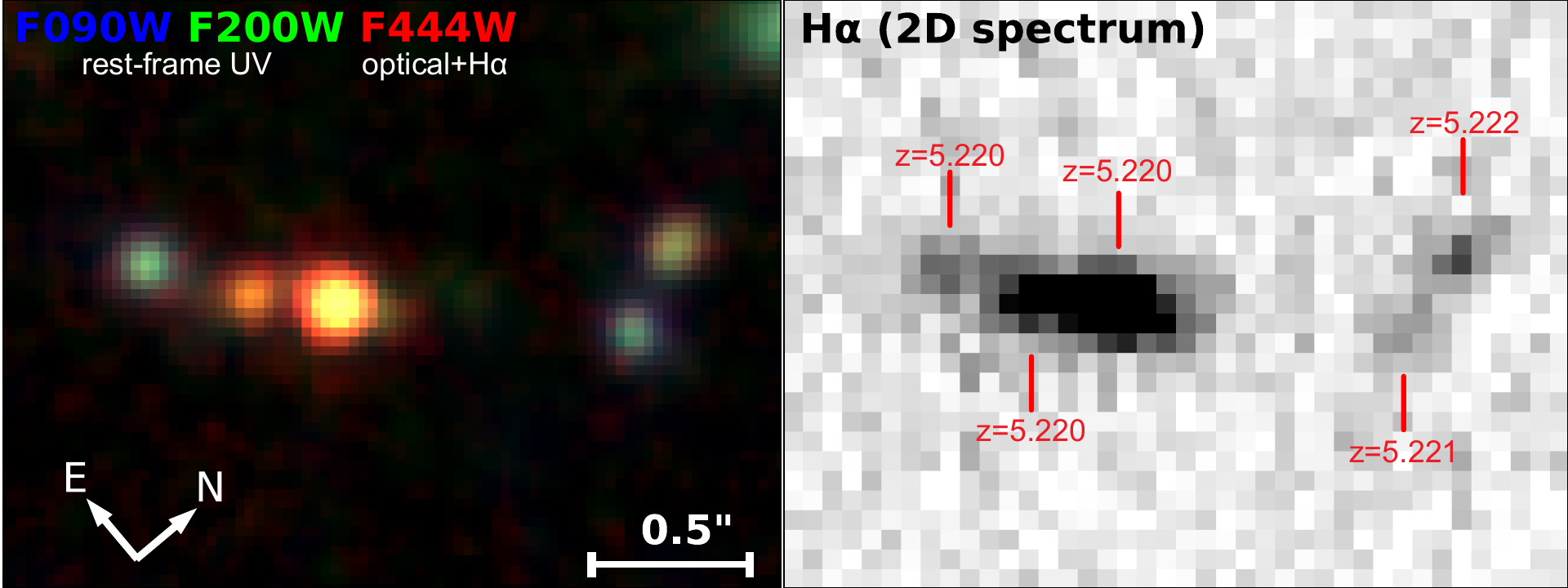}
\caption{NIRCam F444W-F200W-F090W RGB image (left) and F444W grism spectrum (right) of five \ha-emitting galaxies at $z=5.220-5.222$, obtained through the JADES and FRESCO surveys, respectively. The 
NIRCam image is rotated to align with the dispersion direction.
The brightest galaxy in the center (F444W=25.3\,AB mag) is at R.A.$=$12:37:03.699, Decl.$=+$62\arcdeg11\arcmin57\farcs7.
\ha\ redshifts of the five galaxies are noted in the right panel, and the two galaxies on the right (northwest) were spectroscopically confirmed previously at $z_\mathrm{Ly\alpha}=5.224$ \citep{calvi21}.
}
\label{fig:06_core}
\end{figure*}

\subsection{\ha-Emitting Galaxies at Similar Redshifts}
\label{ss:05a_od}

HDF850.1 is known to reside in an overdense environment at $z\sim5.2$, which has been confirmed with ground-based \lya\ spectroscopy \citep{walter12,calvi21,calvi23}.
Before the launch of the JWST, these works reported \textred{22} spectroscopically confirmed galaxies within the overdensity, including a known quasar at $z=5.186$ \citep{barger02}.
To better understand the environment of HDF850.1, we follow a method similar to that described in \citet{helton23} to identify \ha-emitting galaxies at $z=5.1-5.5$, where the \ha\ emission is within the bandwidth of F410M and F444W filters, resulting in F410M flux excess.

Similar to \citet{tacchella23}, we start from the full JADES photometric catalog in the GOODS-N field (B.\ Robertson et al.\ in preparation).
We conduct circular aperture photometry at 0.4--5.0\,\micron\ (obtained with HST/ACS and JWST/NIRCam) with diameters of $0\farcs2$ and point-source aperture correction factors.
Although the flux densities of extended sources can be underestimated, the color information at the centroids of galaxies is preserved, and therefore the following photometric redshift analysis is not affected. 
We use the code \textsc{eazy} \citep{eazy}, which estimates photometric redshifts using a template-fitting approach.
The templates and parameters being used in this step are the same as those described in \citet{helton23} and \citet{hainline23a}.
We select sources with F444W\,$<$\,29\,AB mag, $z_\mathrm{phot} > 4$ and $\Delta z_\mathrm{phot} < 1$ for FRESCO grism spectrum extraction, where the \textsc{eazy} confidence interval ($\Delta z_\mathrm{phot}$) is defined to be the difference between the 16th and 84th percentiles of the photometric redshift posterior distribution.
\textred{We caution that the accuracy of $z_\mathrm{phot}$ estimate can be degraded for intrinsically red sources (e.g., SMGs) because of the use of small apertures and SED templates best suited for the selection of $z>8$ galaxies \citep{hainline23a}, and therefore our survey completeness is expected to decrease towards the galaxy population with higher dust attenuation.
}

For a total of $\sim$4000 galaxies that satisfy our selection criteria, we extract 2D grism spectra and collapse them into 1D spectra using a boxcar aperture with height of five NIRCam LW native pixels (total of 0\farcs31).
We then perform automatic emission line identification with the 1D spectra, finding peaks at S/N\,$\geq 5$ with various wavelength bin sizes (1-8\,nm; corresponding to $\Delta v = 60-600$\,\si{km.s^{-1}}).
\textred{Here the noise is measured from the covariance matrix of Gaussian-profile fitting using 1D scientific and noise spectra.}

Similar to \citet{helton23}, we tentatively assign an emission line solution (\ha\ or \oiii) to each of the detected peaks that minimizes the difference between the estimated photometric redshifts and proposed spectroscopic redshifts. 
\textred{Based on our visual inspection, the miss-identification of \ha\ emitters as \oiii\ emitters at higher redshifts is very rare ($<$1\%). 
This is because \ha\ emitters at $z\sim5.2$ typically also have strong \oiii\ lines in the F277W band and thus color excess to filters consecutive in wavelength space (e.g., F200W and F335M), resulting in very tight photometric redshift constraint.
}
We then perform visual inspection on these solutions to remove spurious detections caused by either noise or contamination, and revise misidentifications for a few cases.
After this step, we also optimally re-extract 1D spectra of confirmed sources using their F444W surface brightness profiles \citep{horne86}, which recovers more of the \ha\ line fluxes for extended sources than boxcar extraction.
We fit the extracted 1D spectra with Gaussian profiles to measure the redshifts and fluxes.
For sources with blended \ha\ emission in grism data, we fit multiple Gaussian profiles to decompose the fluxes and measure the redshifts properly.
The typical redshift uncertainty is $\Delta z = 0.001$, and the typical $5\sigma$ detection limit of \ha\ line is $2\times10^{-18}$\,\si{erg.s.^{-1}.cm^{-2}}, similar to that reported in \citet{helton23} for \ha\ emitters in the GOODS-S field.

We confirm 146 \ha-emitting galaxies at $z=5.1-5.5$ with $\geq5\sigma$ detections of \ha\ lines from the FRESCO spectra, including HDF850.1 itself.
The 16, 50 and 84th percentiles of the difference between photometric and spectroscopic redshifts ($z_\mathrm{phot} - z_\mathrm{spec}$) are $-0.04$, 0.01 and 0.09, respectively.
Figure~\ref{fig:05_sky} shows the on-sky distribution of these 146 galaxies overlaid on the footprint of JADES and FRESCO, including 109 galaxies at $z=5.17-5.30$.
We present the NIRCam cutout images and grism spectra of these galaxies in Figure~\ref{fig:apd_spec} of Appendix~\ref{apd:01_spec}.
We found five point-like sources in our spectroscopic sample with broad \ha\ line emission with FWHM\,$\sim$\,2000\,\si{km.s^{-1}}, which could be interpreted as active galaxy nuclei (AGN) with bolometric luminosities of $L_\mathrm{bol} \simeq 10^{45} - 10^{46}$\,\si{erg.s^{-1}}.
These sources will be presented and discussed in a companion paper (E.\ Egami et al.\ in preparation; \textred{see also \citealt{matthee23b}}).

In the redshift histogram, we identify four peaks at $z=5.185$, 5.222, 5.268 and 5.296.
In the 1D redshift space with bin size of $\Delta z =0.005$, these peaks are $36\pm5$, $17\pm3$, $8\pm2$ and $12\pm2$ times more overdense compared with random field galaxies at $z=5.10-5.17$ and $5.30-5.50$.
We also evaluate the galaxy overdensity $\delta_g = n / \bar{n} - 1$ in the 3D space, where $n$ is the volume density of galaxies in an overdense region and $\bar{n}$ is the volume density of field galaxies.
We derive  $\bar{n} = 10^{-2.8\pm0.1}$\,cMpc$^{-3}$ from 37 field galaxies at $z=5.10-5.17$ and $5.30-5.50$.
Note that the actual volume density of \ha\ emitters at this redshift \citep[c.f.][]{sun22c} should be higher than $\bar{n}$ because of completeness corrections.
Galaxies in our sample share a similar selection function across $z=5.1-5.5$ on account of the \lya\ dropout in the HST F775W band and \ha\ excess in the NIRCam F410M band, and thus the use of $\bar{n}$ is fair for galaxies in and out of the overdense environment.
Assuming a search radius of 5.5\,cMpc ($\Delta V\sim700$\,\si{cMpc^3}) in which we expect one random field \ha\ emitter, the galaxy overdensities at the aforementioned four redshift peaks are found to be $\delta_g=$\,25, 20, 9 and 7, respectively.
The significances of these overdensities assuming Poisson statistics are 10.9, 9.0, 5.2 and 4.3$\sigma$, respectively.
No other peak is found to lie above a significance of 3.5$\sigma$.

Figure~\ref{fig:06_core} shows the core region of the galaxy overdensity corresponding to the second redshift peak at $z=5.22$.
Five galaxies are spectroscopically confirmed at $z=5.220-5.222$ within a separation of 2\arcsec\ ($\sim$\,12 proper kpc, pkpc), making it the most overdense region of $z>5$ galaxies in the joint footprint of JADES and FRESCO in the GOODS-N field. 
Similar compact assemblies of $z\gtrsim5$ galaxies have been reported in recent JWST/NIRCam studies \citep[]{jin23,helton23}, which are suggested as the progenitors of massive galaxies (stellar mass $M_\mathrm{star}\gtrsim10^{11}$\,\msun) seen at lower redshifts.
The two galaxies to the northwest of the bright central galaxy are blended in ground-based images obtained with the SHARDS survey \citep[SHARDS20013448;][]{pg13,ah18}, which were later spectroscopically confirmed at $z=5.224$ through the detection of \lya\ emission \citep{calvi21}.
The velocity offset between \lya\ and \ha\ emission is $120\pm50$\,\si{km.s^{-1}}.

A total of eight galaxies in our sample have been reported as part of the $z\sim5.2$ overdensity through ground-based spectroscopy \citep{walter12,calvi21}.
The remaining \textred{14} galaxies that are known as part of the overdensity are outside of our JADES and FRESCO joint footprint, suggesting that a significant fraction of overdensity member galaxies are not included in our sample \textblue{(see also recent study by \citealt{hd23})}.
Indeed, through the JADES photometric redshift analyses, we have identified galaxies outside of the FRESCO footprint that are likely associated with the HDF850.1 overdensity.
These sources will be presented in a forthcoming clustering analysis paper from the JADES collaboration \textred{(J.\ Helton et al., in preparation)}.


\begin{figure*}[!th]
\centering
\includegraphics[width=\linewidth]{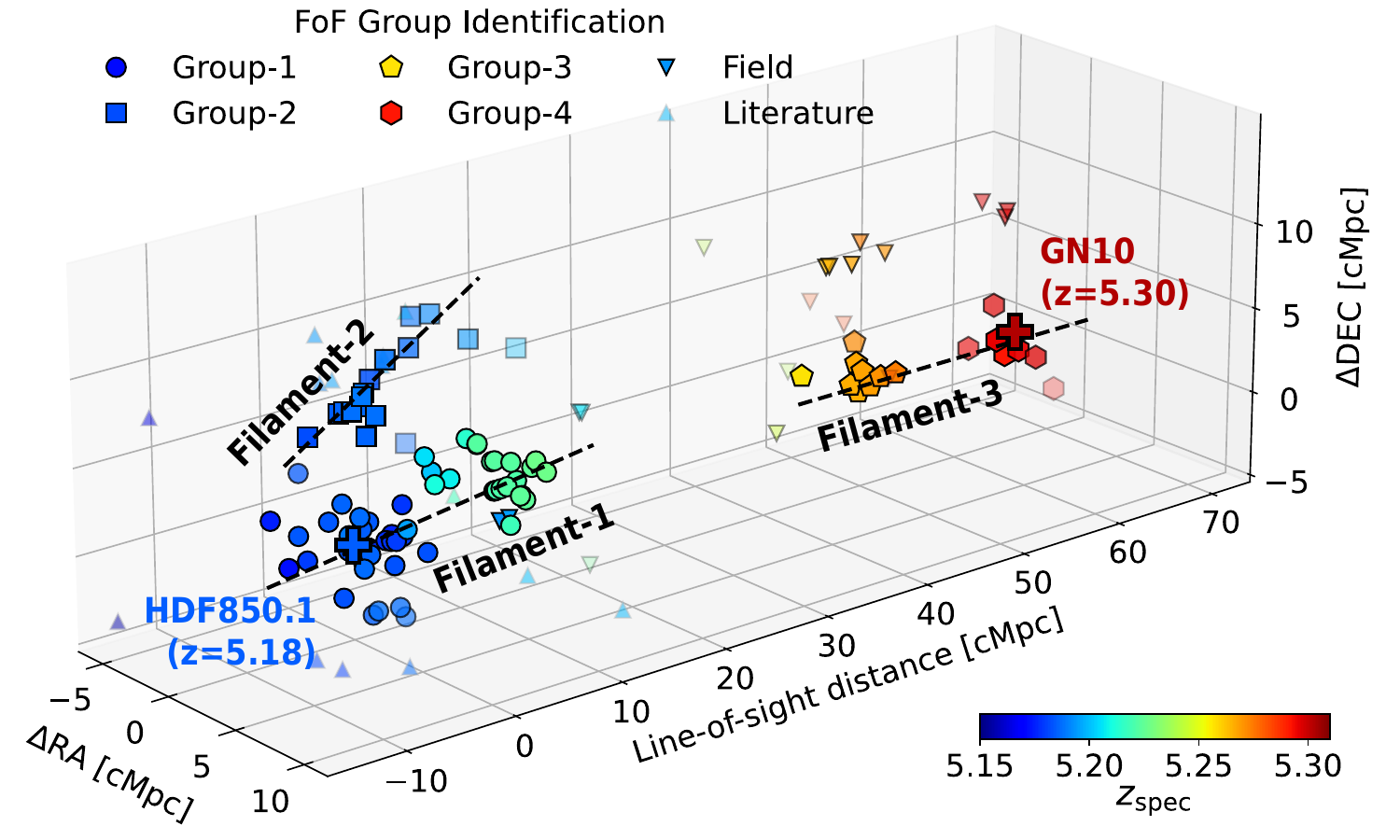}
\caption{3D large-scale structure of the overdense environment in the GOODS-N field at $z=5.17-5.30$. 
Coordinates of galaxies are with respect to that of HDF850.1 in units of comoving Mpc. 
All galaxies are color-coded by their redshifts, and galaxies in less dense regions (suggested by Gaussian kernel density estimation) are shown as transparent symbols.
\textred{Spectroscopically confirmed galaxies in \citet{walter12} and \citet{calvi21} that are not part of our sample are denoted as upward triangles.
}
Through our clustering analysis, we identify four galaxy groups in the overdensities, potentially consisting of three filamentary structures over a volume of 16$\times$18$\times$64\,\si{cMpc^3}.
The massive SMGs HDF850.1 and GN10 (shown as colored plus signs) reside in our proposed Filament-1 and -3, respectively.
}
\label{fig:10_filament}
\end{figure*}

\begin{figure}[!t]
\centering
\includegraphics[width=\linewidth]{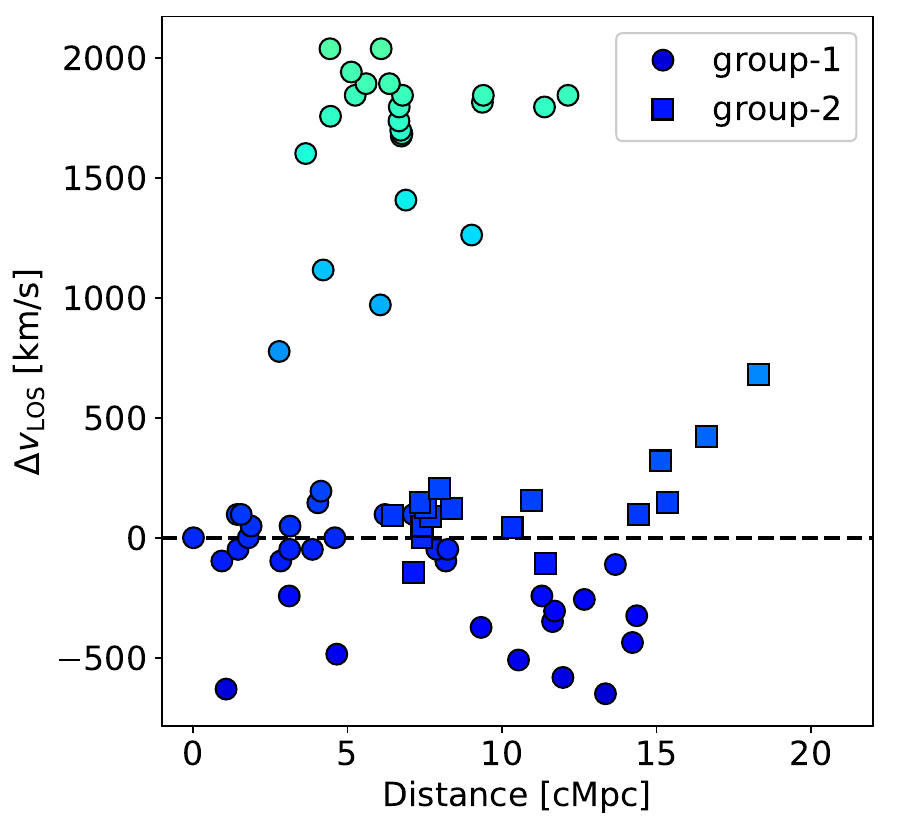}
\caption{Phase-space diagram of galaxies in group-1 (circles) and -2 (squares) classified with FoF algorithm (Section~\ref{ss:05b_cluster}). 
The projected distances and line-of-sight velocity offsets are with respect to HDF850.1.
Galaxies are color-coded by redshifts (same as that in Figure~\ref{fig:05_sky} and \ref{fig:10_filament}).
}
\label{fig:11_phase}
\end{figure}

\subsection{Clustering Analysis}
\label{ss:05b_cluster}

We adopt a Friends-of-Friends (FoF) algorithm to identify groups among 109 galaxies at $z=5.17-5.30$ in 3D space.
Galaxy groups are selected iteratively, consisting of galaxies that have projected separations and line-of-sight (LOS) velocity offsets below the adopted linking parameters of $d_\mathrm{link}=500$\,pkpc and $\Delta v=500$\,\si{km.s^{-1}}, respectively.
These parameters are identical to those adopted in \citet[same overdense environment]{calvi21} and \citet[same selection technique]{helton23}, which are motivated by the typical virial radius and velocity dispersion of galaxy groups.

We identify four groups of galaxies within the overdense environment with numbers of confirmed members at $N\geq7$. Figure~\ref{fig:10_filament} presents the group identification of galaxies in 3D space. 
The largest group found over the full volume of 16$\times$18$\times$64\,\si{cMpc^3} contains 55 member galaxies at $z=5.17-5.23$, including HDF850.1 (group-1).
In the RA, DEC and redshift direction, the coordinates of galaxies are computed as their offsets to HDF850.1 in comoving distance, but we also warn that the redshift-space distortion effect \citep[e.g.,][]{kaiser87} could complicate the actual line-of-sight distance by $\Delta d_\mathrm{LOS} \sim 10$\,cMpc with $\Delta v = 1000$\,\si{km.s^{-1}} ($\Delta z = 0.02$ at $z=5.2$), which is commonly seen in low-redshift clusters \citep{struble99}.
Also as argued in Section~\ref{ss:05a_od}, a significant number of member galaxies at this redshift could be missed because they do not fall in the joint footprint of JADES and FRESCO.
The noticeable gap in our survey area (around R.A.$=$189.16\arcdeg, Decl.$=$+62.26\arcdeg; Figure~\ref{fig:05_sky}) caused by the gap between the two NIRCam modules prevents us from selecting potential overdensity members in this region.
If galaxies at $z=5.26-5.30$ exist in this region, they can potentially connect the galaxies currently classified as field sources to the north of groups -3 \& -4, or even bridge the gap of groups -3 \& -4 in redshift space.
Indeed, three substructures presented by \citet{calvi21} in the same overdense region are now connected in group-1 through our clustering analysis.

The phase-space diagram of groups-1 \& -2 at $z=5.16-5.23$ is shown as Figure~\ref{fig:11_phase}.
In local or low-redshift virialized galaxy clusters, the scattering of peculiar velocity of cluster members decreases at a larger projected distance from the cluster center, which is a natural result with a NFW-like dark matter halo profile \citep{nfw}.
This is not seen in the phase-space diagram of HDF850.1 overdensity, suggesting that the protocluster has not yet evolved into a dynamically relaxed system. 

\subsection{Filamentary Structure}
\label{ss:05c_fila}

Cosmological simulations have suggested that $\sim90$\% of baryonic and dark matter at $z\gtrsim5$ resides in un-virialized filamentary large scale structure \citep[e.g.,][]{haider16}.
We find that the distribution of group member galaxies in 3D space can be potentially interpreted by three filamentary structures (Figure~\ref{fig:10_filament}).
Filament-1 (traced by group-1) and Filament-3 (traced by groups -3 \& -4) are structures elongated primarily along the line-of-sight direction with lengths of $\sim30$\,cMpc.
We warn that the peculiar velocities may lead to the identification of filaments along the LOS that are artificially long, although the peculiar velocities seen in each redshift clustering are rather small ($\sigma_V \lesssim 300$\,\si{km.s^{-1}}, corresponding to $\sigma d_\mathrm{LOS} \lesssim 3$\,cMpc).
Filament-3 is only anchored by two groups of galaxies, and they could be two co-spatial groups infalling to the same gravitational potential with opposite velocities.
Filament-2 (group-2) elongates along both the declination and redshift direction with a length of $\sim15$\,cMpc, with a RMS width in the transverse direction of $\sim1.6$\,cMpc.

It is interesting that the redshift differences between galaxies in Filament-2 and HDF850.1 decline with smaller projected distances to HDF850.1 (Figure~\ref{fig:11_phase}), and therefore the two filaments could be physically related with the region close to HDF850.1 as the potential node of the cosmic web.
Given our current survey volume, the redshift gap between Filament-1 and -3 is real because our selection function (based on photometric redshifts and \ha\ line detections) does not discriminate galaxies at this specific redshift.
However, there is a chance that the two filaments are linked with faint galaxies whose \ha\ line luminosities are below our detection limit or through galaxies outside of our survey area.
It is worth mentioning again that \textred{14} out of \textred{22} galaxies spectroscopically confirmed previously in the $z\sim5.2$ protocluster \citep{walter12,calvi21} are outside of the JADES and FRESCO joint footprint, and we do expect more complicated substructures of the overdensity to be uncovered with NIRCam grism or NIRSpec micro-shutter assembly (MSA) observations (e.g., GO-2674; PI: Arrabal Haro), which may or may not support our current filamenatry interpretation of the large-scale structure.

As one of the most massive galaxies at $z>5$ in the GOODS-N field, HDF850.1 resides in the core region of Filament-1 with 11 galaxies in its 500-pkpc proximity.
Although we do not directly detect the cold gas in this overdensity, the existence of filamentary large-scale structures in the protocluster environment, if real,  suggests efficient gas inflow from the intergalactic medium (IGM) through the cosmic web \citep[e.g.,][]{narayanan15}, which could be responsible for triggering the intense starburst and rapid assembly of the massive stellar component in HDF850.1 \textblue{despite of a short molecular gas depletion time scale} \citep[e.g.,][]{casey16}.
Similarly, another HST-dark SMG, GN10 ($z=5.303$, see NIRCam images and grism spectrum in Appendix~\ref{apd:01_spec}; \citealt{wang04}, \citealt{pope05}, \citealt{riechers20}), resides at the high-redshift end of Filament-3. 
\citet{calvi23} suggested that GN10 also traces an overdense environment through photometric redshift analysis \textblue{by applying the well-established poisson probability method \citep{castignani14,castignani19}}, and our study provides the crucial spectroscopic confirmation with NIRCam grism spectroscopy \citep[see also][]{hd23}.


\subsection{The ubiquity of galaxy overdensities at $z\sim5$}
\label{ss:05d_ubi}

\citet{helton23} reported the discovery of a galaxy protocluster in the GOODS-S field with 43 confirmed members at $z=5.4$ using the same selection method that we have used (NIRCam photometric redshifts and grism redshift confirmation with \ha). 
These galaxies were selected in a 41-arcmin$^2$ joint survey area with JADES and FRESCO, and 53 field galaxies that are not part of the overdensity were confirmed at $z=5.2-5.5$.

It is certainly an interesting fact that both GOODS fields host galaxy overdensities at comparable redshifts where \ha\ lines are within the bandwidth of the F410M filter, \textred{as JADES did not preferentially target both fields for overdensity mapping}. 
Both of the overdensities can be detected and characterized at high fidelity with a comoving survey volume of $\sim4\times10^{4}$\,\si{cMpc^{-3}} in each field, and both could evolve into Coma-like clusters in the present Universe \citep{calvi21,helton23}.
In fact, the expected dark matter halo mass of HDF850.1 itself based on its stellar mass is already $M_\mathrm{halo} = 10^{12.5} - 10^{13}$\,\msun\ using the empirical $M_\mathrm{star} - M_\mathrm{halo}$ relation at $z=5$ from \citet{behroozi19}, which is similar to or greater than the halo mass of the progenitor of Coma cluster at this redshift based on simulations \citep{chiang13}.

\begin{figure}[!t]
\centering
\includegraphics[width=\linewidth]{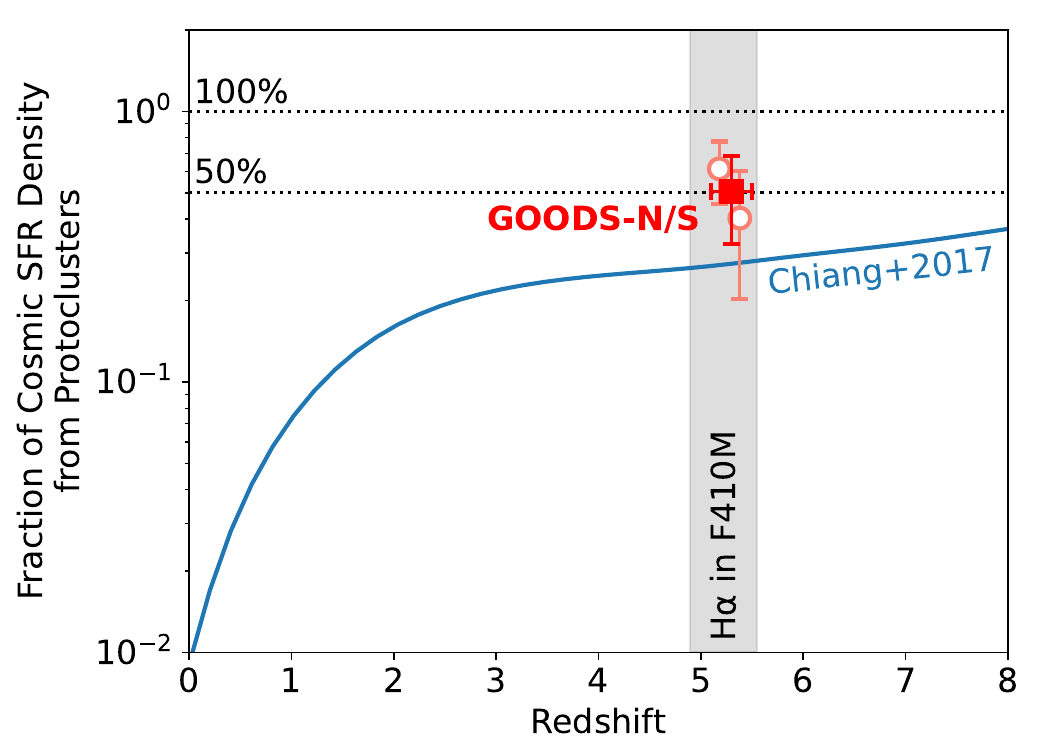}
\caption{Fraction of cosmic SFR density from protoclusters as a function of redshift.
The constraints from \ha\ emitters at $z=5.1-5.5$ in both the GOODS-N (this work) and GOODS-S \citep{helton23} fields are shown as the open circles (for each field) or the filled red square (combined).
Model prediction from \citet{chiang17} is shown as the solid blue line.
}
\label{fig:12_frac}
\end{figure}

These findings may suggest that massive galaxy overdensities in the high-redshift Universe are fairly common.
In fact, similar overdensities at $z\simeq 5-6$ have also been reported along the sight line of $z>6$ quasars in EIGER \citep{kashino22} and ASPIRE \citep{wang23} surveys through JWST/NIRCam slitless spectroscopy.
If we assume that the \ha\ luminosity functions in and out of an overdense environment share the same shape but with different normalizations, we can  roughly estimate the fraction of instantaneous cosmic SFR density hosted in protocluster environments through the number fraction of \ha\ emitters.
With the JADES and FRESCO joint survey in the GOODS-S/N fields, we estimate that $\sim50$\% of cosmic SFR densities are hosted in protoclusters at $z\sim5.2$, which is slightly higher than the fraction of $\sim30$\% as suggested by simulations (\citealt{chiang17}; see Figure~\ref{fig:12_frac}).

The discrepancy between derived and predicted fraction of cosmic SFR densities from protoclusters could arise from the cosmic variance \citep[$\sigma\sim20$\%;][]{trenti08}, the difference in the shapes of \ha\ luminosity functions in and out of protoclusters, and also the fraction of obscured cosmic SFR densities in two distinct environments.
In fact, if we sum the \ha\ luminosities of galaxies in and out of overdensities, then $\sim$60\%\ of \ha\ luminosity density above our detection limit ($L_\mathrm{H\alpha}\gtrsim 6 \times10^{41}$\,\si{erg.s^{-1}}, $5\sigma$) is from protocluster member galaxies.
A large cosmic variance is indeed observed, as the protocluster in the GOODS-N field is more overdense than the one in the GOODS-S field \citep{helton23} by a factor of $\sim$2.

\section{Summary}
\label{sec:05_sum}

In this paper, we have studied the stellar component and overdense environment of HDF850.1 at $z=5.18$, the brightest submillimeter galaxy in the Hubble Deep Field.
By combining JWST/NIRCam 0.8--5.0\,\micron\ imaging observations obtained with JADES and 3.9--5.0\,\micron\ WFSS observation obtained with FRESCO, we detect and resolve the rest-frame optical stellar component of HDF850.1 for the first time since its discovery \citep{hughes98}.
We also detect the \ha\ line emission from HDF850.1 through grism spectroscopy. In addition, we identify  109 galaxies in the redshift range of $z=5.17-5.30$, where only eight galaxies were spectroscopically confirmed previously. 
Our main conclusions are summarized as follows:

\begin{enumerate}

\item HDF850.1 is split into two components in NIRCam images because of heavy dust attenuation in the center. 
Through pixelated SED modeling, we reconstruct the map of \ha\ emission, dust attenuation ($A_V$) and the stellar mass distribution.
The northern component is higher in $A_V$ with lower \ha\ surface brightness, and the southern component is lower in $A_V$ and leaks rest-frame UV and \ha\ photons.
The location of the high-$A_V$ region matches well with that of dust emission as observed with PdBI \citep{neri14}.

\item After correcting for a lensing magnification of \textblue{$\bar{\mu}=2.5$}, we derive the stellar mass  \textblue{ ($\log(M_\mathrm{star}/\mathrm{M}_\mathrm{\odot})=10.8\pm0.1$), star-formation rate ($\log[\mathrm{SFR} / (\mathrm{M}_\mathrm{\odot}\mathrm{yr}^{-1})]=2.8\pm0.2$} \textred{based on mid-to-far-IR SED modeling}) and dust attenuation ($A_V = 4.6\pm0.7$) for HDF850.1. 
As one of the most massive galaxies at $z>5$, most of the star-formation of HDF850.1 is dust-obscured.
This places HDF850.1 at the massive end of the star-forming main sequence at $z\sim5$.

\item After correcting for dust attenuation, we find that the morphology of HDF850.1 can be well described by a single galaxy. 
In the image plane (lensing-uncorrected), we measure a half-mass radius of 0\farcs27$\pm$0\farcs02 from the 1D profile, which is smaller than the half-light radius in rest-frame $V$-band (0\farcs49$\pm$0\farcs02).
This is because the high dust attenuation in the galaxy center flattens out the light profile.

\item\ The \ha\ emission of HDF850.1 is detected spectroscopically with NIRCam WFSS. 
The \ha\ redshift is consistent with that of \cii\ measured with PdBI \citep{walter12,neri14}, and the morphology of \ha\ emission in the 2D grism spectrum is also consistent with that derived from pixelated fitting of imaging data.
We also resolve the kinematics of HDF850.1 through the \ha\ line, and we find a velocity offset of $330\pm70$\,\si{km.s^{-1}} from the \ha\ centroid of the southern component to the galaxy center.

\item Through morphological and kinematic information, we conclude that HDF850.1 is not necessarily a major merger system in the final coalescence phase. However, our observations cannot rule out the potential existence of a minor merger, or even a major merger in the past star-formation history.

\item Despite the high dust attenuation and HST-dark nature ($>$28\,AB mag below 1.6\,\micron), the dust attenuation of HDF850.1 is inhomogeneous, placing it $\sim100$ times above the empirical IRX--$\beta_\mathrm{UV}$ relation at its UV continuum slope ($-0.8\pm0.2$). 
HDF850.1 is compact in both stellar mass and dust continuum emission (half-mass/light radius of \textblue{$1.0\pm0.1$\,kpc} in the source plane), which is likely the cause of the high dust attenuation seen for the system.

\item\ Leveraging off JADES high-accuracy  photometric redshifts, we spectroscopically confirm 146 galaxies in the joint footprint of JADES and FRESCO at $z=5.1-5.5$ with $\geq 5\sigma$ detections of \ha\ lines.
109 of them are in the narrower redshift range of $z=5.17-5.30$, in which four peaks of redshift clustering are detected at $z=5.185$, 5.222, 5.268 and 5.296 with significance greater than $4\sigma$.
Among this sample, only eight sources were spectroscopically confirmed previously as members of the $z=5.2$ protocluster \citep{walter12,calvi21}.
\textred{Fourteen} galaxies that were confirmed as members of this protocluster are outside of our survey area, suggesting that a significant fraction of member galaxies and substructures are missed with this study.

\item Through a Friends-of-Friends clustering analysis, we identify four $N\geq 7$ groups of galaxies at $z=5.17-5.30$. 
The grouped galaxies in 3D space can be interpreted in terms of  three filamentary structures with lengths of 15--30\,cMpc.
HDF850.1 resides in one filament, and it potentially serves as a node of the cosmic web by connecting to another filament. 
Another SMG (GN10) at $z=5.30$ is associated with the third proposed filament.
If these filamentary structures are real, the efficient cold gas inflow through the cosmic web can be responsible for the vigorous starburst and rapid mass assembly of luminous SMGs at this epoch.

\item Both the GOODS-N/S fields are now confirmed to contain galaxy protoclusters at $z=5.2-5.4$ \citep[][and this work]{calvi21,helton23}. 
This may suggest that Coma-progenitor-like galaxy overdensities in the high-redshift Universe are fairly common. 
From our observed fraction of \ha\ emitters associated with galaxy protoclusters at $z=5.1-5.5$, we estimate that $50\pm20\%$ of cosmic SFR density occurs in the protocluster environment at this epoch. 
This may be slightly higher than previous simulation predictions \citep[$\sim$30\%;][]{chiang17}, but our estimate is subject to a variety of uncertainties, e.g., cosmic variance and the difference in \ha\ luminosity function shapes in and out of overdense environments.

\end{enumerate}


\section*{Acknowledgement}

We sincerely thank Roberto Neri and Fabian Walter for kindly sharing their PdBI millimeter continuum and \cii\ line emission data of HDF850.1, and Pablo G.\ P\'{e}rez-Gonz\'{a}lez for sharing his measurements of HST-dark galaxies with the CEERS survey.
FS\ thanks Christina Williams for helpful early discussions on this work.
\textblue{FS\ also thanks Caitlin Casey, Xiaohui Fan and Ian Smail for helpful comments.}
\textred{We thank the anonymous referee for a helpful report.}

FS, EE, GHR, CNAW, BR, ST, DJE, ZJ, MJR acknowledge JWST/NIRCam contract to the University of Arizona NAS5-02015.
DJE is also supported as a Simons Investigator.
AJB and JC acknowledge funding from the ``FirstGalaxies" Advanced Grant from the European Research Council (ERC) under the European Union’s Horizon 2020 research and innovation programme (Grant agreement No. 789056).
ECL acknowledges support of an STFC Webb Fellowship (ST/W001438/1).
RM, WB, LS, JW acknowledge support by the Science and Technology Facilities Council (STFC) and by the ERC through Advanced Grant 695671 ``QUENCH". RM also acknowledges funding from a research professorship from the Royal Society.
H{\"U} gratefully acknowledges support by the Isaac Newton Trust and by the Kavli Foundation through a Newton-Kavli Junior Fellowship.
KB is supported in part by the Australian Research Council Centre of Excellence for All Sky Astrophysics in 3 Dimensions (ASTRO 3D), through project number CE170100013.
ALD thanks the University of Cambridge Harding Distinguished Postgraduate Scholars Programme and Technology Facilities Council (STFC) Center for Doctoral Training (CDT) in Data intensive science at the University of Cambridge (STFC grant number 2742605) for a PhD studentship.

This work is based on observations made with the NASA/ESA/CSA James Webb Space Telescope. The data were obtained from the Mikulski Archive for Space Telescopes at the Space Telescope Science Institute, which is operated by the Association of Universities for Research in Astronomy, Inc., under NASA contract NAS 5-03127 for JWST. These observations are associated with program \#1181 and 1895. The authors sincerely thank the FRESCO team (PI: Pascal Oesch) for developing their observing program with a zero-exclusive-access period. 
This research is based (in part) on observations made with the NASA/ESA Hubble Space Telescope obtained from the Space Telescope Science Institute, which is operated by the Association of Universities for Research in Astronomy, Inc., under NASA contract NAS 5–26555. 
All the HST data used in this paper can be found in MAST: \dataset[10.17909/T91019]{http://dx.doi.org/10.17909/T91019}.
Additionally, this work made use of the {\it lux} supercomputer at UC Santa Cruz which is funded by NSF MRI grant AST 1828315, as well as the High Performance Computing (HPC) resources at the University of Arizona which is funded by the Office of Research Discovery and Innovation (ORDI), Chief Information Officer (CIO), and University Information Technology Services (UITS).


%

\vspace{5mm}
\facilities{JWST(NIRCam), HST(ACS), PdBI.}


\software{\textsc{astropy} \citep{2013A&A...558A..33A,2018AJ....156..123A}, 
\textsc{bagpipes} \citep{bagpipes},
\textsc{cigale} \citep{noll09,cigale},
\textsc{eazy} \citep{eazy},
\textsc{galfit} \citep{galfit},
\textsc{lenstool} \citep{lenstool}
\textsc{photutils} \citep{photutils},
}



\begin{deluxetable*}{@{\extracolsep{2pt}}rrrrr}
\tablecaption{Summary of the properties of HDF850.1 at $z=5.18$
\label{tab:01_hdf}}
\tablewidth{0pt}
\tabletypesize{\footnotesize}
\tablehead{
\colhead{} & \colhead{HDF850.1-N} & 
\colhead{HDF850.1-S} & \colhead{HDF850.1} 
}
\startdata
R.A.\  [deg] & 189.21655 & 189.21661 & 189.21658 \\
Decl.\ [deg] &  62.20723 &  62.20706 &  62.20715 \\
\hline\multicolumn4l{Photometric measurements (with 5\% noise floor)}\\\hline
F090W  [nJy] & $  5.0\pm 3.1$ & $ 10.0\pm 2.3$ & $ 15.0\pm 3.9$ \\
F115W  [nJy] & $  2.8\pm 2.3$ & $ 13.1\pm 1.8$ & $ 15.8\pm 2.9$ \\
F150W  [nJy] & $  2.0\pm 2.8$ & $ 19.0\pm 2.2$ & $ 20.9\pm 3.5$ \\
F200W  [nJy] & $  2.5\pm 2.3$ & $ 31.7\pm 2.3$ & $ 34.2\pm 3.2$ \\
F277W  [nJy] & $ 48.7\pm 2.8$ & $ 95.7\pm 4.9$ & $144.4\pm 5.6$ \\
F335M  [nJy] & $172.2\pm 8.8$ & $ 85.3\pm 4.4$ & $257.4\pm 9.8$ \\
F356W  [nJy] & $203.0\pm10.2$ & $ 99.1\pm 5.0$ & $302.1\pm11.4$ \\
F410M  [nJy] & $500.5\pm25.1$ & $266.7\pm13.4$ & $767.2\pm28.5$ \\
F444W  [nJy] & $630.9\pm31.6$ & $242.5\pm12.2$ & $873.4\pm33.9$ \\
\hline\multicolumn4l{Spectroscopic measurements}\\\hline
$z_\mathrm{spec}$ & $5.179\pm0.005$ & $5.192\pm0.001$ & $5.184\pm0.002$ \\
H$\alpha$ flux [\si{10^{-18}.erg.s^{-1}.cm^{-2}}] & $7.7\pm0.2$ & $9.3\pm0.3$ & $17.7\pm1.2$ \\
\hline\multicolumn4l{Physical properties (corrected for lensing magnification $\mu$)}\\\hline
$\mu$                             & 2.7 & 1.9 & 2.5 \\
$\log[M_\mathrm{star}$/\msun]     & $10.7\pm0.4$ & $9.7\pm0.1$ & $10.8\pm0.1$ \\
$\log[\mathrm{SFR}_\mathrm{100Myr}$/\smpy] & $2.1\pm0.2$ & $1.8\pm0.1$ & $2.4\pm0.3$ \\
$\log[\mathrm{SFR}_\mathrm{UV}$/\smpy] & $-1.2\pm0.2$ & $-0.7\pm0.1$ & $-0.7\pm0.1$ \\
$A_V$ [mag]                       & $4.9\pm0.4$ & $4.3\pm0.1$ & $4.6\pm0.7$ \\
$\log(\mathrm{IRX})=\log(L_\mathrm{IR}/L_\mathrm{UV})$            & \nodata & \nodata & $3.6\pm0.2$ \\
$\beta_\mathrm{UV}$               & \nodata & $-0.44\pm0.18$ & $-0.77\pm0.23$ \\
$f_\mathrm{gas}=M_\mathrm{gas}/(M_\mathrm{star}+M_\mathrm{gas})$  & \nodata & \nodata & $0.18\pm0.10$ \\
\enddata
\tablecomments{\ha\ line fluxes for HDF850.1-N/S are measured from the \ha\ line image segments, while the flux for the whole system is measured from the grism spectrum.
SFRs derived from SED modeling are averaged over 100\,Myr of the most recent star-formation history.
UV SFRs are derived from the best-fit UV continuum flux densities with conversion factor in \citet{ke12} and are not corrected for dust attenuation.
}
\end{deluxetable*}

\bibliography{00_main}{}
\bibliographystyle{aasjournal}

\appendix

\section{NIRCam Cutout Images and Grism Spectra of Confirmed Galaxies at $z=5.1-5.5$}
\label{apd:01_spec}

Figure~\ref{fig:apd_spec} shows the NIRCam cutout images and extracted 2D and 1D grism spectra for 140 galaxies at $z=5.1-5.5$ that we spectroscopically confirmed with $\geq5\sigma$ detection of \ha\ line emission.
The properties of galaxies in this sample is presented in Table~\ref{tab:02_hae}.
We note that the spectrum of HDF850.1 is not shown in this figure because it is presented in Figure~\ref{fig:04_spec}.
The cutout images and spectra of five AGN candidates identified through broad \ha\ emission lines are also not shown in the table. 
They will be presented and studied in details with a companion paper (E.\ Egami et al.\ in preparation; also \citealt{matthee23b}).

GN10 ($z=5.303$; R.A.$=$12:36:33.398, Decl.$=+$62\arcdeg14\arcmin08\farcs4) is another luminous SMG that is within the overdense environment. 
This source is not included in our photometric-redshift parent sample because of heavy dust obscuration, which leads to a large uncertainty of $z_\mathrm{phot}$.
Figure~\ref{fig:gn10} shows the NIRCam images and grism spectrum of GN10.
GN10 is also an HST-dark SMG and only detected at above 2\,\micron.
Multiple components could be identified from the image, suggesting a hint of merger nature.
In the NIRCam grism spectrum, we detect the \ha\ ($(9.5\pm0.9)\times10^{-18}$\,\si{erg.s^{-1}.cm^{-2}}) and \nii\,$\lambda$6583 ($(4.9\pm0.8)\times10^{-18}$\,\si{erg.s^{-1}.cm^{-2}}) line emission at $z=5.303$, the same redshift reported previously through CO lines \citep{riechers20}.

\textred{For completeness, we also show the full NIRCam 1D grism spectra (3.90--4.85\,\micron) of HDF850.1 and GN10 in Figure~\ref{fig:full_spec}.
We note that the spectrum of GN10 is contaminated by a bright galaxy at $z=1.381$, whose \feii\,$\lambda$1.646\,\micron\ and Paschen\,$\alpha$ lines are visible in the 1D spectrum. 
These interloping emission lines are marked for clarity.
}

\startlongtable
\begin{deluxetable*}{@{\extracolsep{2pt}}ccccccccc}
\tablecaption{Summary of spectroscopically confirmed \ha-emitting galaxies at $z=5.1-5.5$
\label{tab:02_hae}}
\tablewidth{0pt}
\tabletypesize{\footnotesize}
\tablehead{
\colhead{Index} & \colhead{Name} & 
\colhead{R.A.} & \colhead{Decl.} &
\colhead{$z_\mathrm{spec}$} & \colhead{$M_\mathrm{UV}$}  & \colhead{F444W} & \colhead{\ha\ line flux} & \colhead{Group} \\ 
 \colhead{}  &  \colhead{} & \colhead{(deg)} & \colhead{(deg)} & \colhead{} & \colhead{(mag)} 
 & \colhead{(mag)}  & \colhead{($10^{-18}$\,\si{erg.s^{-1}.cm^{-2}})}  & \colhead{}  
}
\startdata
  1 & JADES-GN-189.06961+62.27808 & 189.06961 & $+62.27808$ & 5.115 & $-18.71\pm0.08$ & 27.80 & $ 2.06\pm0.27$ & F \\
  2 & JADES-GN-189.09098+62.27600 & 189.09098 & $+62.27600$ & 5.126 & $-18.38\pm0.12$ & 27.71 & $ 1.61\pm0.19$ & F \\
  3 & JADES-GN-189.08706+62.26738 & 189.08706 & $+62.26738$ & 5.131 & $-18.72\pm0.06$ & 27.14 & $ 5.64\pm0.47$ & F \\
  4 & JADES-GN-189.15434+62.24967 & 189.15434 & $+62.24967$ & 5.141 & $-19.71\pm0.03$ & 27.25 & $ 3.07\pm0.48$ & F \\
  5 & JADES-GN-189.08801+62.26874 & 189.08801 & $+62.26874$ & 5.141 & $-20.39\pm0.03$ & 25.92 & $ 4.97\pm0.56$ & F \\
  6 & JADES-GN-189.06429+62.27429 & 189.06429 & $+62.27429$ & 5.145 & $-19.45\pm0.03$ & 26.44 & $ 4.82\pm0.32$ & F \\
  7 & JADES-GN-189.11978+62.22501 & 189.11978 & $+62.22501$ & 5.172 & $-19.30\pm0.11$ & 27.25 & $ 2.73\pm0.42$ & 1 \\
  8 & JADES-GN-189.20867+62.20743 & 189.20867 & $+62.20743$ & 5.172 & $-20.30\pm0.03$ & 25.52 & $12.34\pm1.07$ & 1 \\
  9 & JADES-GN-189.13071+62.22752 & 189.13071 & $+62.22752$ & 5.173 & $-18.92\pm0.15$ & 27.50 & $ 1.96\pm0.26$ & 1 \\
 10 & JADES-GN-189.14055+62.22303 & 189.14055 & $+62.22303$ & 5.175 & $-19.12\pm0.06$ & 27.18 & $ 3.00\pm0.42$ & 1 \\
 11 & JADES-GN-189.24900+62.21834 & 189.24900 & $+62.21834$ & 5.175 & $-18.94\pm0.03$ & 27.24 & $ 2.93\pm0.43$ & 1 \\
 12 & JADES-GN-189.11531+62.23411 & 189.11531 & $+62.23411$ & 5.176 & $-19.24\pm0.09$ & 27.02 & $ 4.36\pm0.33$ & 1 \\
 13 & JADES-GN-189.14948+62.22198 & 189.14948 & $+62.22198$ & 5.177 & $-20.11\pm0.03$ & 26.32 & $ 3.07\pm0.49$ & 1 \\
 14 & JADES-GN-189.13328+62.22751 & 189.13328 & $+62.22751$ & 5.178 & $-20.17\pm0.03$ & 25.03 & $17.08\pm0.73$ & 1 \\
 15 & JADES-GN-189.11761+62.24433 & 189.11761 & $+62.24433$ & 5.178 & $-19.52\pm0.10$ & 26.84 & $ 3.00\pm0.36$ & 1 \\
 16 & JADES-GN-189.13261+62.22674 & 189.13261 & $+62.22674$ & 5.179 & $-19.66\pm0.03$ & 26.32 & $ 5.66\pm0.36$ & 1 \\
 17 & JADES-GN-189.12590+62.22916 & 189.12590 & $+62.22916$ & 5.180 & $-20.27\pm0.03$ & 25.54 & $ 9.21\pm0.47$ & 1 \\
 18 & JADES-GN-189.13539+62.22536 & 189.13539 & $+62.22536$ & 5.180 & $-20.64\pm0.03$ & 25.74 & $10.63\pm0.49$ & 1 \\
 19 & JADES-GN-189.23848+62.20022 & 189.23848 & $+62.20022$ & 5.180 & $-18.92\pm0.04$ & 26.48 & $ 3.25\pm0.47$ & 1 \\
 20 & JADES-GN-189.24868+62.24843 & 189.24868 & $+62.24843$ & 5.182 & $-19.57\pm0.03$ & 25.21 & $13.00\pm0.71$ & 2 \\
 21 & JADES-GN-189.11695+62.22208 & 189.11695 & $+62.22208$ & 5.183 & $-19.73\pm0.03$ & 26.38 & $ 5.00\pm0.39$ & 1 \\
 22 & JADES-GN-189.18074+62.28224 & 189.18074 & $+62.28224$ & 5.183 & $-20.02\pm0.06$ & 25.49 & $ 9.84\pm0.45$ & 2 \\
 23 & JADES-GN-189.15632+62.21000 & 189.15632 & $+62.21000$ & 5.183 & $-19.89\pm0.03$ & 25.84 & $ 5.57\pm0.43$ & 1 \\
 24 & JADES-GN-189.21582+62.18648 & 189.21582 & $+62.18648$ & 5.183 & $-19.55\pm0.03$ & 25.46 & $14.99\pm0.40$ & 1 \\
 25 & JADES-GN-189.20968+62.20726 & 189.20968 & $+62.20726$ & 5.183 & $-19.75\pm0.03$ & 26.65 & $ 4.77\pm0.52$ & 1 \\
 26 & JADES-GN-189.27453+62.20191 & 189.27453 & $+62.20191$ & 5.184 & $-18.50\pm0.06$ & 27.97 & $ 1.56\pm0.25$ & 1 \\
 27 & JADES-GN-189.20777+62.21324 & 189.20777 & $+62.21324$ & 5.184 & $-19.99\pm0.03$ & 25.38 & $10.43\pm0.82$ & 1 \\
 28 & JADES-GN-189.23869+62.21371 & 189.23869 & $+62.21371$ & 5.184 & $-20.33\pm0.03$ & 26.14 & $ 3.01\pm0.51$ & 1 \\
 29 & JADES-GN-189.19199+62.22137 & 189.19199 & $+62.22137$ & 5.184 & $-18.71\pm0.05$ & 27.16 & $ 3.11\pm0.41$ & 1 \\
 30 & JADES-GN-189.27348+62.22835 & 189.27348 & $+62.22835$ & 5.184 & $-20.01\pm0.03$ & 26.13 & $ 3.72\pm0.62$ & 1 \\
 31 & JADES-GN-189.20629+62.21533 & 189.20629 & $+62.21533$ & 5.185 & $-19.32\pm0.03$ & 26.22 & $ 5.37\pm0.37$ & 1 \\
 32 & JADES-GN-189.19472+62.18163 & 189.19472 & $+62.18163$ & 5.185 & $-19.08\pm0.03$ & 27.40 & $ 3.13\pm0.41$ & 1 \\
 33 & JADES-GN-189.23137+62.25908 & 189.23137 & $+62.25908$ & 5.185 & $-20.09\pm0.03$ & 25.82 & $23.78\pm0.59$ & 2 \\
 34 & JADES-GN-189.16011+62.25755 & 189.16011 & $+62.25755$ & 5.186 & $-19.20\pm0.05$ & 26.08 & $ 5.49\pm0.61$ & 2 \\
 35 & JADES-GN-189.23545+62.22043 & 189.23545 & $+62.22043$ & 5.186 & $-19.66\pm0.03$ & 26.65 & $ 4.52\pm0.45$ & 1 \\
 36 & JADES-GN-189.23124+62.25913 & 189.23124 & $+62.25913$ & 5.186 & $-20.24\pm0.03$ & 25.89 & $24.48\pm0.59$ & 2 \\
 37 & JADES-GN-189.23103+62.25912 & 189.23103 & $+62.25912$ & 5.186 & $-19.67\pm0.03$ & 26.47 & $15.82\pm0.59$ & 2 \\
 38 & JADES-GN-189.20293+62.20467 & 189.20293 & $+62.20467$ & 5.186 & $-19.73\pm0.03$ & 24.56 & $40.83\pm1.19$ & 1 \\
 39 & JADES-GN-189.20621+62.25295 & 189.20621 & $+62.25295$ & 5.186 & $-17.63\pm0.21$ & 26.65 & $ 1.20\pm0.39$ & 2 \\
 40 & JADES-GN-189.21754+62.26311 & 189.21754 & $+62.26311$ & 5.187 & $-19.27\pm0.10$ & 27.38 & $ 2.01\pm0.36$ & 2 \\
 41 & JADES-GN-189.21680+62.19577 & 189.21680 & $+62.19577$ & 5.187 & $-19.40\pm0.03$ & 26.09 & $ 8.14\pm0.77$ & 1 \\
 42 & JADES-GN-189.22121+62.21652 & 189.22121 & $+62.21652$ & 5.187 & $-19.57\pm0.03$ & 26.32 & $ 4.42\pm0.66$ & 1 \\
 43 & JADES-GN-189.17570+62.18671 & 189.17570 & $+62.18671$ & 5.187 & $-19.95\pm0.03$ & 25.42 & $ 7.37\pm0.71$ & 1 \\
 44 & JADES-GN-189.16961+62.18386 & 189.16961 & $+62.18386$ & 5.187 & $-19.40\pm0.04$ & 26.75 & $ 5.13\pm0.52$ & 1 \\
 45 & JADES-GN-189.21809+62.26809 & 189.21809 & $+62.26809$ & 5.188 & $-20.65\pm0.03$ & 24.96 & $ 5.18\pm0.63$ & 2 \\
 46 & JADES-GN-189.20512+62.26072 & 189.20512 & $+62.26072$ & 5.188 & $-21.09\pm0.03$ & 25.45 & $16.10\pm0.44$ & 2 \\
 47 & JADES-GN-189.20744+62.17914 & 189.20744 & $+62.17914$ & 5.188 & $-19.57\pm0.04$ & 26.42 & $ 4.41\pm0.33$ & 1 \\
 48 & JADES-GN-189.23482+62.25750 & 189.23482 & $+62.25750$ & 5.188 & $-20.21\pm0.03$ & 25.51 & $10.84\pm0.38$ & 2 \\
 49 & JADES-GN-189.16549+62.30689 & 189.16549 & $+62.30689$ & 5.188 & $-19.96\pm0.03$ & 26.21 & $ 3.02\pm0.30$ & 2 \\
 50 & JADES-GN-189.19692+62.28451 & 189.19692 & $+62.28451$ & 5.188 & $-20.21\pm0.03$ & 26.00 & $ 3.52\pm0.27$ & 2 \\
 51 & JADES-GN-189.24689+62.20341 & 189.24689 & $+62.20341$ & 5.189 & $-19.02\pm0.04$ & 25.98 & $ 5.32\pm0.77$ & 1 \\
 52 & JADES-GN-189.23078+62.26355 & 189.23078 & $+62.26355$ & 5.189 & $-19.87\pm0.03$ & 26.71 & $ 5.21\pm0.35$ & 2 \\
 53 & JADES-GN-189.16658+62.30560 & 189.16658 & $+62.30560$ & 5.192 & $-20.20\pm0.04$ & 24.48 & $11.78\pm0.56$ & 2 \\
 54 & JADES-GN-189.10025+62.23082 & 189.10025 & $+62.23082$ & 5.193 & $-19.53\pm0.08$ & 26.73 & $ 4.24\pm0.69$ & F \\
 55 & JADES-GN-189.13667+62.29864 & 189.13667 & $+62.29864$ & 5.194 & $-18.67\pm0.09$ & 27.72 & $ 1.19\pm0.18$ & 2 \\
 56 & JADES-GN-189.10034+62.23089 & 189.10034 & $+62.23089$ & 5.195 & $-19.21\pm0.05$ & 26.61 & $ 5.75\pm0.62$ & F \\
 57 & JADES-GN-189.11534+62.29500 & 189.11534 & $+62.29500$ & 5.199 & $-19.16\pm0.12$ & 26.96 & $ 3.30\pm0.31$ & 2 \\
 58 & JADES-GN-189.22478+62.22575 & 189.22478 & $+62.22575$ & 5.201 & $-18.31\pm0.08$ & 26.77 & $ 1.74\pm0.21$ & 1 \\
 59 & JADES-GN-189.25771+62.22433 & 189.25771 & $+62.22433$ & 5.205 & $-18.81\pm0.04$ & 26.77 & $ 2.24\pm0.41$ & 1 \\
 60 & JADES-GN-189.08401+62.26984 & 189.08401 & $+62.26984$ & 5.205 & $-18.42\pm0.19$ & 27.76 & $ 1.92\pm0.28$ & F \\
 61 & JADES-GN-189.08302+62.26969 & 189.08302 & $+62.26969$ & 5.206 & $-18.98\pm0.05$ & 27.20 & $ 2.57\pm0.31$ & F \\
 62 & JADES-GN-189.24659+62.21494 & 189.24659 & $+62.21494$ & 5.208 & $-19.87\pm0.03$ & 25.25 & $15.67\pm0.82$ & 1 \\
 63 & JADES-GN-189.28294+62.20472 & 189.28294 & $+62.20472$ & 5.211 & $-19.23\pm0.03$ & 27.33 & $ 3.10\pm0.28$ & 1 \\
 64 & JADES-GN-189.26381+62.22539 & 189.26381 & $+62.22539$ & 5.214 & $-20.42\pm0.03$ & 25.49 & $19.94\pm0.55$ & 1 \\
 65 & JADES-GN-189.23731+62.19031 & 189.23731 & $+62.19031$ & 5.218 & $-19.03\pm0.03$ & 27.25 & $ 2.42\pm0.33$ & 1 \\
 66 & JADES-GN-189.26554+62.19931 & 189.26554 & $+62.19931$ & 5.220 & $-18.49\pm0.06$ & 26.58 & $ 7.01\pm1.09$ & 1 \\
 67 & JADES-GN-189.26569+62.19927 & 189.26569 & $+62.19927$ & 5.220 & $-19.23\pm0.03$ & 27.04 & $ 3.49\pm1.19$ & 1 \\
 68 & JADES-GN-189.26541+62.19937 & 189.26541 & $+62.19937$ & 5.220 & $-19.39\pm0.03$ & 25.27 & $13.83\pm1.11$ & 1 \\
 69 & JADES-GN-189.26502+62.19955 & 189.26502 & $+62.19955$ & 5.221 & $-19.06\pm0.04$ & 27.49 & $ 2.44\pm0.25$ & 1 \\
 70 & JADES-GN-189.24928+62.20532 & 189.24928 & $+62.20532$ & 5.221 & $-18.66\pm0.04$ & 25.66 & $23.07\pm5.11$ & 1 \\
 71 & JADES-GN-189.26510+62.19963 & 189.26510 & $+62.19963$ & 5.222 & $-18.94\pm0.04$ & 27.13 & $ 3.43\pm0.25$ & 1 \\
 72 & JADES-GN-189.28549+62.20682 & 189.28549 & $+62.20682$ & 5.222 & $-19.57\pm0.03$ & 25.91 & $ 6.30\pm0.88$ & 1 \\
 73 & JADES-GN-189.28570+62.20676 & 189.28570 & $+62.20676$ & 5.223 & $-19.40\pm0.03$ & 26.44 & $ 1.97\pm1.25$ & 1 \\
 74 & JADES-GN-189.26643+62.20915 & 189.26643 & $+62.20915$ & 5.223 & $-21.12\pm0.03$ & 24.77 & $ 9.20\pm0.66$ & 1 \\
 75 & JADES-GN-189.25382+62.19701 & 189.25382 & $+62.19701$ & 5.223 & $-18.47\pm0.06$ & 26.83 & $ 4.71\pm0.26$ & 1 \\
 76 & JADES-GN-189.30578+62.21089 & 189.30578 & $+62.21089$ & 5.223 & $-20.81\pm0.03$ & 25.81 & $14.82\pm0.56$ & 1 \\
 77 & JADES-GN-189.26168+62.19490 & 189.26168 & $+62.19490$ & 5.224 & $-19.58\pm0.05$ & 26.40 & $ 5.96\pm0.39$ & 1 \\
 78 & JADES-GN-189.25565+62.19426 & 189.25565 & $+62.19426$ & 5.224 & $-19.93\pm0.04$ & 25.75 & $ 6.30\pm0.42$ & 1 \\
 79 & JADES-GN-189.25428+62.20769 & 189.25428 & $+62.20769$ & 5.225 & $-19.34\pm0.03$ & 26.51 & $ 3.82\pm0.48$ & 1 \\
 80 & JADES-GN-189.19396+62.17536 & 189.19396 & $+62.17536$ & 5.226 & $-18.43\pm0.06$ & 26.55 & $ 3.26\pm0.33$ & F \\
 81 & JADES-GN-189.26137+62.20801 & 189.26137 & $+62.20801$ & 5.227 & $-19.80\pm0.03$ & 25.71 & $19.04\pm0.61$ & 1 \\
 82 & JADES-GN-189.24913+62.20519 & 189.24913 & $+62.20519$ & 5.227 & $-20.60\pm0.03$ & 24.89 & $24.33\pm0.78$ & 1 \\
 83 & JADES-GN-189.15686+62.30453 & 189.15686 & $+62.30453$ & 5.242 & $-19.42\pm0.10$ & 27.18 & $ 3.81\pm0.61$ & F \\
 84 & JADES-GN-189.09366+62.23480 & 189.09366 & $+62.23480$ & 5.245 & $-19.03\pm0.05$ & 27.16 & $ 3.55\pm0.35$ & F \\
 85 & JADES-GN-189.13369+62.24499 & 189.13369 & $+62.24499$ & 5.257 & $-18.98\pm0.06$ & 27.00 & $ 3.33\pm0.39$ & 3 \\
 86 & JADES-GN-189.14244+62.28573 & 189.14244 & $+62.28573$ & 5.264 & $-20.04\pm0.03$ & 26.14 & $ 6.07\pm1.16$ & F \\
 87 & JADES-GN-189.14260+62.28570 & 189.14260 & $+62.28570$ & 5.264 & $-18.94\pm0.05$ & 26.71 & $ 5.12\pm0.64$ & F \\
 88 & JADES-GN-189.12617+62.23717 & 189.12617 & $+62.23717$ & 5.266 & $-19.58\pm0.04$ & 26.83 & $ 3.40\pm0.49$ & 3 \\
 89 & JADES-GN-189.11794+62.23552 & 189.11794 & $+62.23552$ & 5.266 & $-19.25\pm0.05$ & 27.25 & $ 3.27\pm0.32$ & 3 \\
 90 & JADES-GN-189.13100+62.28684 & 189.13100 & $+62.28684$ & 5.267 & $-19.54\pm0.08$ & 25.48 & $ 6.44\pm0.58$ & F \\
 91 & JADES-GN-189.12574+62.24596 & 189.12574 & $+62.24596$ & 5.267 & $-19.19\pm0.06$ & 27.16 & $ 3.42\pm0.45$ & 3 \\
 92 & JADES-GN-189.12415+62.24222 & 189.12415 & $+62.24222$ & 5.268 & $-19.26\pm0.05$ & 26.73 & $ 5.78\pm0.55$ & 3 \\
 93 & JADES-GN-189.11906+62.23624 & 189.11906 & $+62.23624$ & 5.269 & $-20.47\pm0.03$ & 25.99 & $ 8.29\pm0.40$ & 3 \\
 94 & JADES-GN-189.11309+62.29239 & 189.11309 & $+62.29239$ & 5.270 & $-18.07\pm0.12$ & 25.24 & $ 3.21\pm0.35$ & F \\
 95 & JADES-GN-189.12004+62.23867 & 189.12004 & $+62.23867$ & 5.271 & $-20.13\pm0.05$ & 25.94 & $ 7.46\pm0.41$ & 3 \\
 96 & JADES-GN-189.15458+62.24755 & 189.15458 & $+62.24755$ & 5.272 & $-19.24\pm0.05$ & 26.33 & $ 4.87\pm0.45$ & 3 \\
 97 & JADES-GN-189.11960+62.23855 & 189.11960 & $+62.23855$ & 5.274 & $-19.72\pm0.03$ & 26.59 & $ 7.90\pm0.75$ & 3 \\
 98 & JADES-GN-189.15980+62.28796 & 189.15980 & $+62.28796$ & 5.274 & $-19.56\pm0.03$ & 27.07 & $ 2.50\pm0.41$ & F \\
 99 & JADES-GN-189.26349+62.24229 & 189.26349 & $+62.24229$ & 5.283 & $-17.13\pm0.22$ & 28.02 & $ 1.84\pm0.34$ & F \\
100 & JADES-GN-189.23581+62.23532 & 189.23581 & $+62.23532$ & 5.285 & $-18.85\pm0.11$ & 26.88 & $ 3.20\pm0.28$ & F \\
101 & JADES-GN-189.09470+62.23936 & 189.09470 & $+62.23936$ & 5.292 & $-19.34\pm0.05$ & 26.13 & $ 9.64\pm0.43$ & 4 \\
102 & JADES-GN-189.05259+62.22990 & 189.05259 & $+62.22990$ & 5.294 & $-19.51\pm0.05$ & 26.18 & $ 5.97\pm0.21$ & 4 \\
103 & JADES-GN-189.14939+62.23187 & 189.14939 & $+62.23187$ & 5.295 & $-20.12\pm0.05$ & 25.41 & $ 7.92\pm0.43$ & 4 \\
104 & JADES-GN-189.09492+62.23948 & 189.09492 & $+62.23948$ & 5.295 & $-20.12\pm0.05$ & 25.71 & $ 7.20\pm0.42$ & 4 \\
105 & JADES-GN-189.11943+62.23994 & 189.11943 & $+62.23994$ & 5.295 & $-20.58\pm0.03$ & 25.01 & $ 7.18\pm0.68$ & 4 \\
106 & JADES-GN-189.12993+62.25251 & 189.12993 & $+62.25251$ & 5.296 & $-20.14\pm0.03$ & 26.44 & $ 5.14\pm0.31$ & 4 \\
107 & JADES-GN-189.08389+62.23713 & 189.08389 & $+62.23713$ & 5.296 & $-19.86\pm0.06$ & 26.51 & $ 5.08\pm0.27$ & 4 \\
108 & JADES-GN-189.12192+62.29101 & 189.12192 & $+62.29101$ & 5.297 & $-19.62\pm0.08$ & 26.31 & $ 2.51\pm0.29$ & F \\
109 & JADES-GN-189.12124+62.29358 & 189.12124 & $+62.29358$ & 5.297 & $-18.21\pm0.12$ & 27.32 & $ 2.49\pm0.20$ & F \\
110 & JADES-GN-189.14903+62.29347 & 189.14903 & $+62.29347$ & 5.297 & $-18.20\pm0.13$ & 27.55 & $ 1.50\pm0.25$ & F \\
111 & JADES-GN-189.25780+62.19996 & 189.25780 & $+62.19996$ & 5.299 & $-18.24\pm0.07$ & 26.96 & $ 1.26\pm0.23$ & F \\
112 & JADES-GN-189.22631+62.21190 & 189.22631 & $+62.21190$ & 5.311 & $-20.27\pm0.03$ & 24.98 & $11.82\pm0.94$ & F \\
113 & JADES-GN-189.11031+62.22544 & 189.11031 & $+62.22544$ & 5.344 & $-20.54\pm0.04$ & 25.47 & $12.84\pm1.45$ & F \\
114 & JADES-GN-189.11044+62.22533 & 189.11044 & $+62.22533$ & 5.346 & $-18.53\pm0.38$ & 27.06 & $ 2.81\pm0.73$ & F \\
115 & JADES-GN-189.25501+62.21625 & 189.25501 & $+62.21625$ & 5.348 & $-19.02\pm0.07$ & 26.89 & $ 2.67\pm0.49$ & F \\
116 & JADES-GN-189.21696+62.23168 & 189.21696 & $+62.23168$ & 5.354 & $-19.10\pm0.12$ & 26.68 & $ 4.35\pm0.48$ & F \\
117 & JADES-GN-189.17252+62.23584 & 189.17252 & $+62.23584$ & 5.355 & $-19.13\pm0.11$ & 27.55 & $ 3.00\pm0.44$ & F \\
118 & JADES-GN-189.12134+62.22823 & 189.12134 & $+62.22823$ & 5.356 & $-19.15\pm0.14$ & 26.66 & $ 3.64\pm0.35$ & F \\
119 & JADES-GN-189.06016+62.25651 & 189.06016 & $+62.25651$ & 5.360 & $-19.42\pm0.05$ & 25.98 & $ 8.22\pm0.63$ & F \\
120 & JADES-GN-189.06947+62.25945 & 189.06947 & $+62.25945$ & 5.362 & $-19.81\pm0.03$ & 26.10 & $ 9.04\pm0.68$ & F \\
121 & JADES-GN-189.04592+62.25416 & 189.04592 & $+62.25416$ & 5.362 & $-19.29\pm0.08$ & 26.39 & $ 6.85\pm0.39$ & F \\
122 & JADES-GN-189.16922+62.23847 & 189.16922 & $+62.23847$ & 5.418 & $-18.94\pm0.06$ & 26.05 & $ 4.09\pm0.52$ & F \\
123 & JADES-GN-189.12883+62.24377 & 189.12883 & $+62.24377$ & 5.420 & $-18.99\pm0.18$ & 27.40 & $ 2.56\pm0.40$ & F \\
124 & JADES-GN-189.21086+62.18661 & 189.21086 & $+62.18661$ & 5.421 & $-19.05\pm0.08$ & 26.86 & $ 3.23\pm0.45$ & F \\
125 & JADES-GN-189.18858+62.21303 & 189.18858 & $+62.21303$ & 5.426 & $-19.28\pm0.06$ & 26.94 & $ 3.06\pm0.38$ & F \\
126 & JADES-GN-189.04727+62.22963 & 189.04727 & $+62.22963$ & 5.428 & $-18.62\pm0.23$ & 27.75 & $ 1.22\pm0.21$ & F \\
127 & JADES-GN-189.26070+62.21626 & 189.26070 & $+62.21626$ & 5.432 & $-18.67\pm0.10$ & 26.97 & $ 5.23\pm0.47$ & F \\
128 & JADES-GN-189.15602+62.23872 & 189.15602 & $+62.23872$ & 5.435 & $-19.10\pm0.12$ & 27.76 & $ 1.75\pm0.31$ & F \\
129 & JADES-GN-189.12828+62.27297 & 189.12828 & $+62.27297$ & 5.435 & $-18.76\pm0.06$ & 27.77 & $ 1.67\pm0.27$ & F \\
130 & JADES-GN-189.04962+62.24405 & 189.04962 & $+62.24405$ & 5.442 & $-20.57\pm0.03$ & 25.24 & $19.43\pm0.64$ & F \\
131 & JADES-GN-189.05777+62.23927 & 189.05777 & $+62.23927$ & 5.443 & $-20.33\pm0.05$ & 25.60 & $ 7.50\pm0.45$ & F \\
132 & JADES-GN-189.08862+62.23949 & 189.08862 & $+62.23949$ & 5.443 & $-19.83\pm0.03$ & 26.05 & $10.85\pm0.45$ & F \\
133 & JADES-GN-189.14305+62.21142 & 189.14305 & $+62.21142$ & 5.446 & $-18.34\pm0.08$ & 26.37 & $ 4.37\pm0.69$ & F \\
134 & JADES-GN-189.18792+62.29720 & 189.18792 & $+62.29720$ & 5.450 & $-19.18\pm0.10$ & 27.51 & $ 1.84\pm0.27$ & F \\
135 & JADES-GN-189.15552+62.20012 & 189.15552 & $+62.20012$ & 5.456 & $-19.12\pm0.03$ & 25.94 & $ 4.15\pm1.05$ & F \\
136 & JADES-GN-189.15888+62.19685 & 189.15888 & $+62.19685$ & 5.456 & $-19.17\pm0.03$ & 27.07 & $ 3.65\pm0.39$ & F \\
137 & JADES-GN-189.15551+62.20002 & 189.15551 & $+62.20002$ & 5.457 & $-19.30\pm0.06$ & 26.07 & $ 8.81\pm1.21$ & F \\
138 & JADES-GN-189.06882+62.26887 & 189.06882 & $+62.26887$ & 5.470 & $-19.87\pm0.03$ & 26.85 & $ 2.26\pm0.33$ & F \\
139 & JADES-GN-189.23458+62.19048 & 189.23458 & $+62.19048$ & 5.484 & $-19.17\pm0.03$ & 27.86 & $ 3.43\pm0.37$ & F \\
140 & JADES-GN-189.14849+62.21166 & 189.14849 & $+62.21166$ & 5.490 & $-18.89\pm0.05$ & 27.18 & $ 3.86\pm0.58$ & F 
\enddata
\tablecomments{Coordinates are in ICRS (J2000.0). The typical uncertainty of grism spectroscopic redshifts ($z_\mathrm{spec}$) is $\Delta z=0.001$. The uncertainty of F444W magnitude is dominated by flux calibration and aperture correction and we adopt a $5\%$ noise floor.
In the group ID column, ``F" denotes field galaxies while other number IDs are assigned through our FoF clustering analysis (Section~\ref{ss:05b_cluster}).
HDF850.1 and five AGN candidates with broad \ha\ emission lines are not included in this table. 
The coordinates and redshifts of AGN candidates will be presented in a companion paper (E.\ Egami et al.\ in preparation).
}
\end{deluxetable*}


\setcounter{figure}{0}
\renewcommand{\thefigure}{\thesection\arabic{figure}}

\begin{figure*}[!ht] 
 \centering
\includegraphics[width=0.49\linewidth]{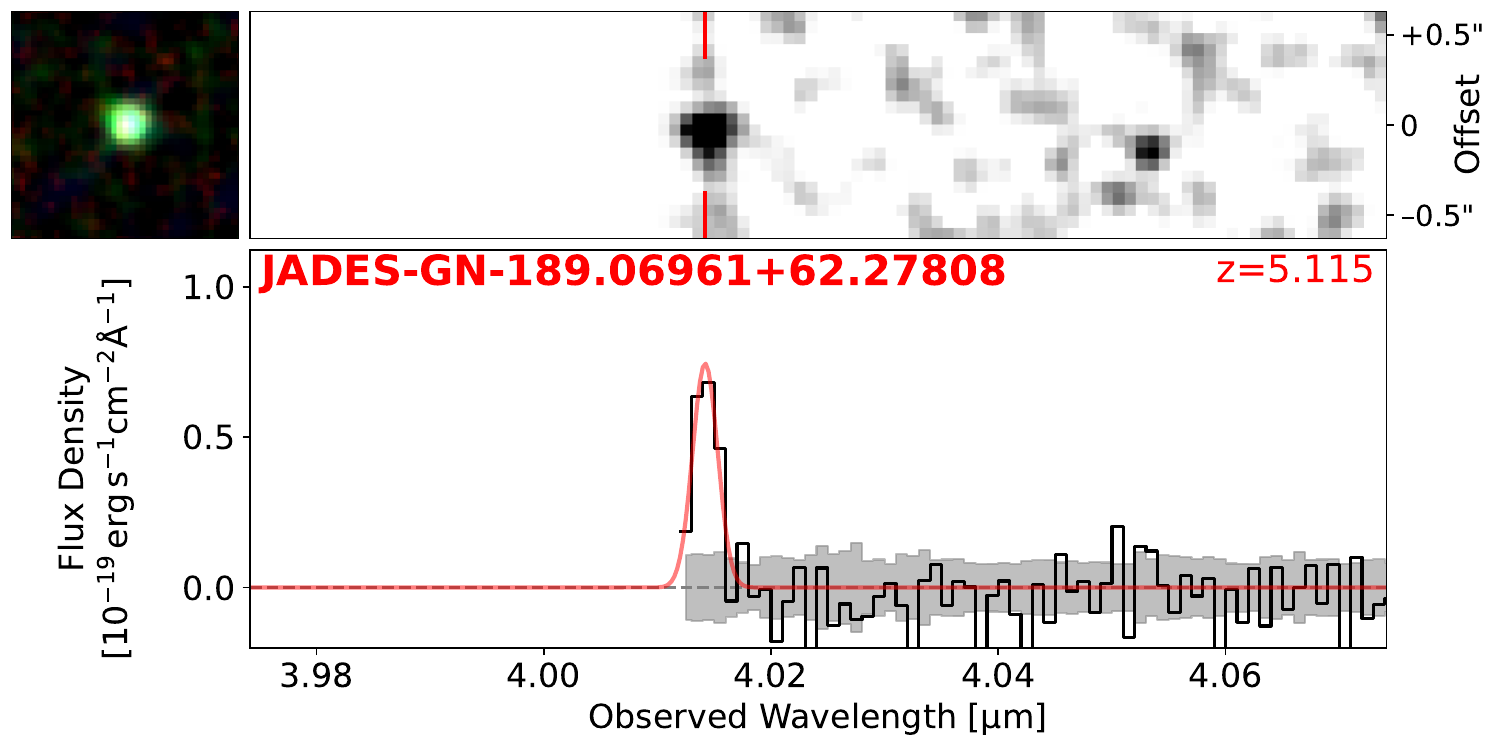}
\includegraphics[width=0.49\linewidth]{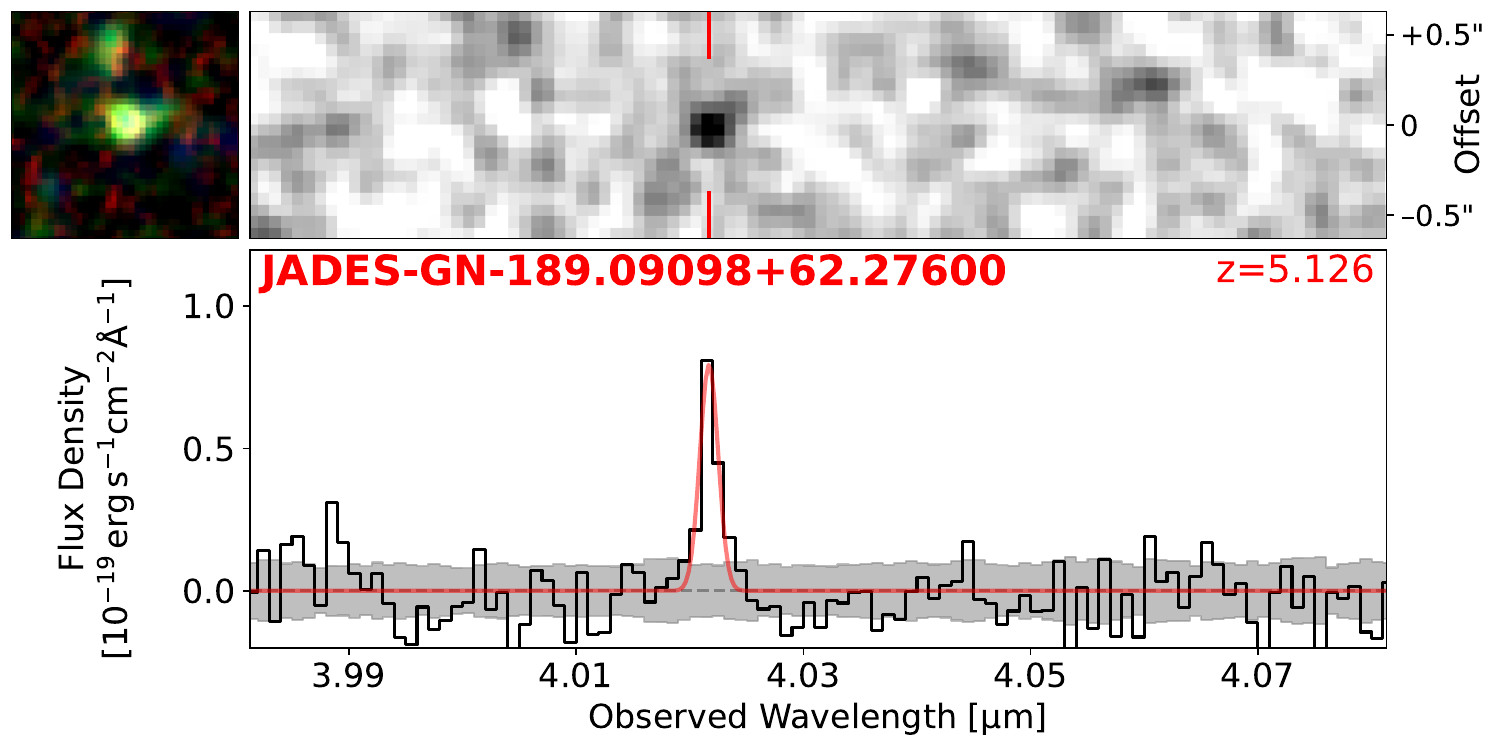}
\includegraphics[width=0.49\linewidth]{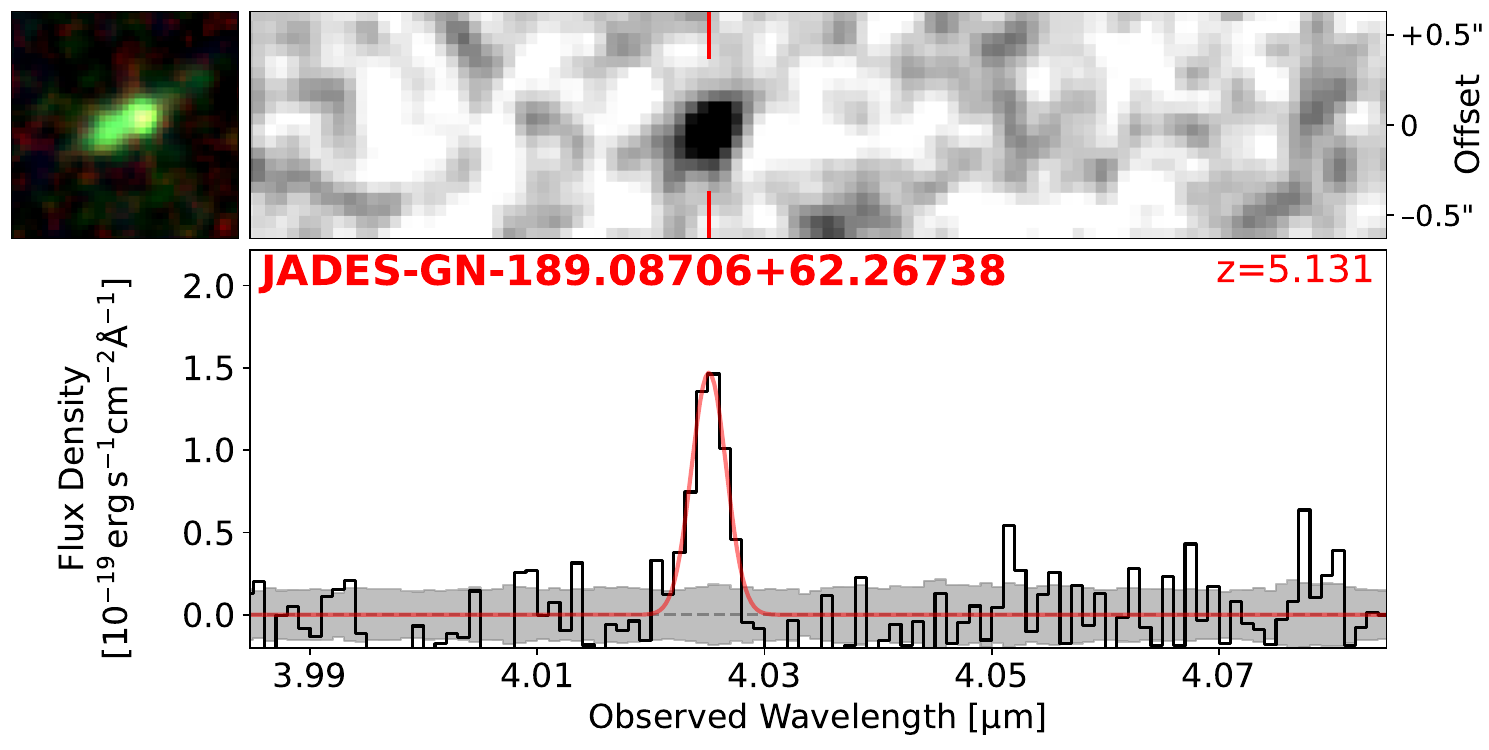}
\includegraphics[width=0.49\linewidth]{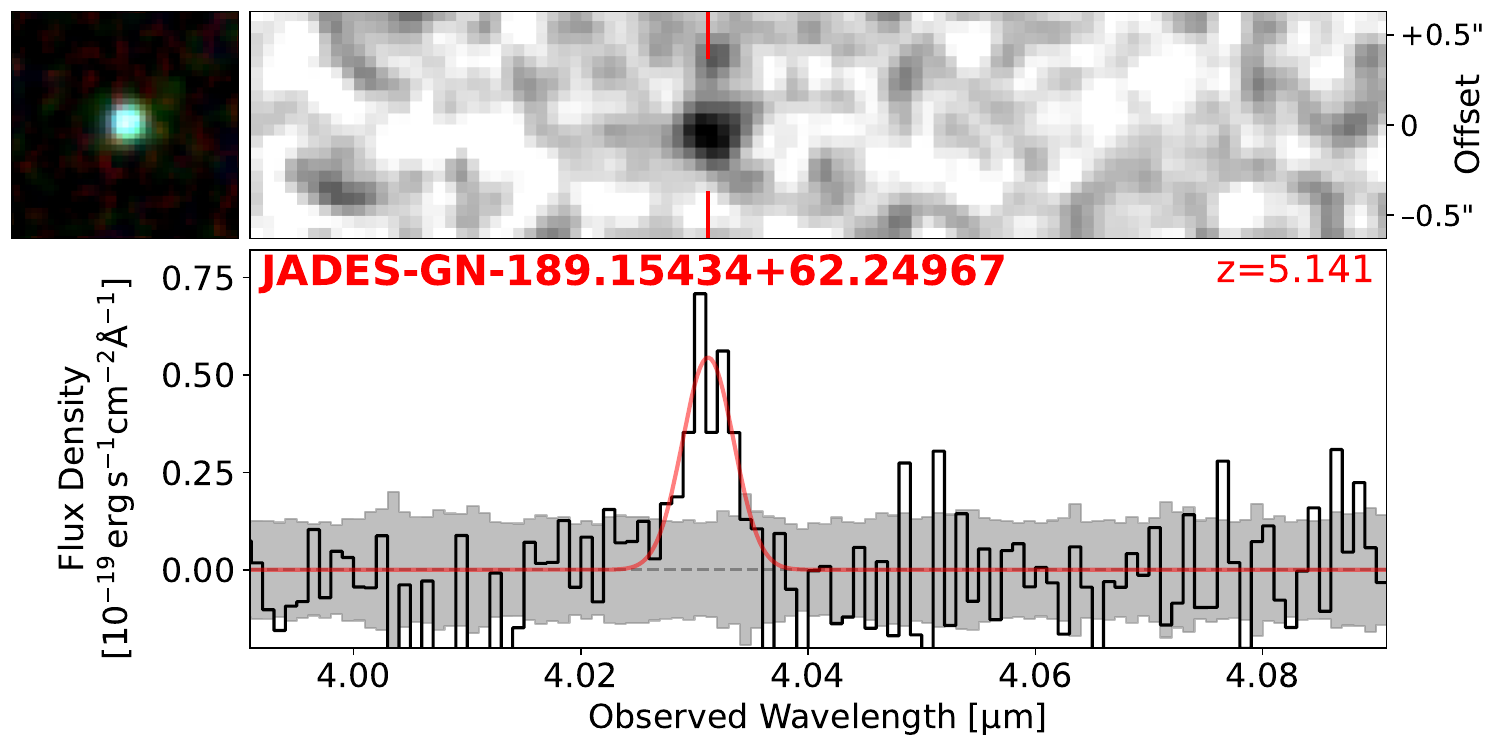}
\includegraphics[width=0.49\linewidth]{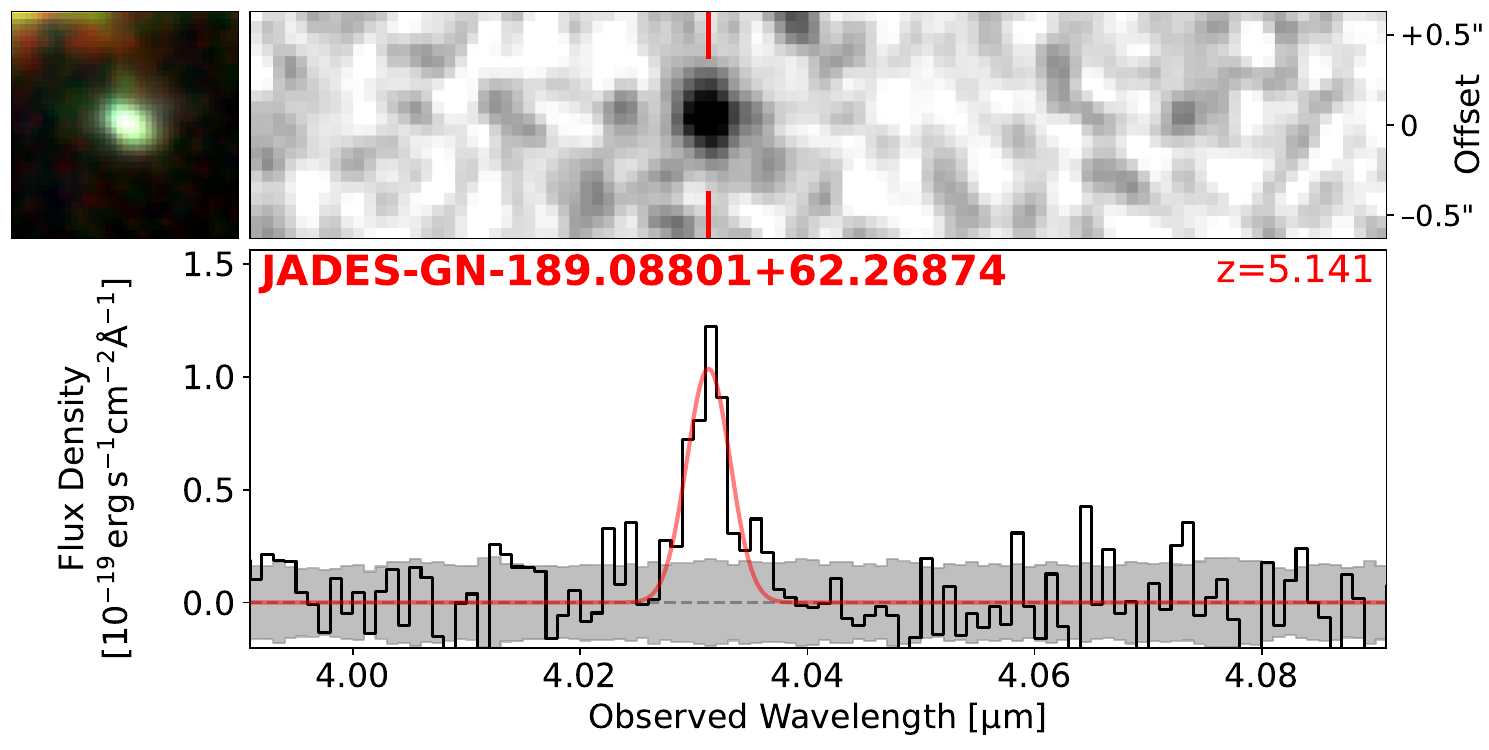}
\includegraphics[width=0.49\linewidth]{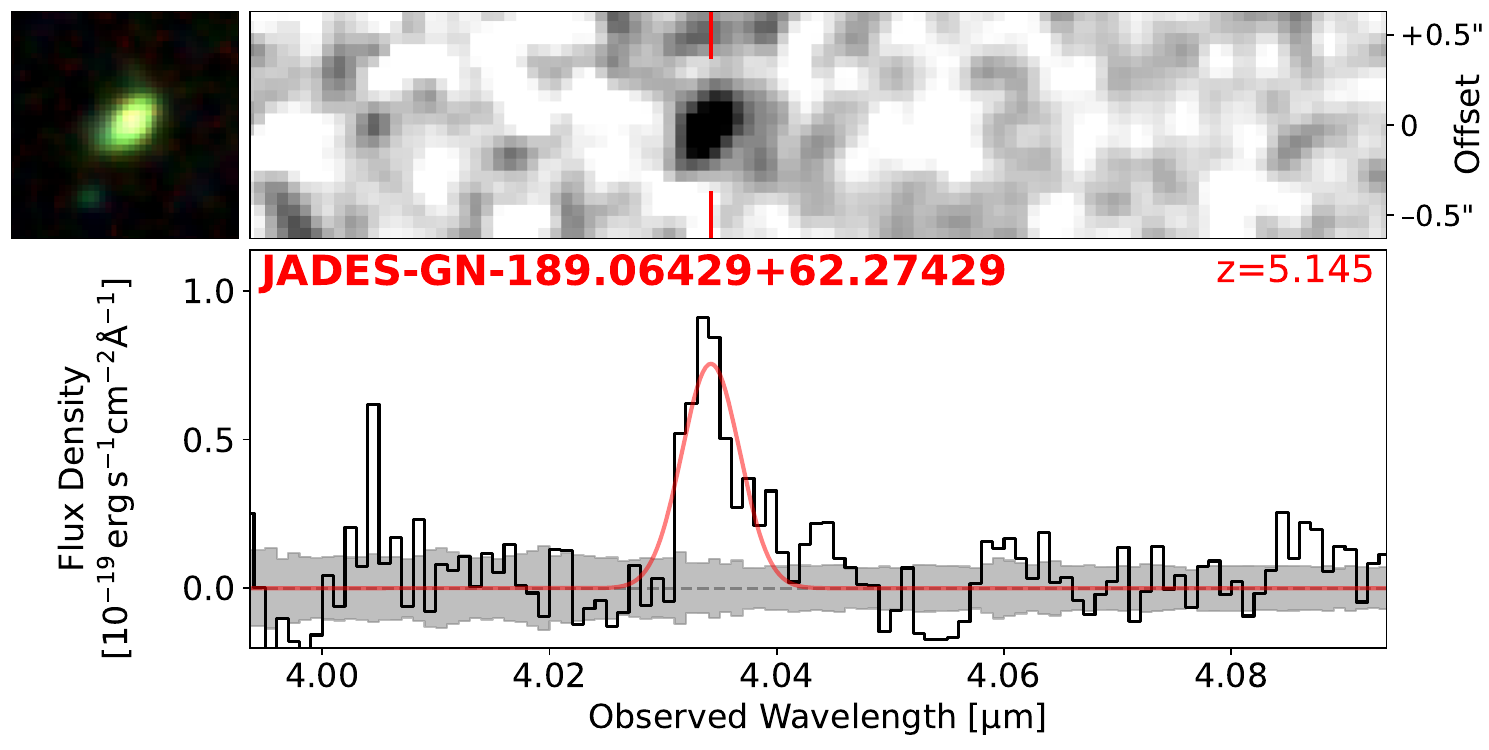}
\includegraphics[width=0.49\linewidth]{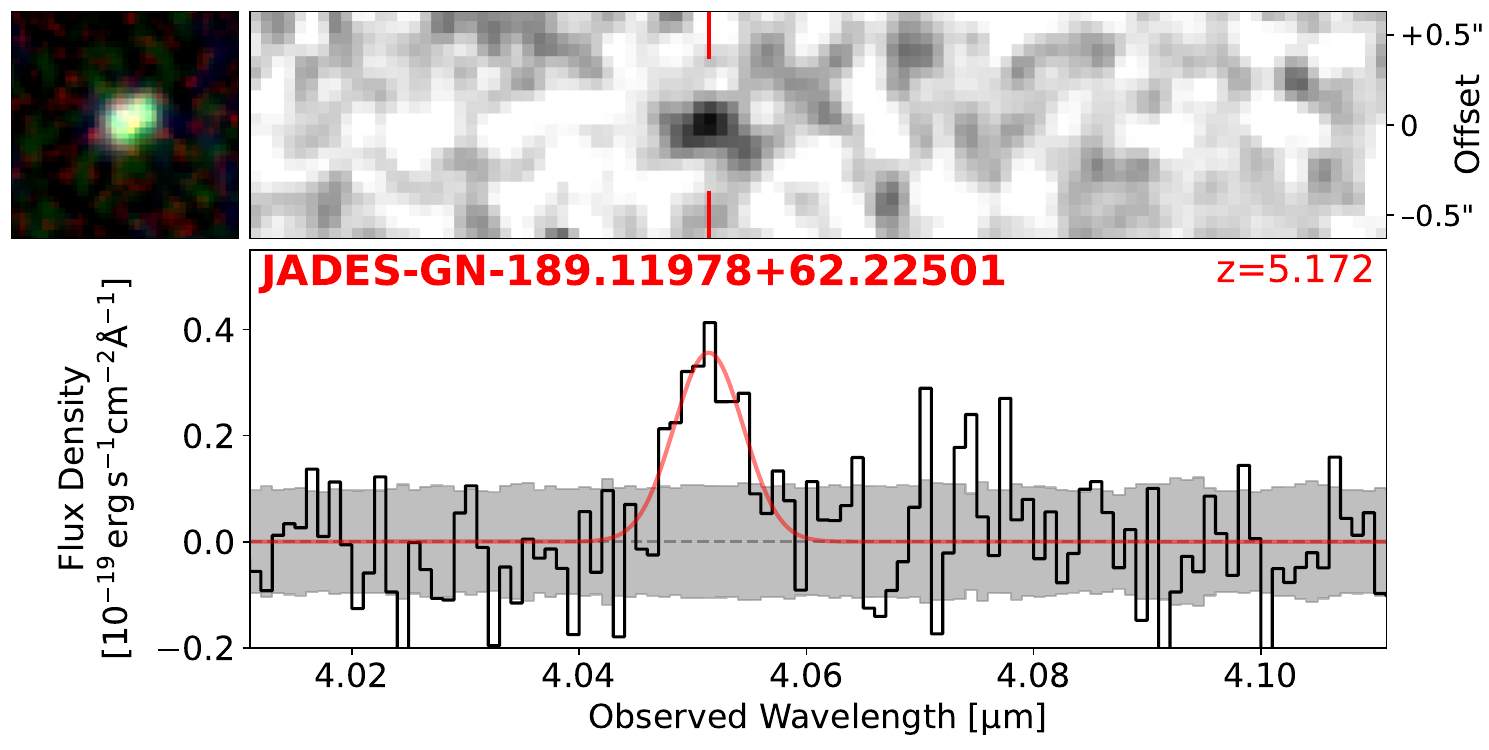}
\includegraphics[width=0.49\linewidth]{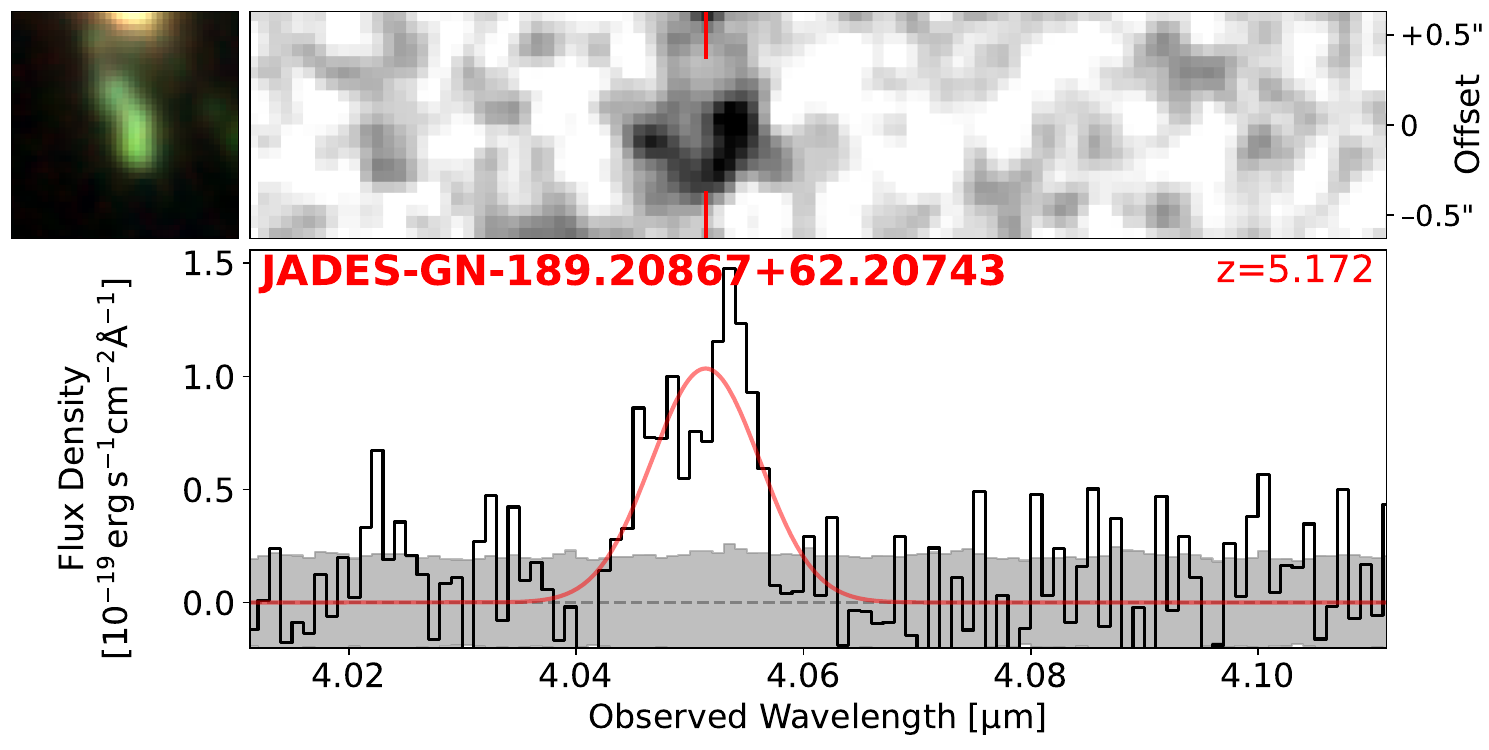}
\includegraphics[width=0.49\linewidth]{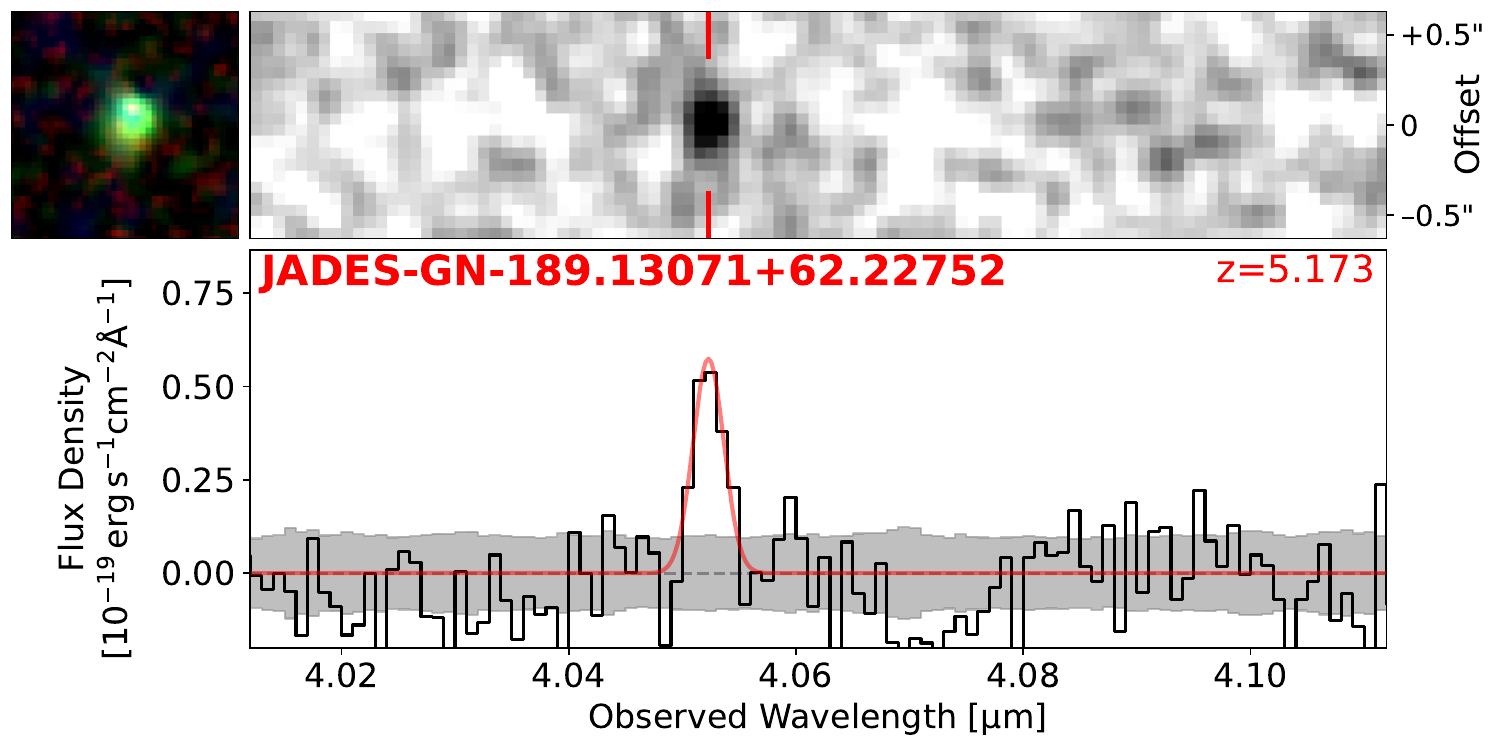}
\includegraphics[width=0.49\linewidth]{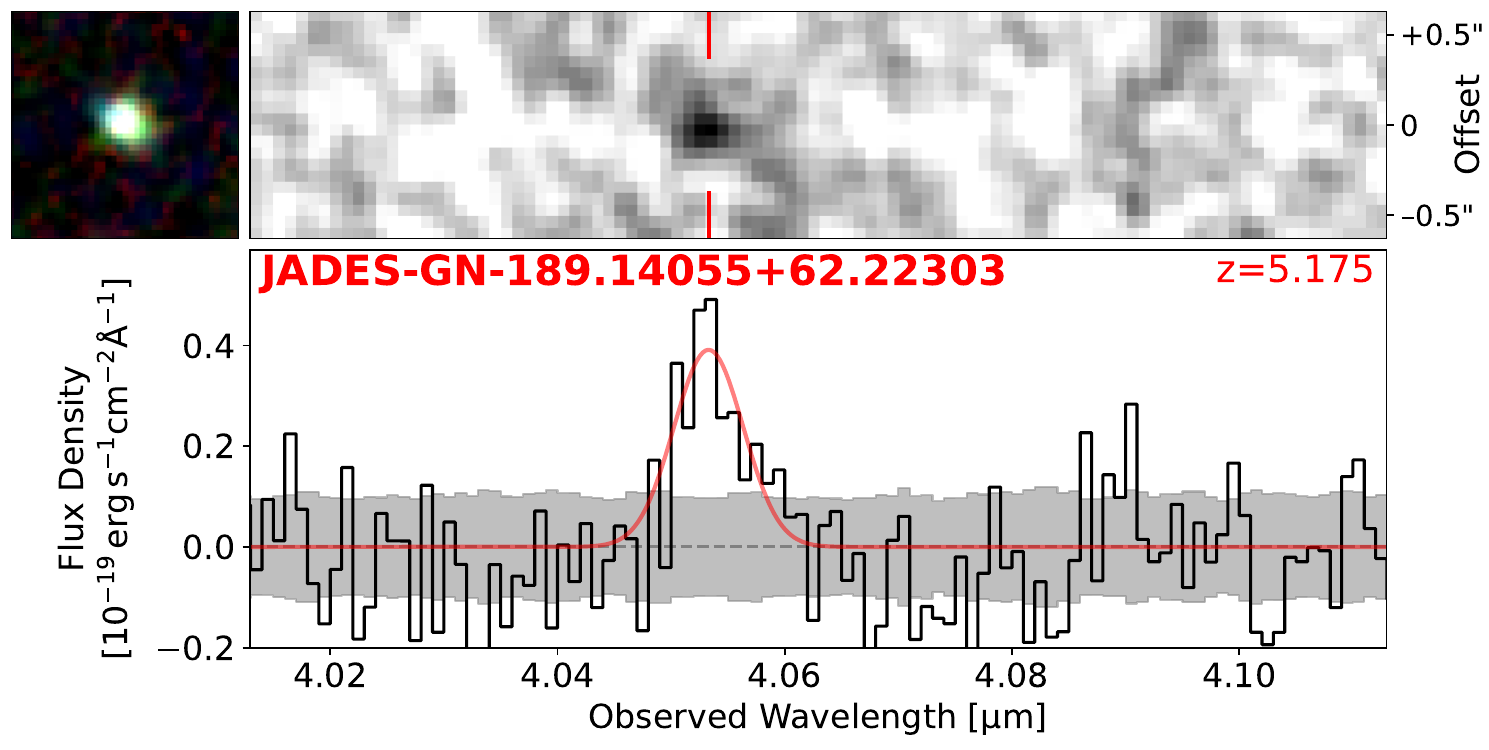}
\caption{NIRCam images, 2D and 1D grism spectra of 140 spectroscopically confirmed galaxies at $z=5.1-5.5$ (not including HDF850.1 and five AGN candidates in E.\ Egami et al.\ in preparation). 
For each galaxy, the upper-left panel shows the 1\farcs2$\times$1\farcs2 F444W--F277W--F150W RGB thumbnail. 
Images are rotated to align with the dispersion direction.
The upper-right panel shows the continuum-subtracted 2D spectrum around the \ha\ emission line detection, indicated by the solid red line. 
The lower-right panel shows the optimally extracted 1D spectrum with the best-fit Gaussian profile indicated by the solid red line. 
The name and confirmed spectroscopic redshift are given in the lower-right panel for each galaxy.}
\label{fig:apd_spec}
\end{figure*} 

 \addtocounter{figure}{-1} 
 \begin{figure*}[!ht] 
 \centering
\includegraphics[width=0.49\linewidth]{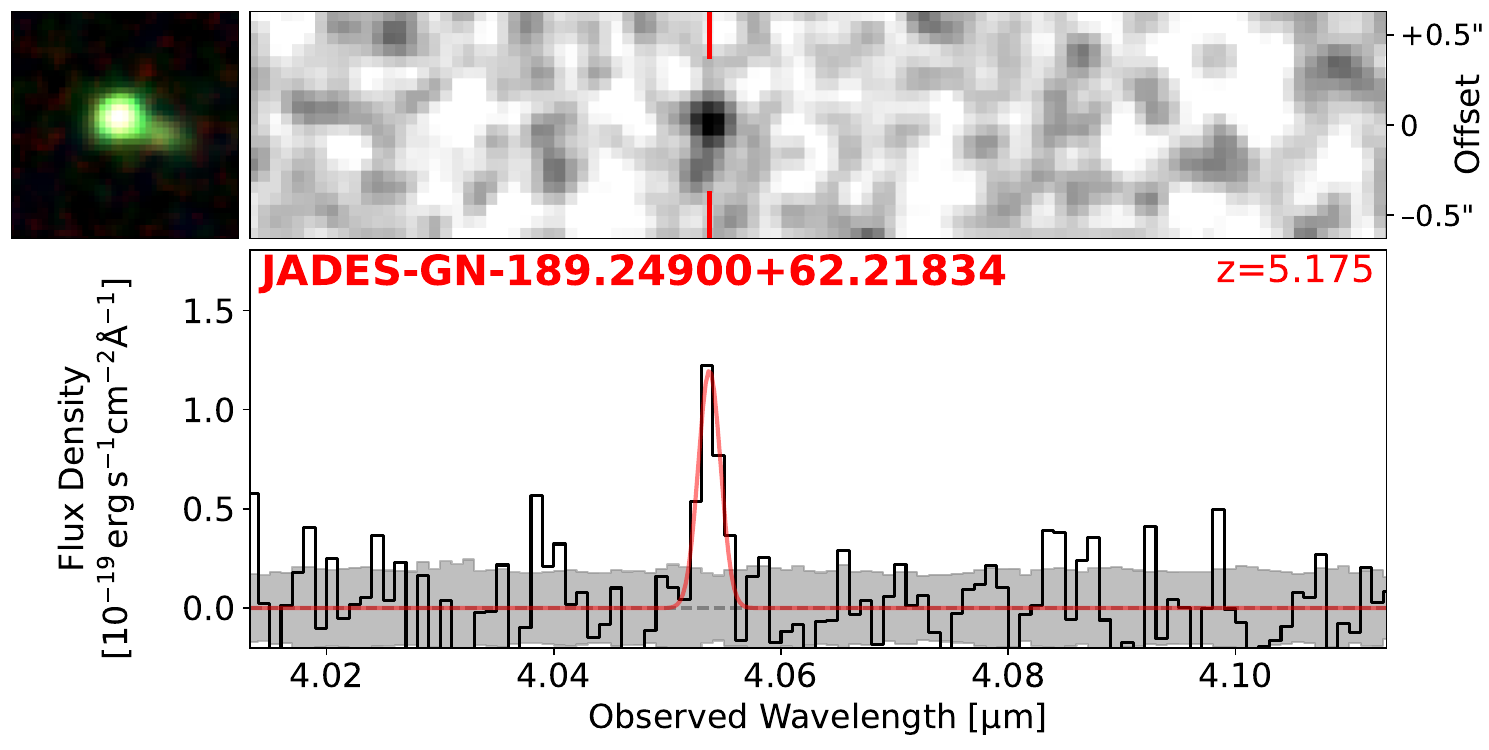}
\includegraphics[width=0.49\linewidth]{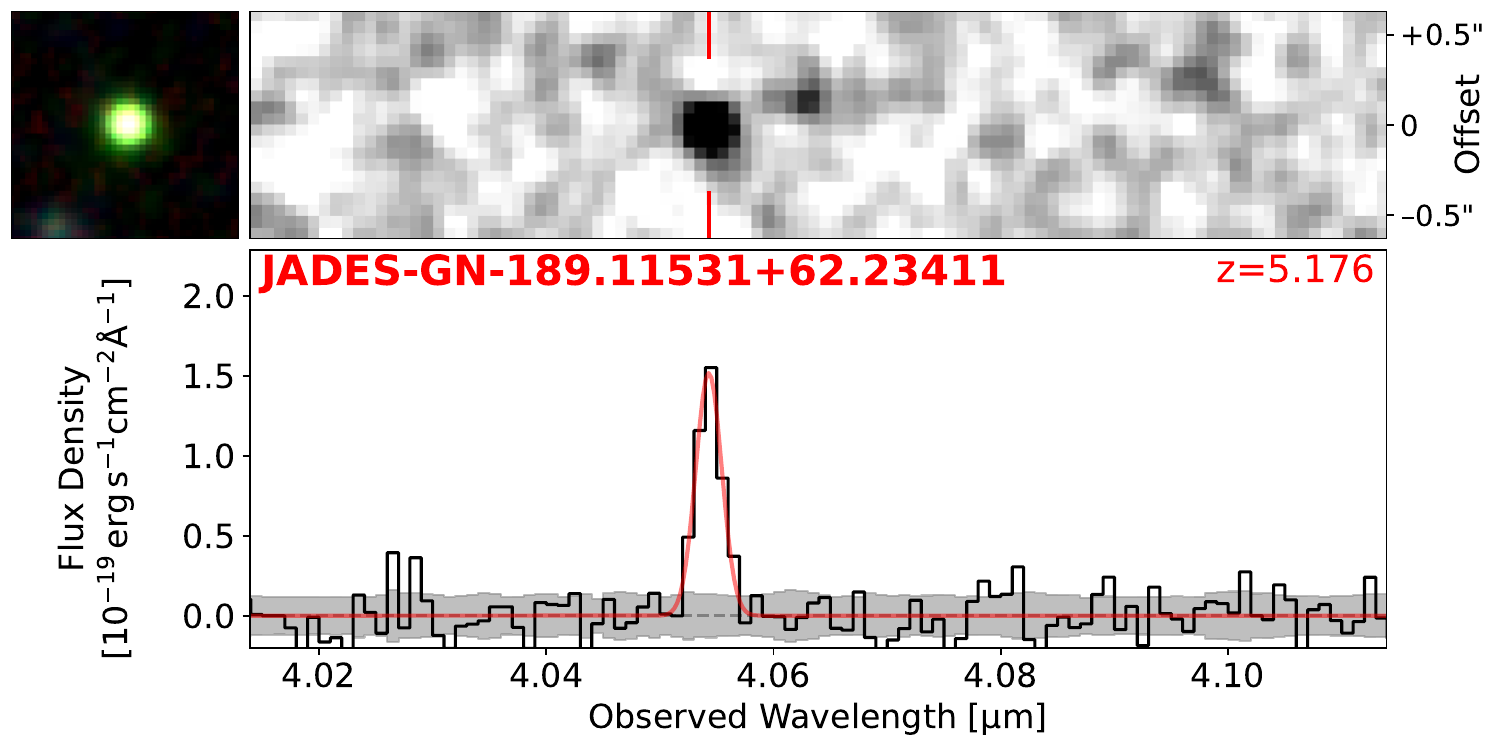}
\includegraphics[width=0.49\linewidth]{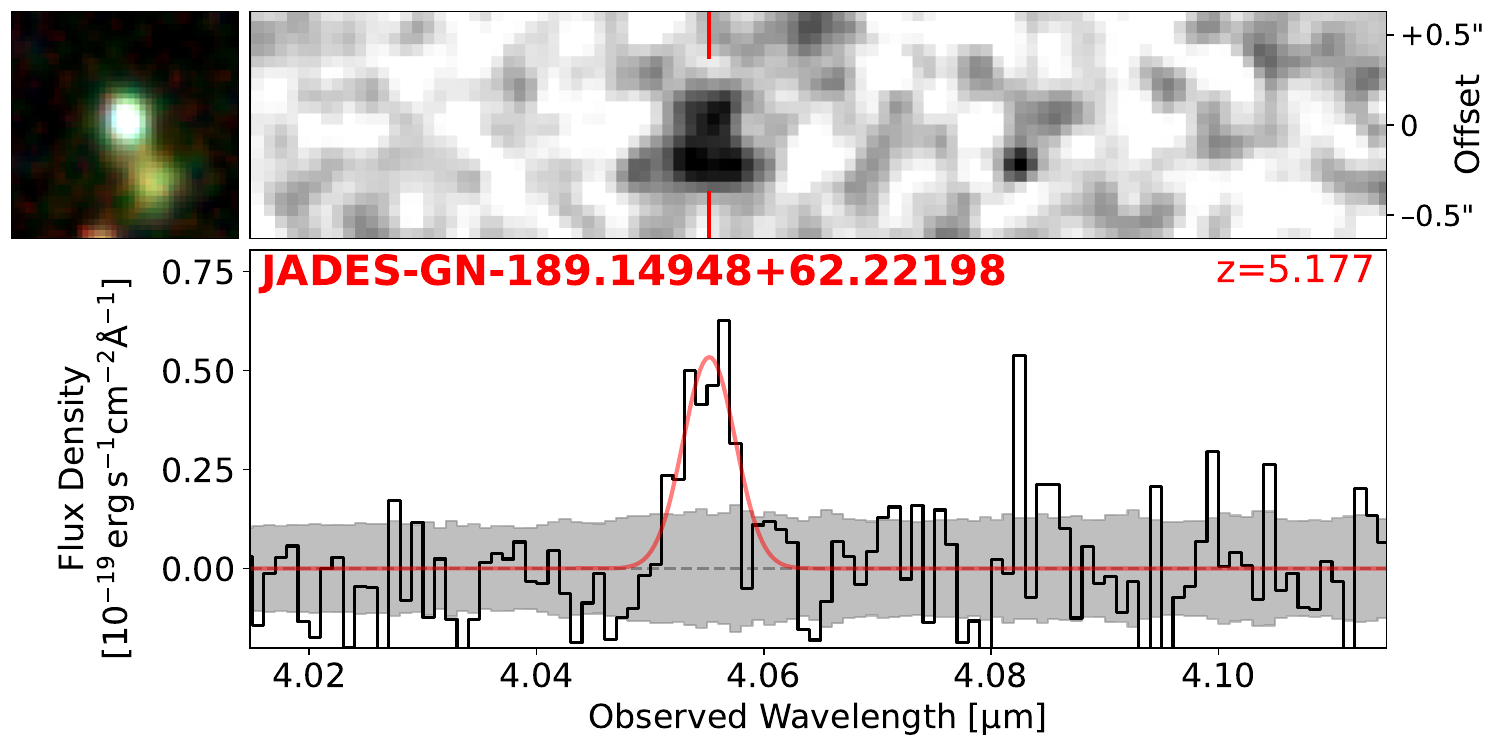}
\includegraphics[width=0.49\linewidth]{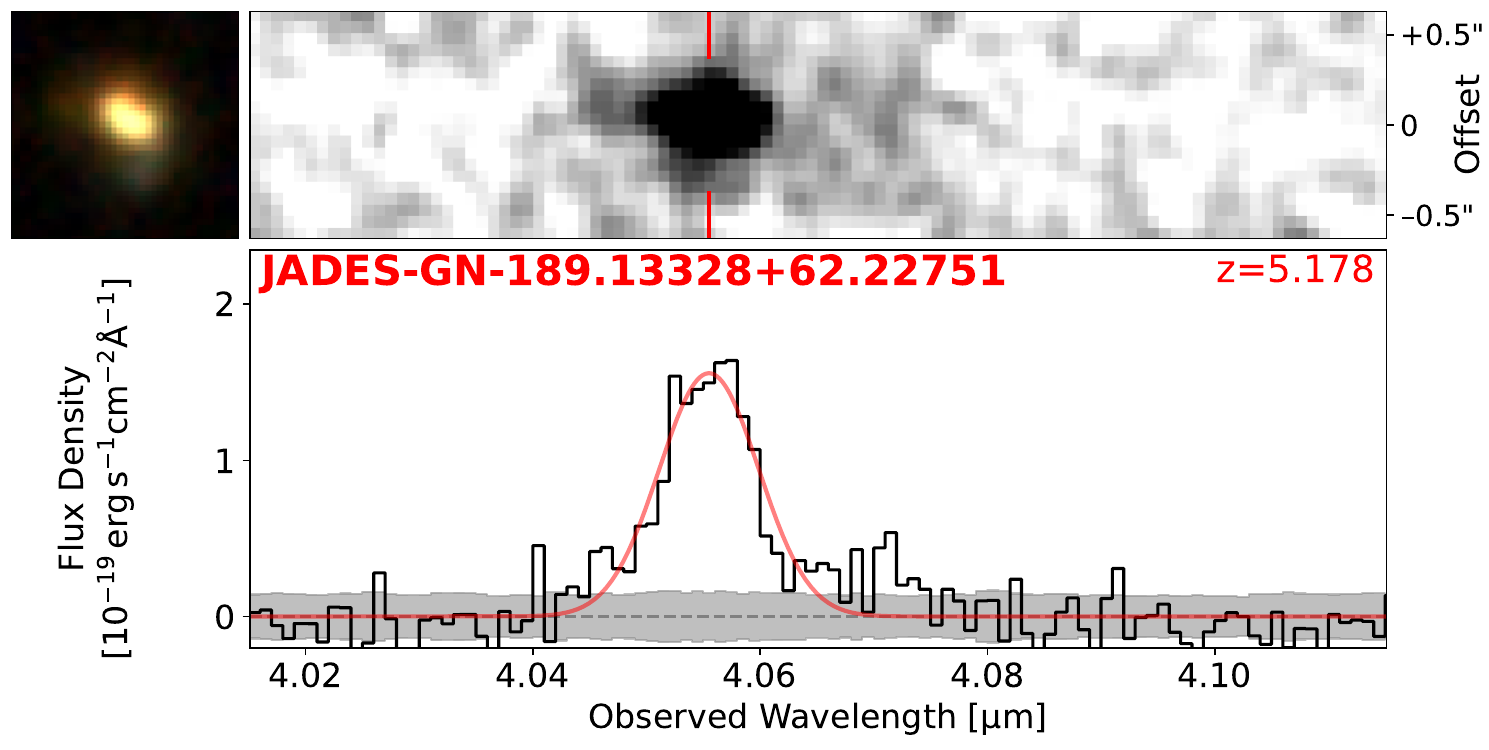}
\includegraphics[width=0.49\linewidth]{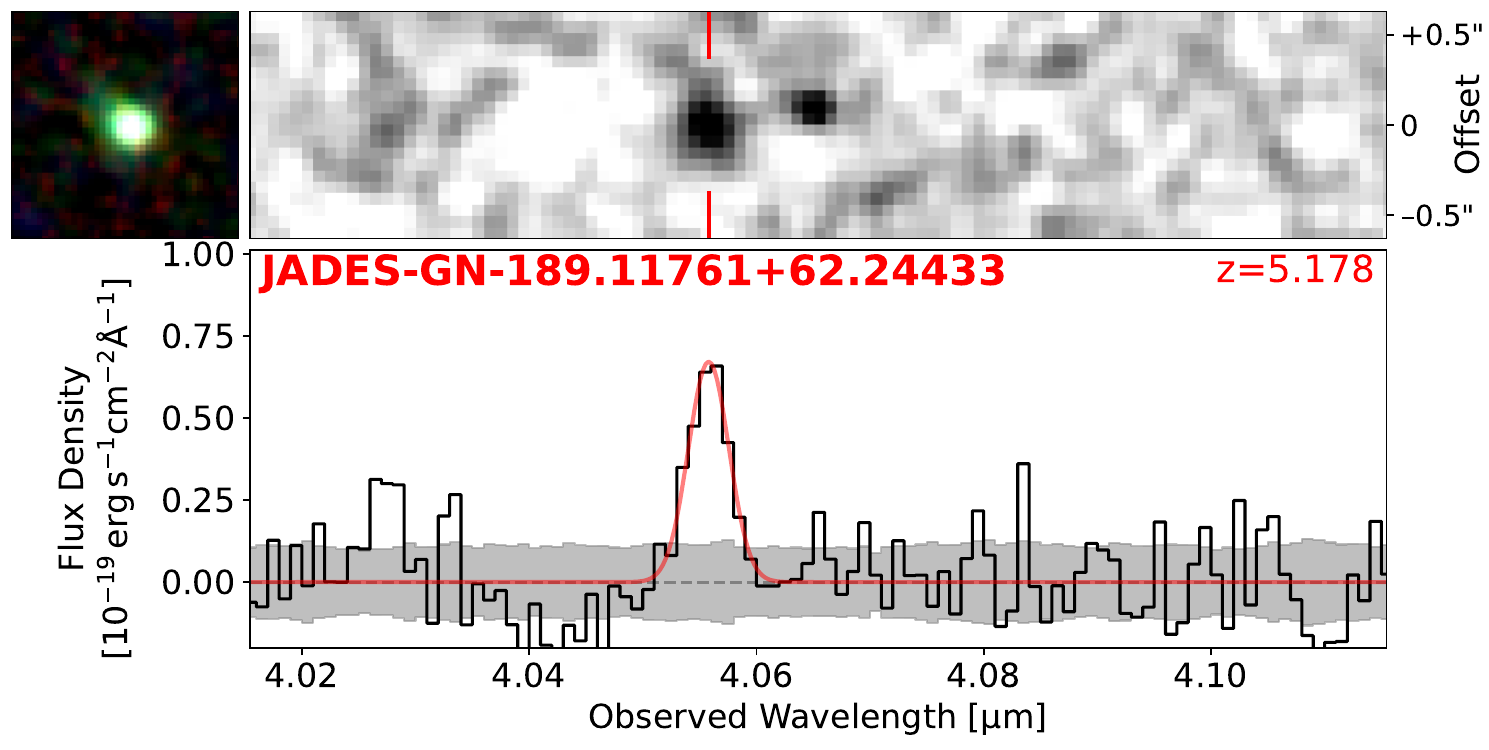}
\includegraphics[width=0.49\linewidth]{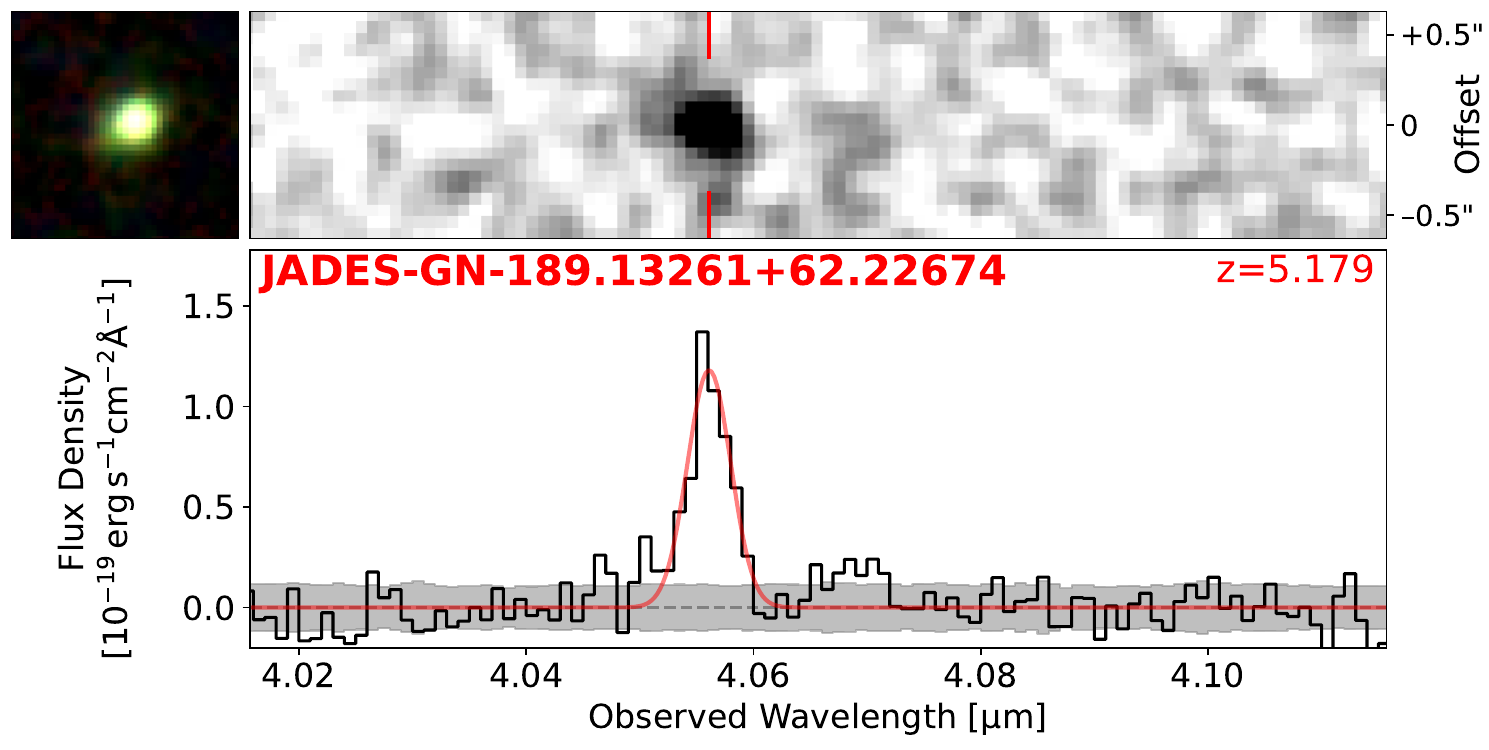}
\includegraphics[width=0.49\linewidth]{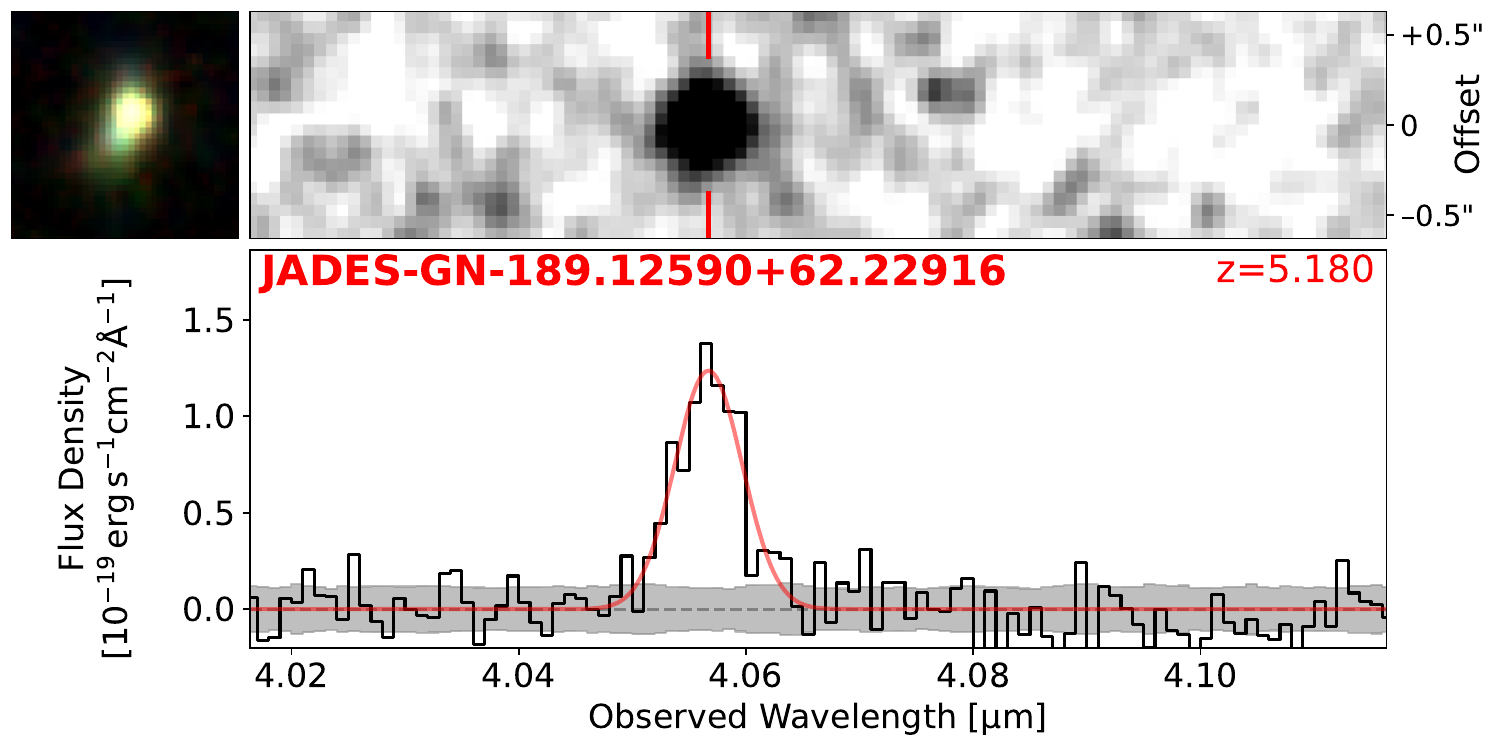}
\includegraphics[width=0.49\linewidth]{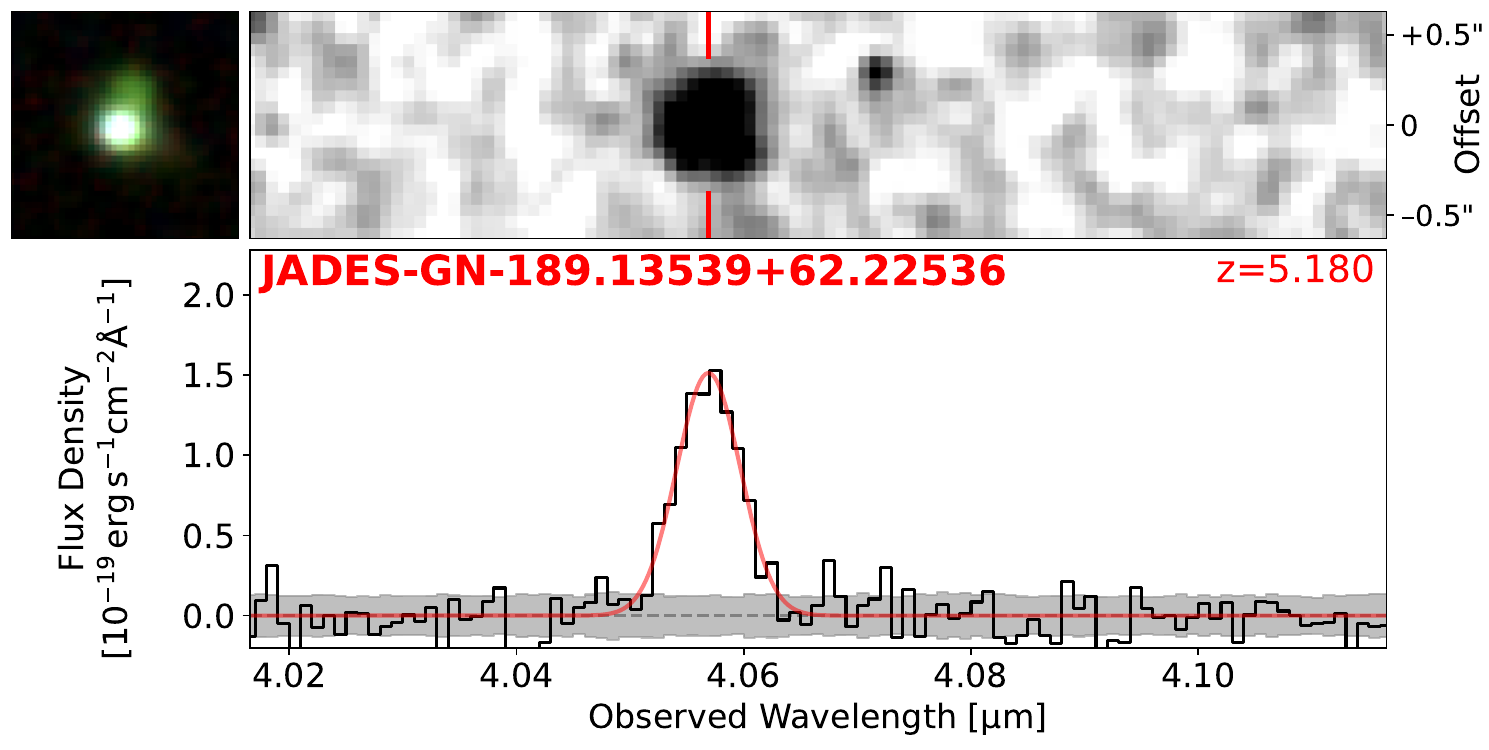}
\includegraphics[width=0.49\linewidth]{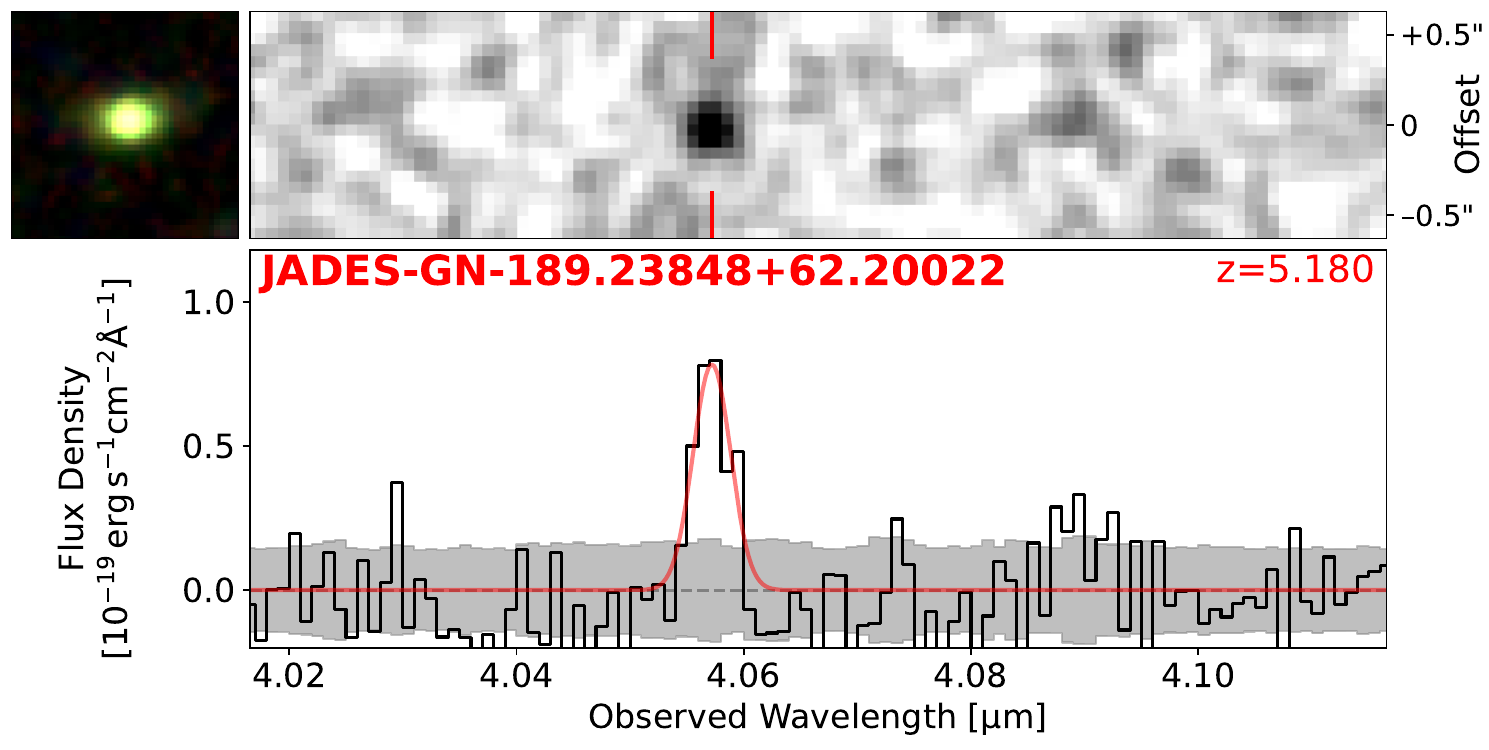}
\includegraphics[width=0.49\linewidth]{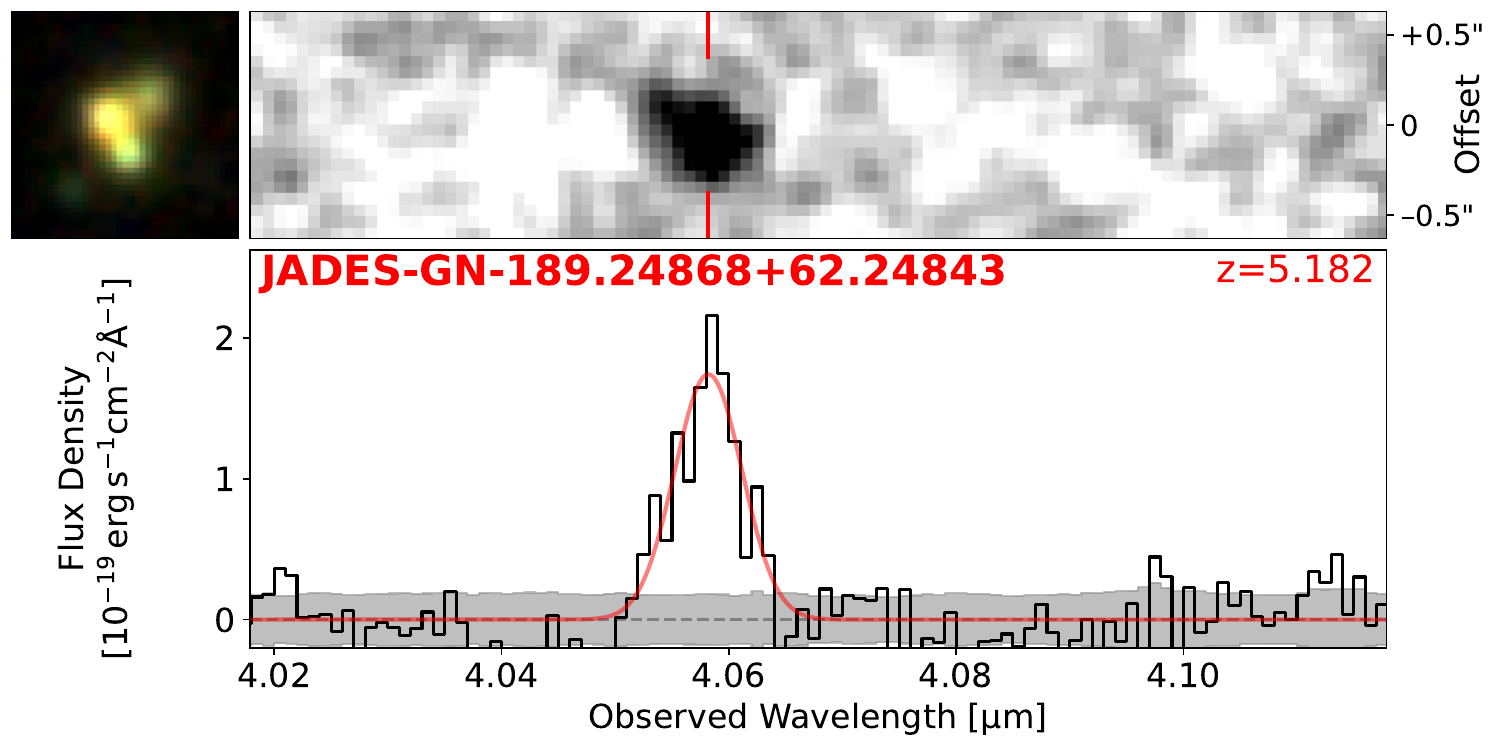}
\caption{Continued.} 
 \end{figure*} 

 \addtocounter{figure}{-1} 
 \begin{figure*}[!ht] 
 \centering
\includegraphics[width=0.49\linewidth]{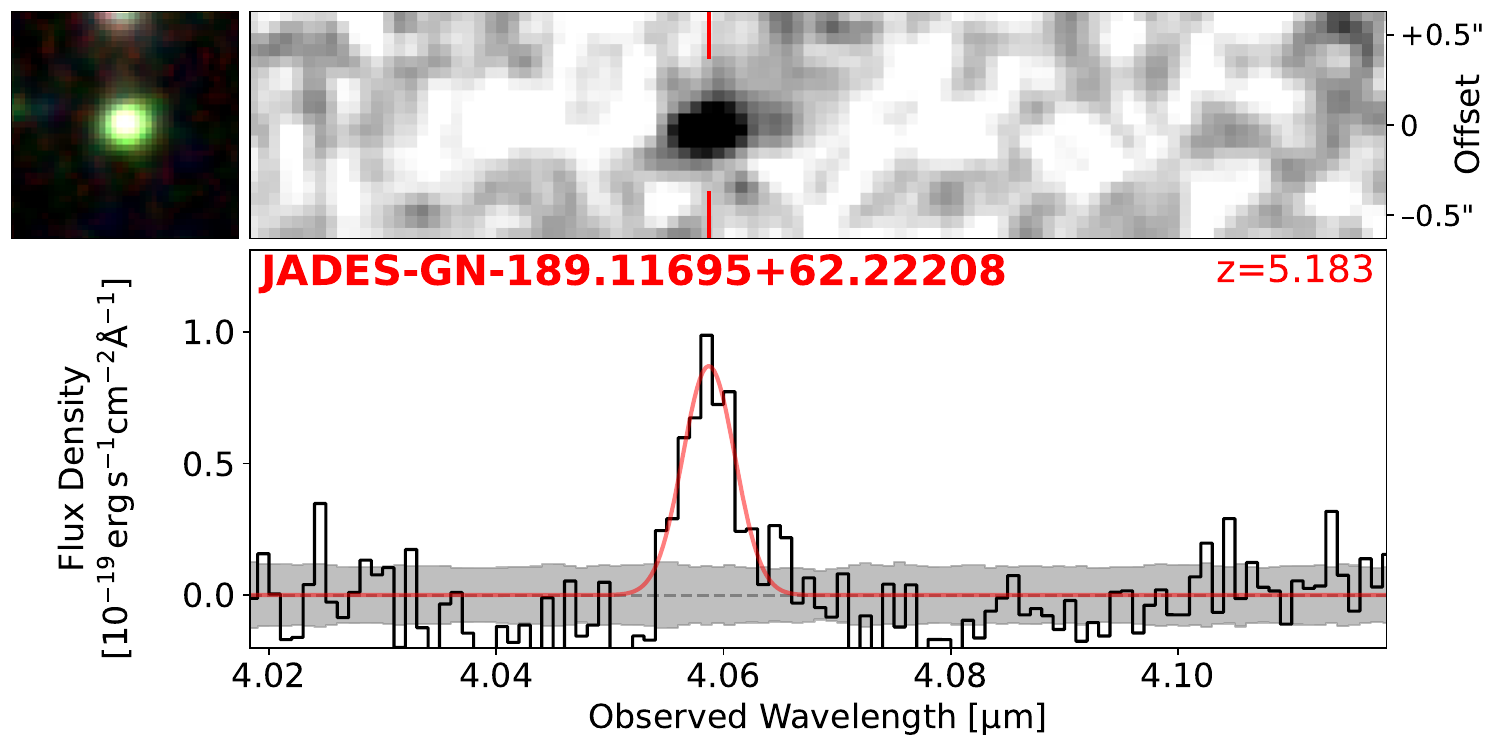}
\includegraphics[width=0.49\linewidth]{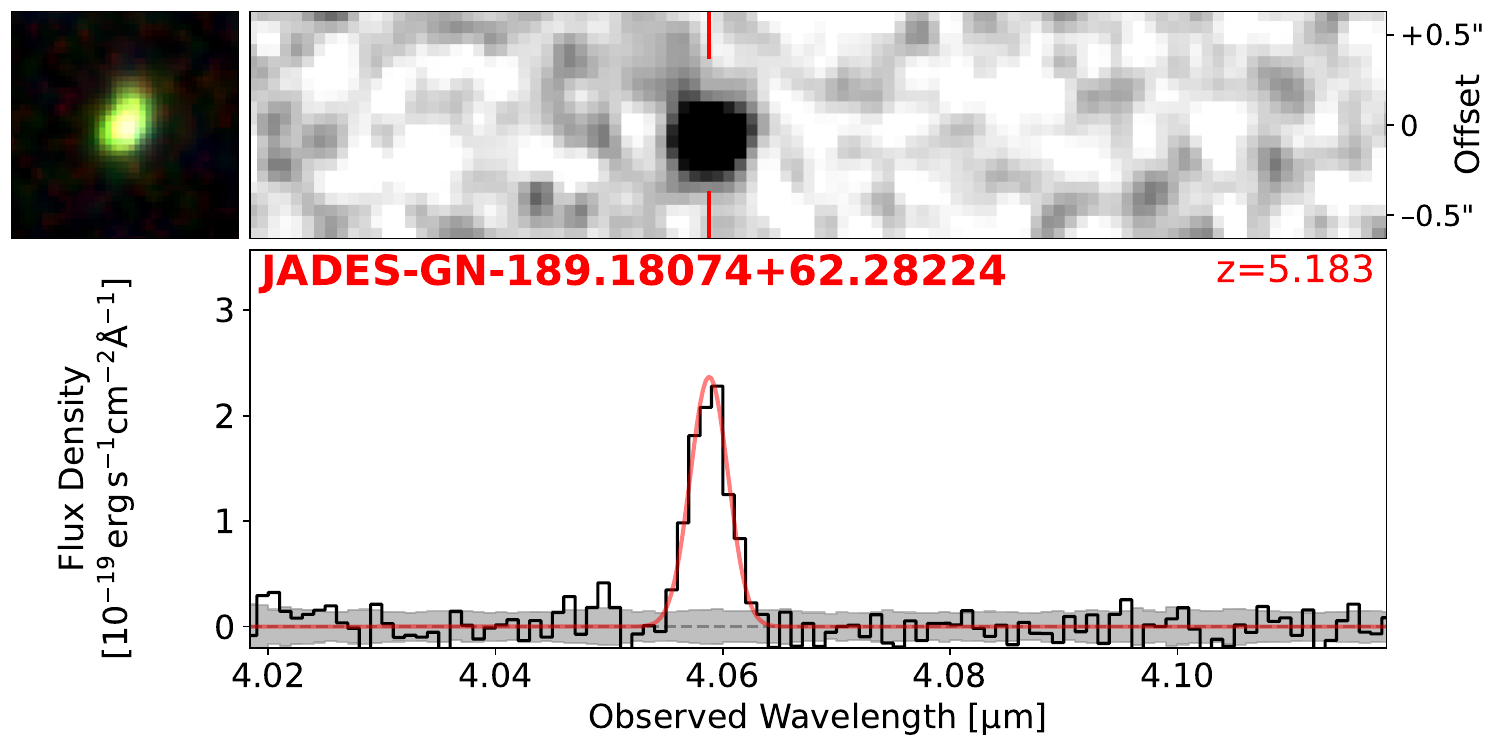}
\includegraphics[width=0.49\linewidth]{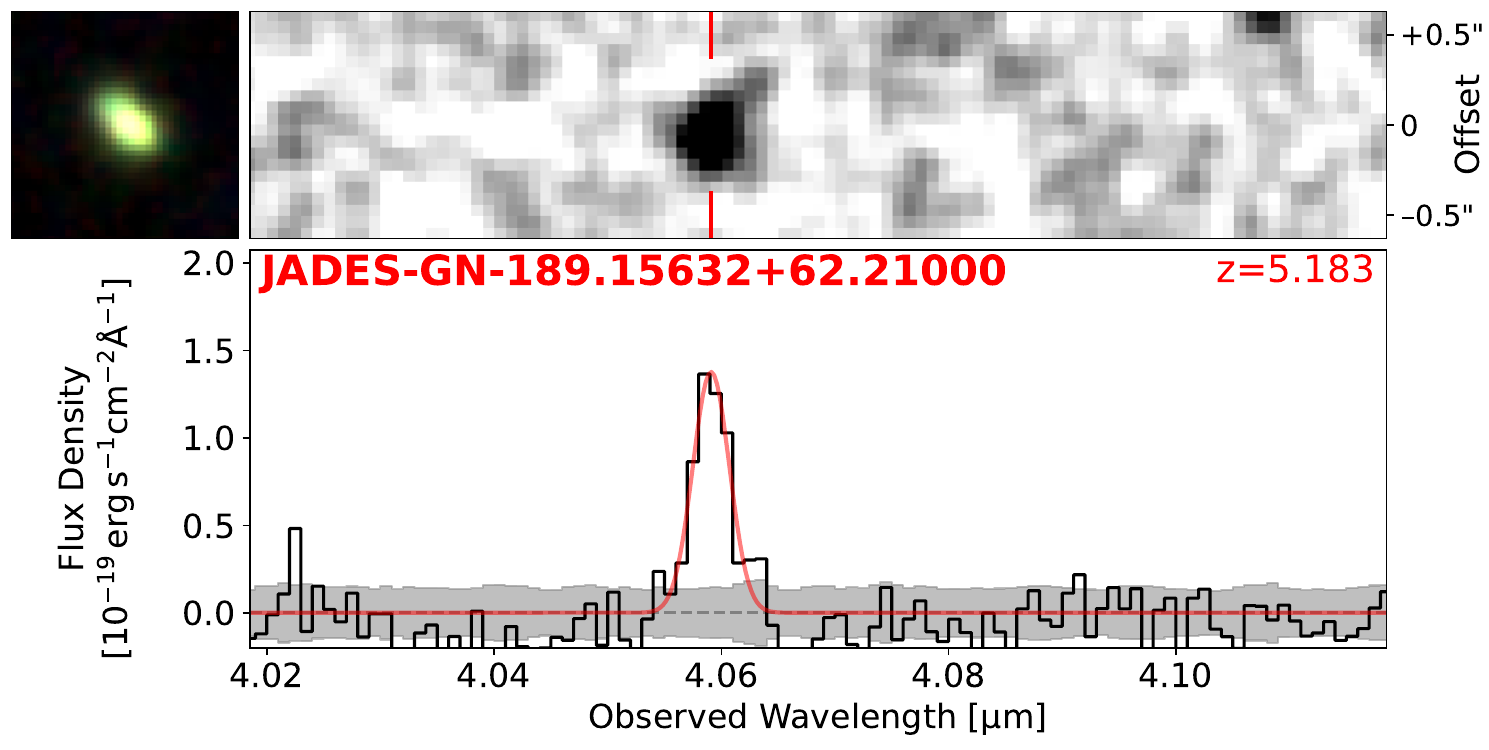}
\includegraphics[width=0.49\linewidth]{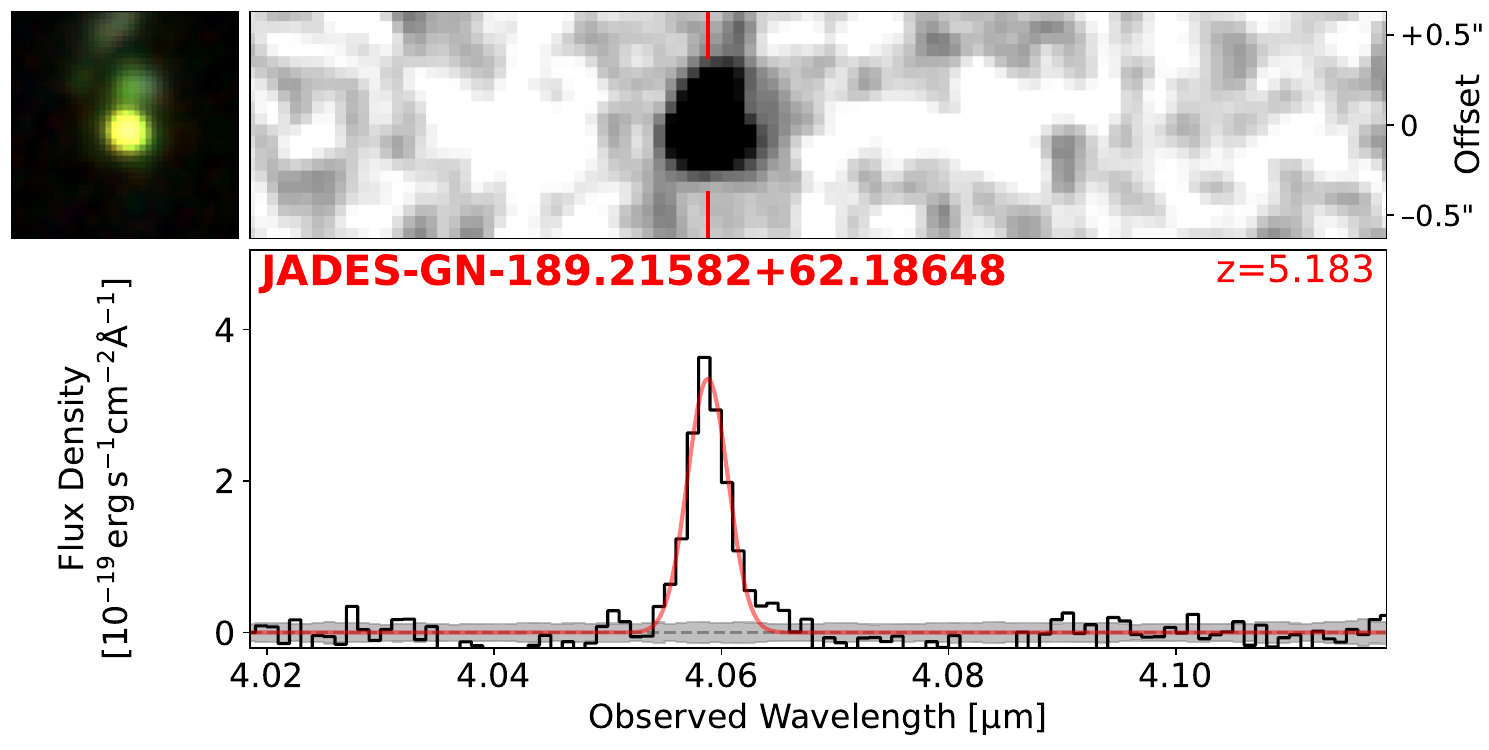}
\includegraphics[width=0.49\linewidth]{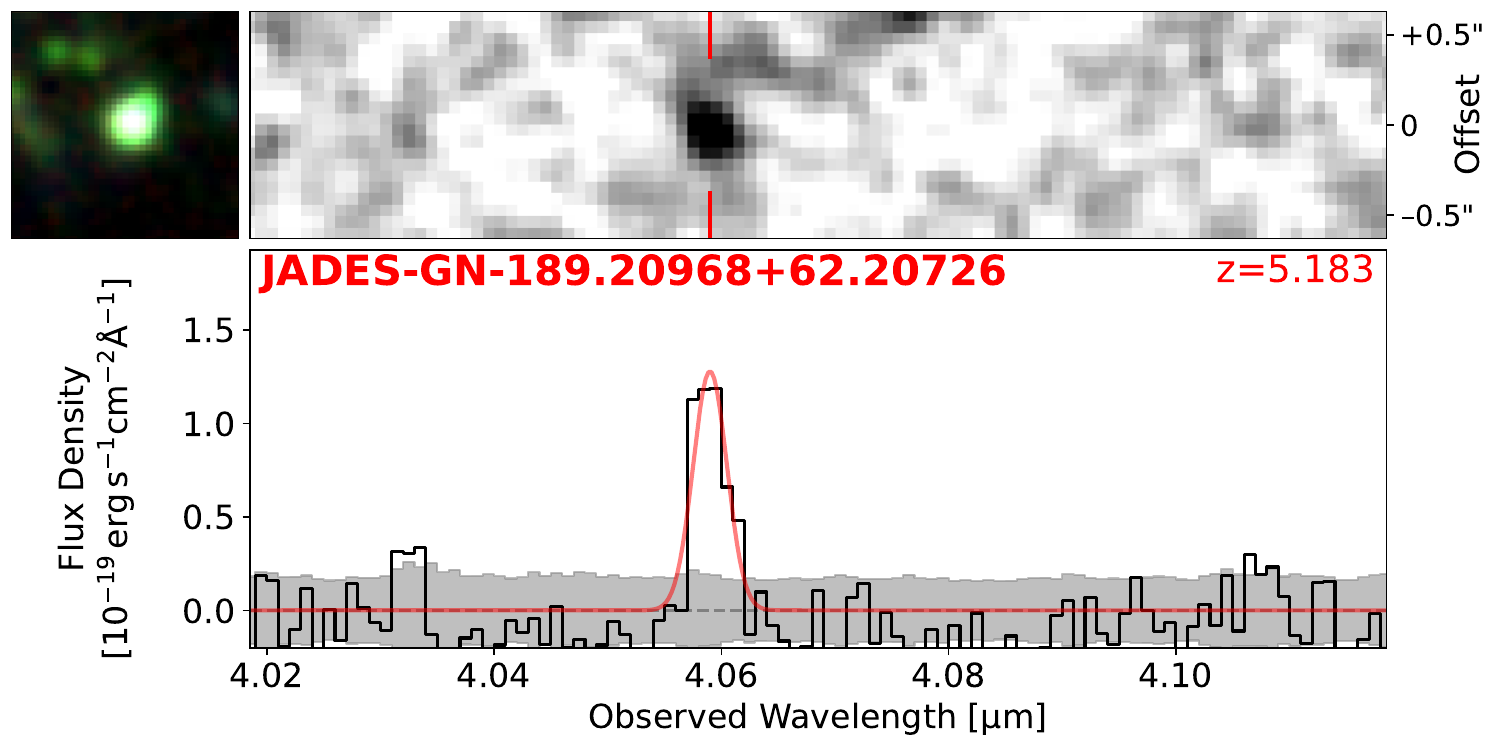}
\includegraphics[width=0.49\linewidth]{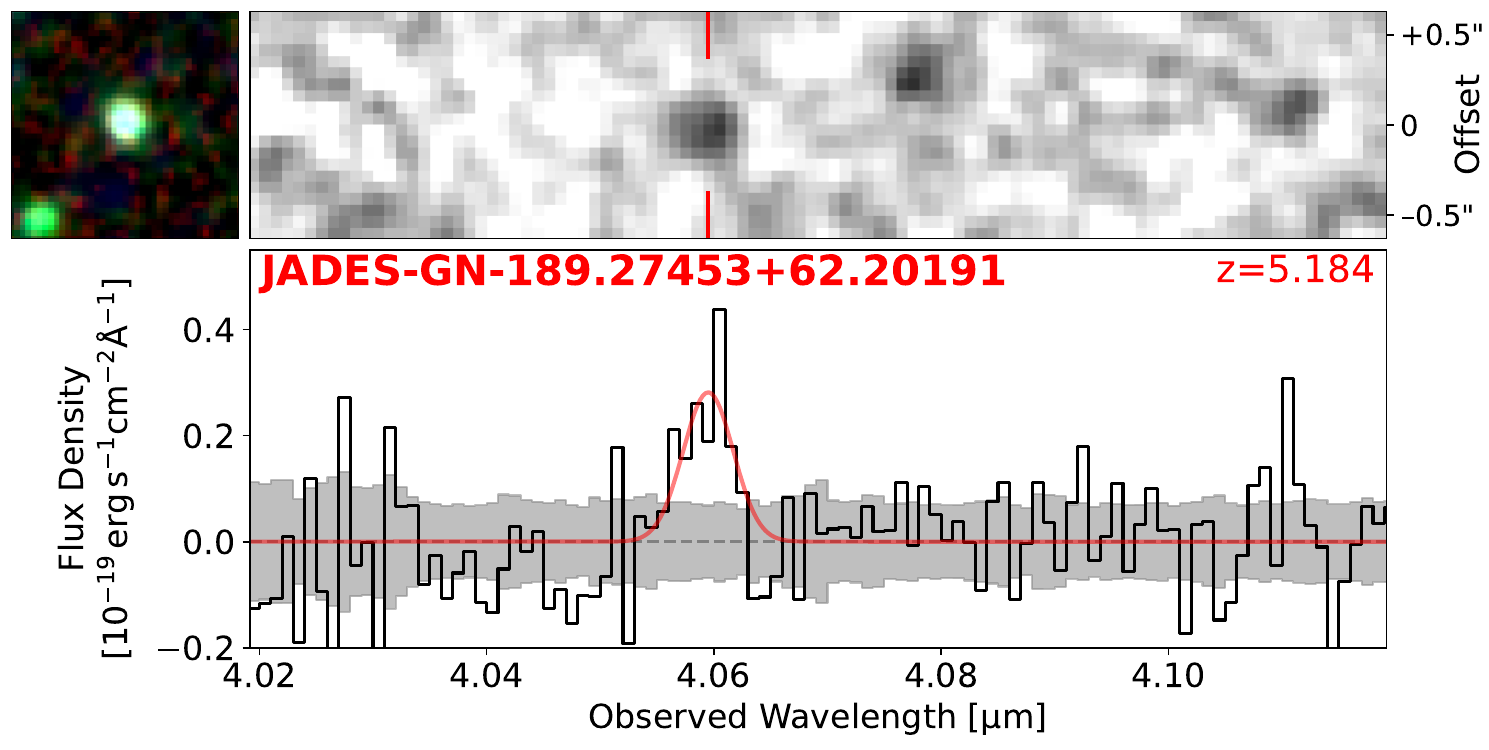}
\includegraphics[width=0.49\linewidth]{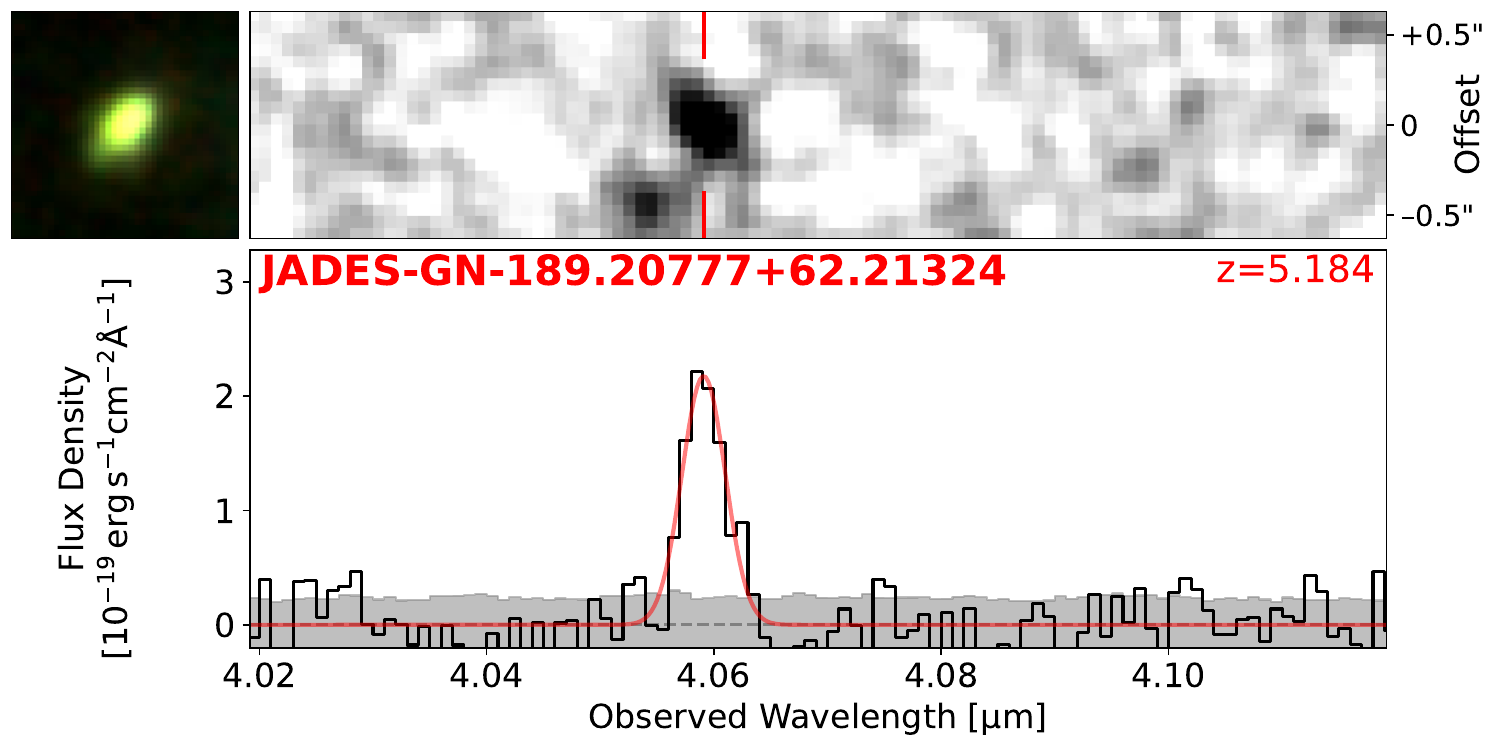}
\includegraphics[width=0.49\linewidth]{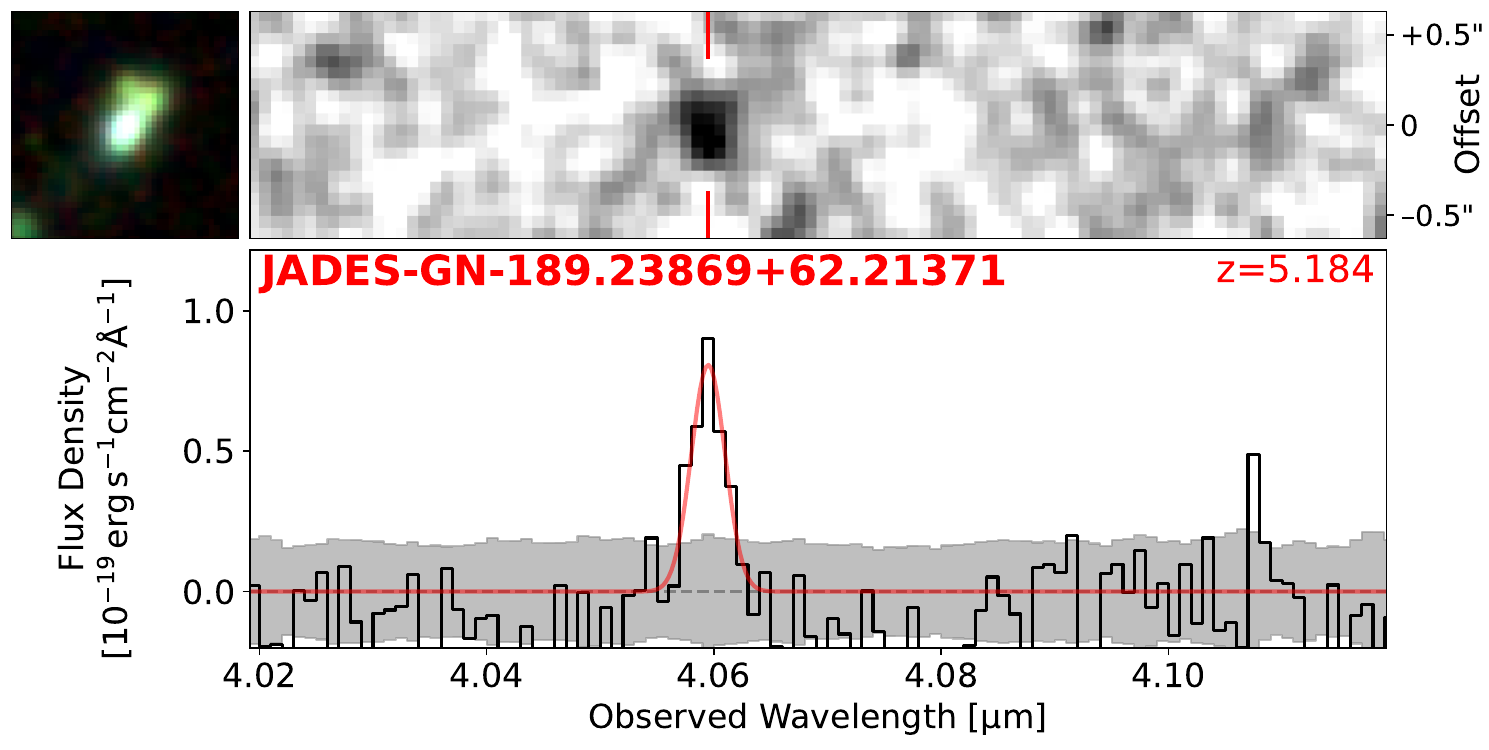}
\includegraphics[width=0.49\linewidth]{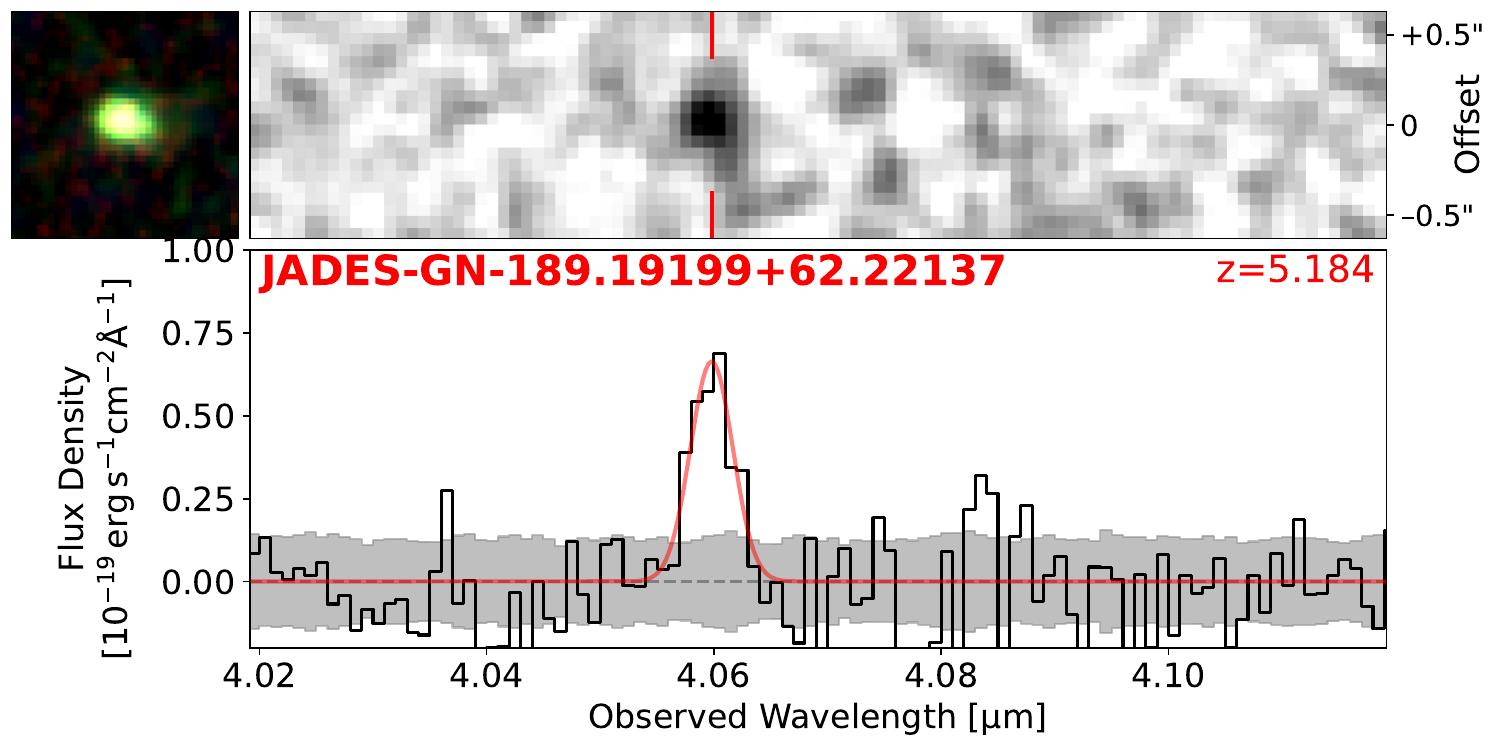}
\includegraphics[width=0.49\linewidth]{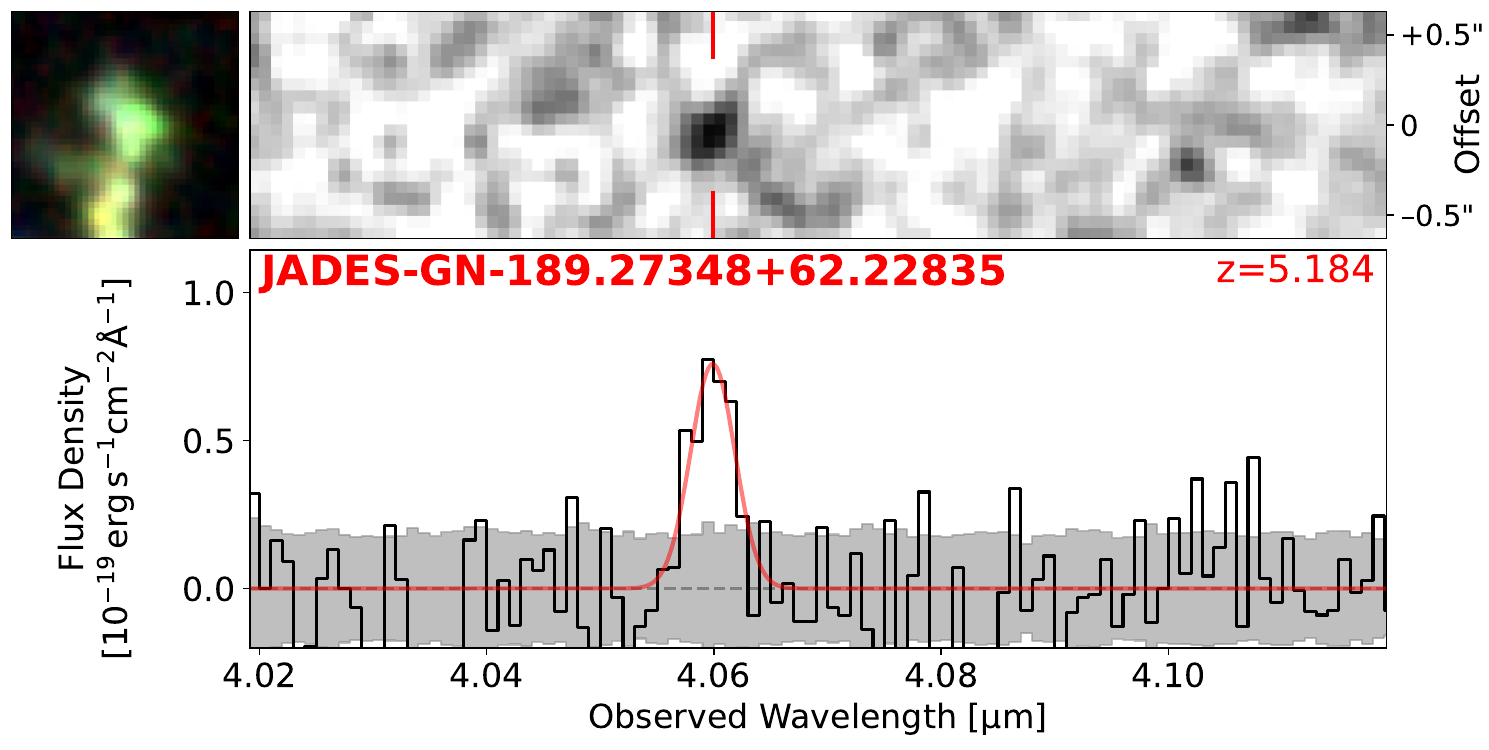}
\caption{Continued.} 
 \end{figure*} 

 \addtocounter{figure}{-1} 
 \begin{figure*}[!ht] 
 \centering
\includegraphics[width=0.49\linewidth]{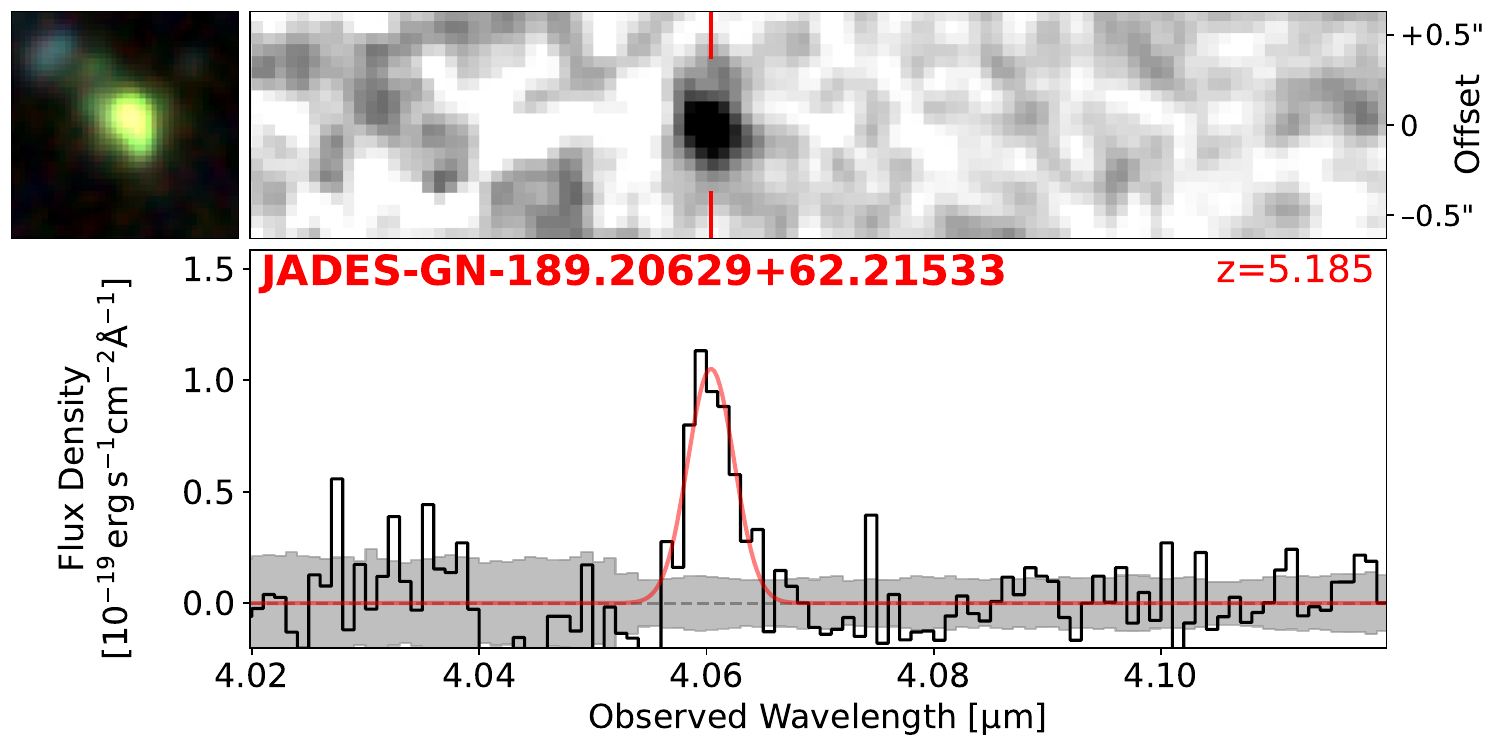}
\includegraphics[width=0.49\linewidth]{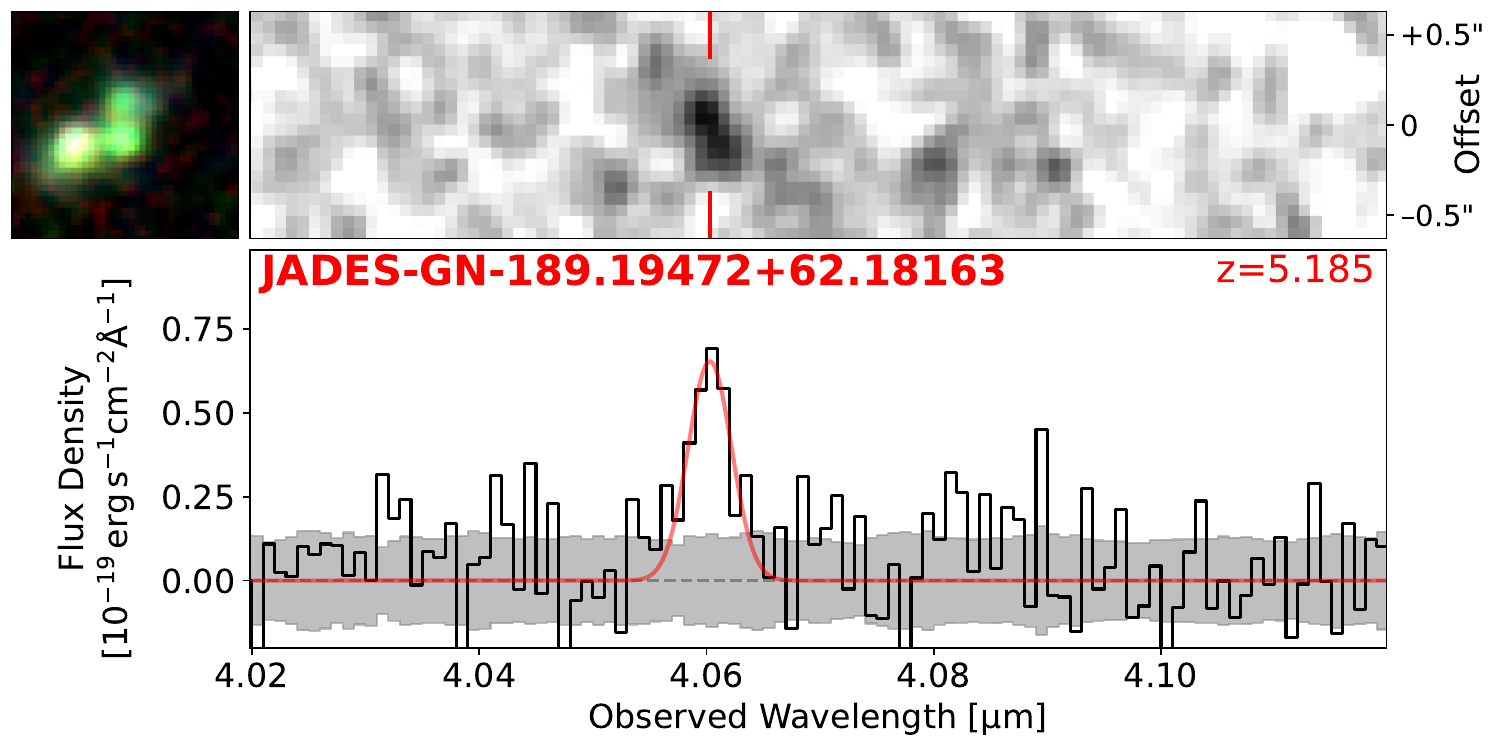}
\includegraphics[width=0.49\linewidth]{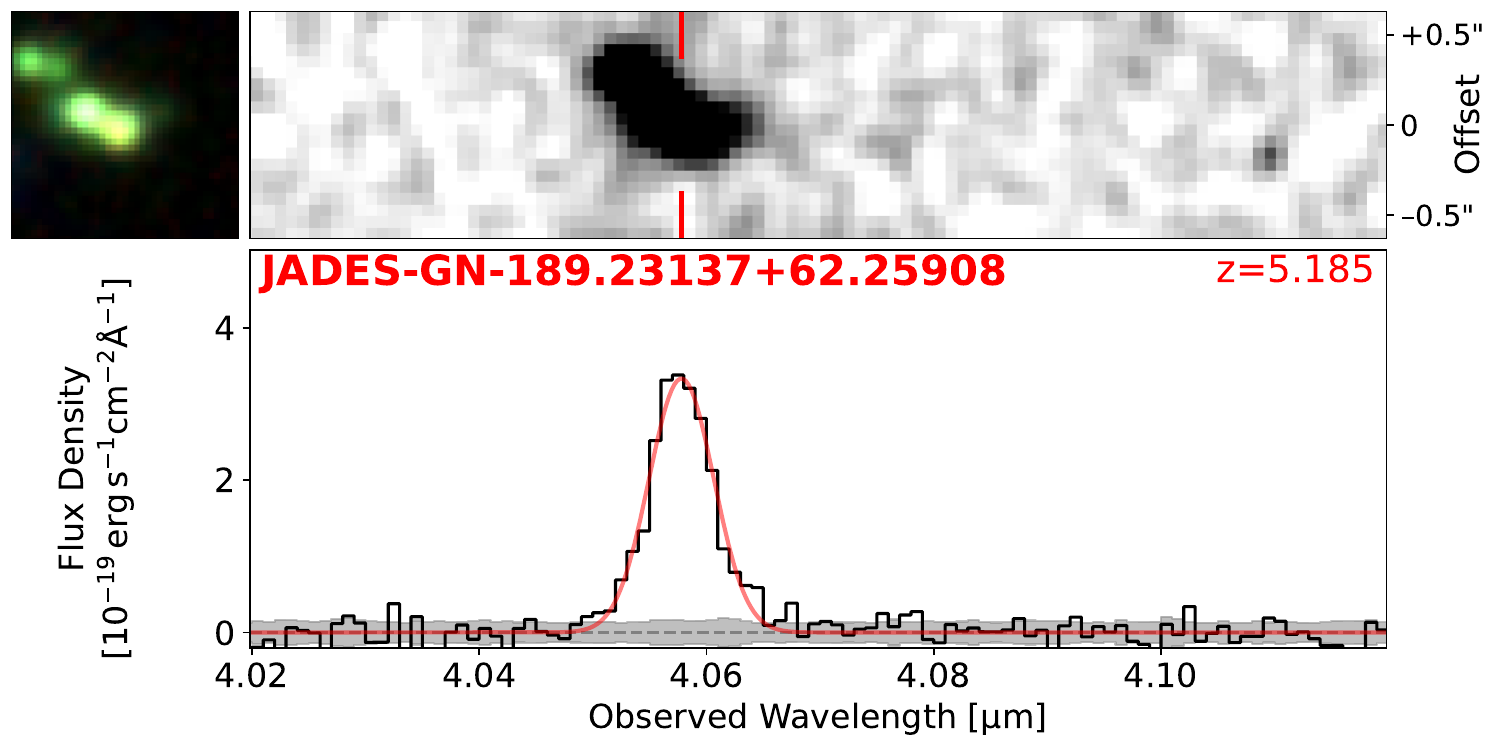}
\includegraphics[width=0.49\linewidth]{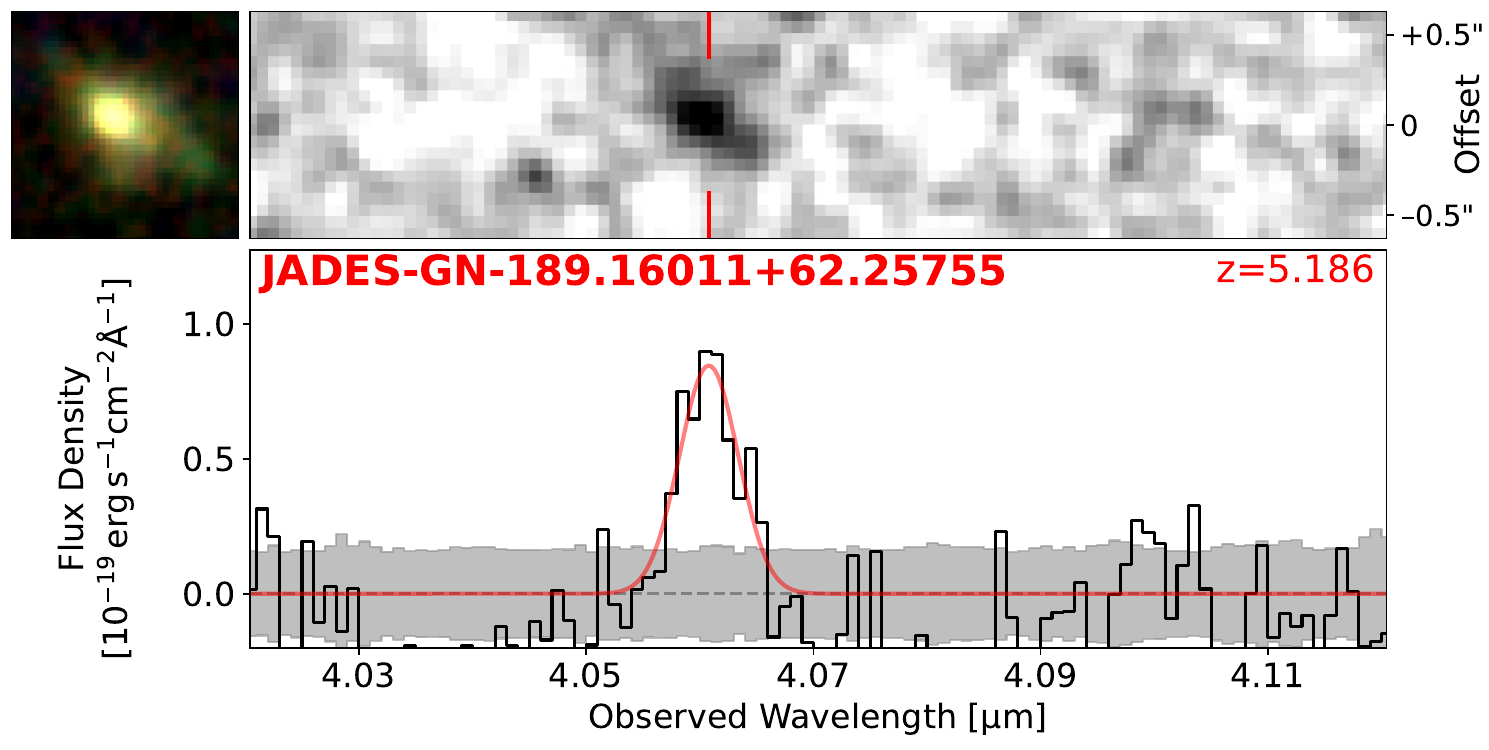}
\includegraphics[width=0.49\linewidth]{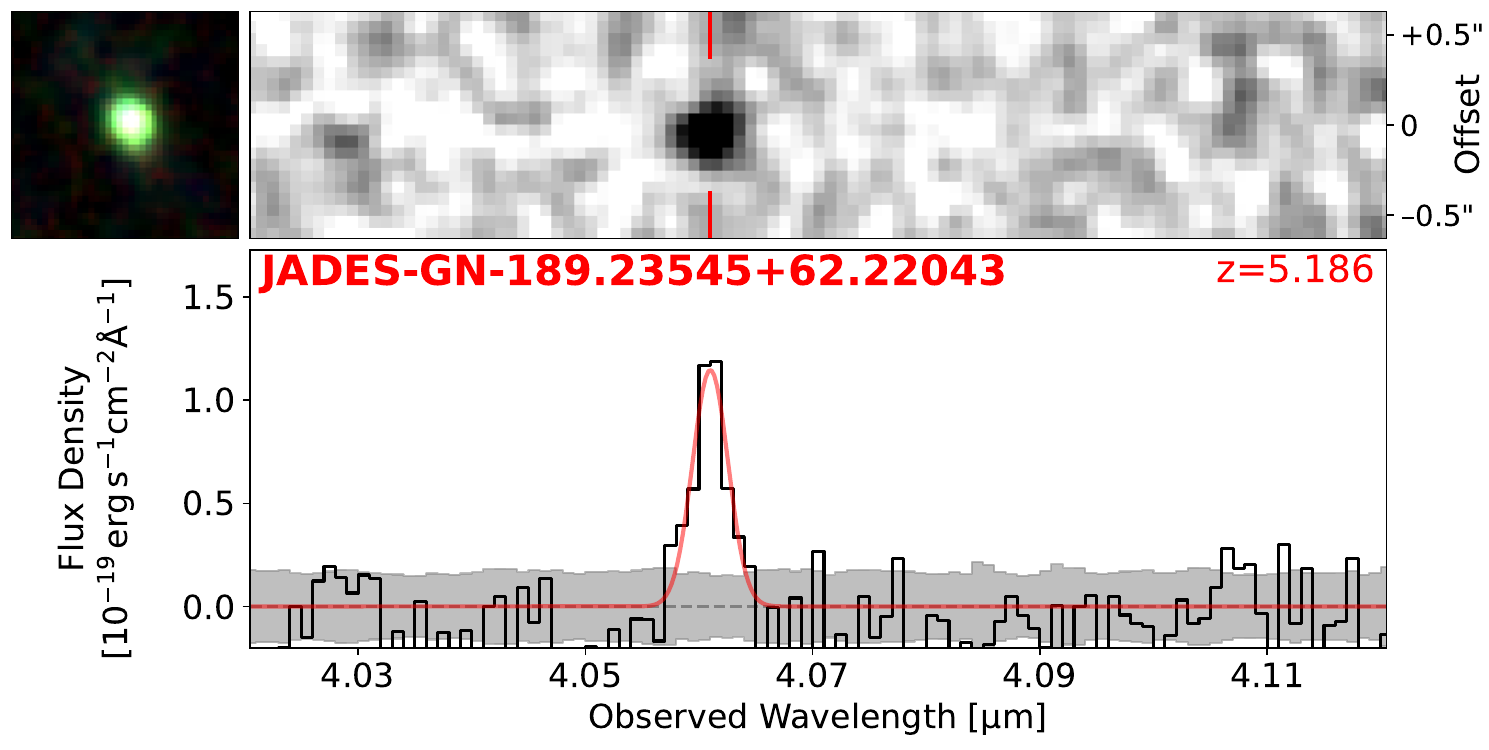}
\includegraphics[width=0.49\linewidth]{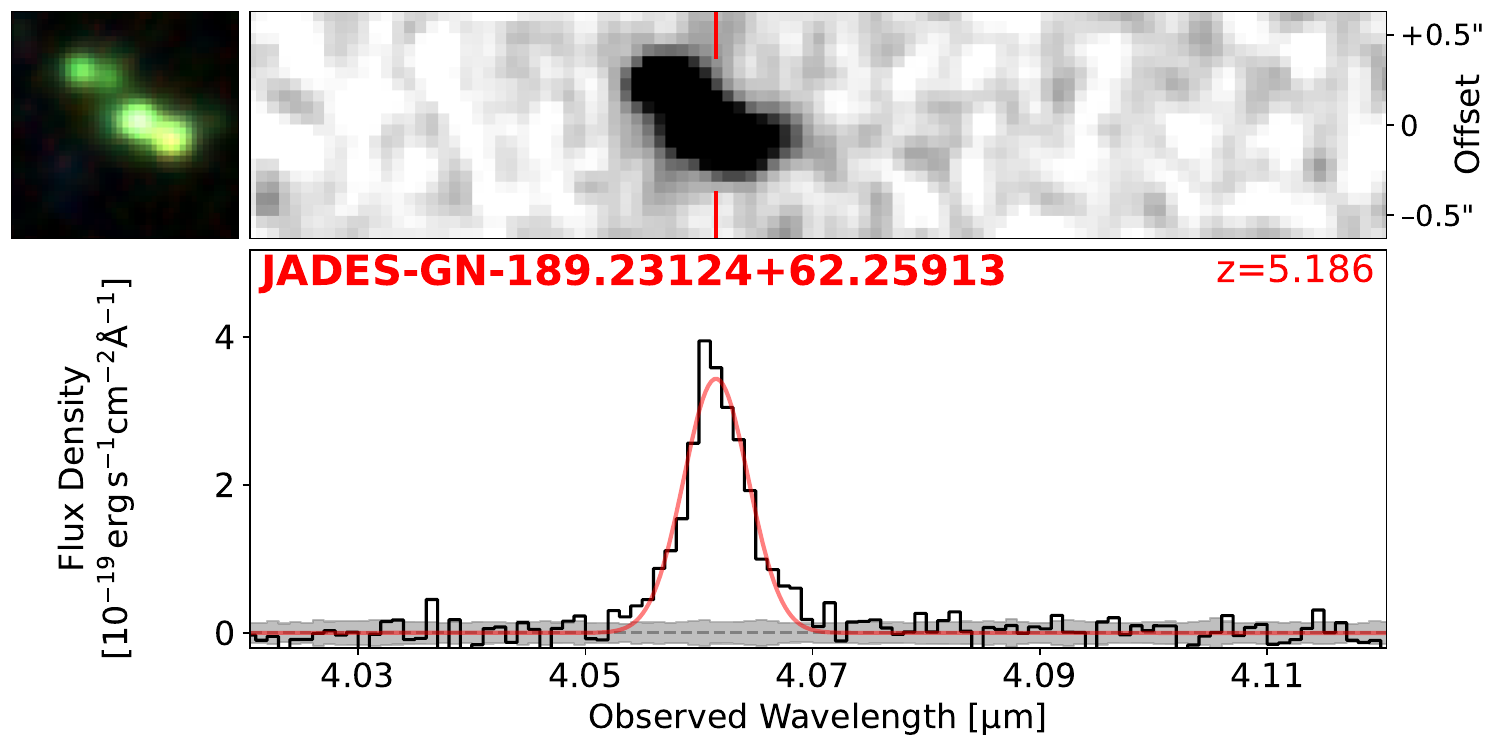}
\includegraphics[width=0.49\linewidth]{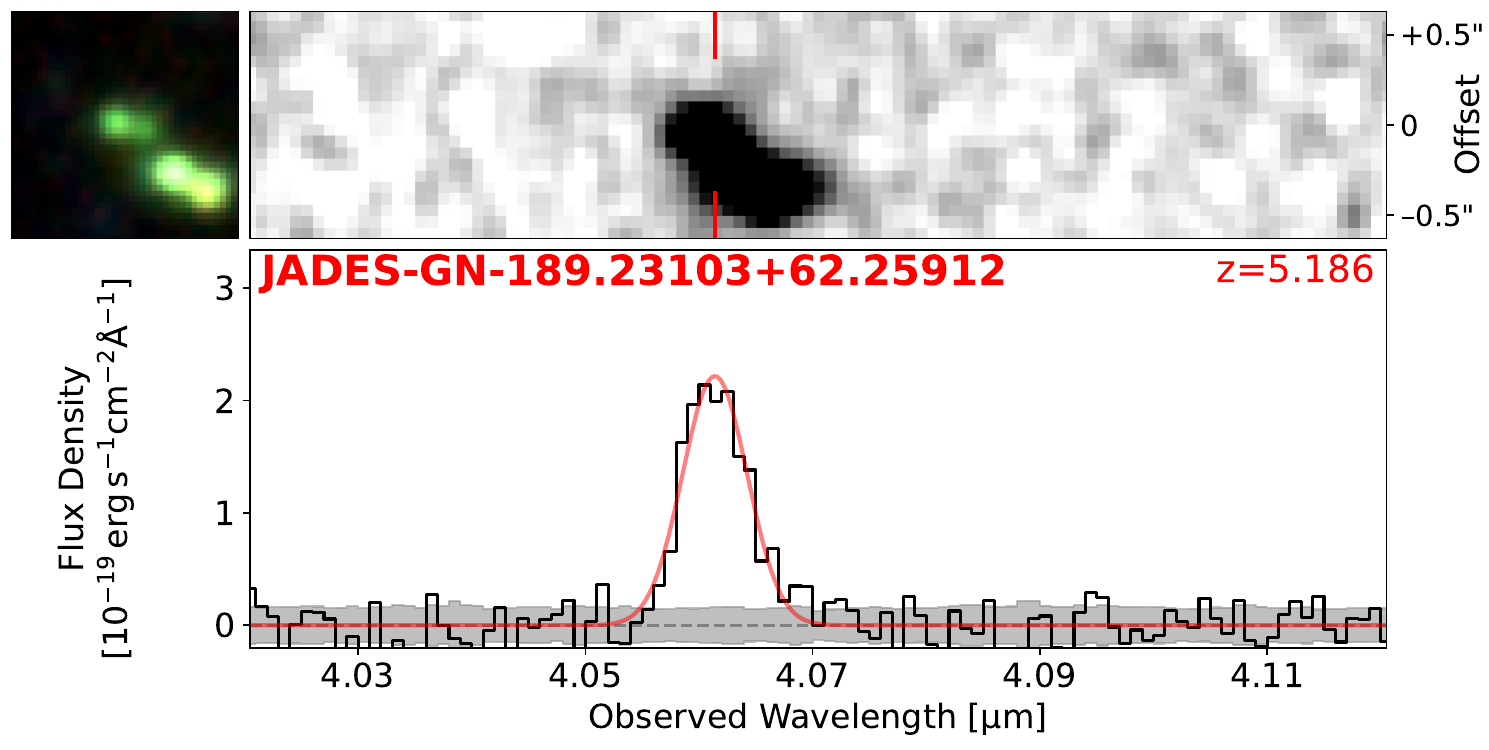}
\includegraphics[width=0.49\linewidth]{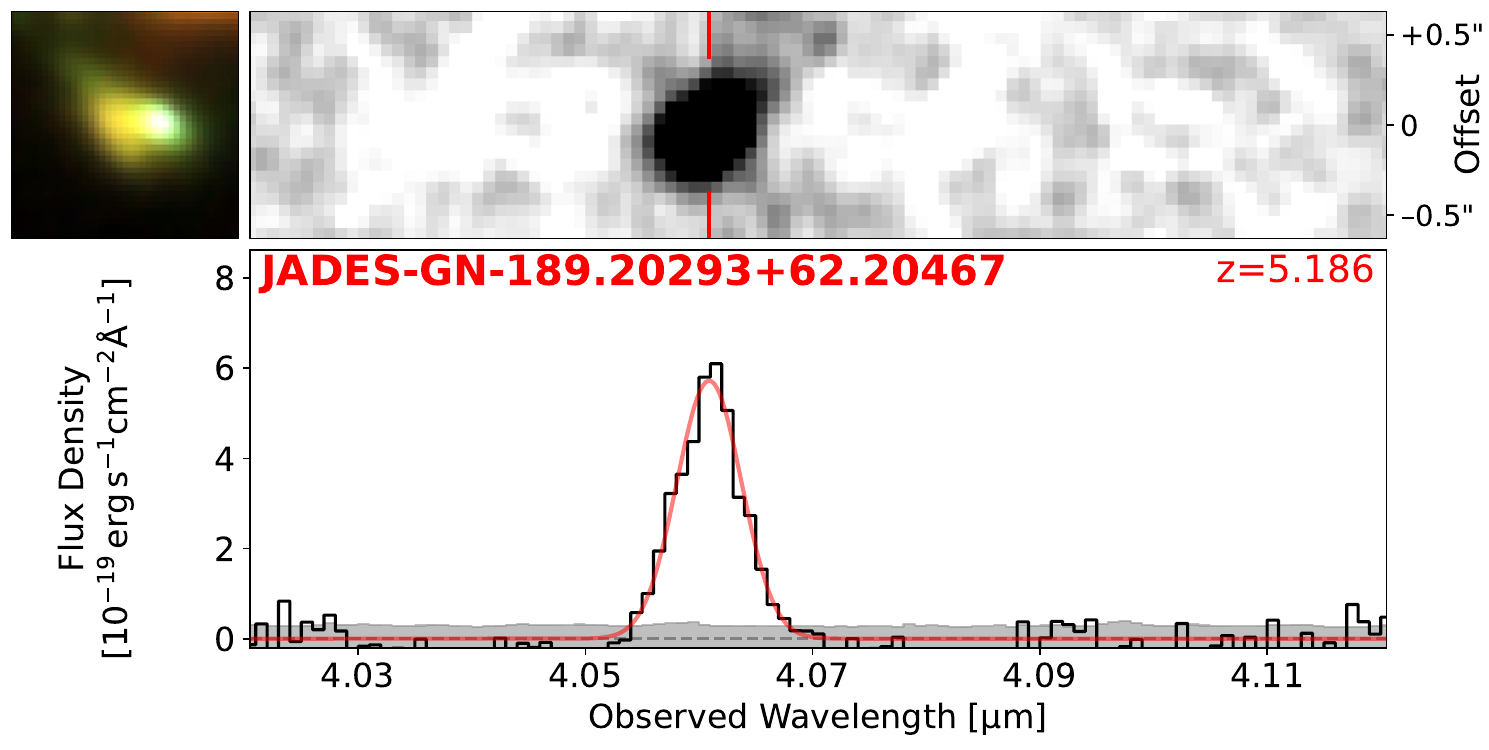}
\includegraphics[width=0.49\linewidth]{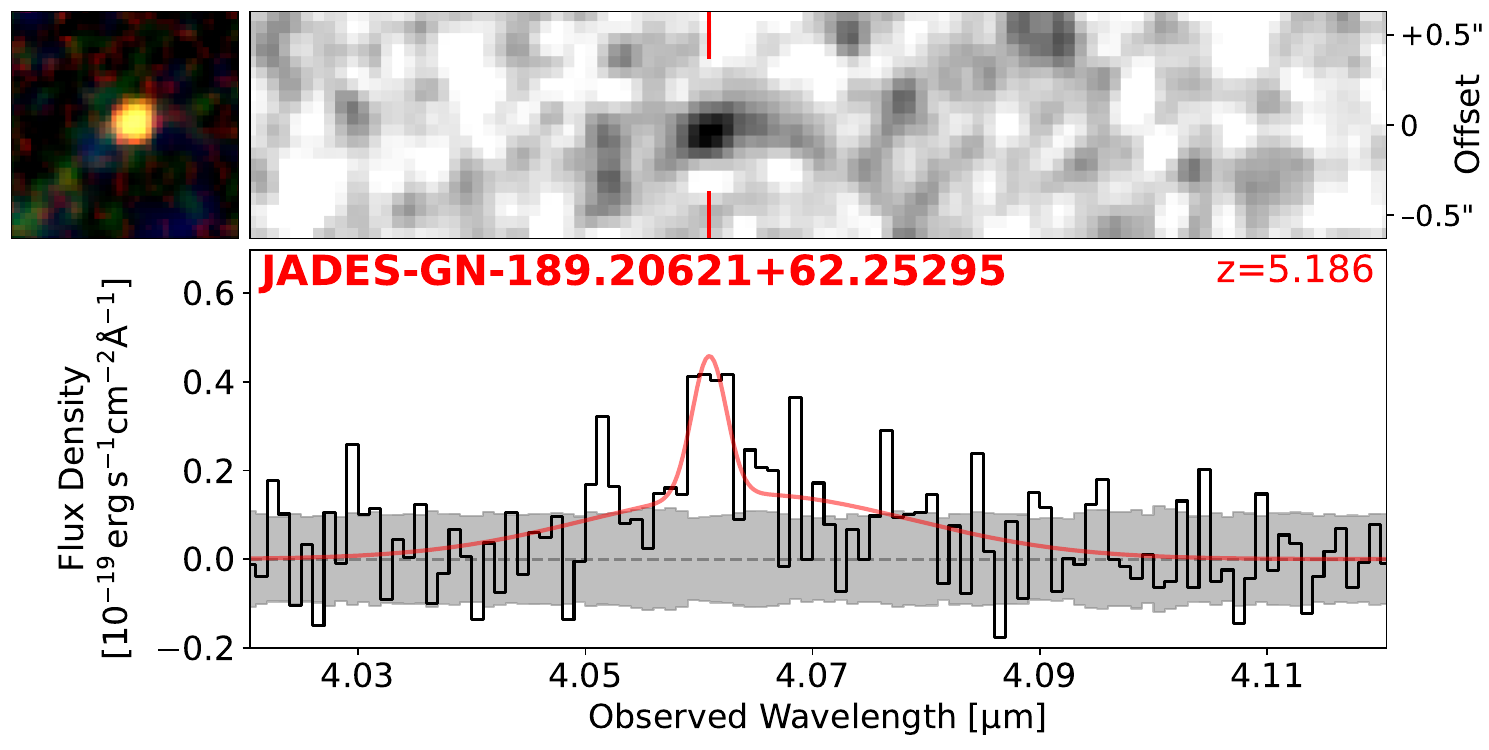}
\includegraphics[width=0.49\linewidth]{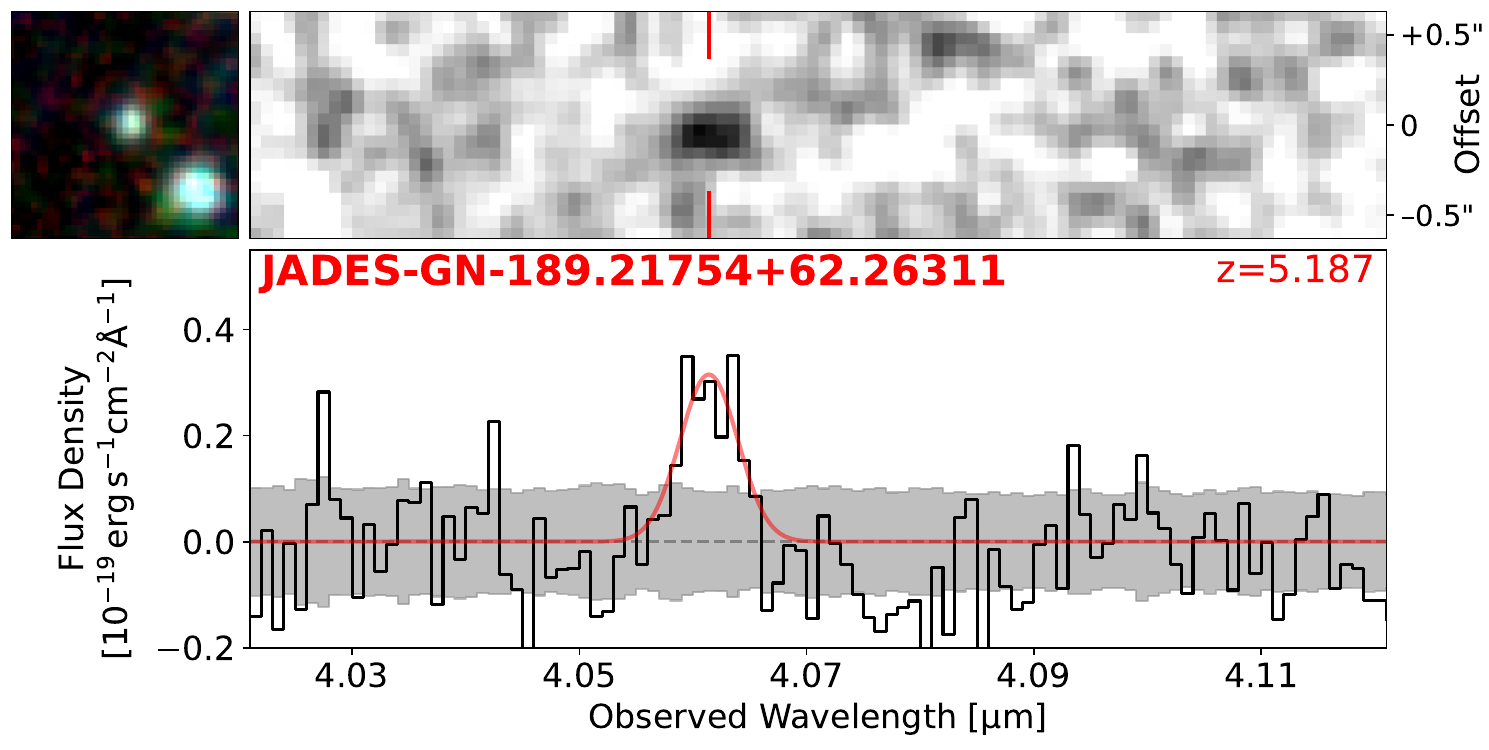}
\caption{Continued.} 
 \end{figure*} 

 \addtocounter{figure}{-1} 
 \begin{figure*}[!ht] 
 \centering
\includegraphics[width=0.49\linewidth]{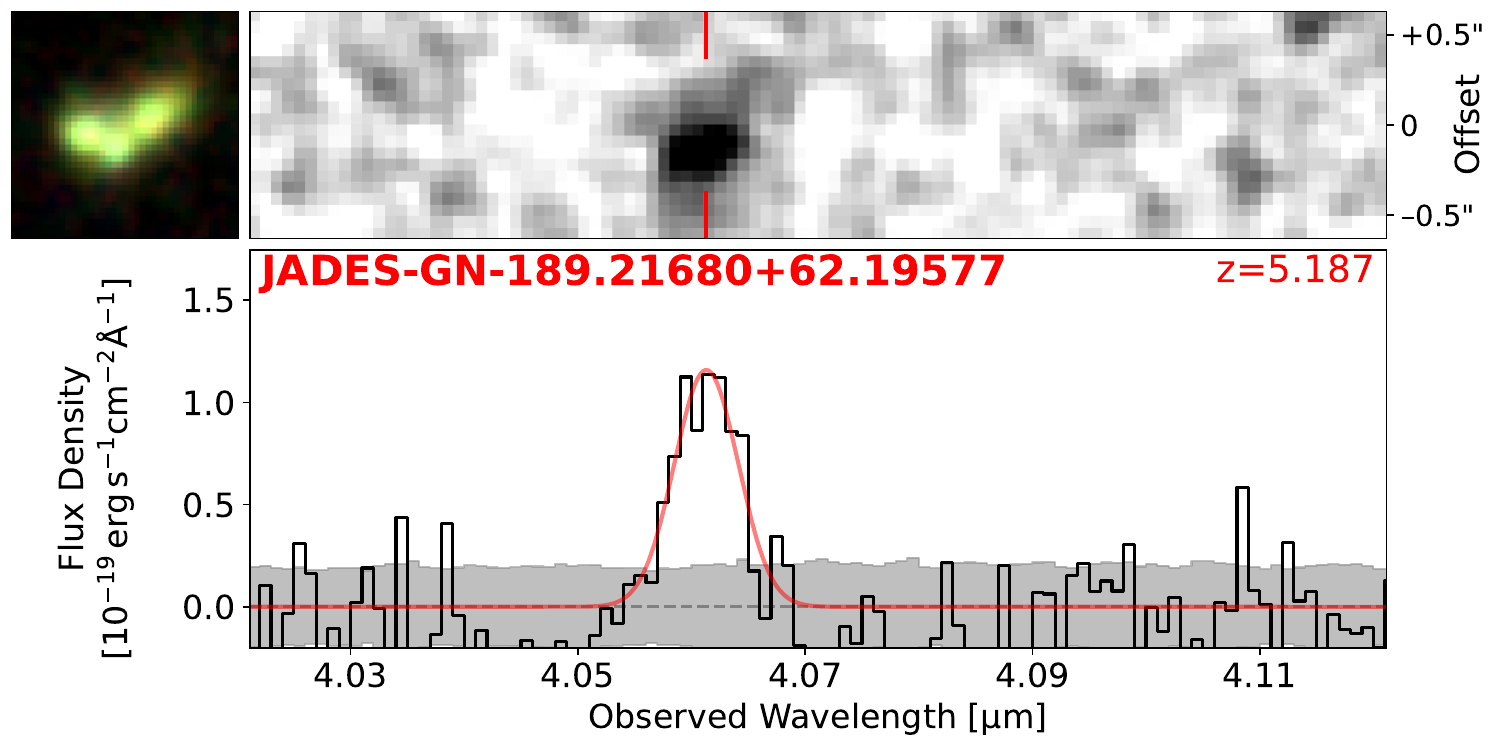}
\includegraphics[width=0.49\linewidth]{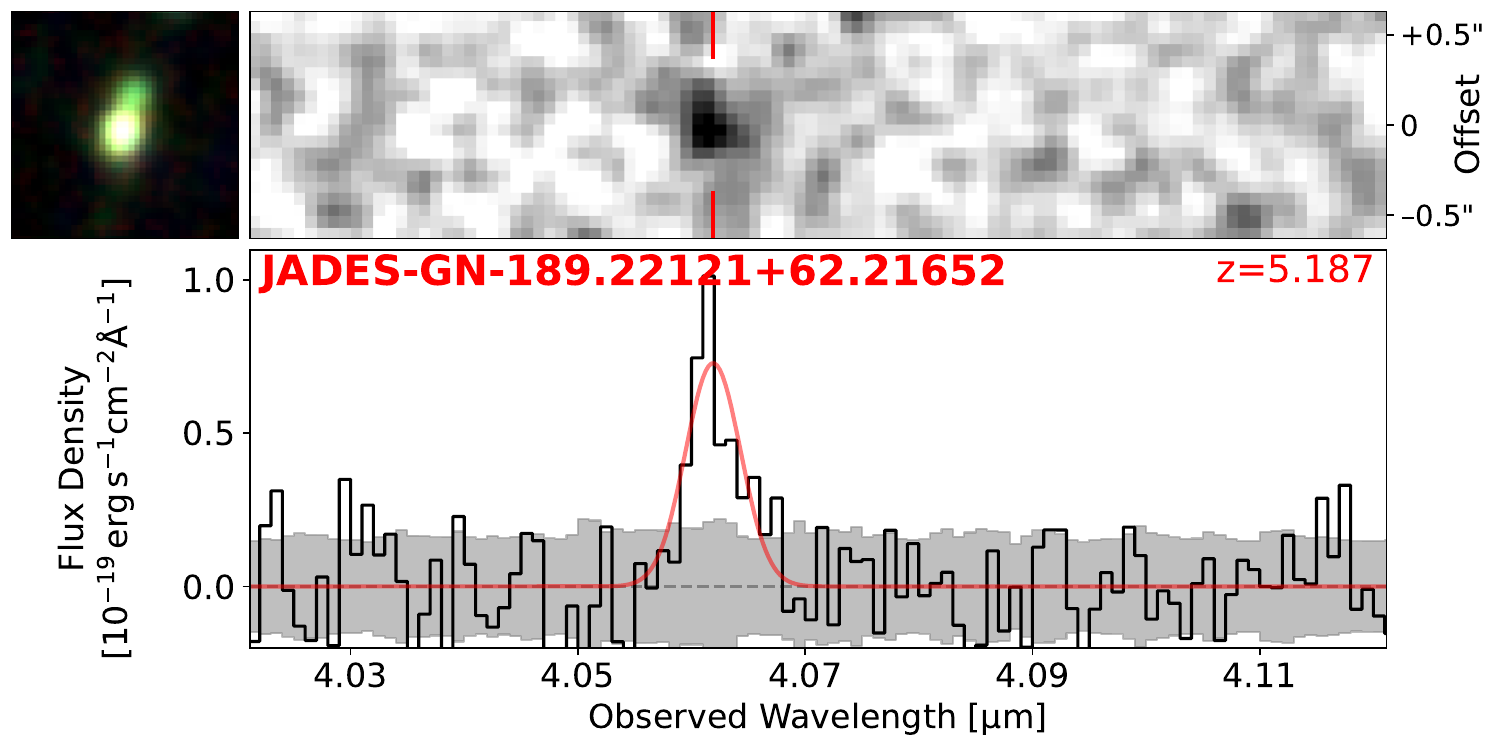}
\includegraphics[width=0.49\linewidth]{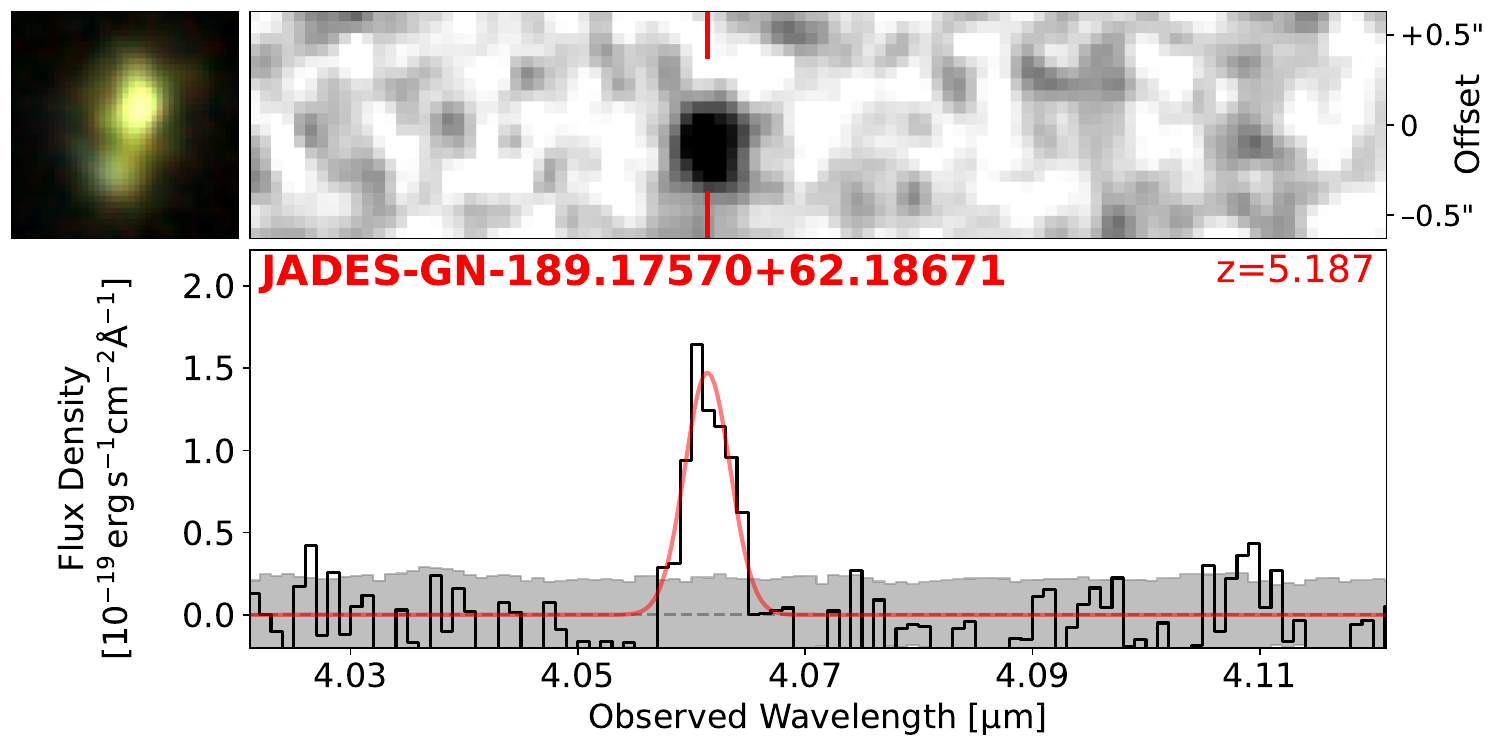}
\includegraphics[width=0.49\linewidth]{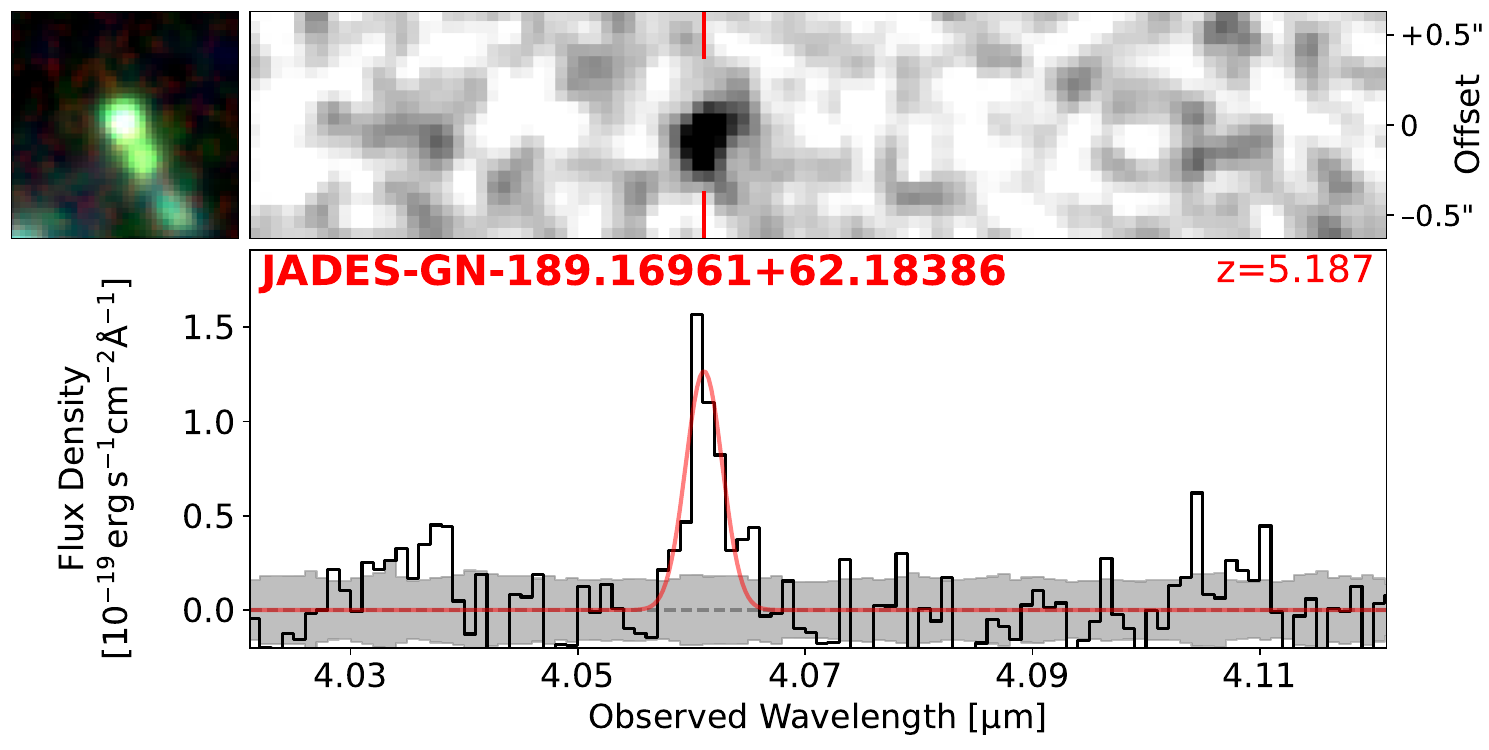}
\includegraphics[width=0.49\linewidth]{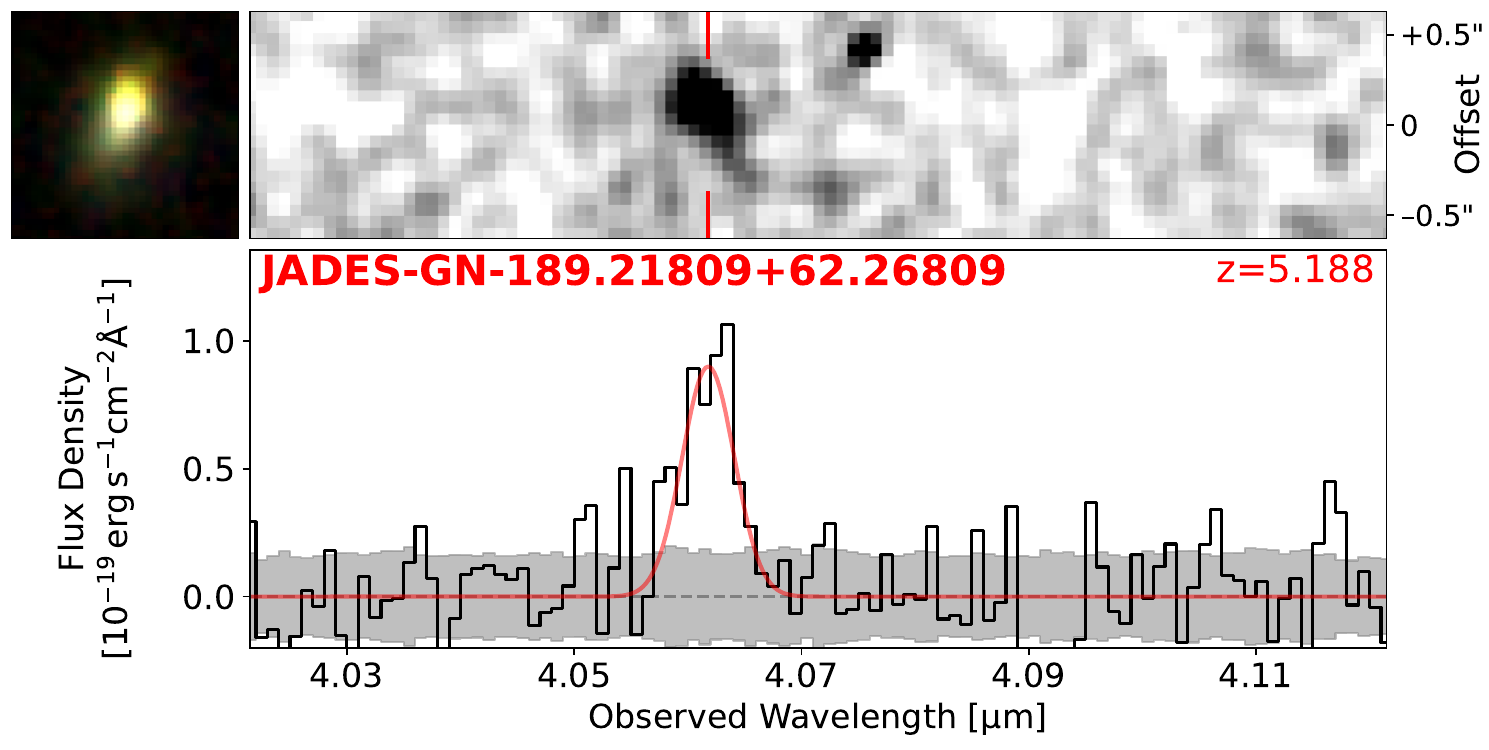}
\includegraphics[width=0.49\linewidth]{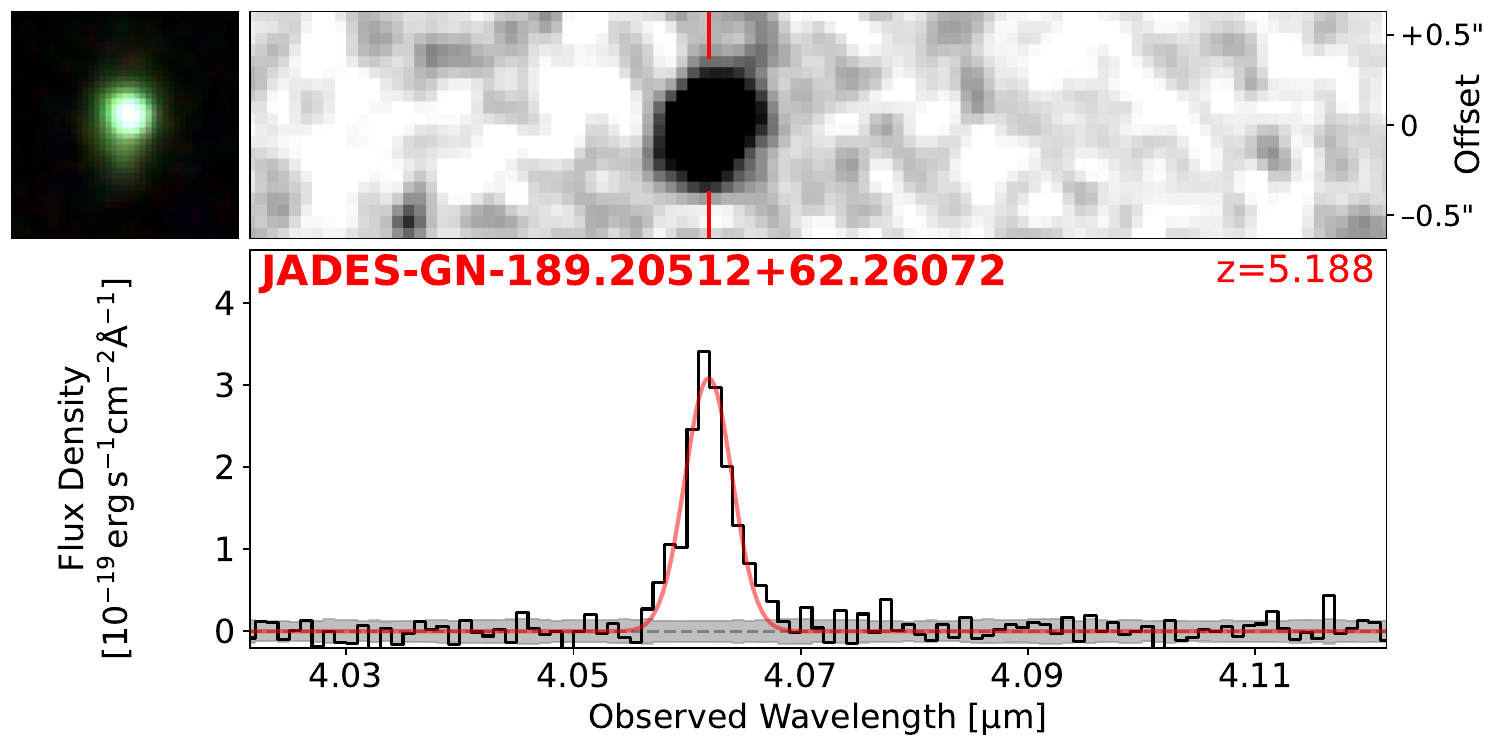}
\includegraphics[width=0.49\linewidth]{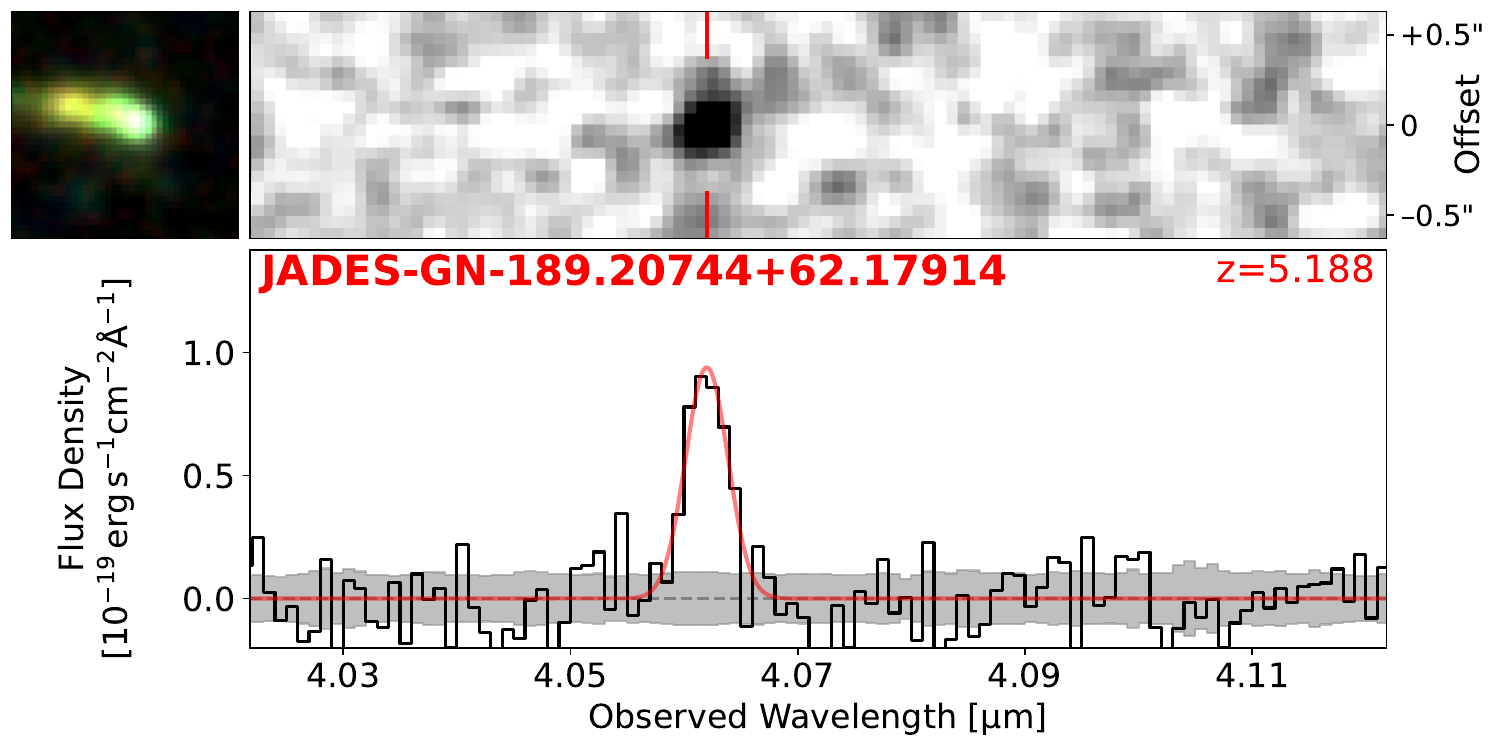}
\includegraphics[width=0.49\linewidth]{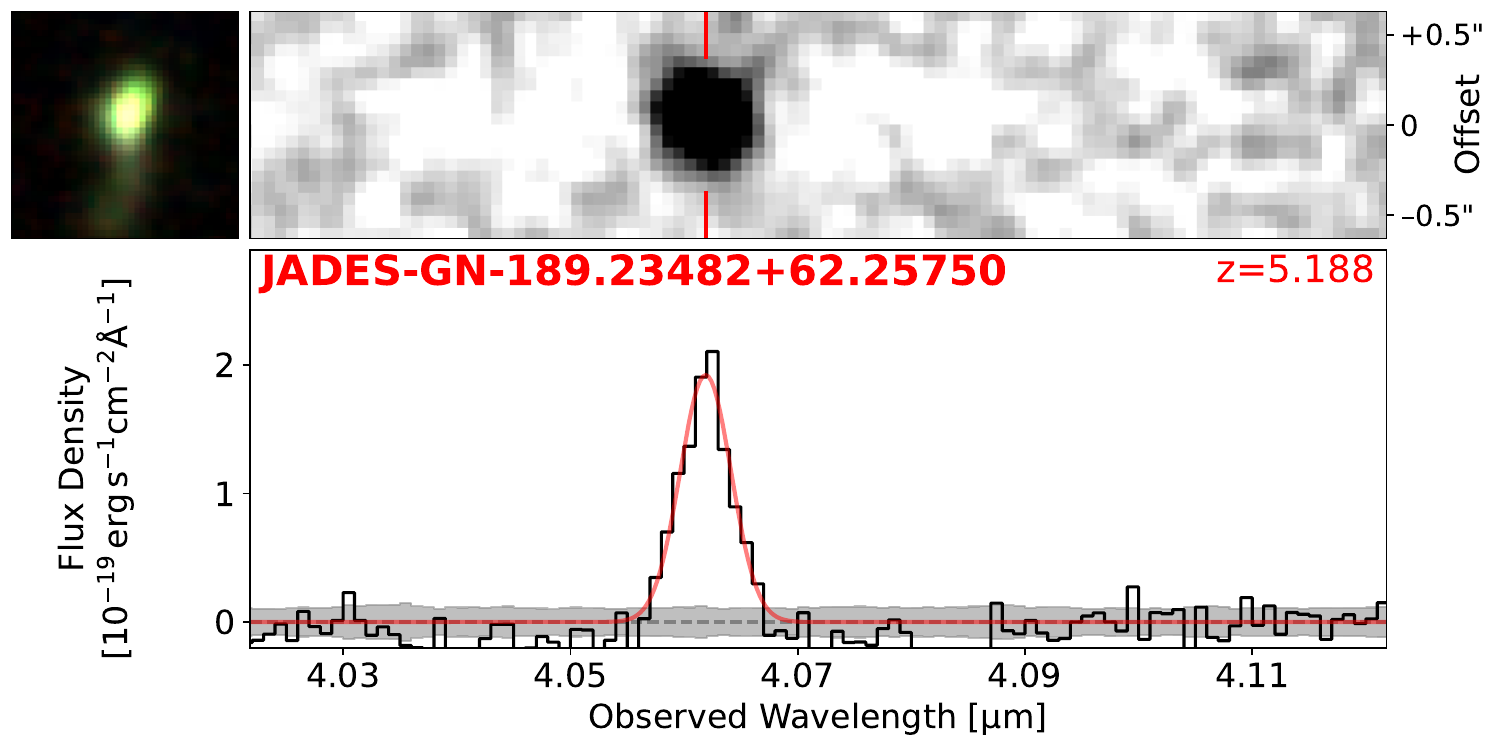}
\includegraphics[width=0.49\linewidth]{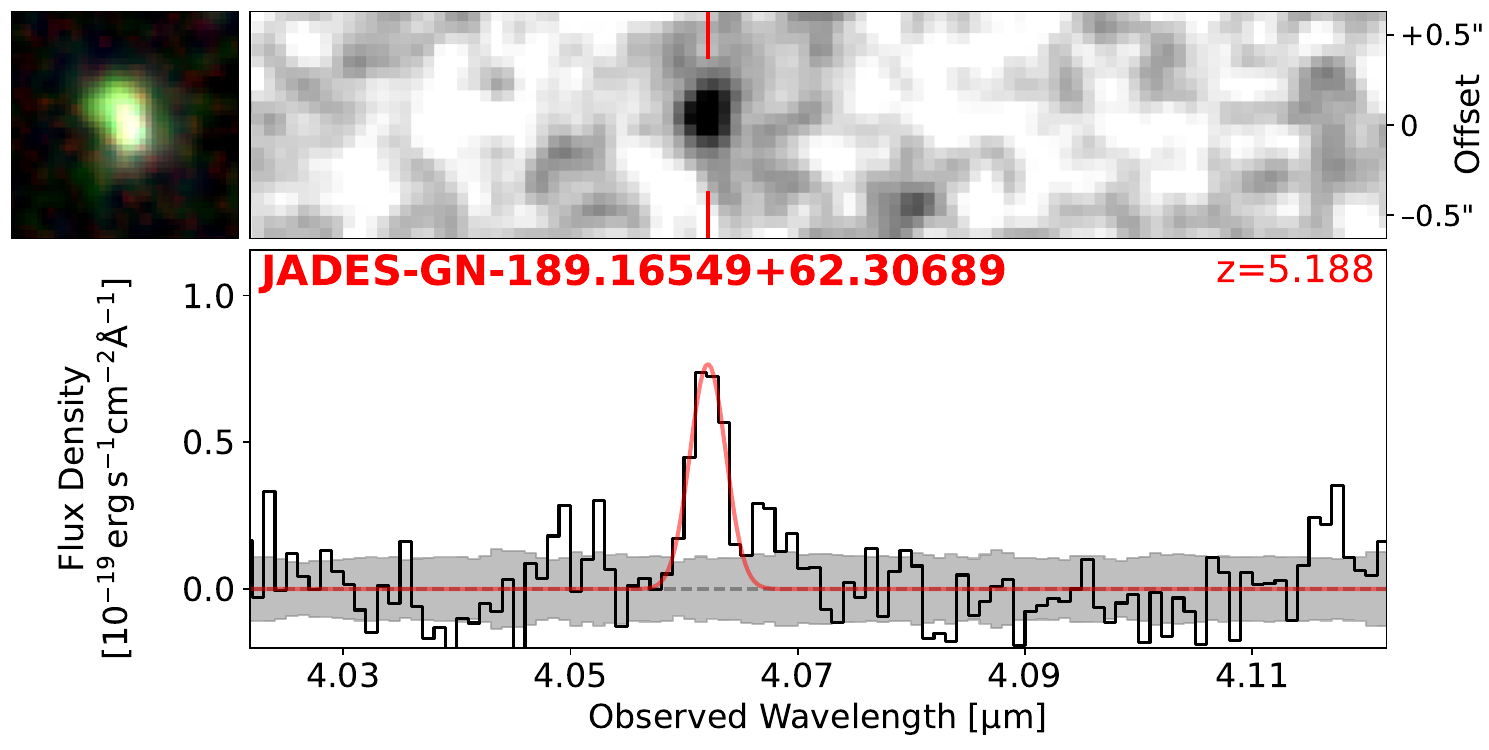}
\includegraphics[width=0.49\linewidth]{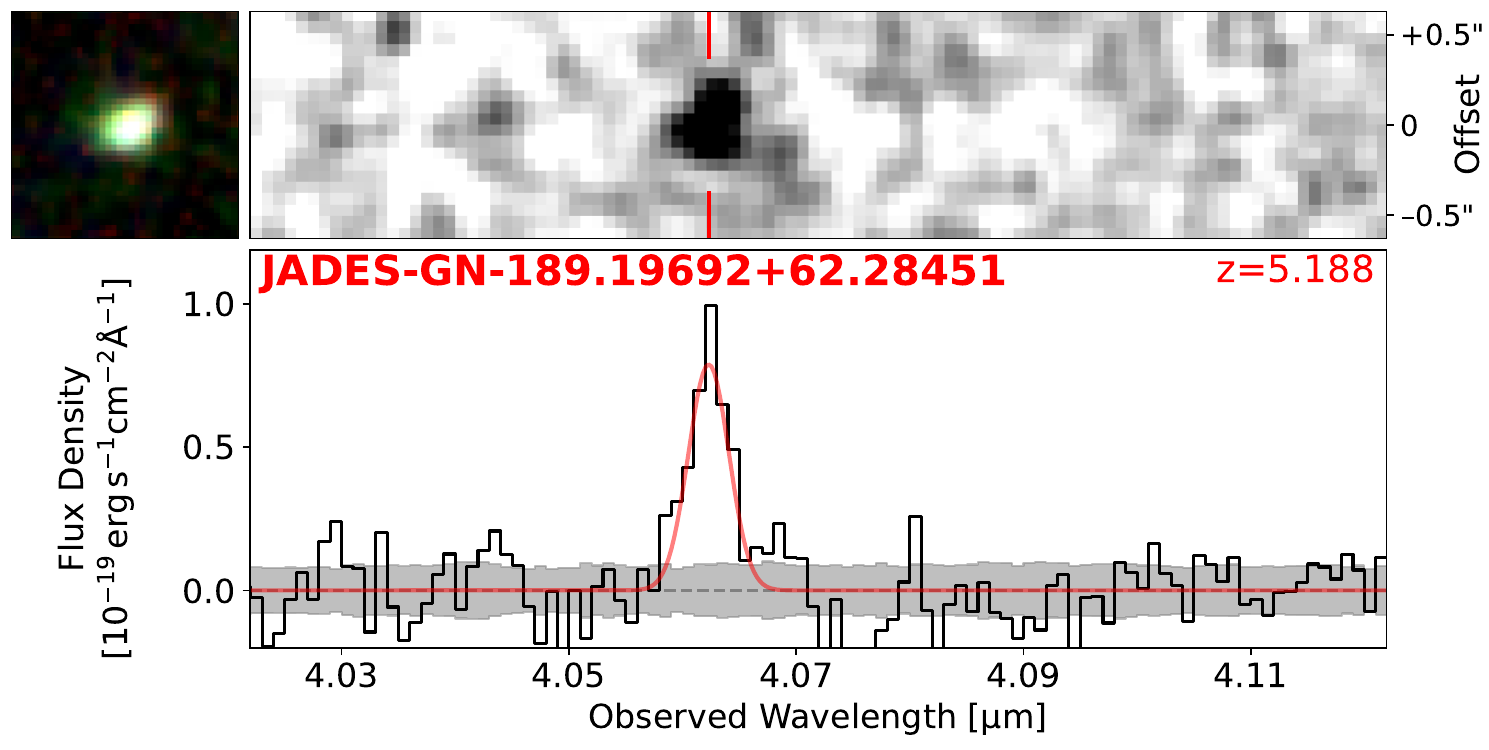}
\caption{Continued.} 
 \end{figure*} 

 \addtocounter{figure}{-1} 
 \begin{figure*}[!ht] 
 \centering
\includegraphics[width=0.49\linewidth]{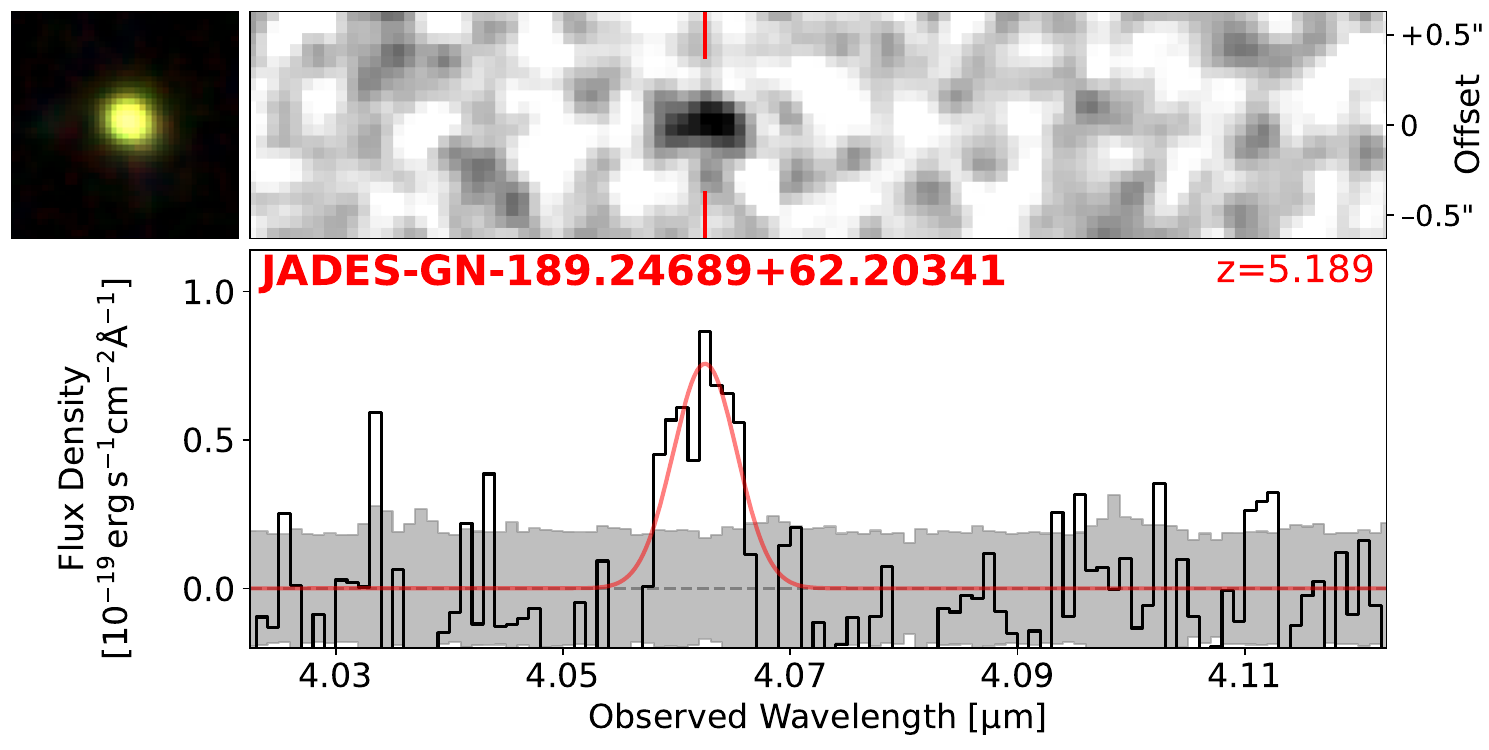}
\includegraphics[width=0.49\linewidth]{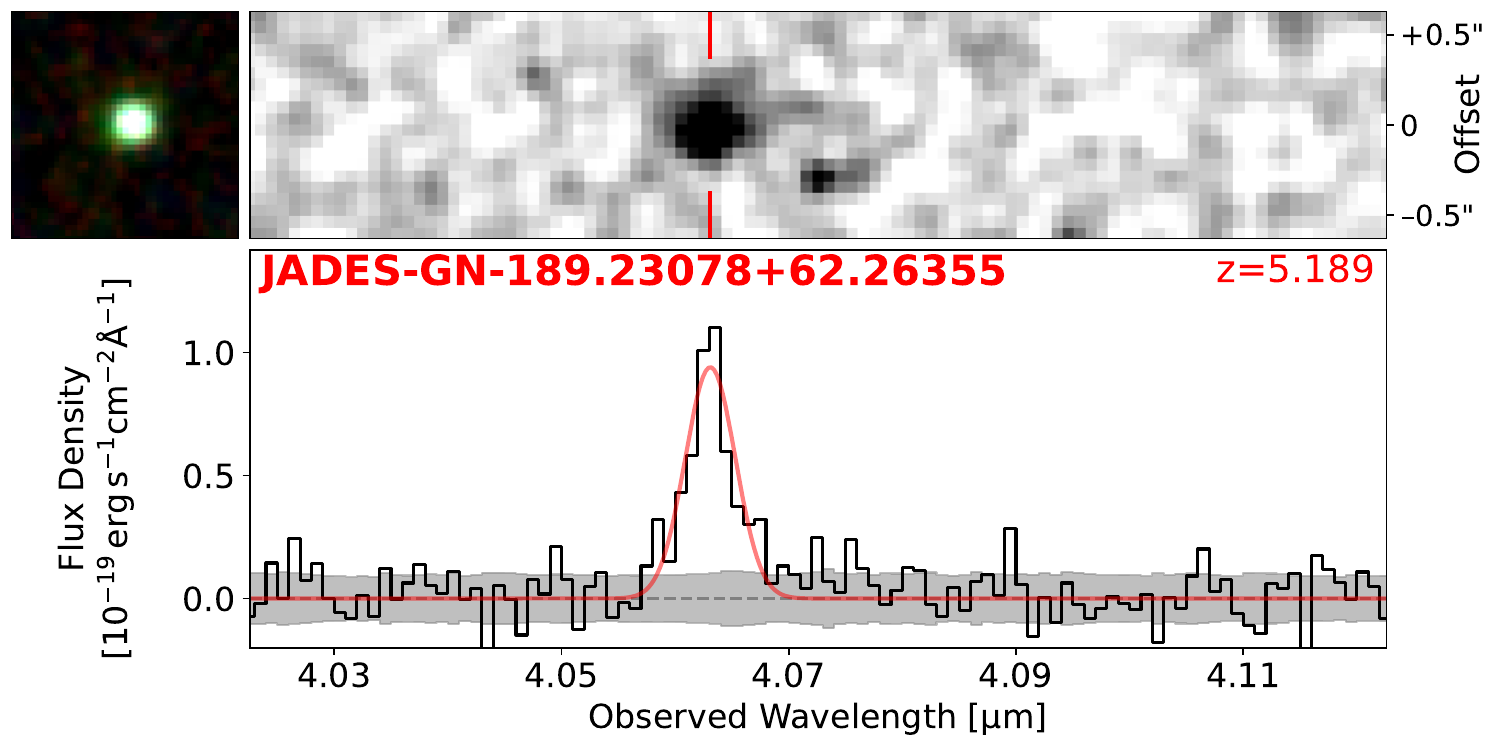}
\includegraphics[width=0.49\linewidth]{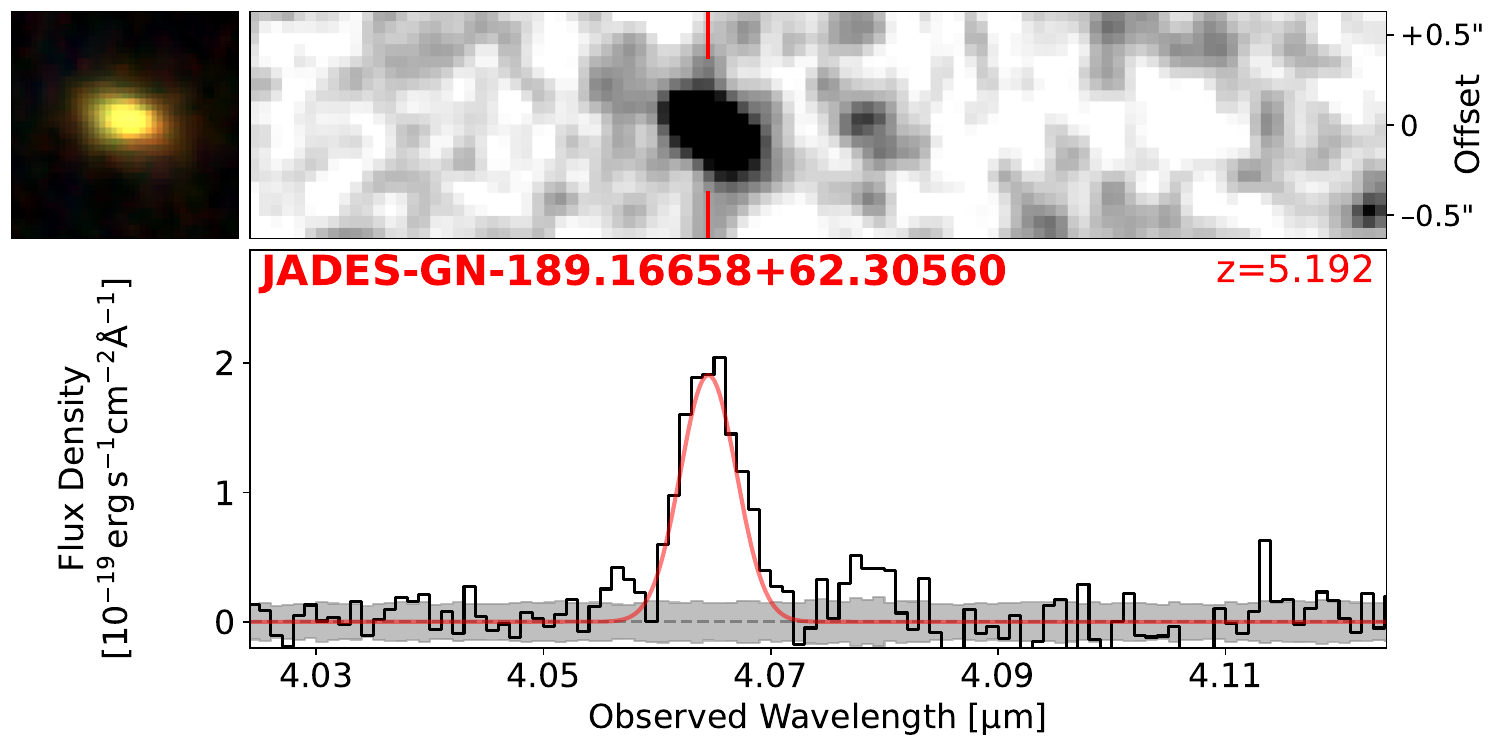}
\includegraphics[width=0.49\linewidth]{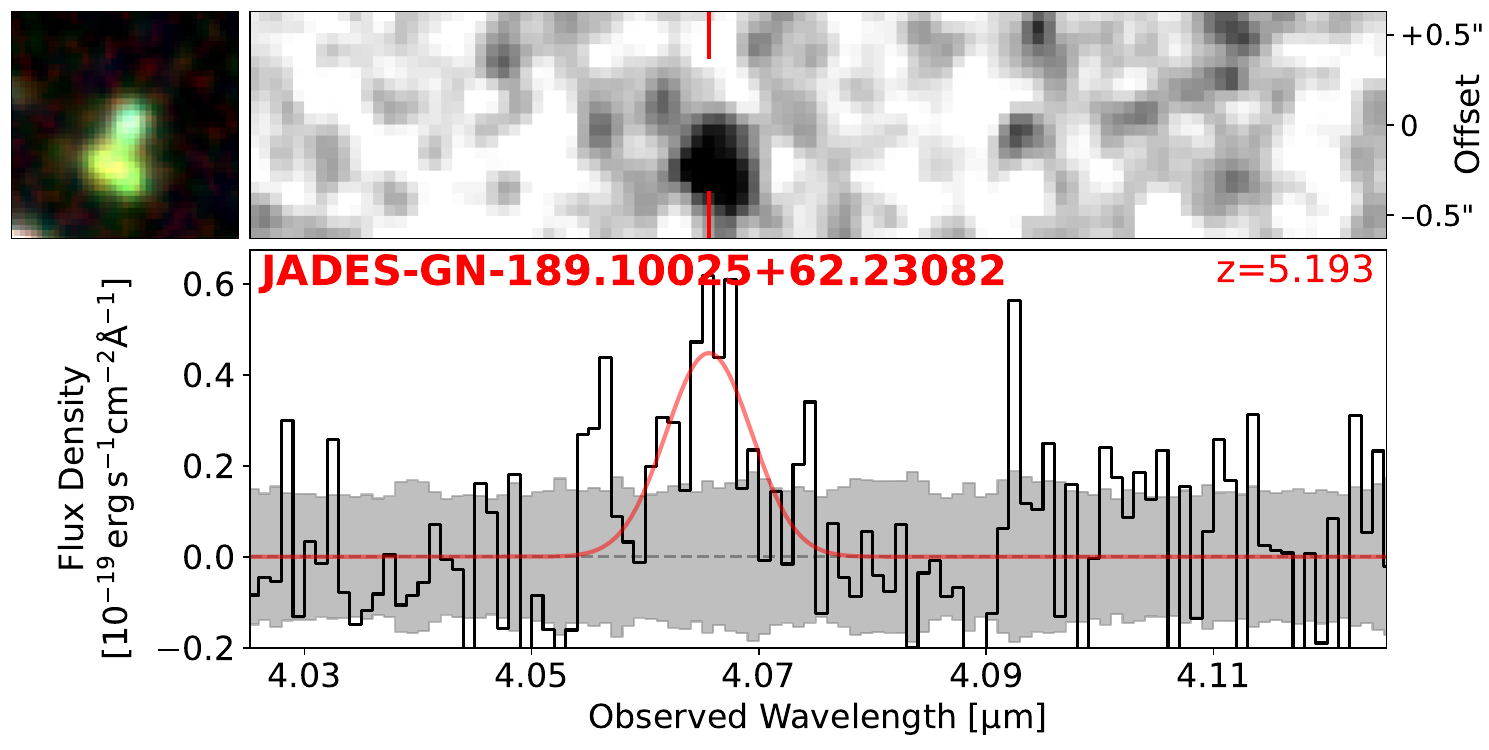}
\includegraphics[width=0.49\linewidth]{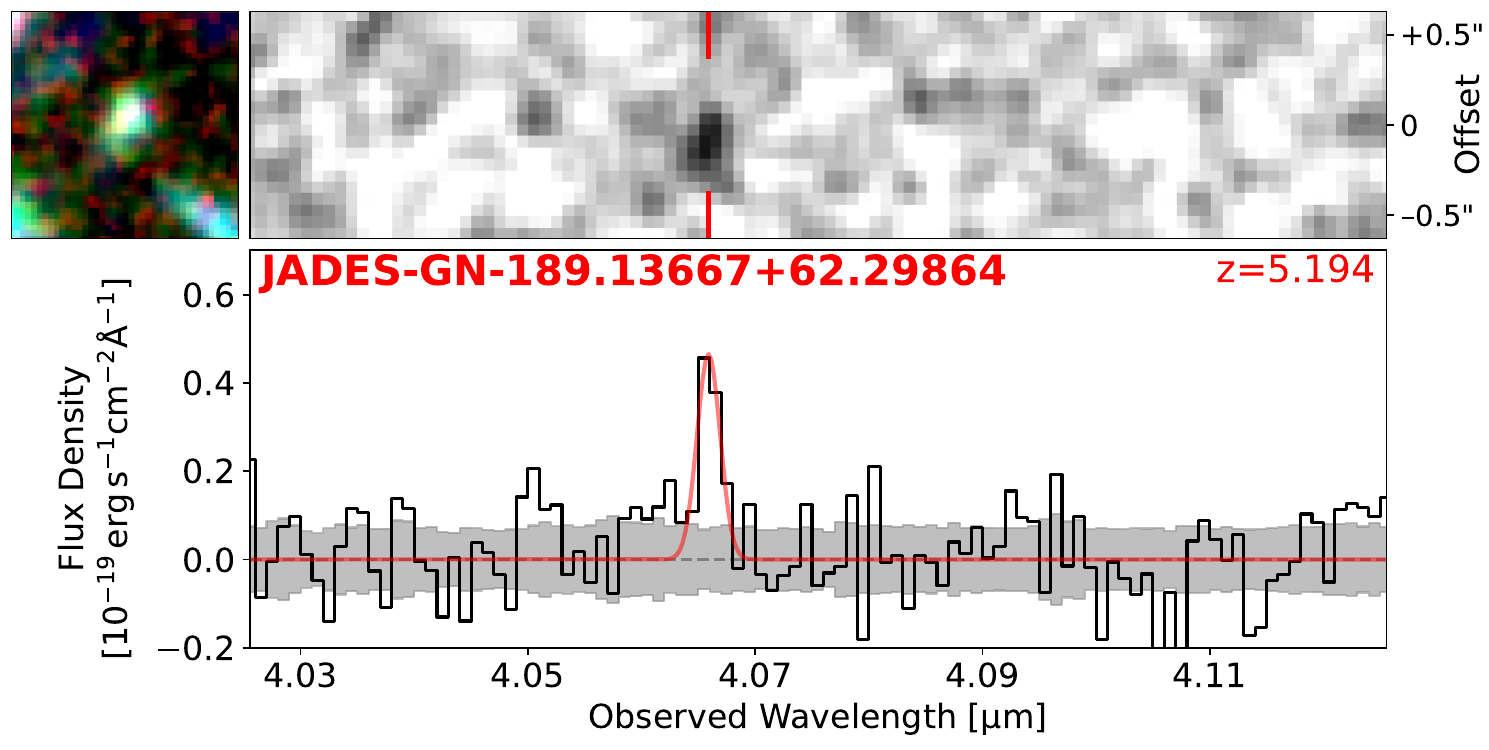}
\includegraphics[width=0.49\linewidth]{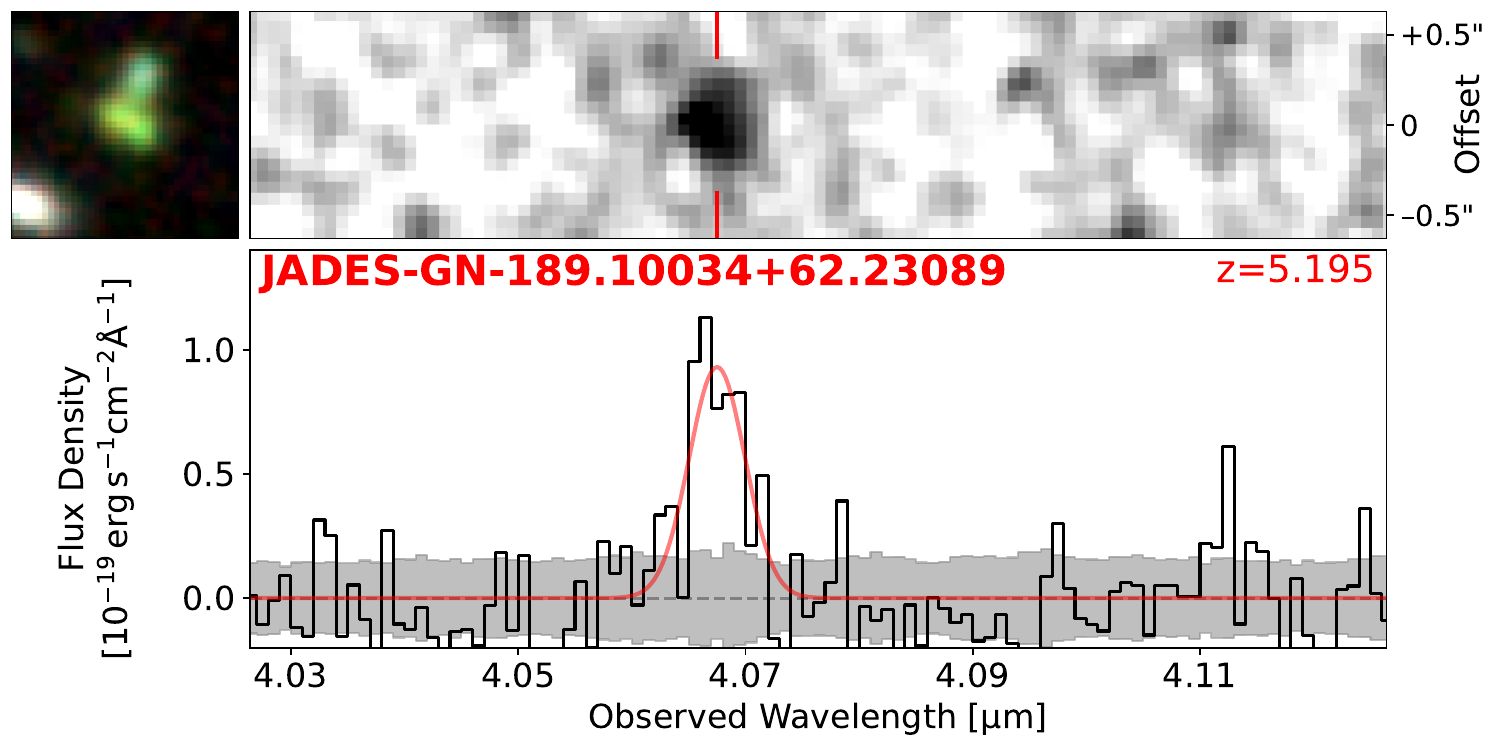}
\includegraphics[width=0.49\linewidth]{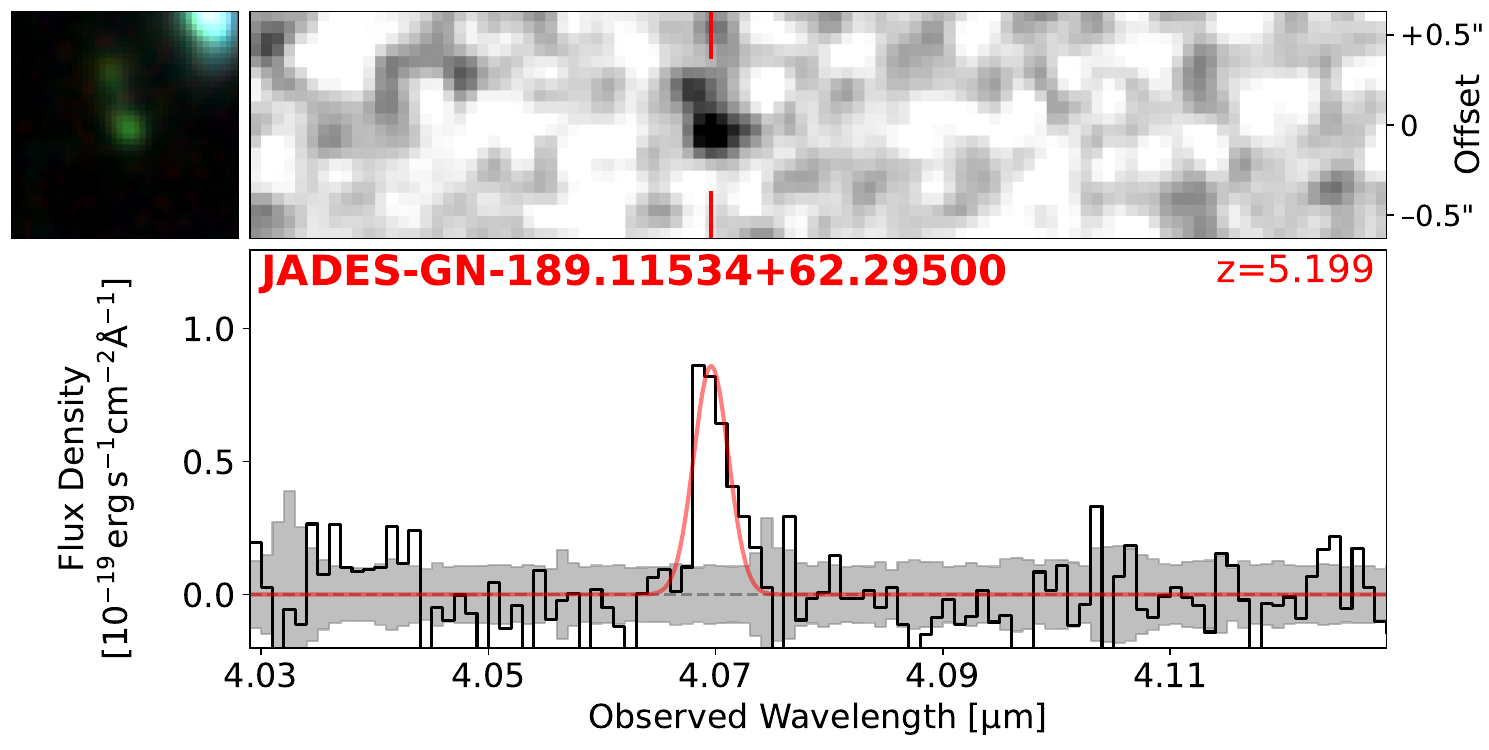}
\includegraphics[width=0.49\linewidth]{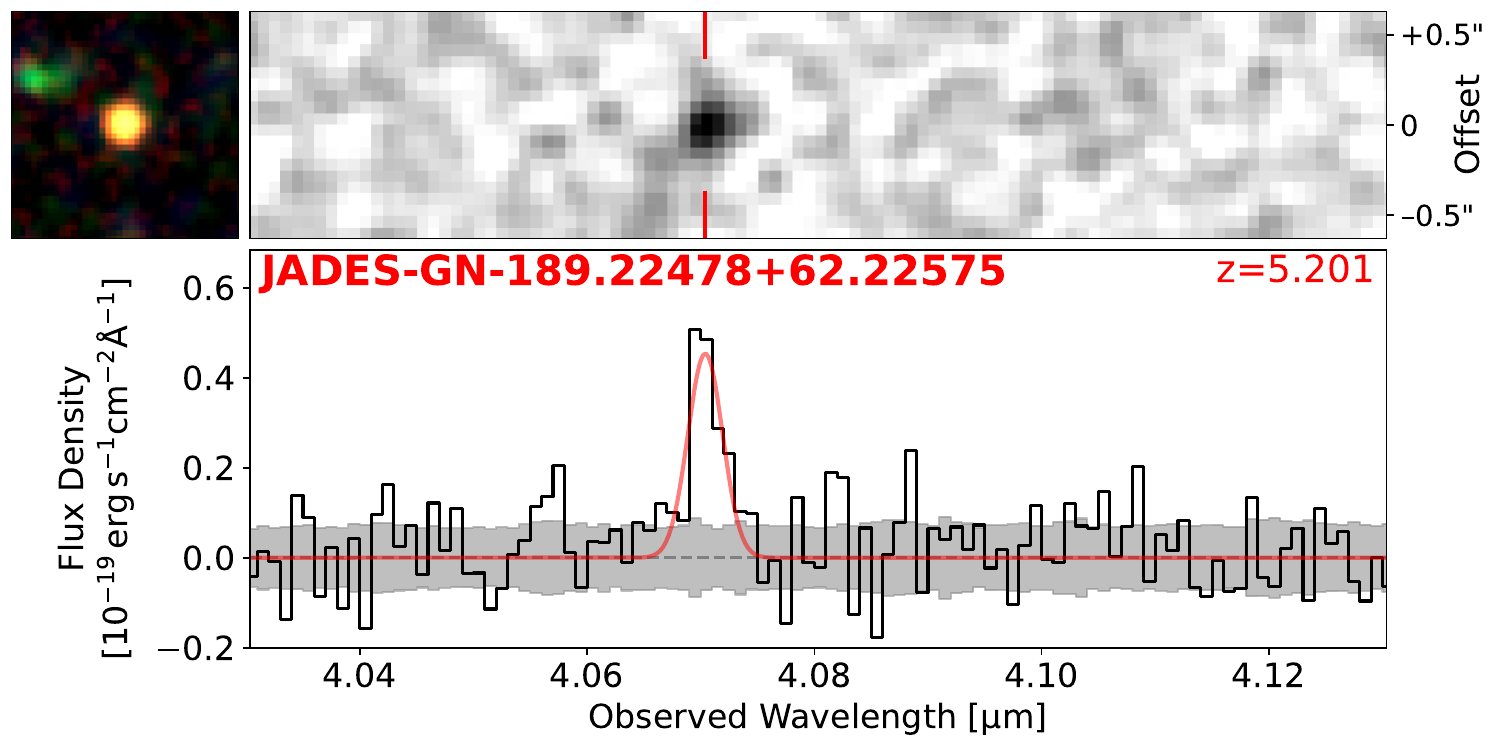}
\includegraphics[width=0.49\linewidth]{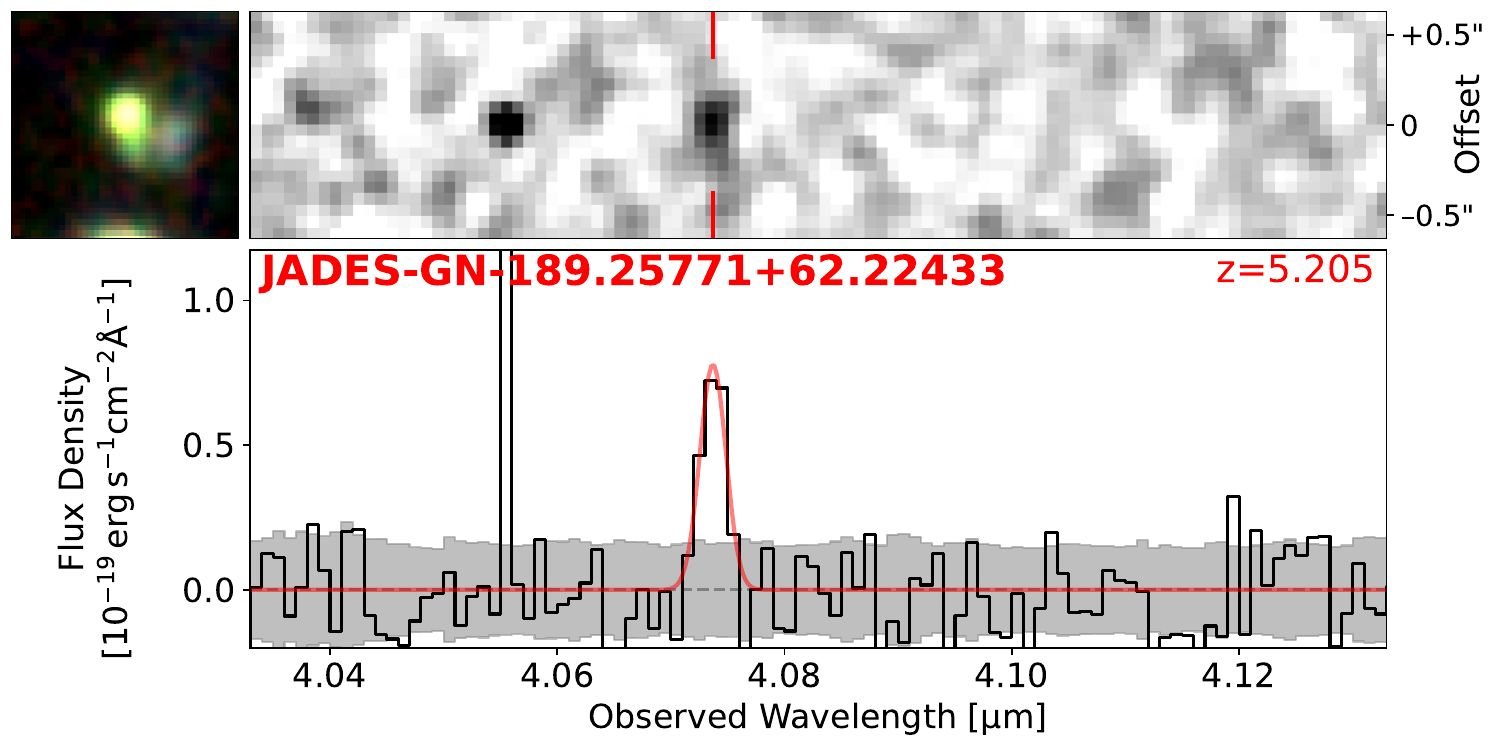}
\includegraphics[width=0.49\linewidth]{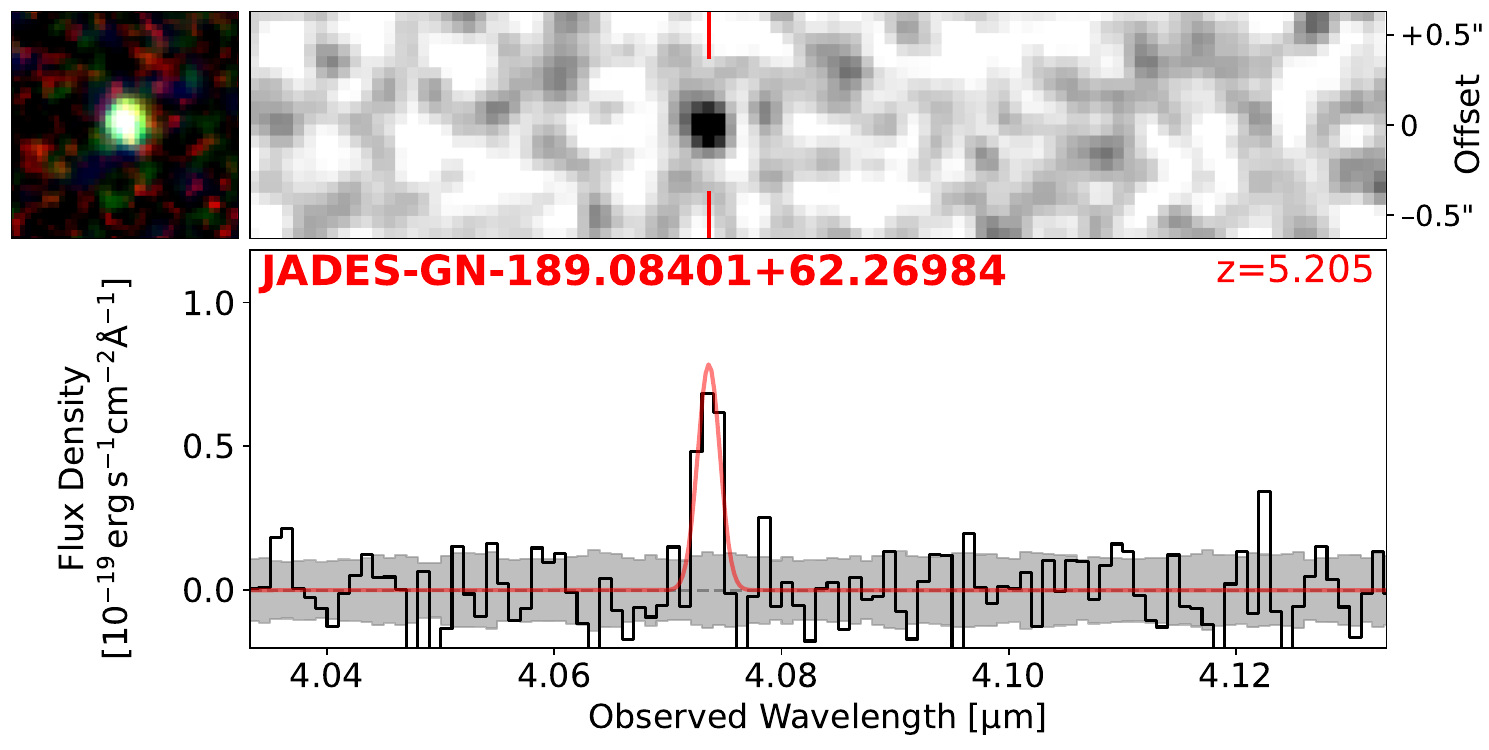}
\caption{Continued.} 
 \end{figure*} 

 \addtocounter{figure}{-1} 
 \begin{figure*}[!ht] 
 \centering
\includegraphics[width=0.49\linewidth]{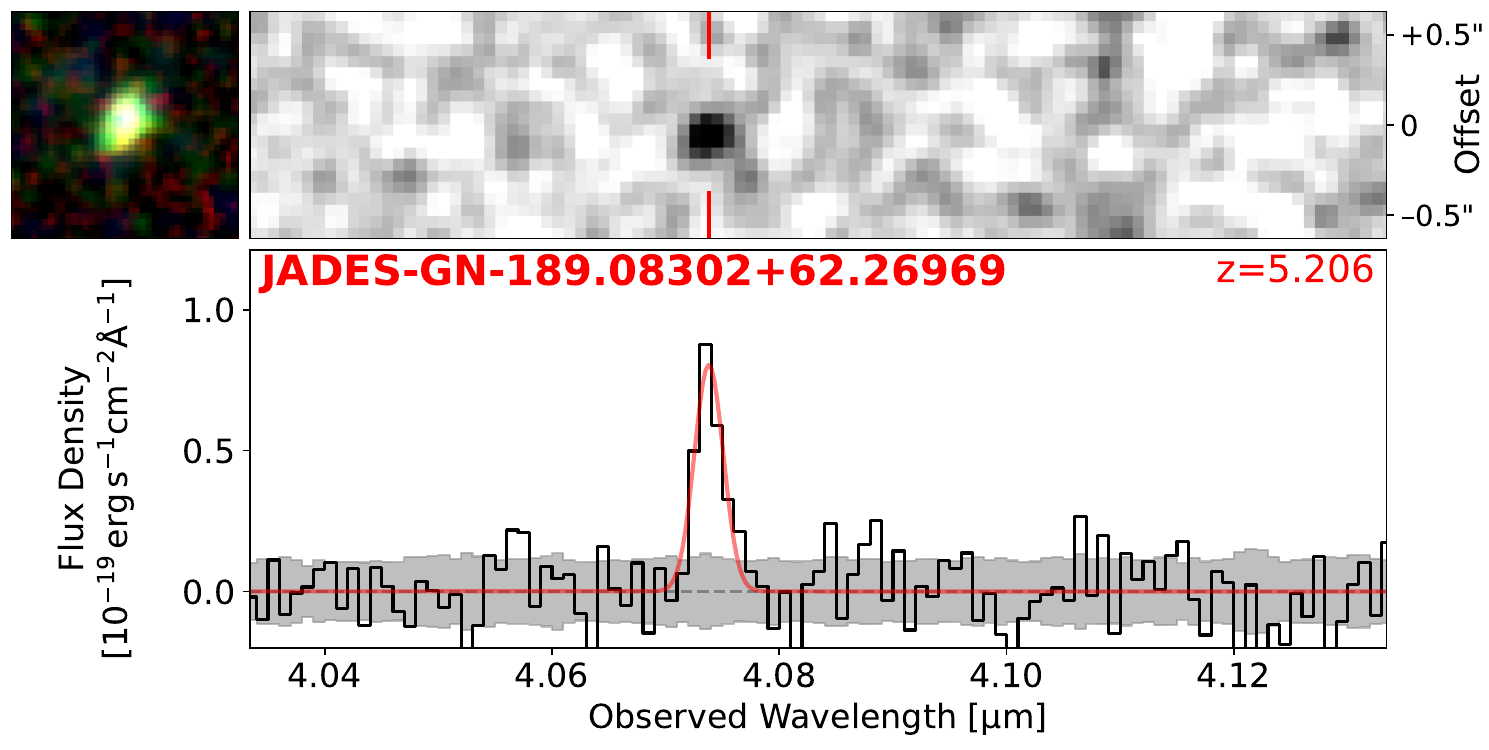}
\includegraphics[width=0.49\linewidth]{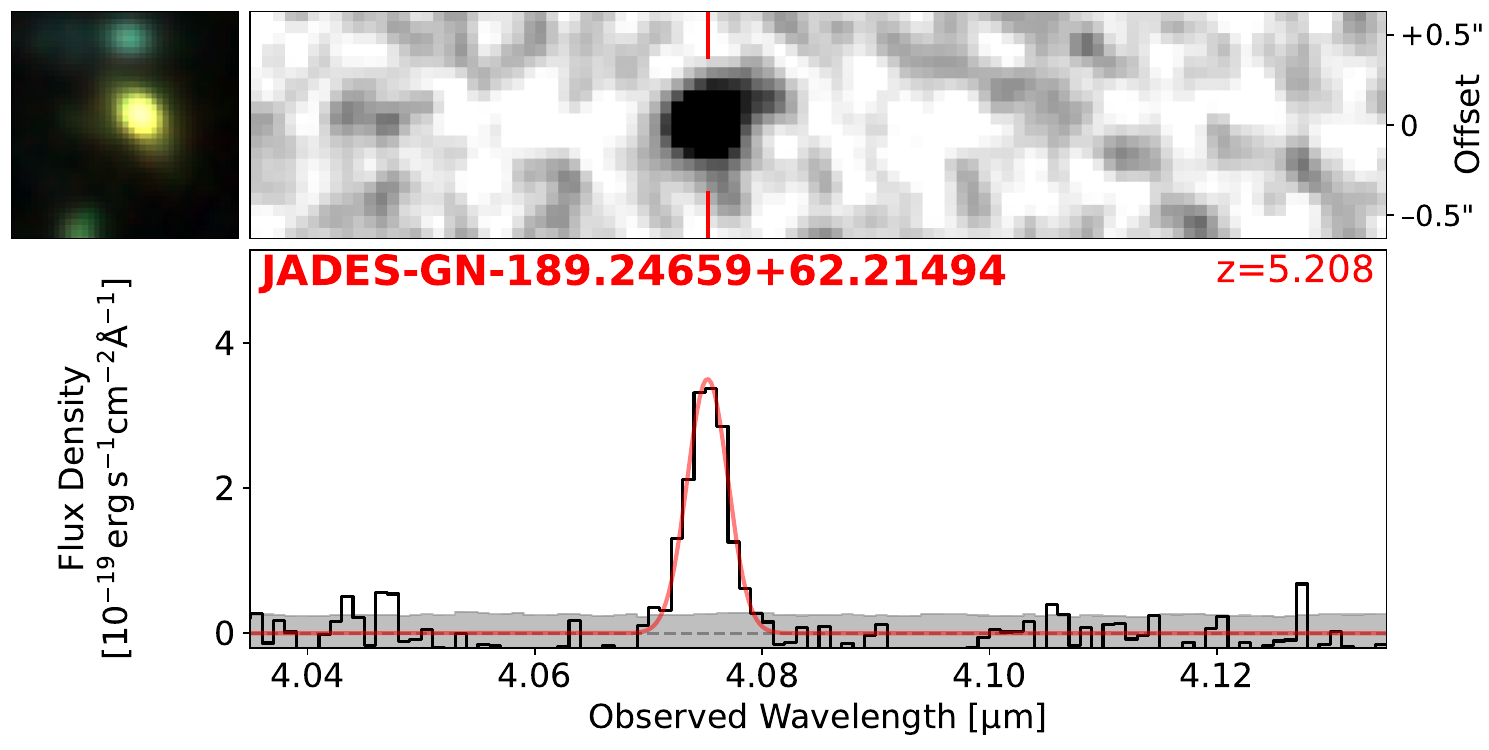}
\includegraphics[width=0.49\linewidth]{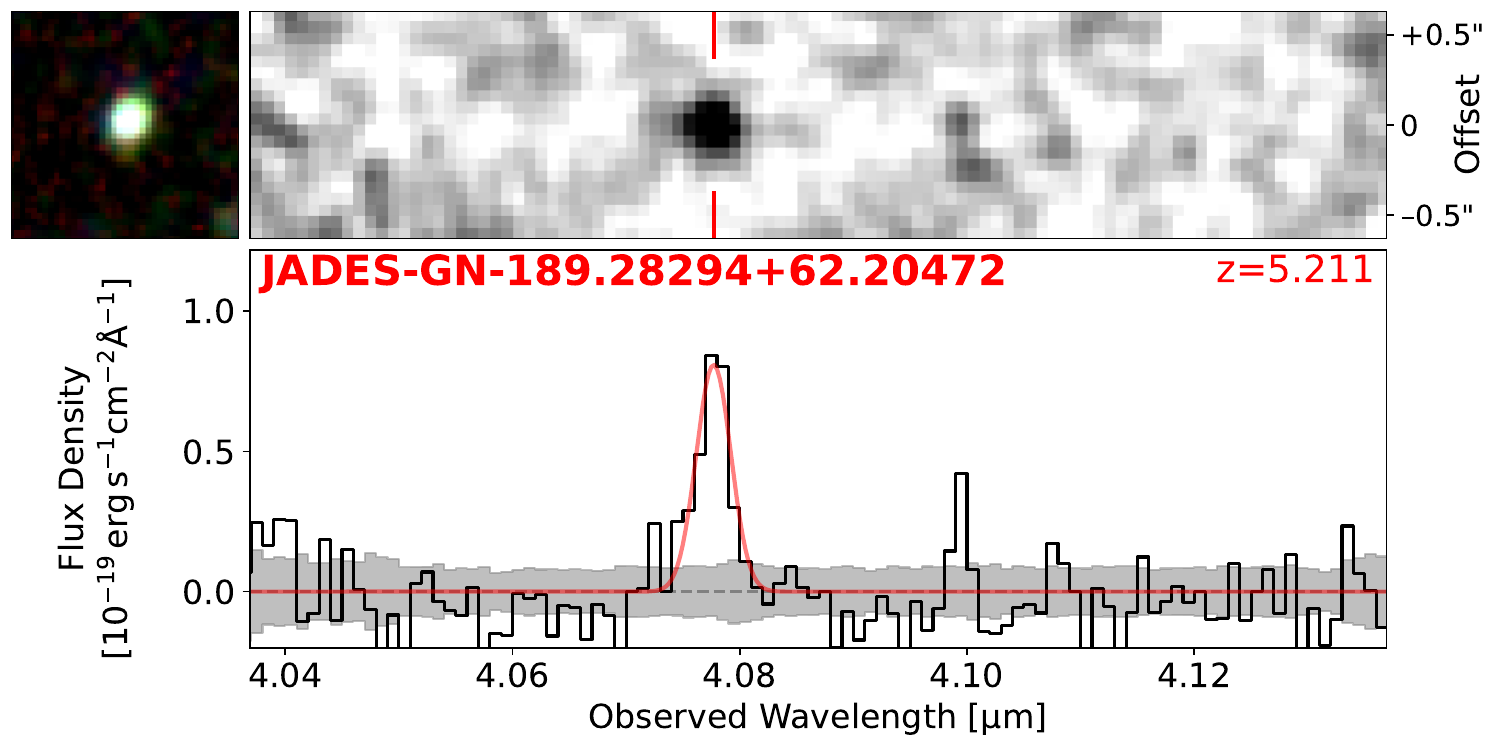}
\includegraphics[width=0.49\linewidth]{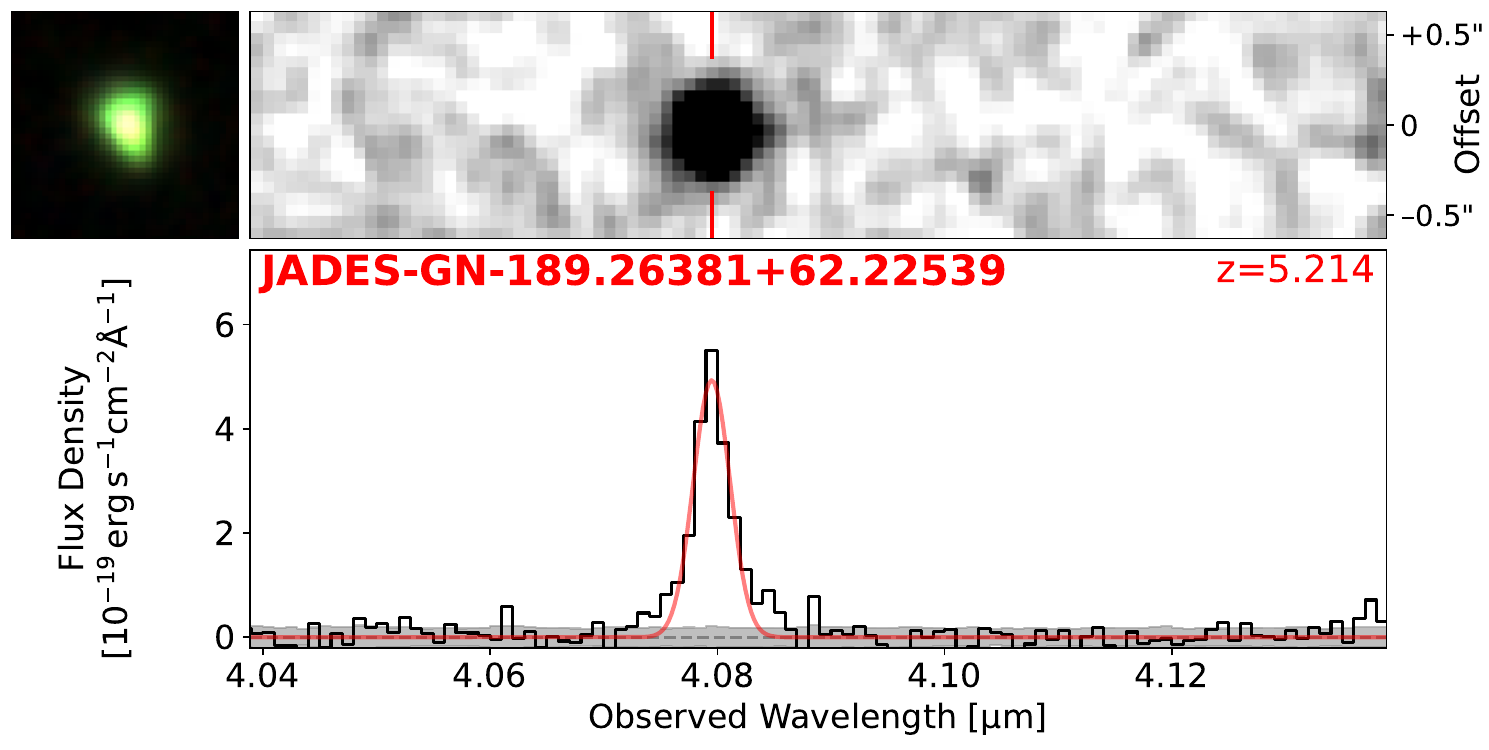}
\includegraphics[width=0.49\linewidth]{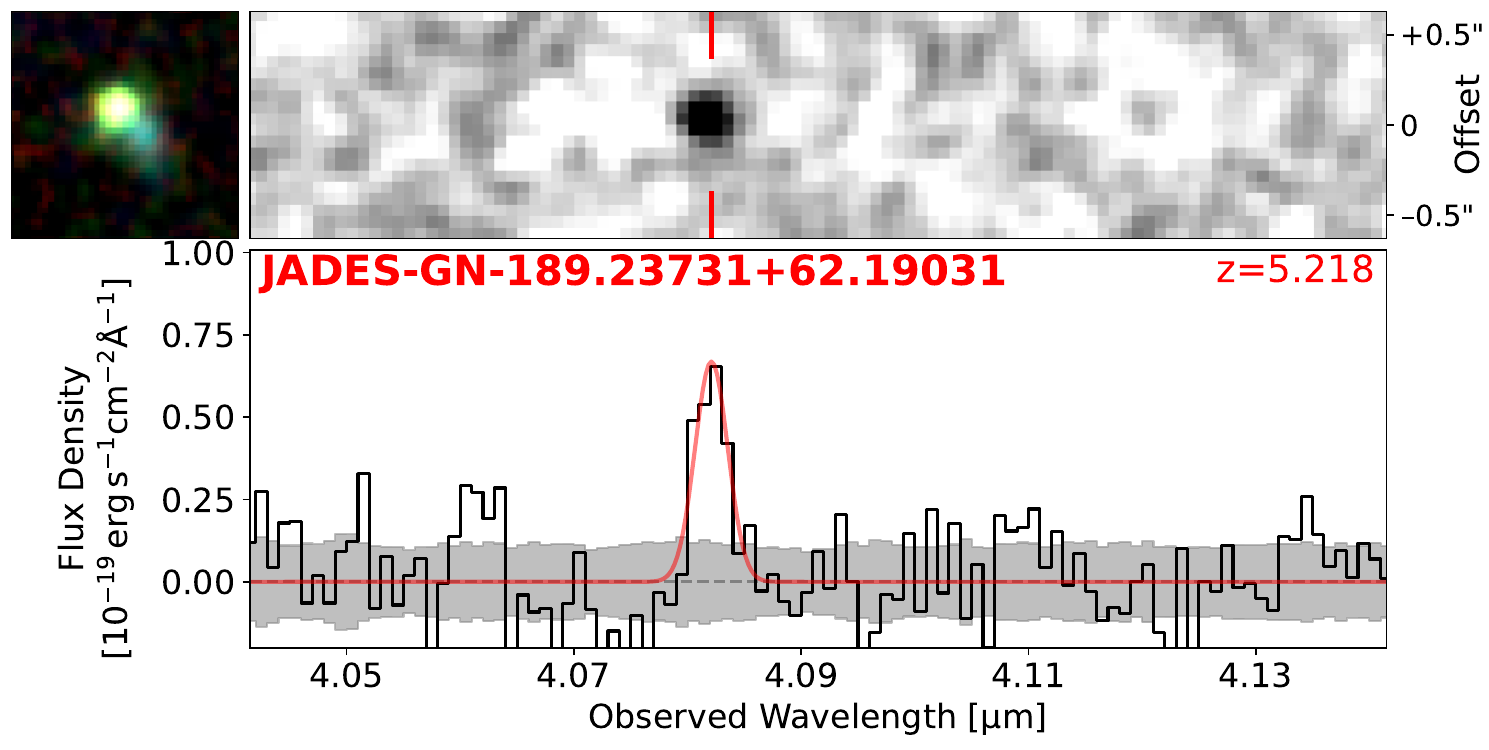}
\includegraphics[width=0.49\linewidth]{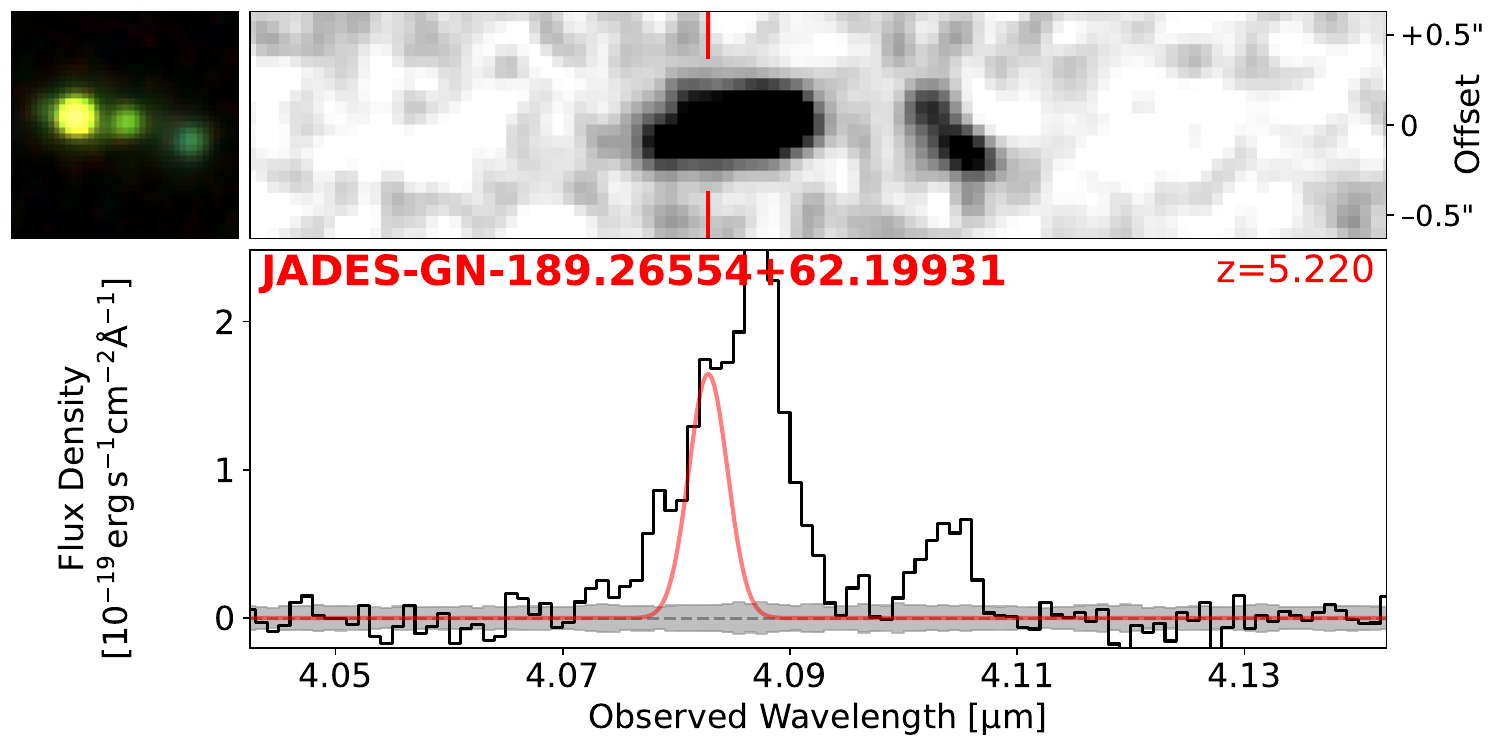}
\includegraphics[width=0.49\linewidth]{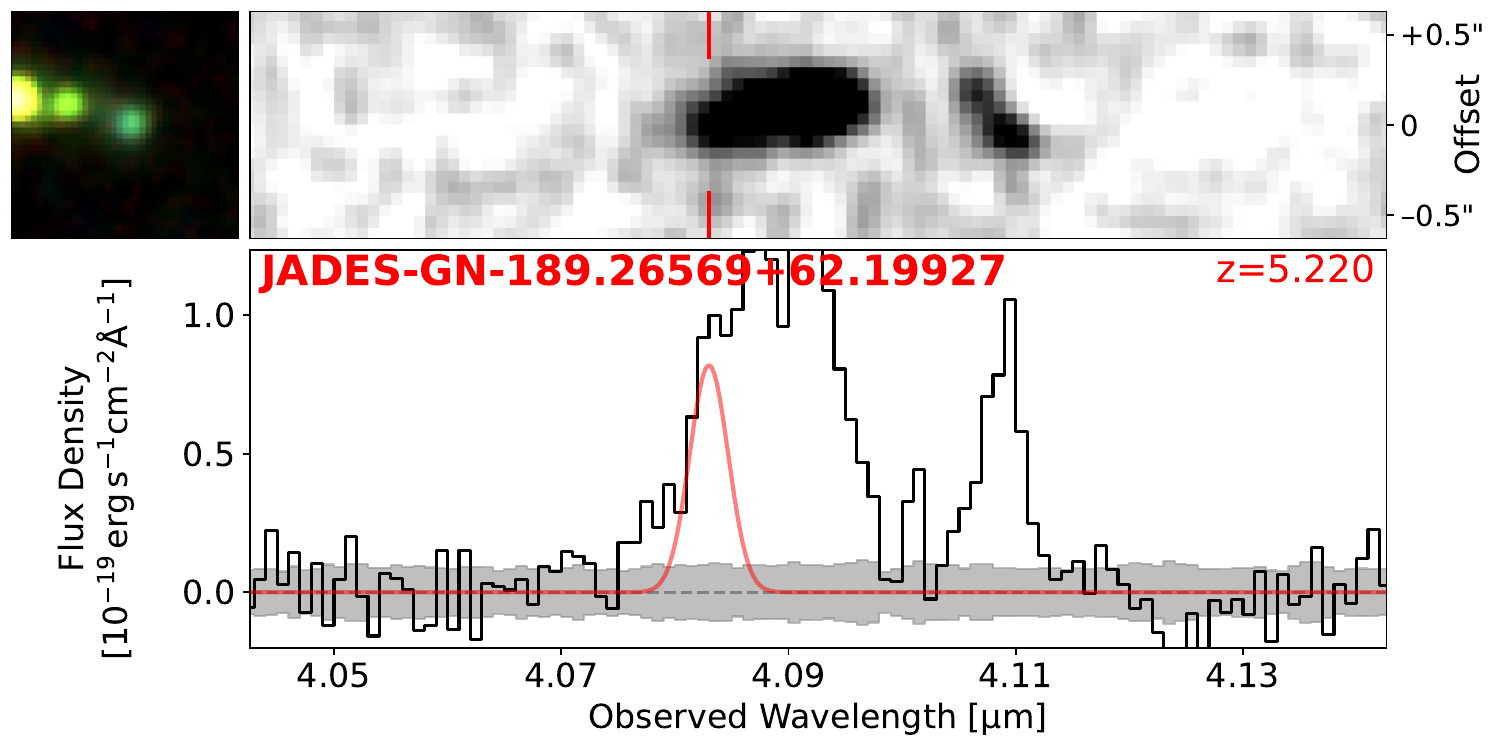}
\includegraphics[width=0.49\linewidth]{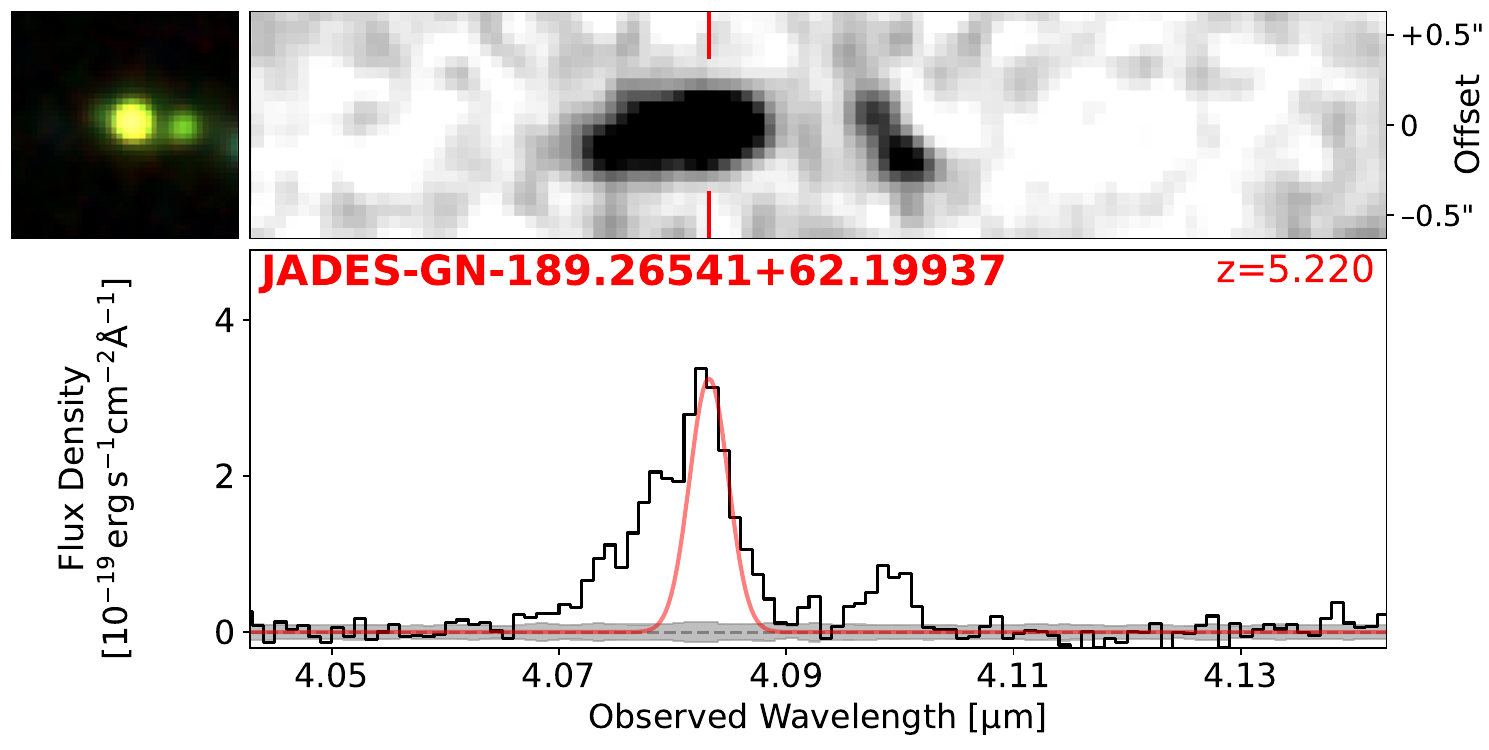}
\includegraphics[width=0.49\linewidth]{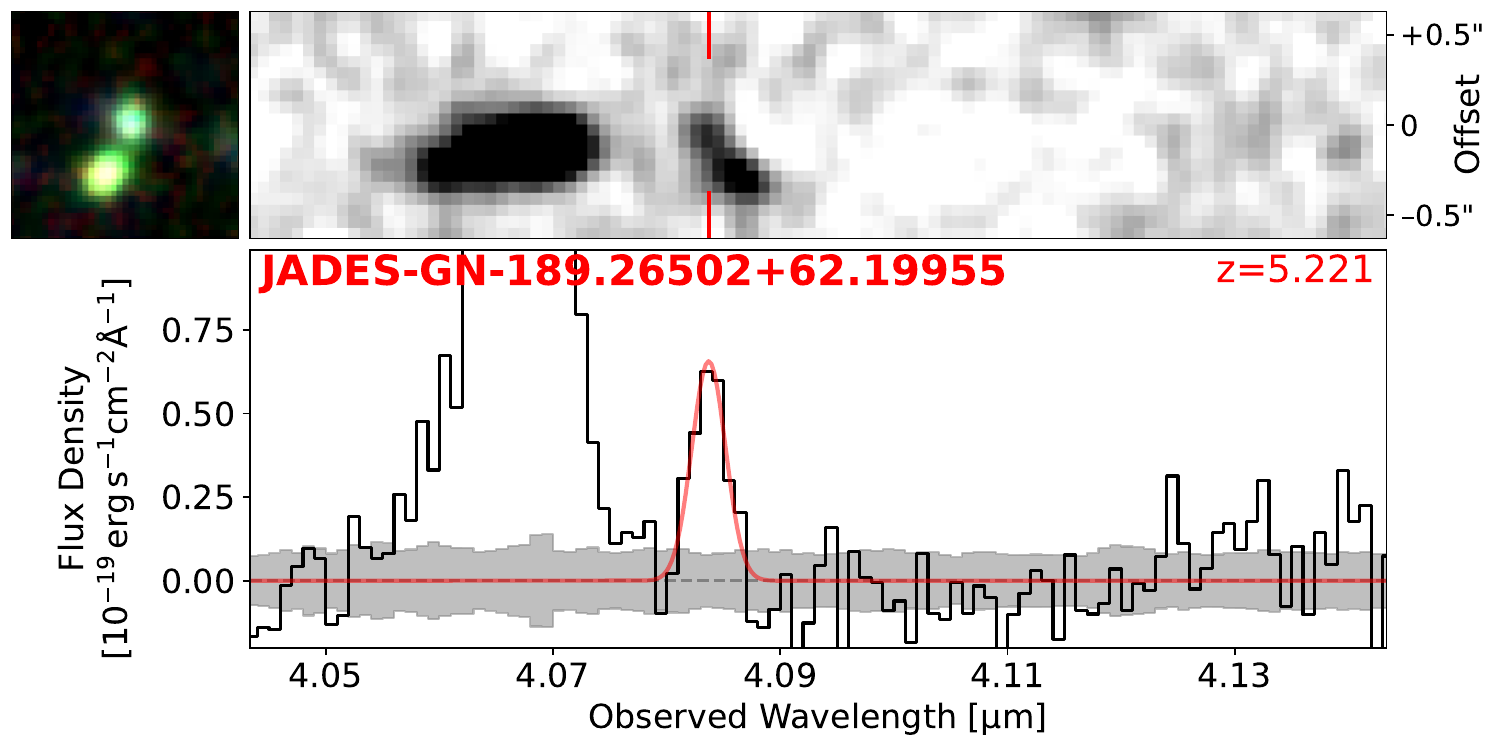}
\includegraphics[width=0.49\linewidth]{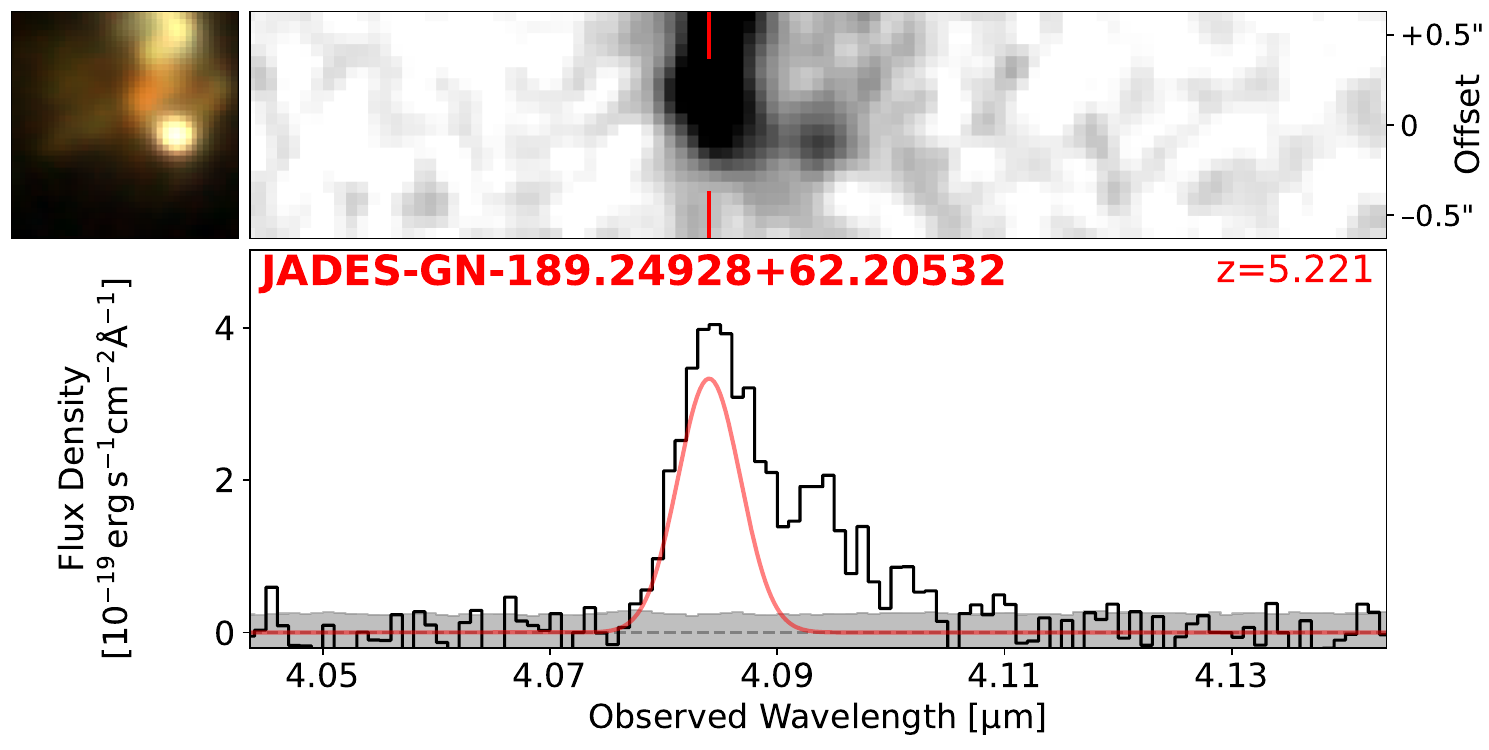}
\caption{Continued.} 
 \end{figure*} 

 \addtocounter{figure}{-1} 
 \begin{figure*}[!ht] 
 \centering
\includegraphics[width=0.49\linewidth]{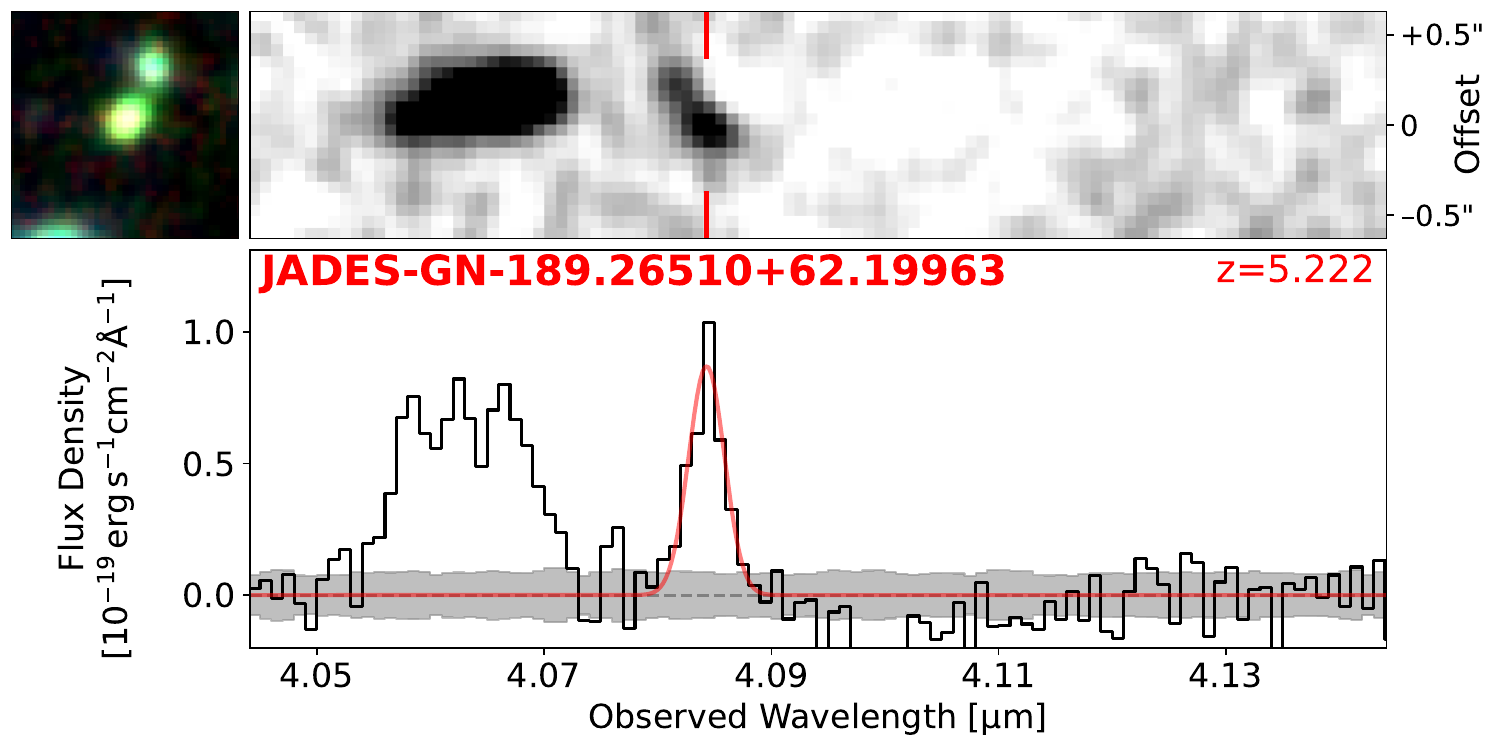}
\includegraphics[width=0.49\linewidth]{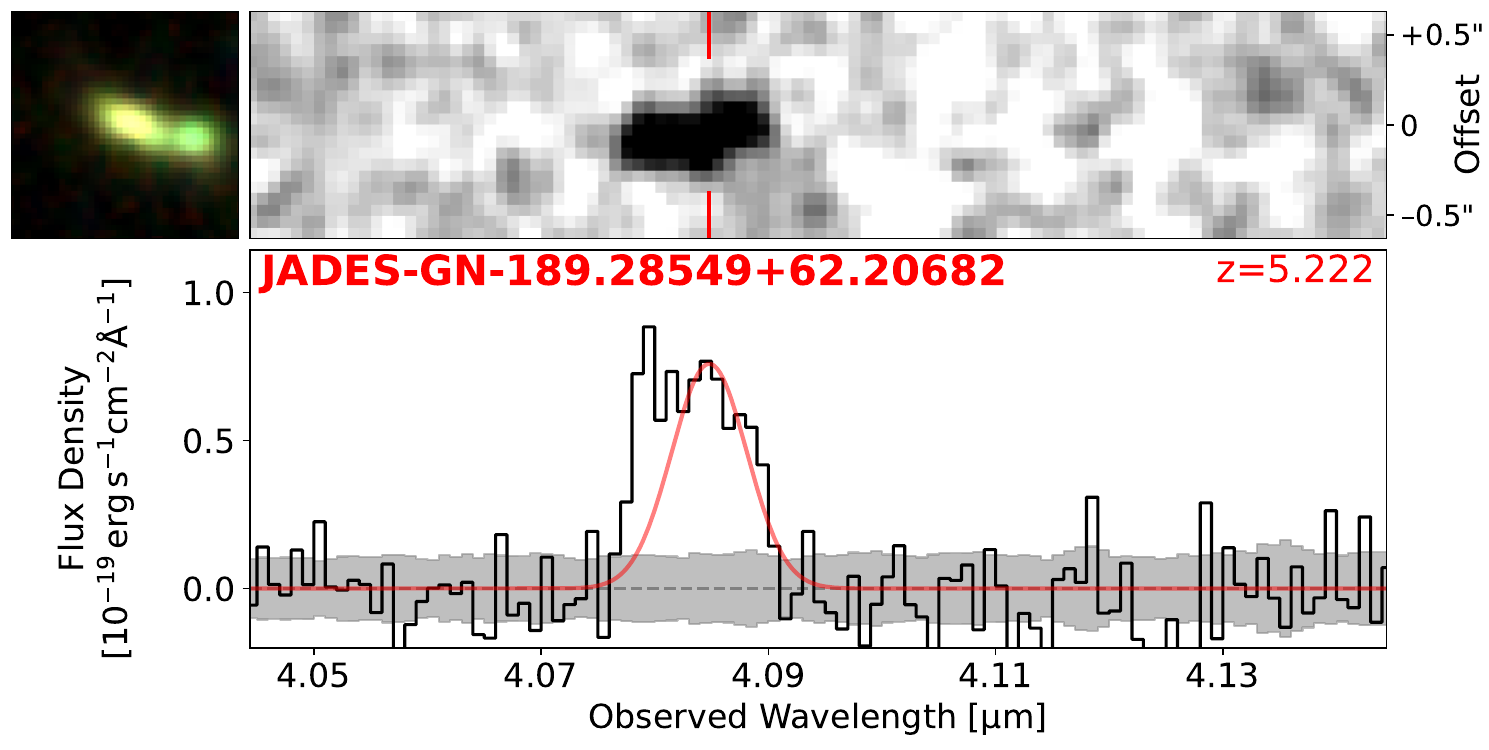}
\includegraphics[width=0.49\linewidth]{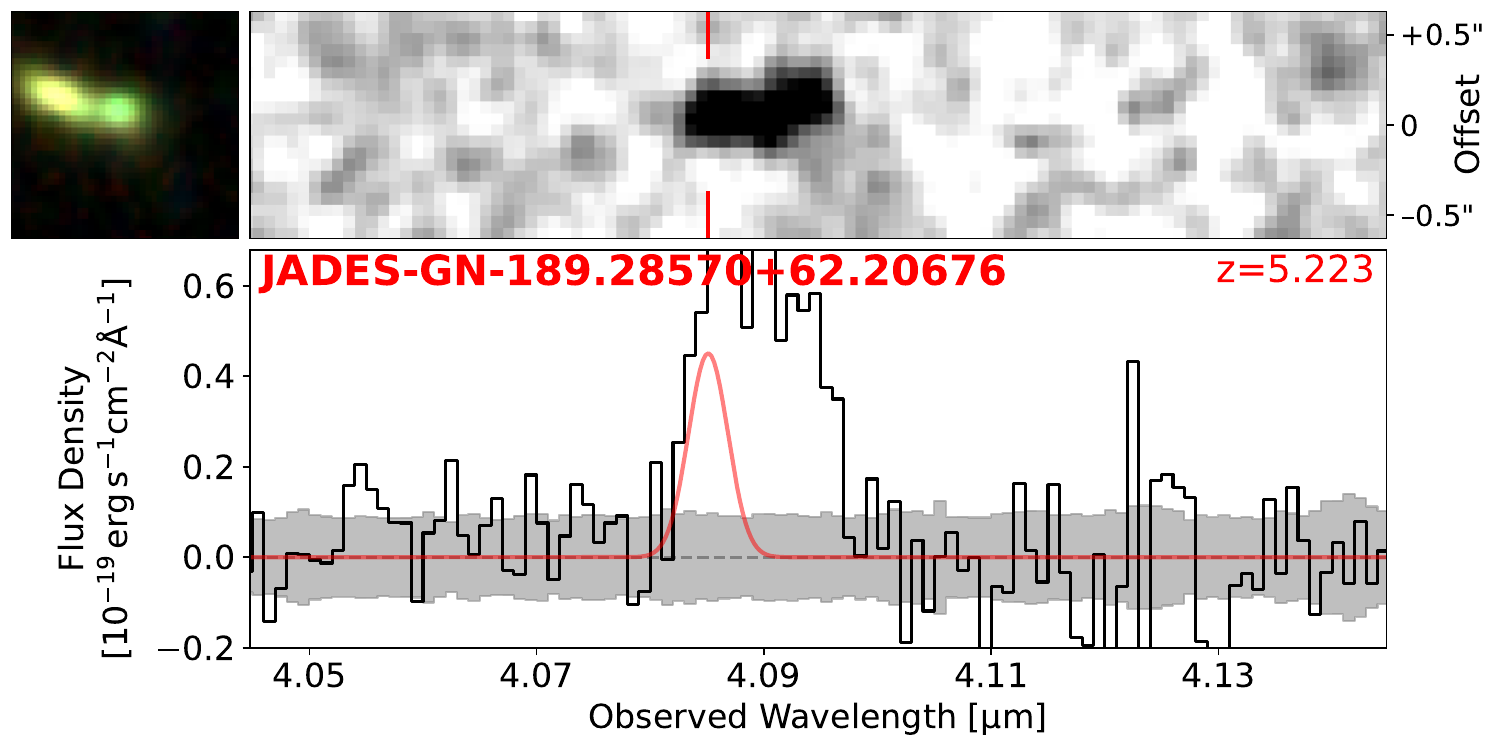}
\includegraphics[width=0.49\linewidth]{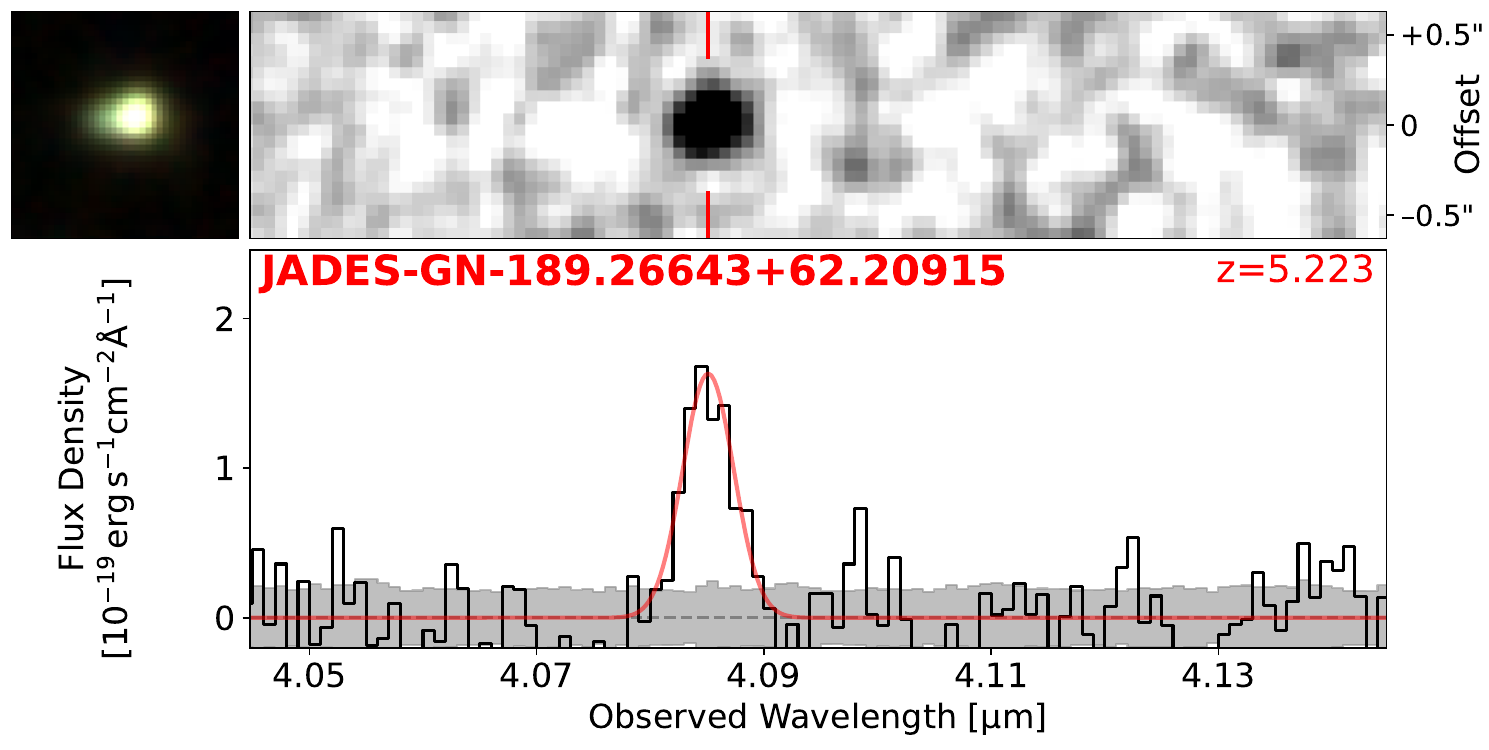}
\includegraphics[width=0.49\linewidth]{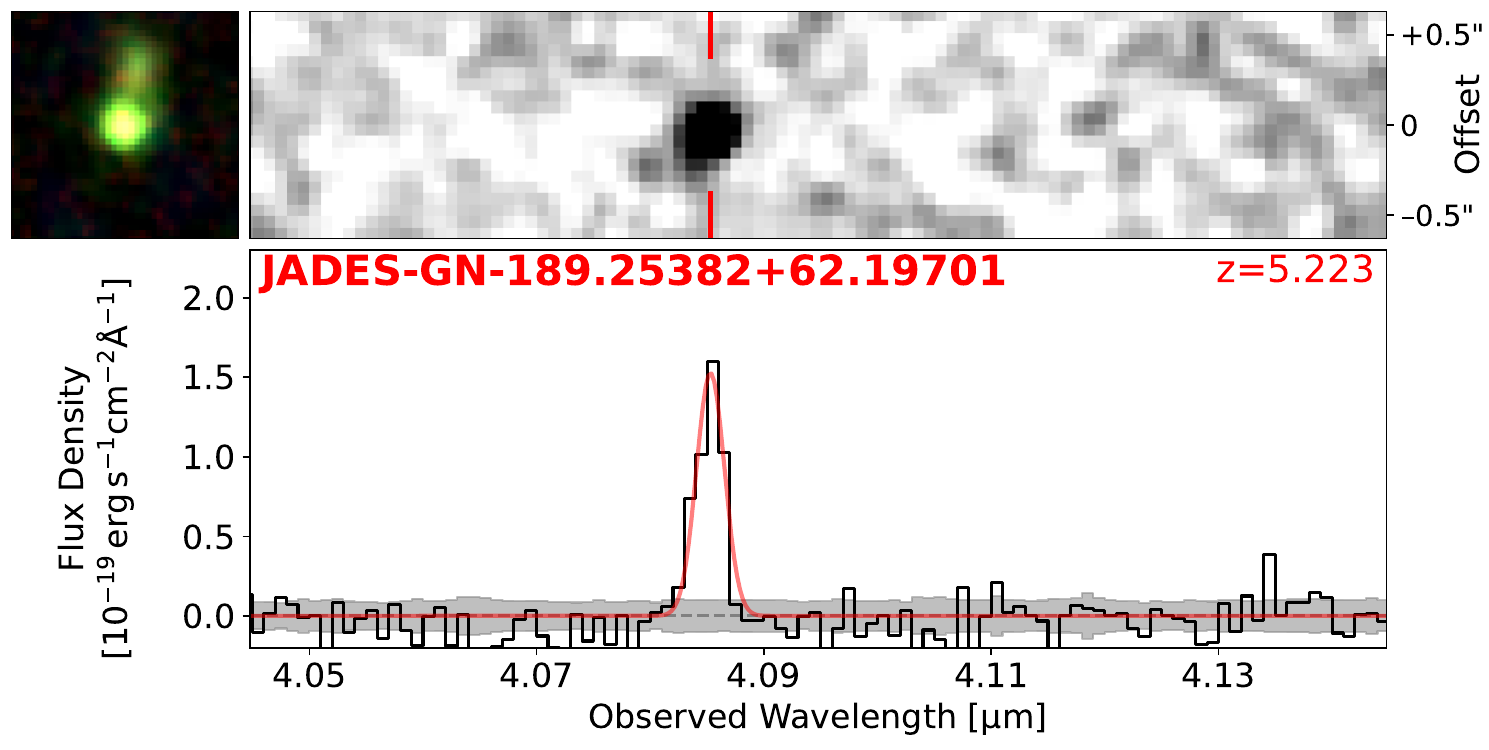}
\includegraphics[width=0.49\linewidth]{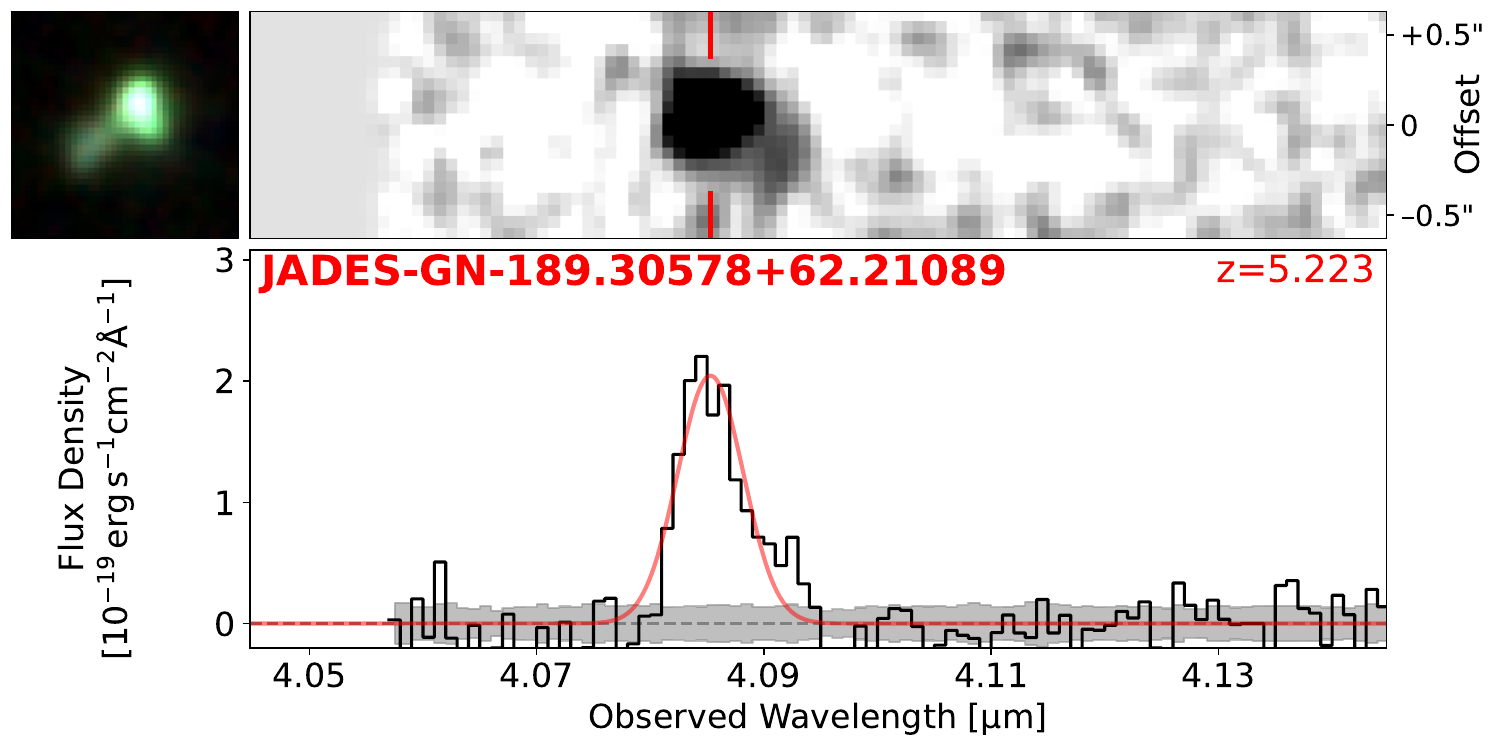}
\includegraphics[width=0.49\linewidth]{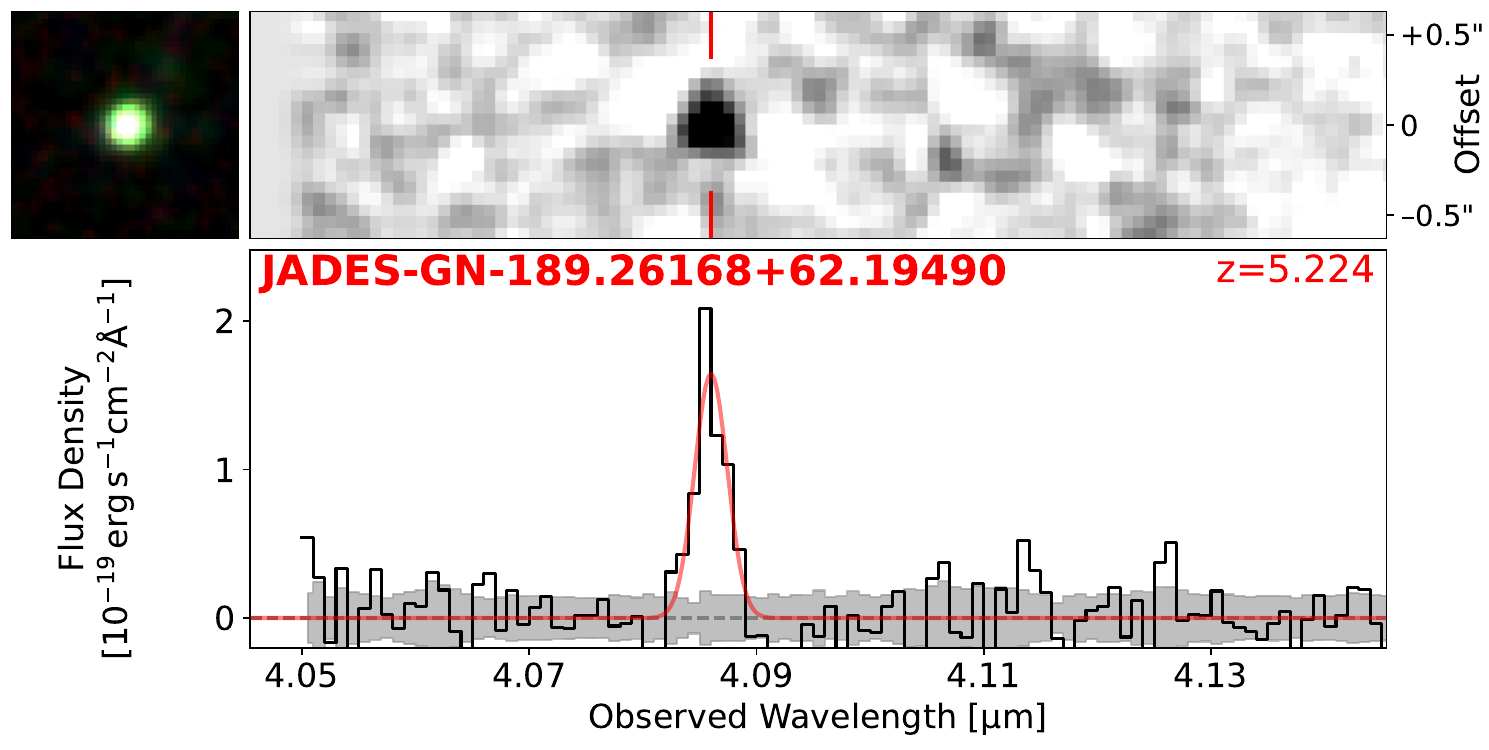}
\includegraphics[width=0.49\linewidth]{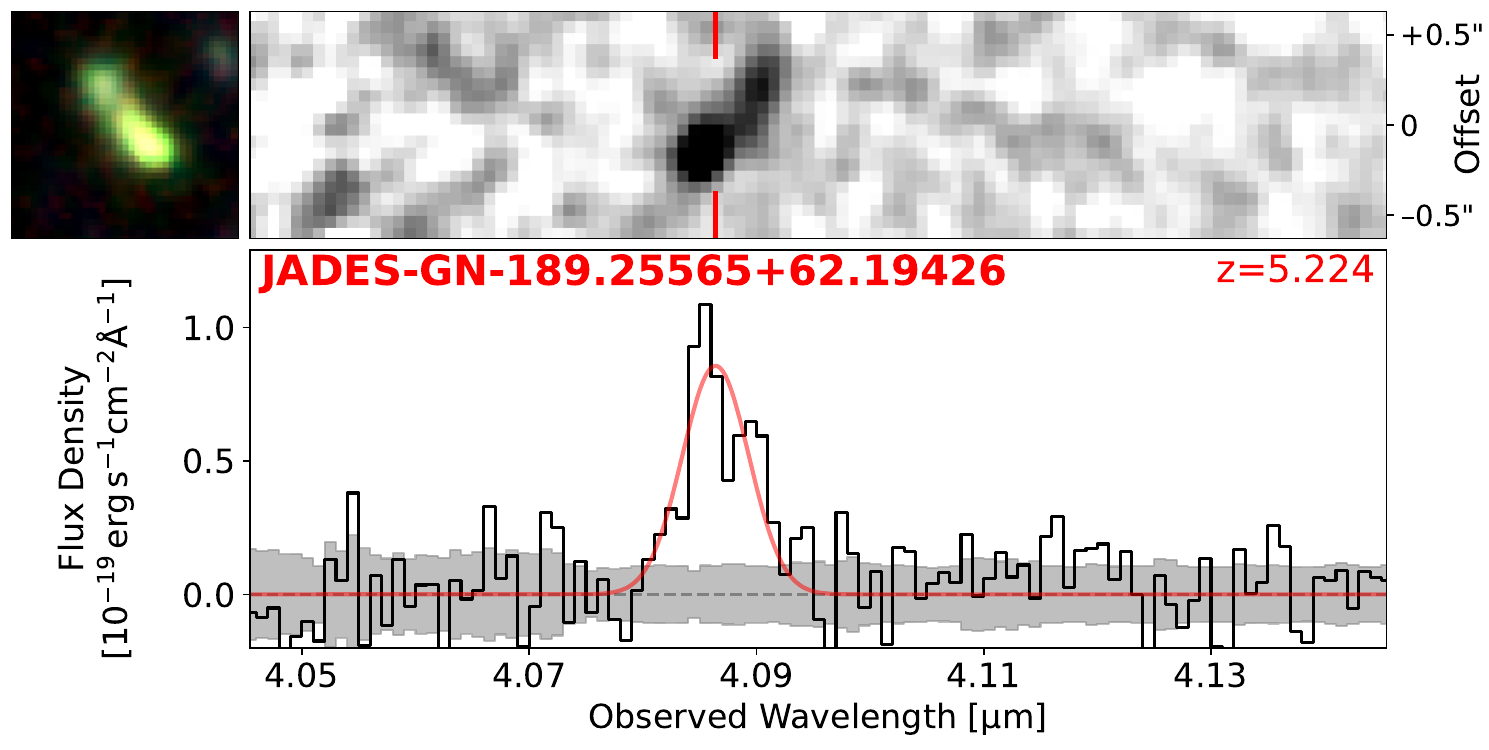}
\includegraphics[width=0.49\linewidth]{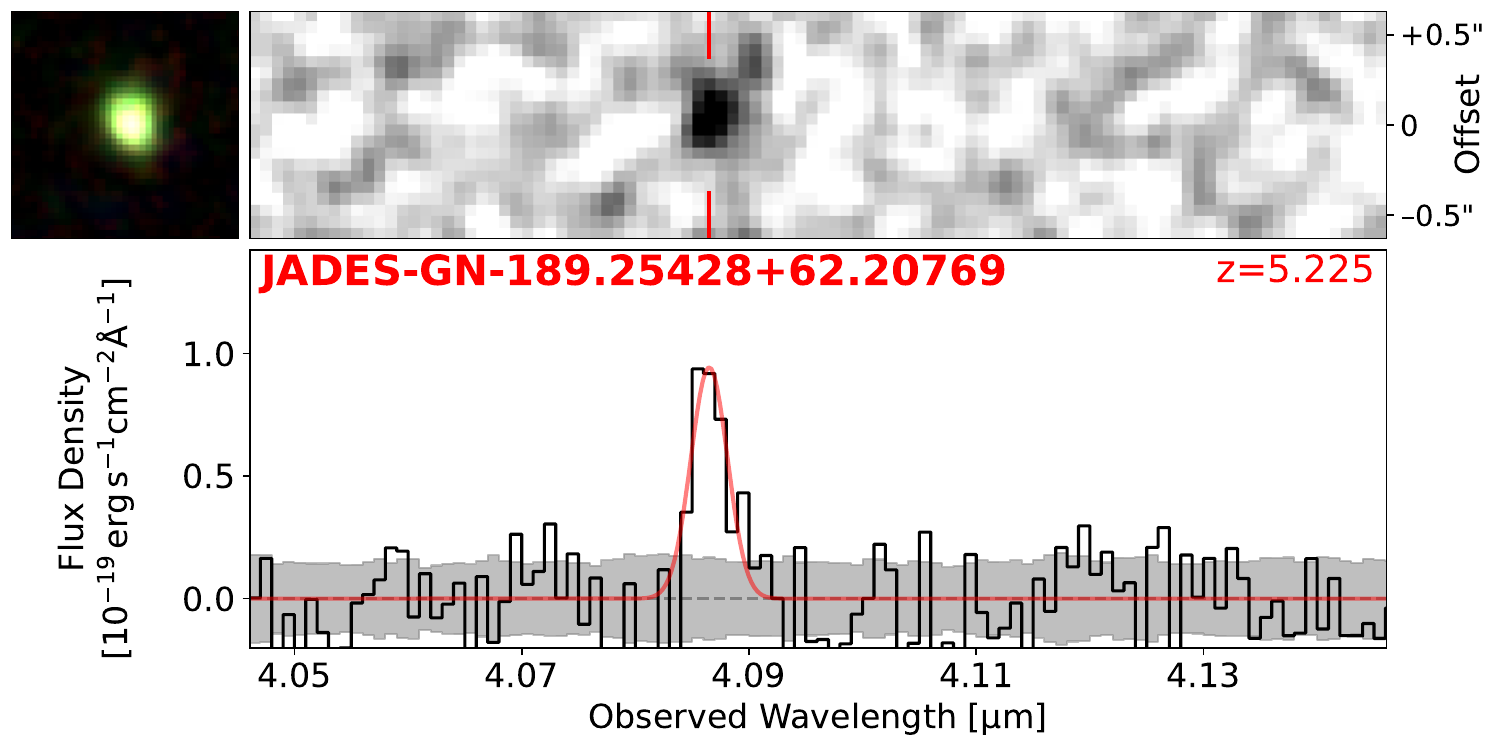}
\includegraphics[width=0.49\linewidth]{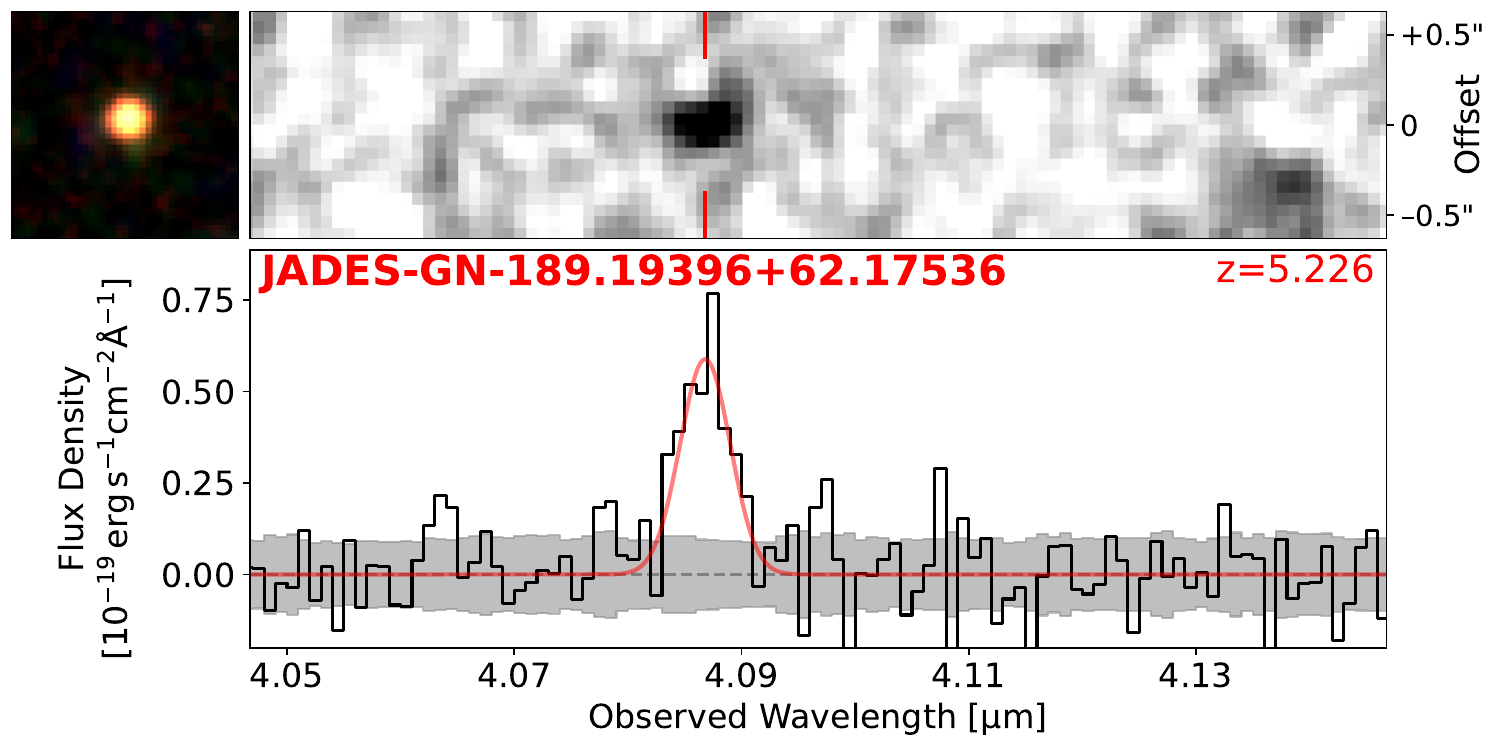}
\caption{Continued.} 
 \end{figure*} 

 \addtocounter{figure}{-1} 
 \begin{figure*}[!ht] 
 \centering
\includegraphics[width=0.49\linewidth]{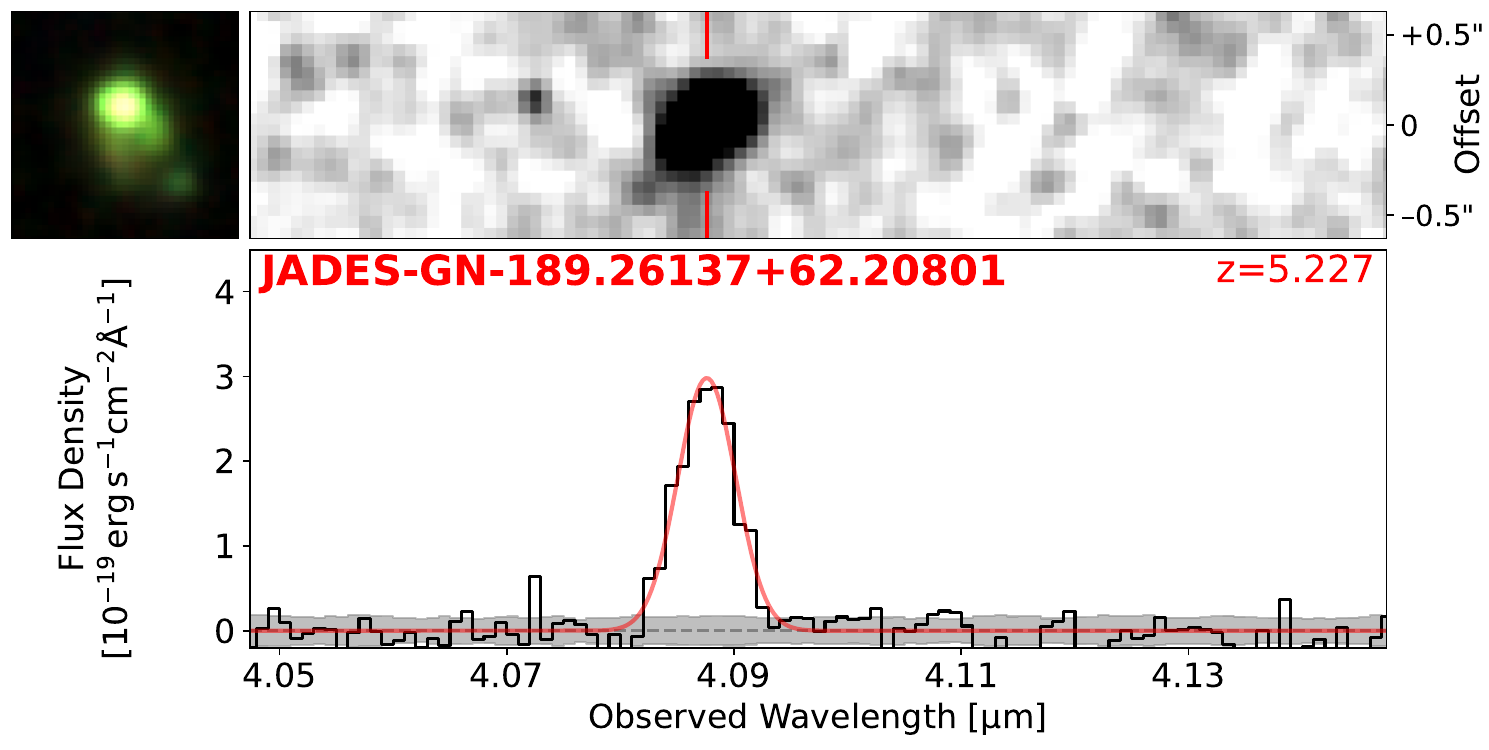}
\includegraphics[width=0.49\linewidth]{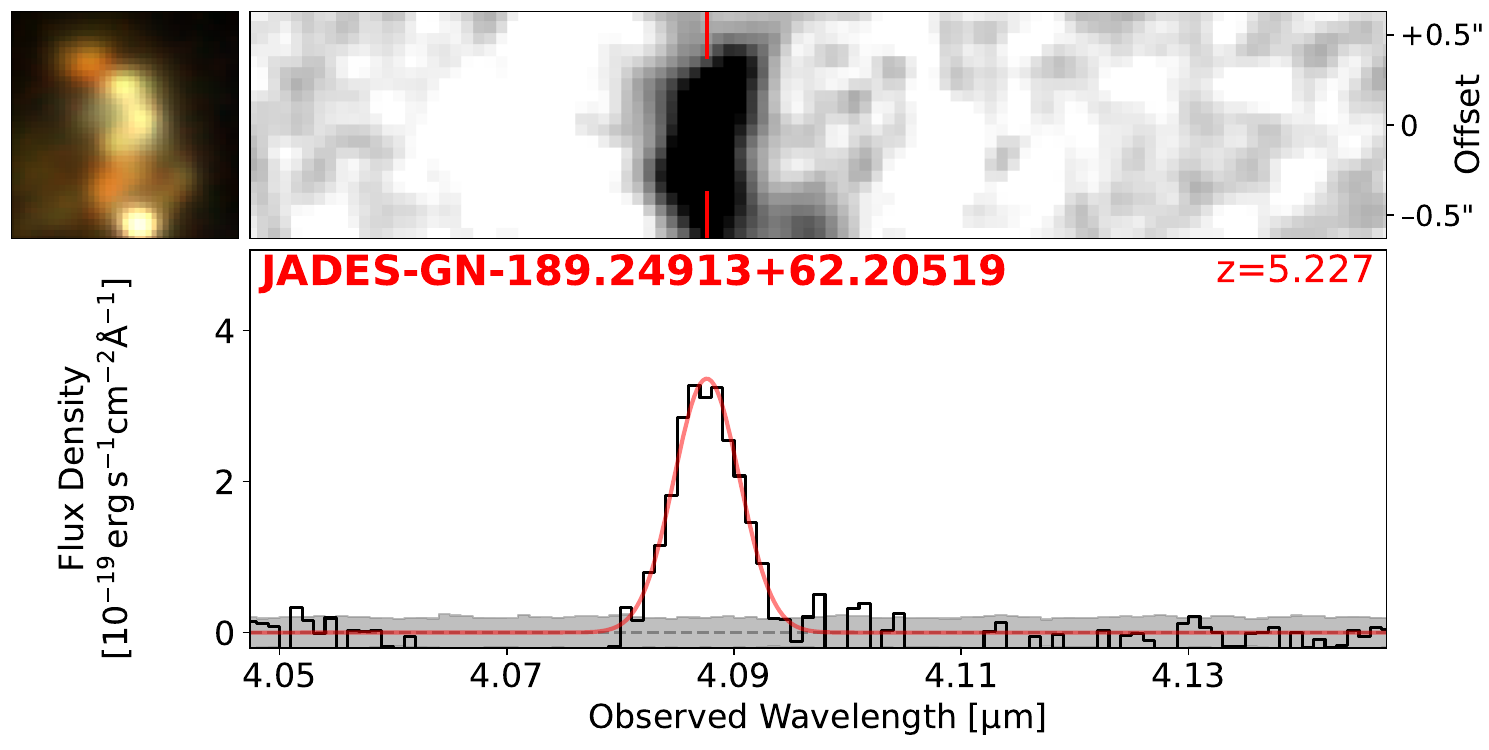}
\includegraphics[width=0.49\linewidth]{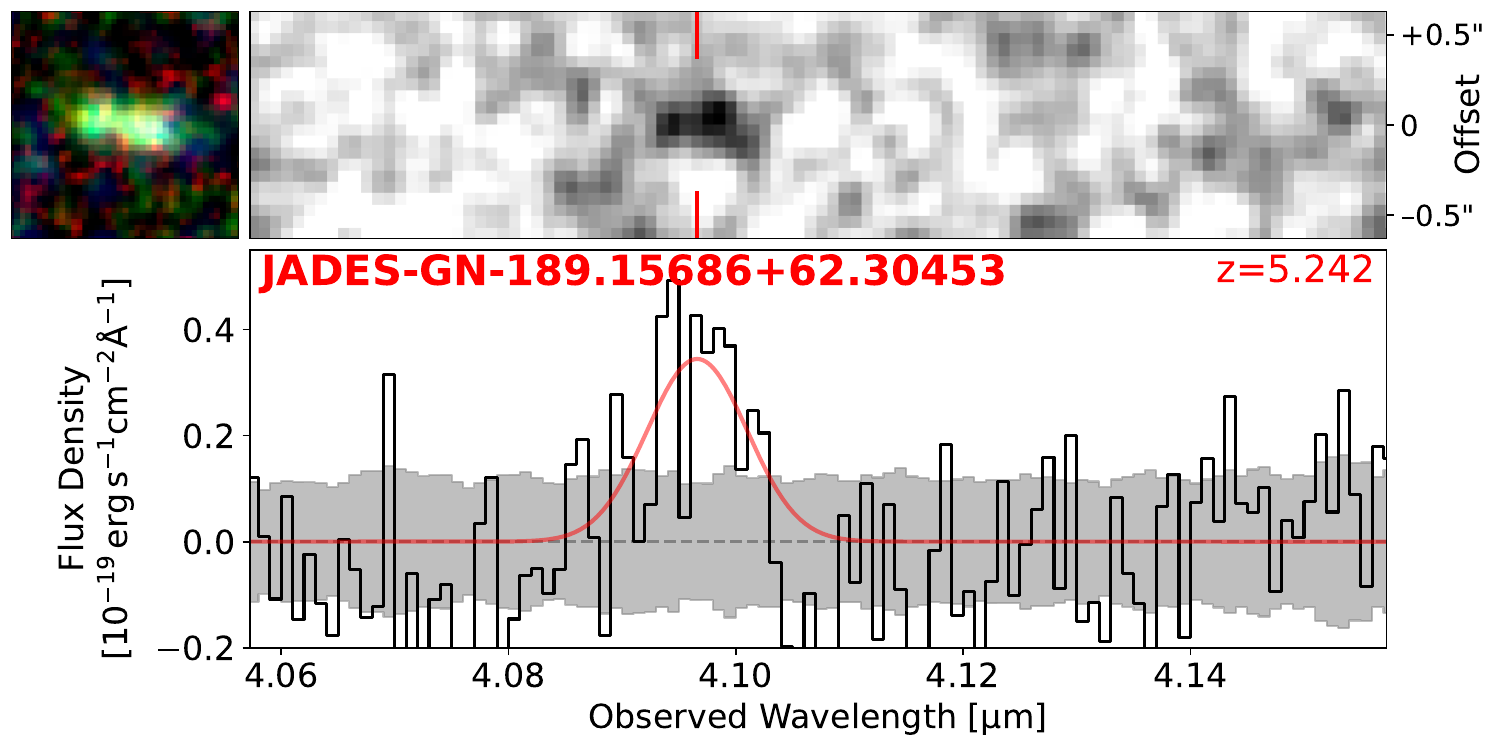}
\includegraphics[width=0.49\linewidth]{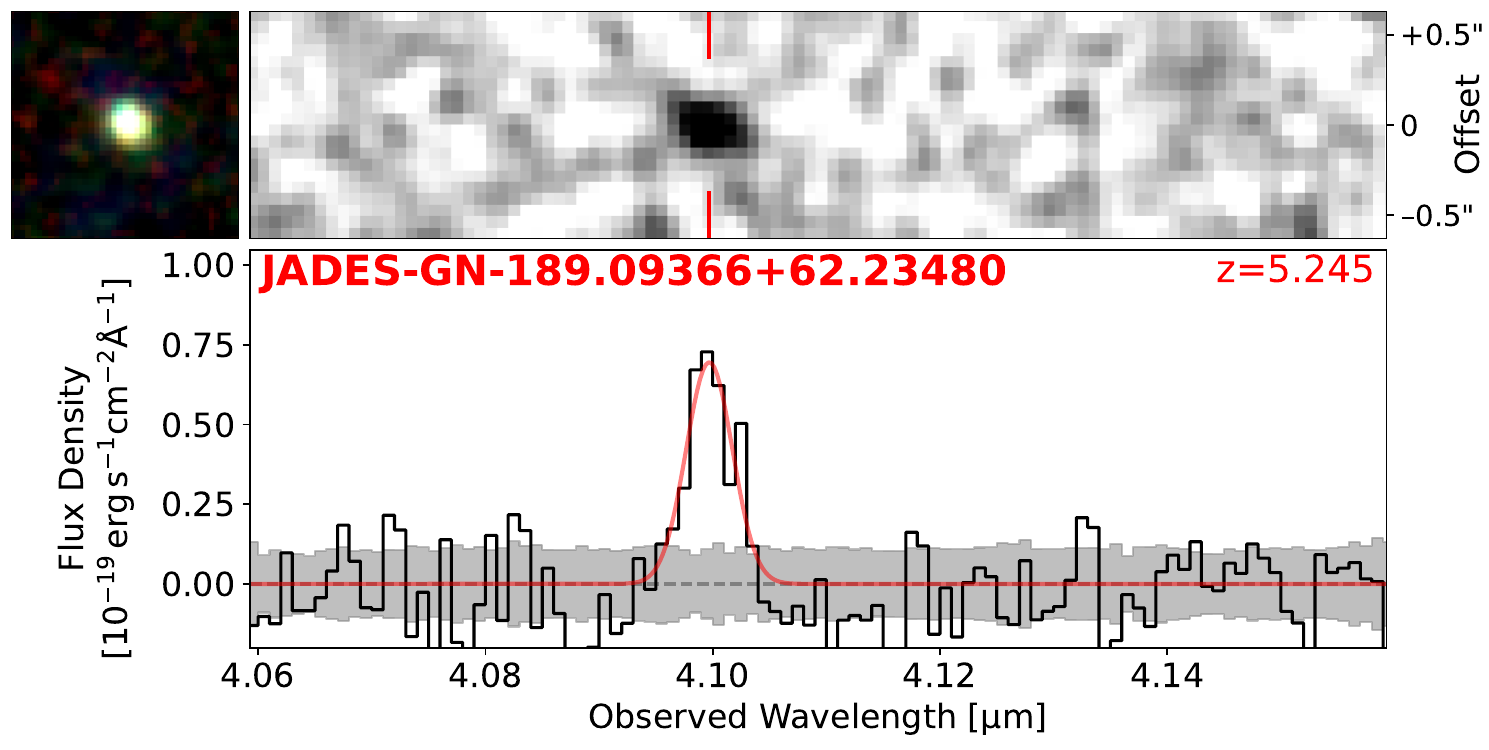}
\includegraphics[width=0.49\linewidth]{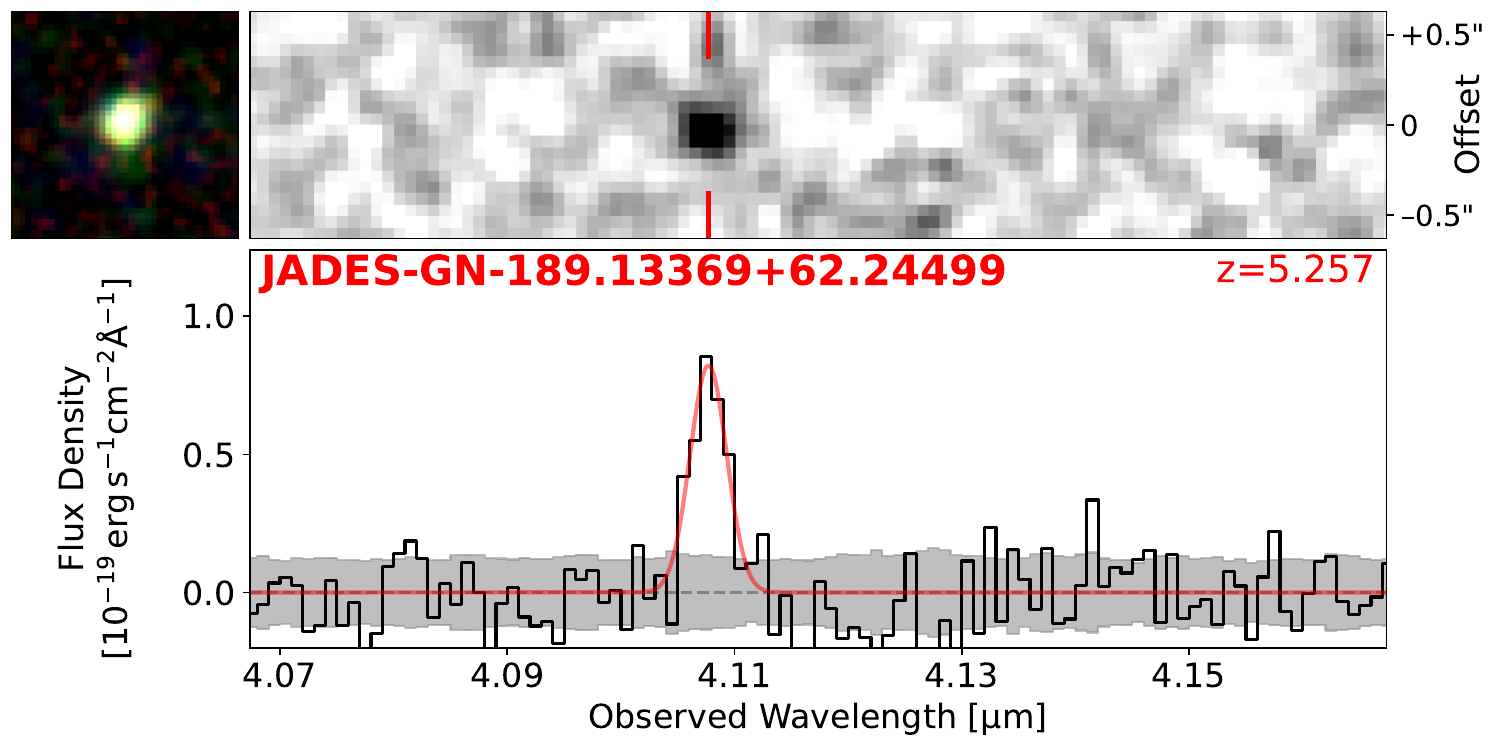}
\includegraphics[width=0.49\linewidth]{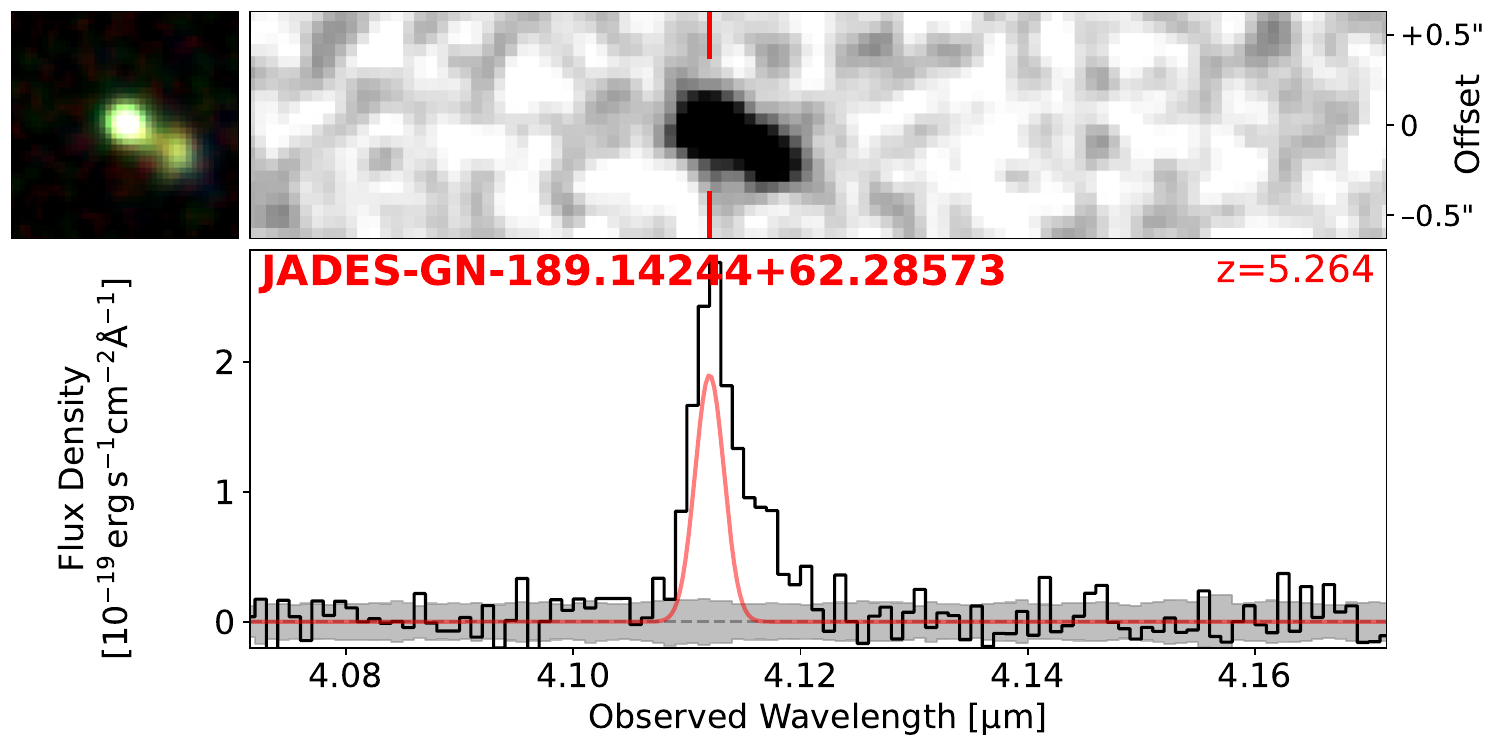}
\includegraphics[width=0.49\linewidth]{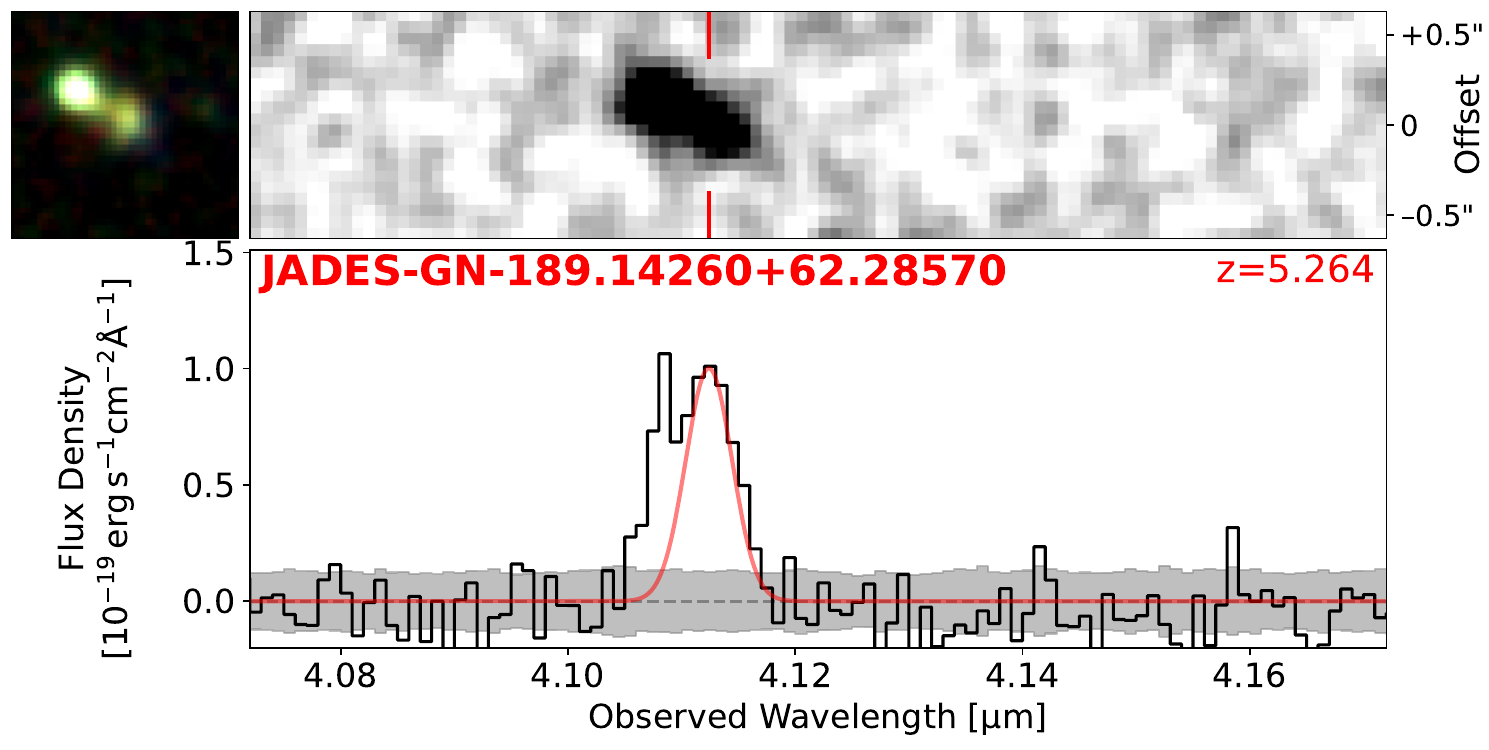}
\includegraphics[width=0.49\linewidth]{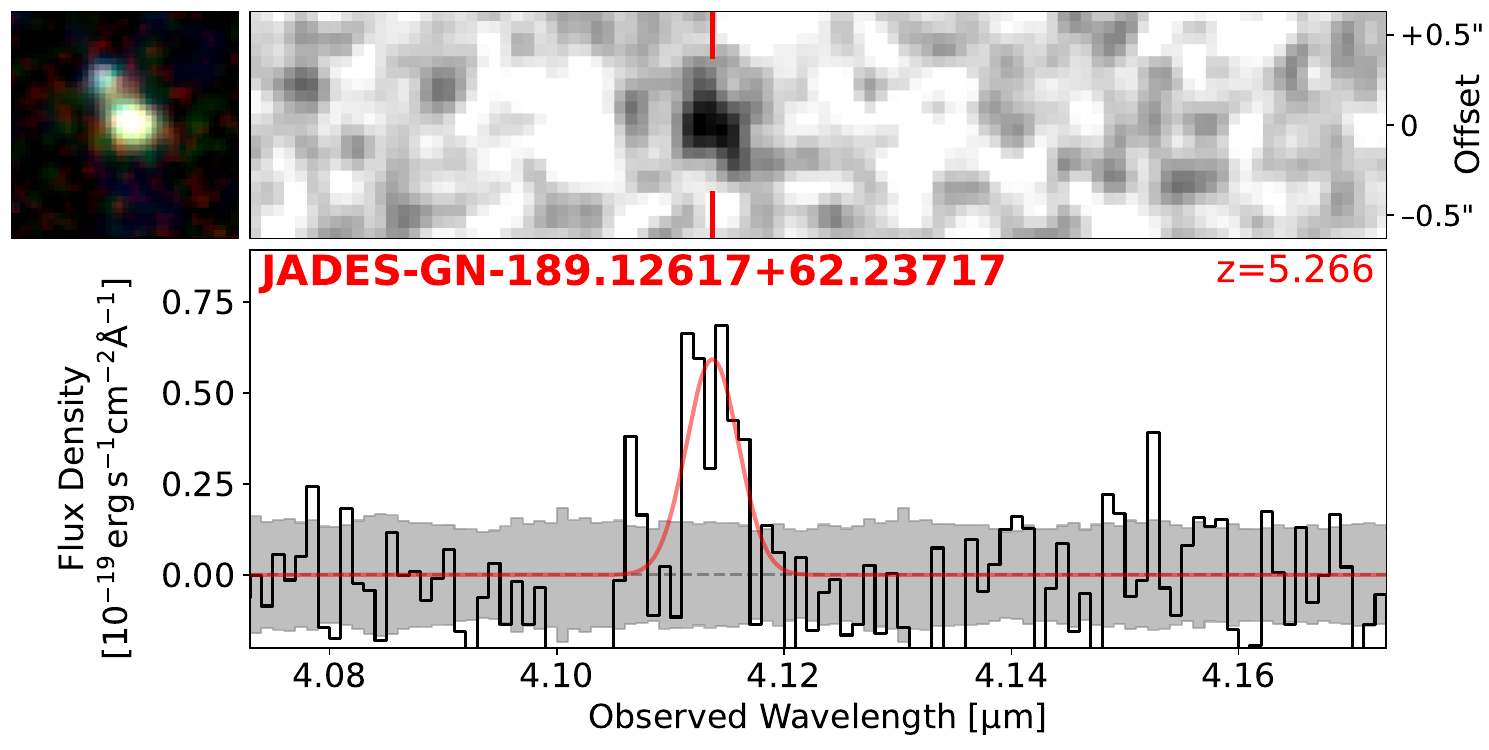}
\includegraphics[width=0.49\linewidth]{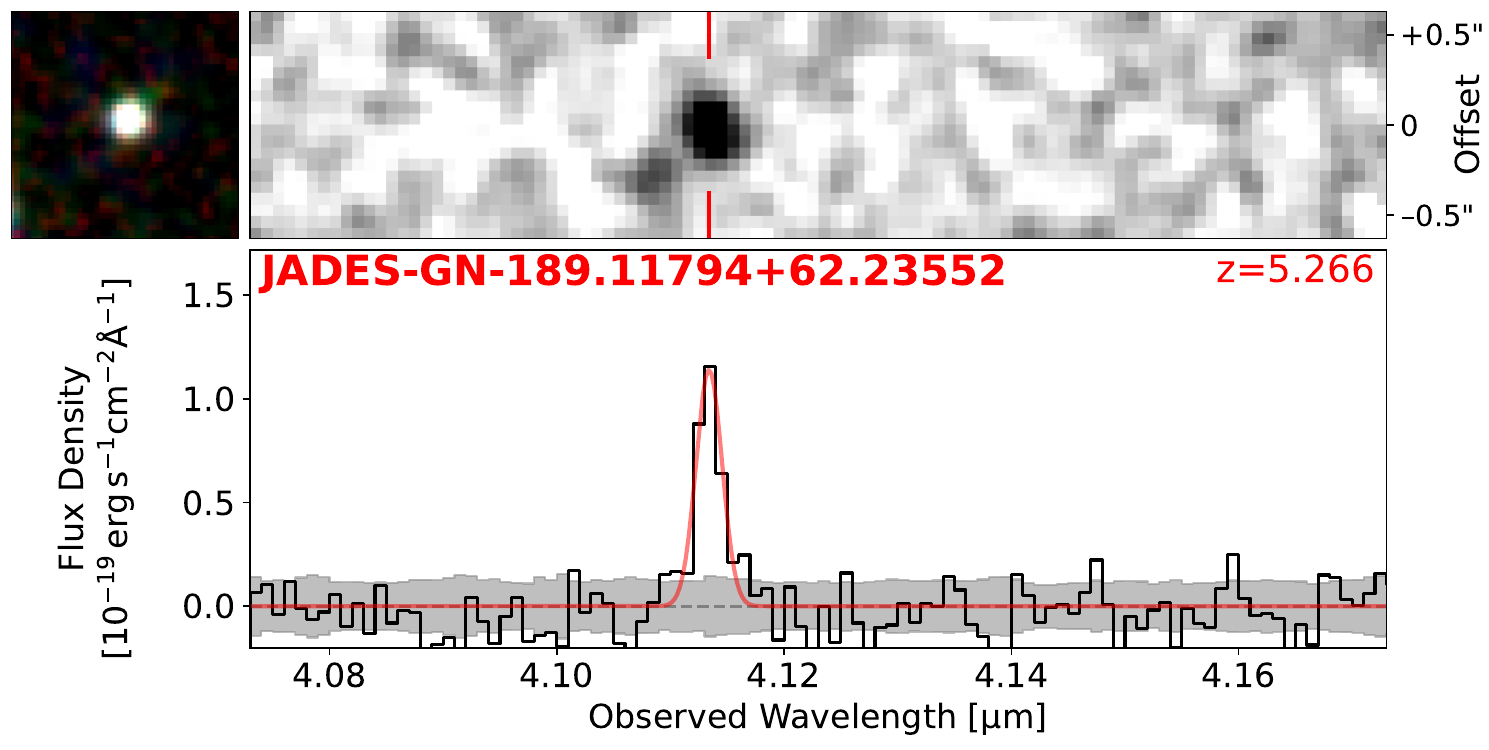}
\includegraphics[width=0.49\linewidth]{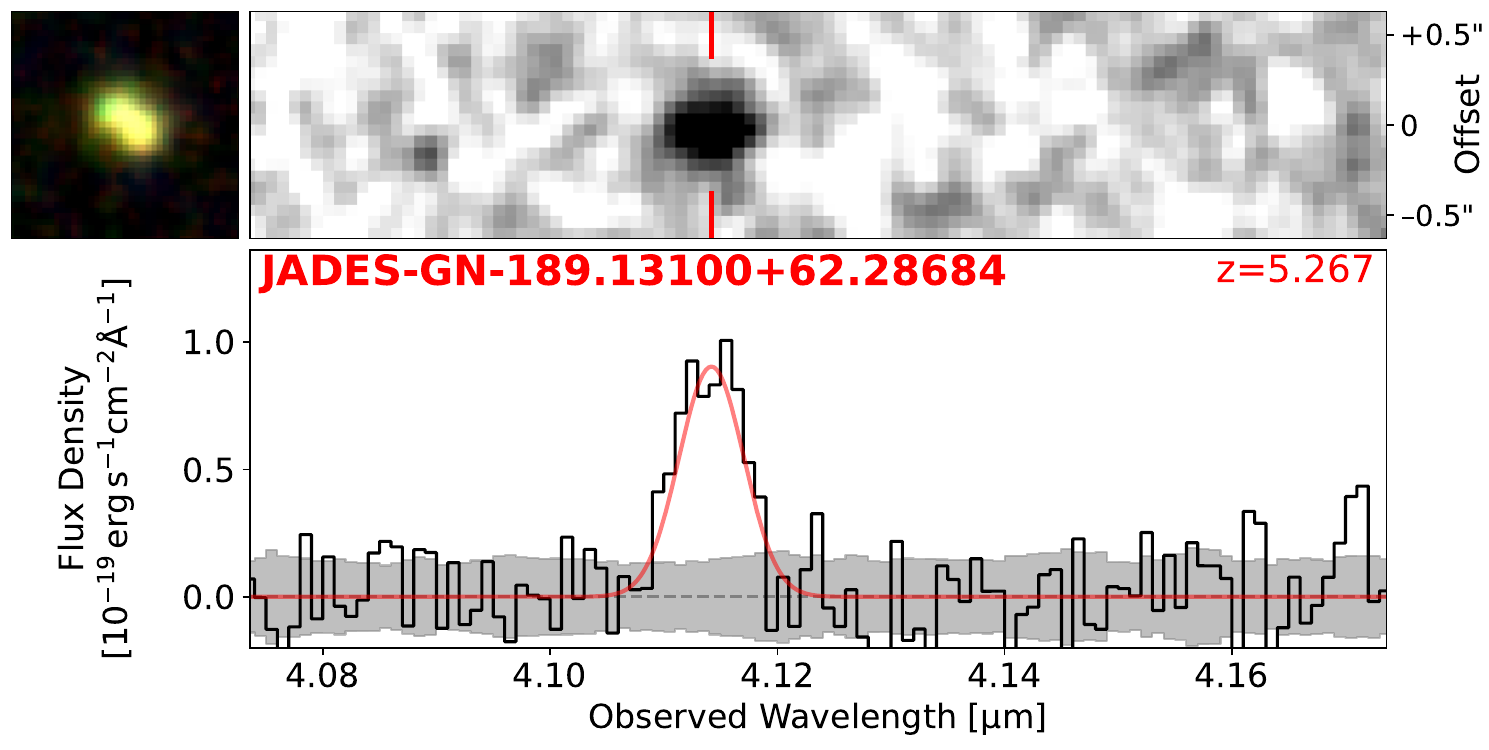}
\caption{Continued.} 
 \end{figure*} 

 \addtocounter{figure}{-1} 
 \begin{figure*}[!ht] 
 \centering
\includegraphics[width=0.49\linewidth]{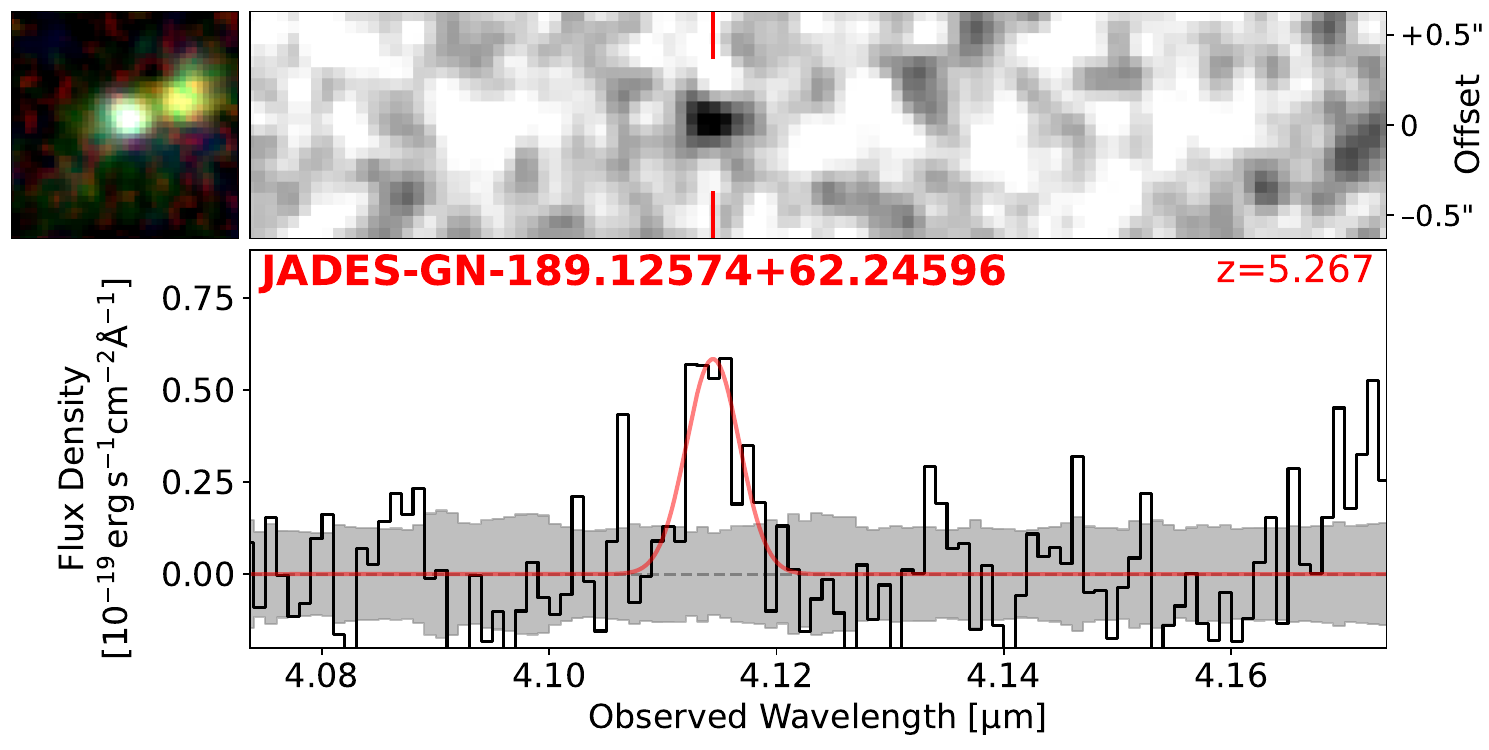}
\includegraphics[width=0.49\linewidth]{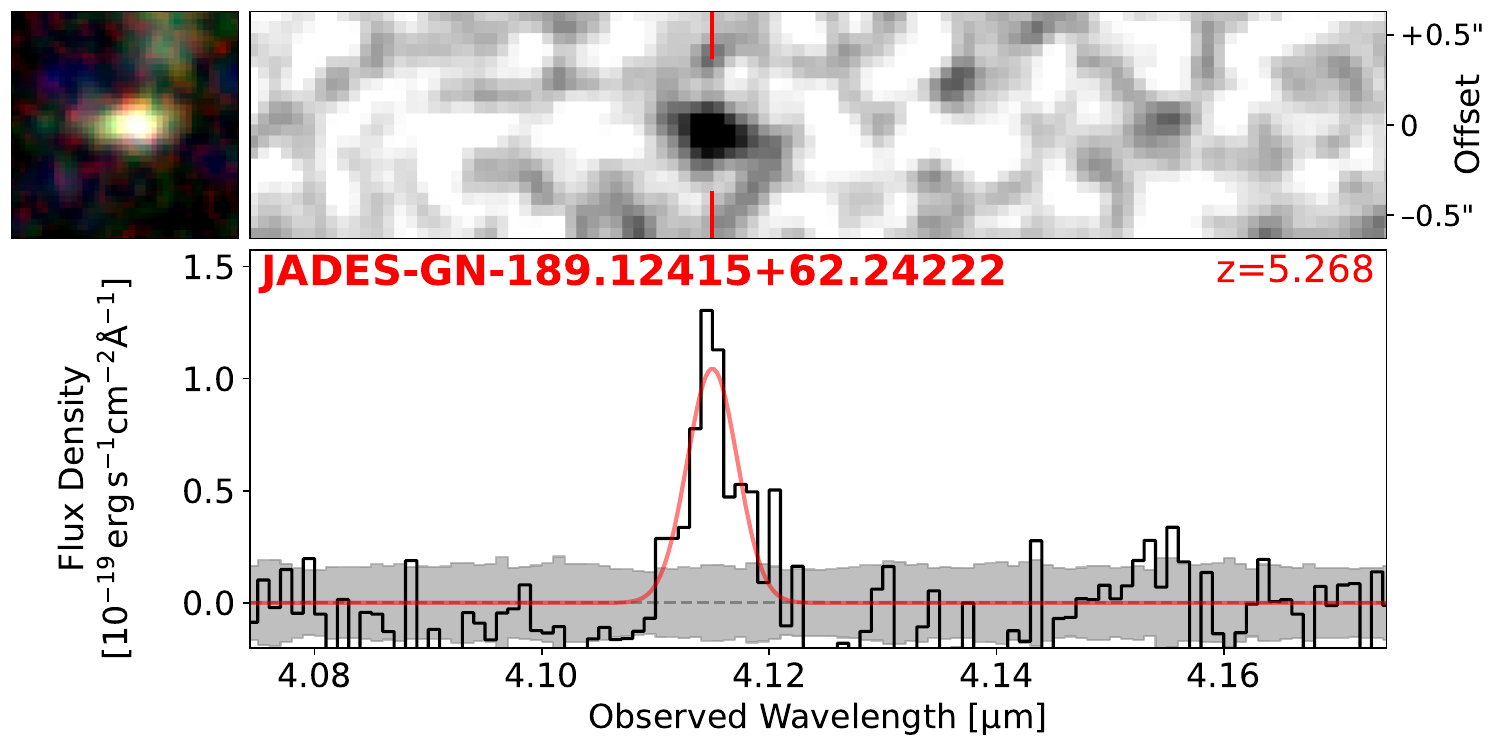}
\includegraphics[width=0.49\linewidth]{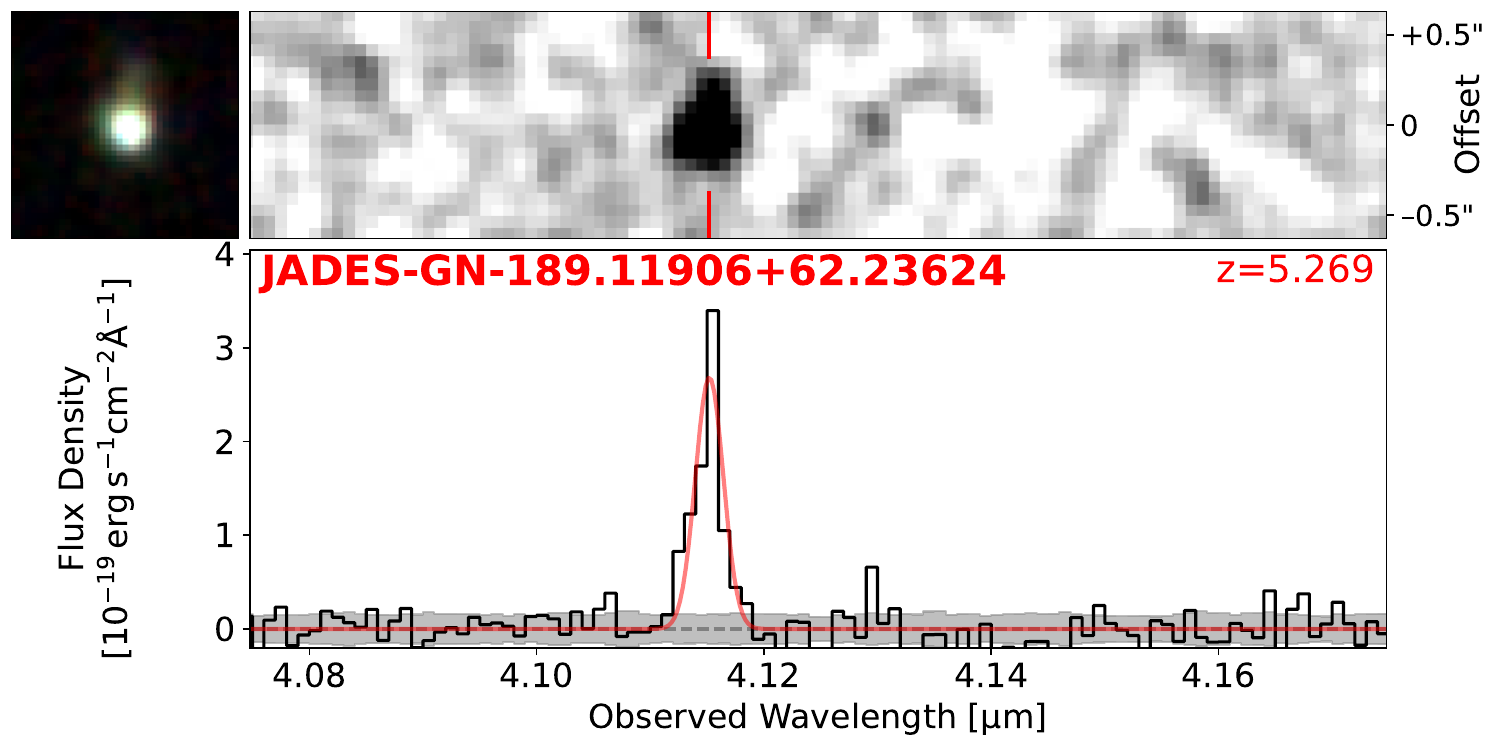}
\includegraphics[width=0.49\linewidth]{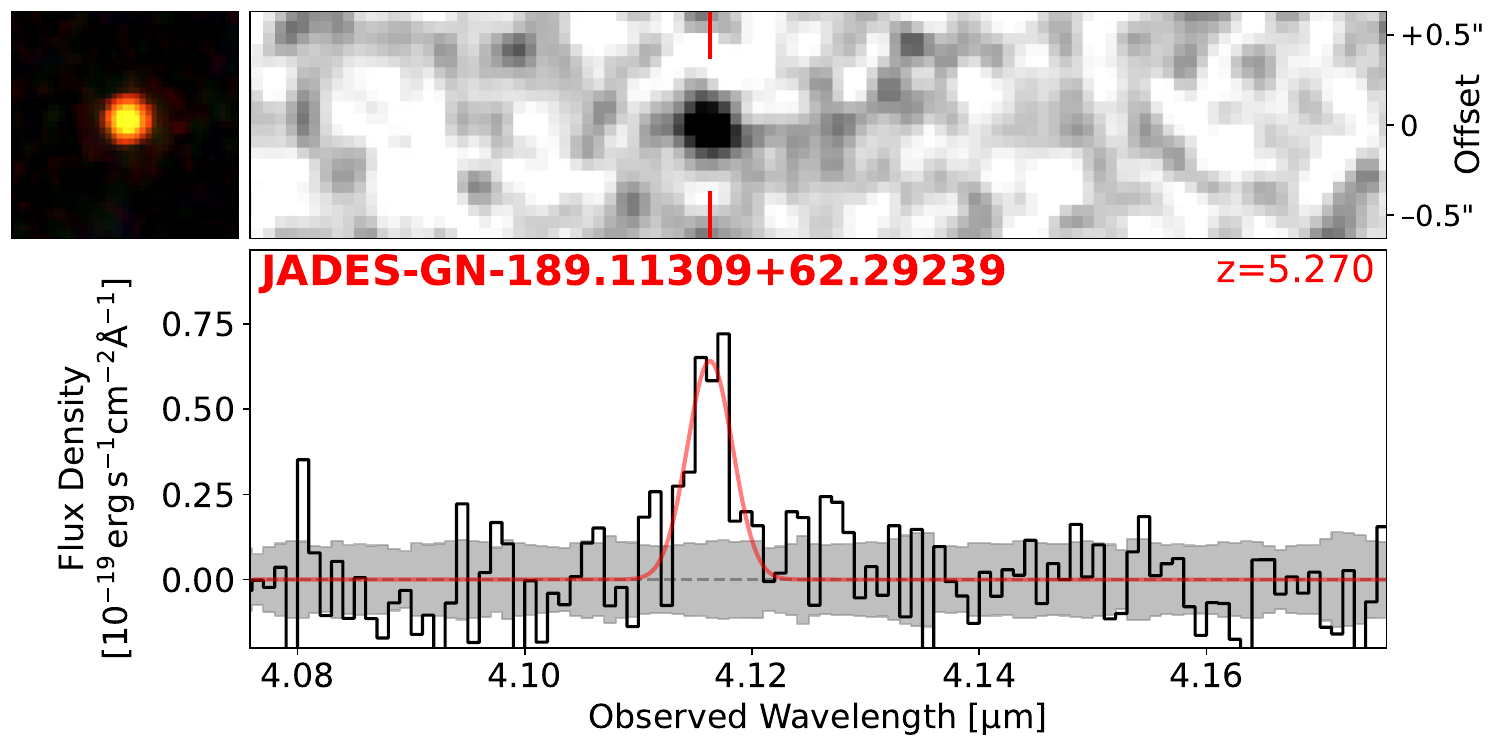}
\includegraphics[width=0.49\linewidth]{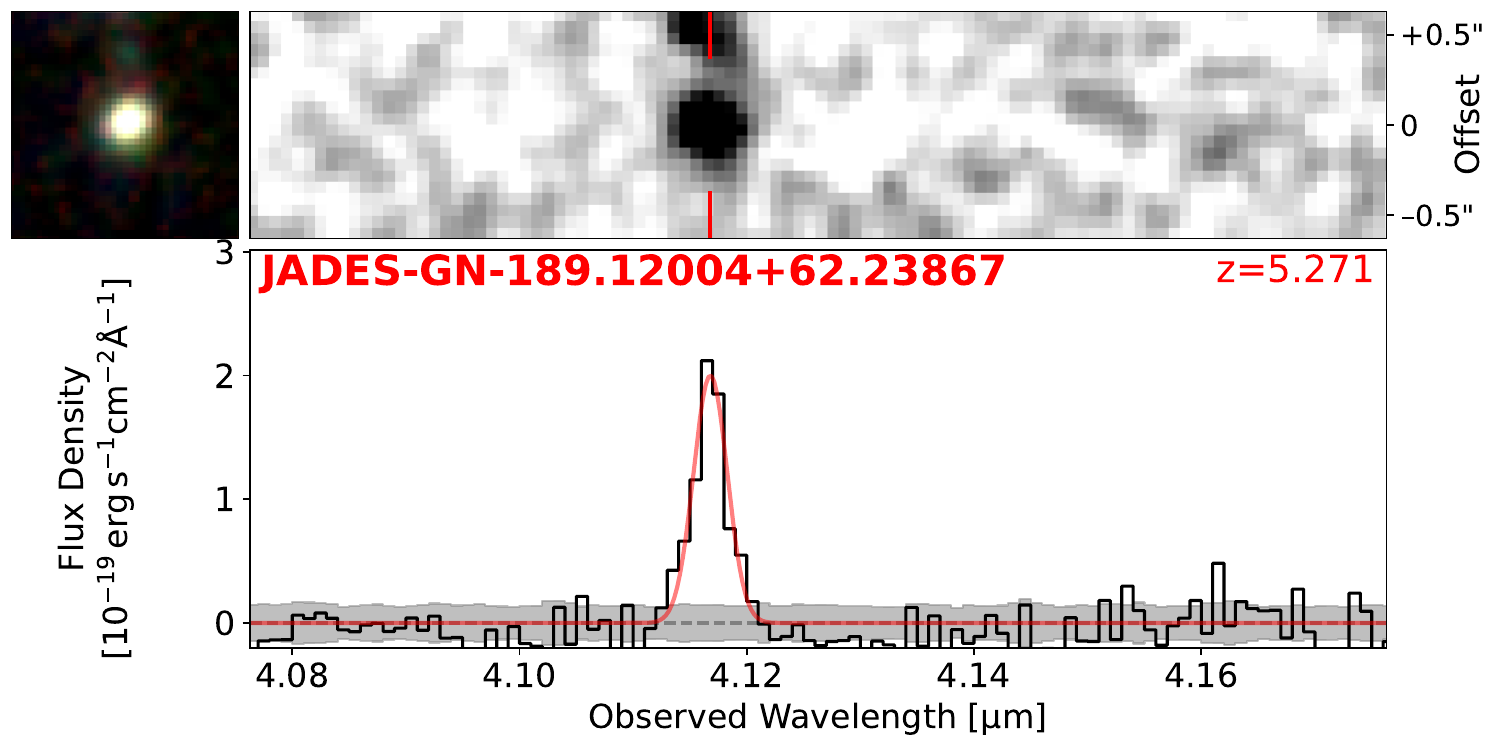}
\includegraphics[width=0.49\linewidth]{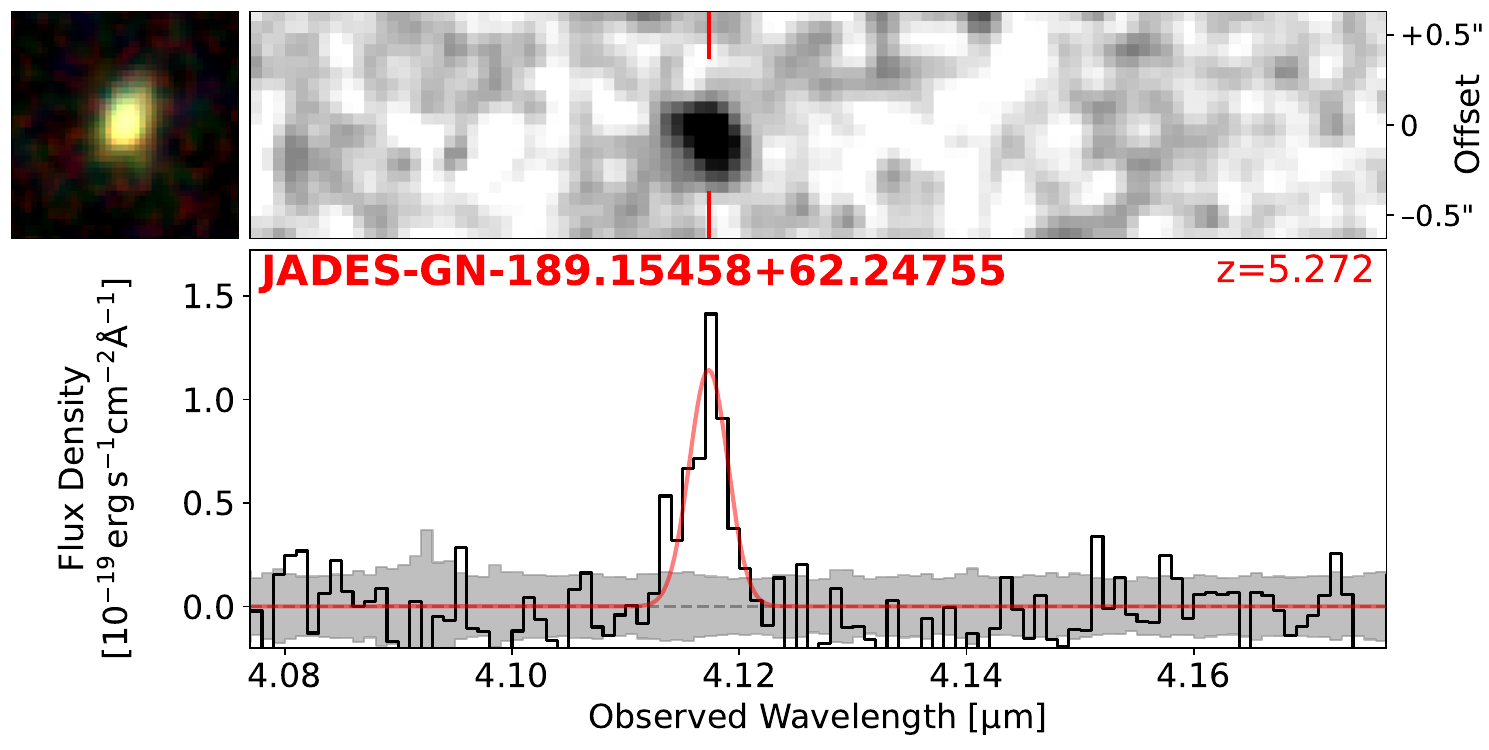}
\includegraphics[width=0.49\linewidth]{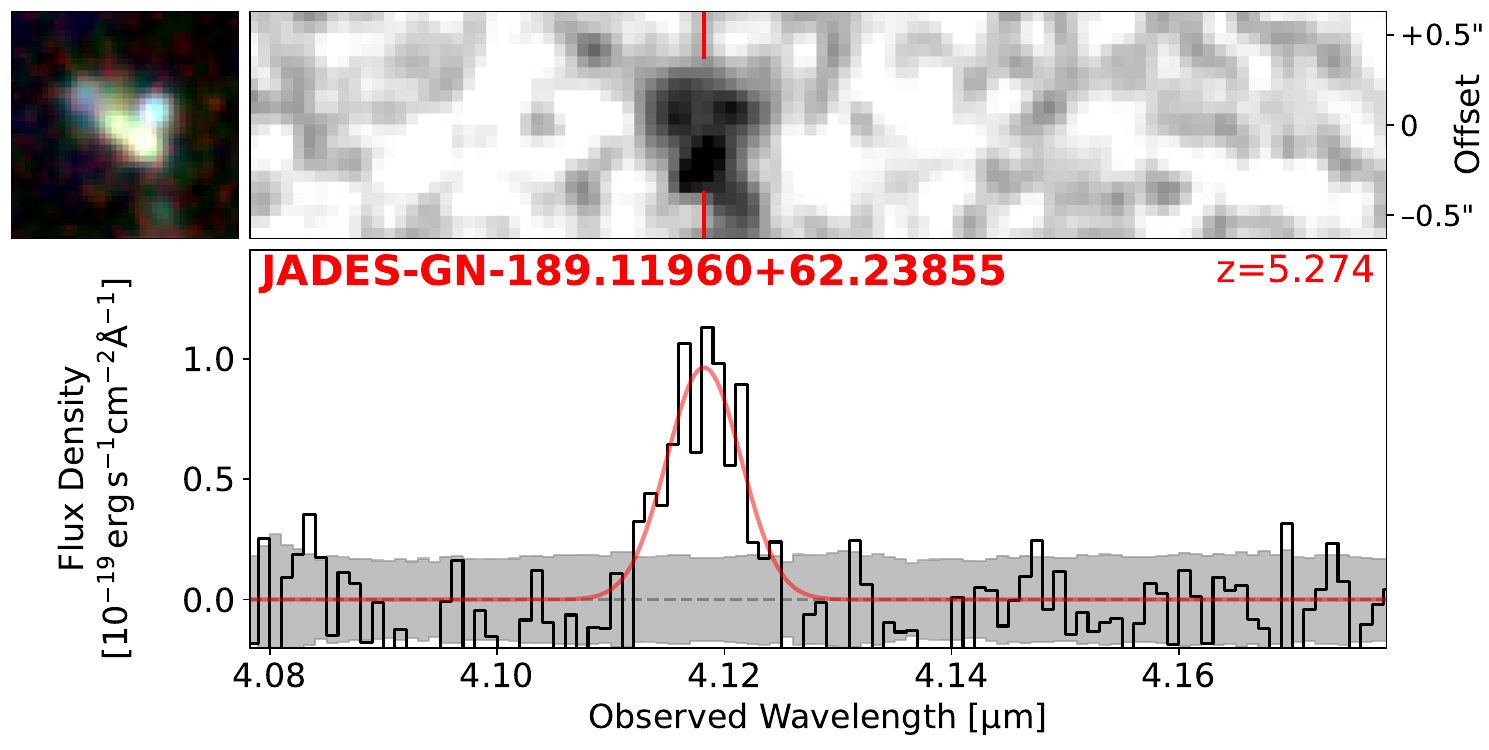}
\includegraphics[width=0.49\linewidth]{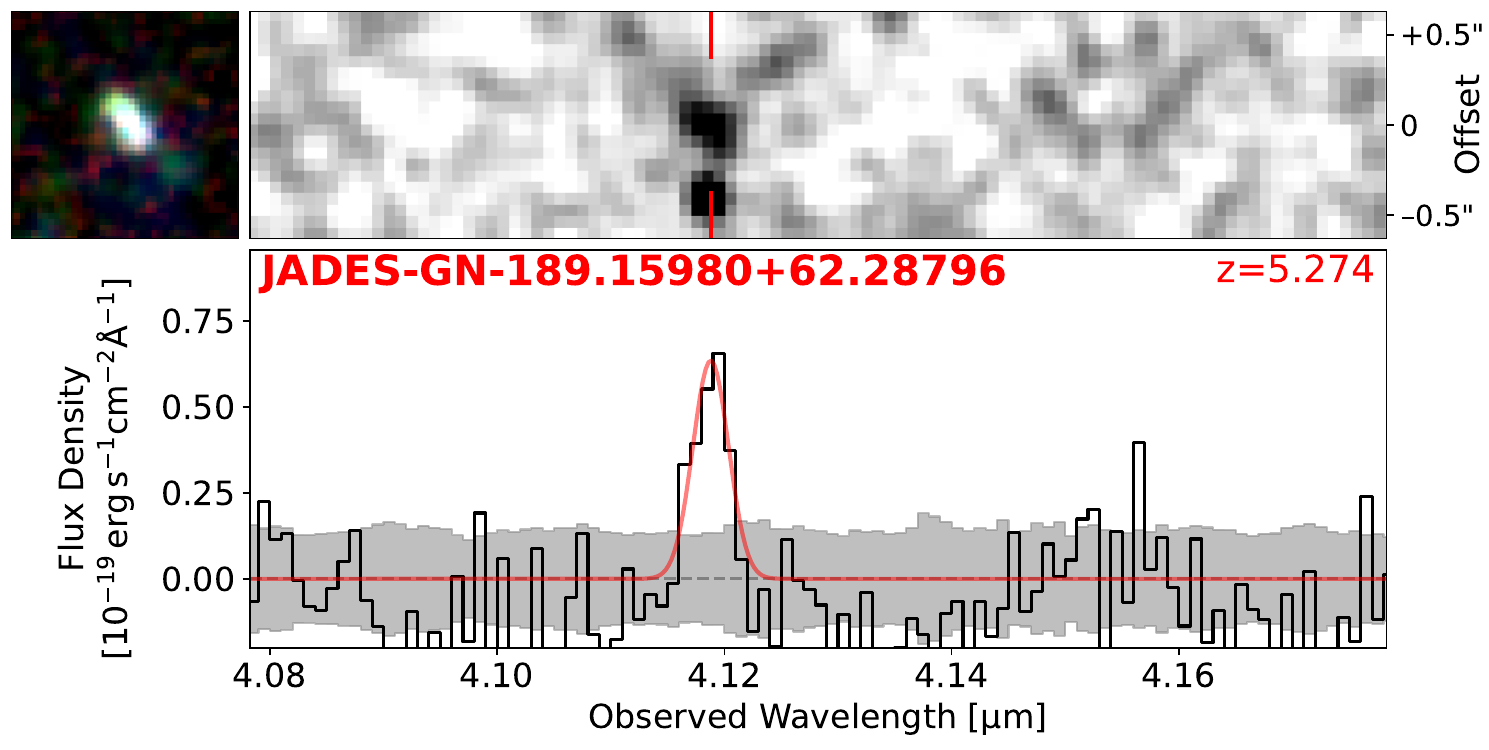}
\includegraphics[width=0.49\linewidth]{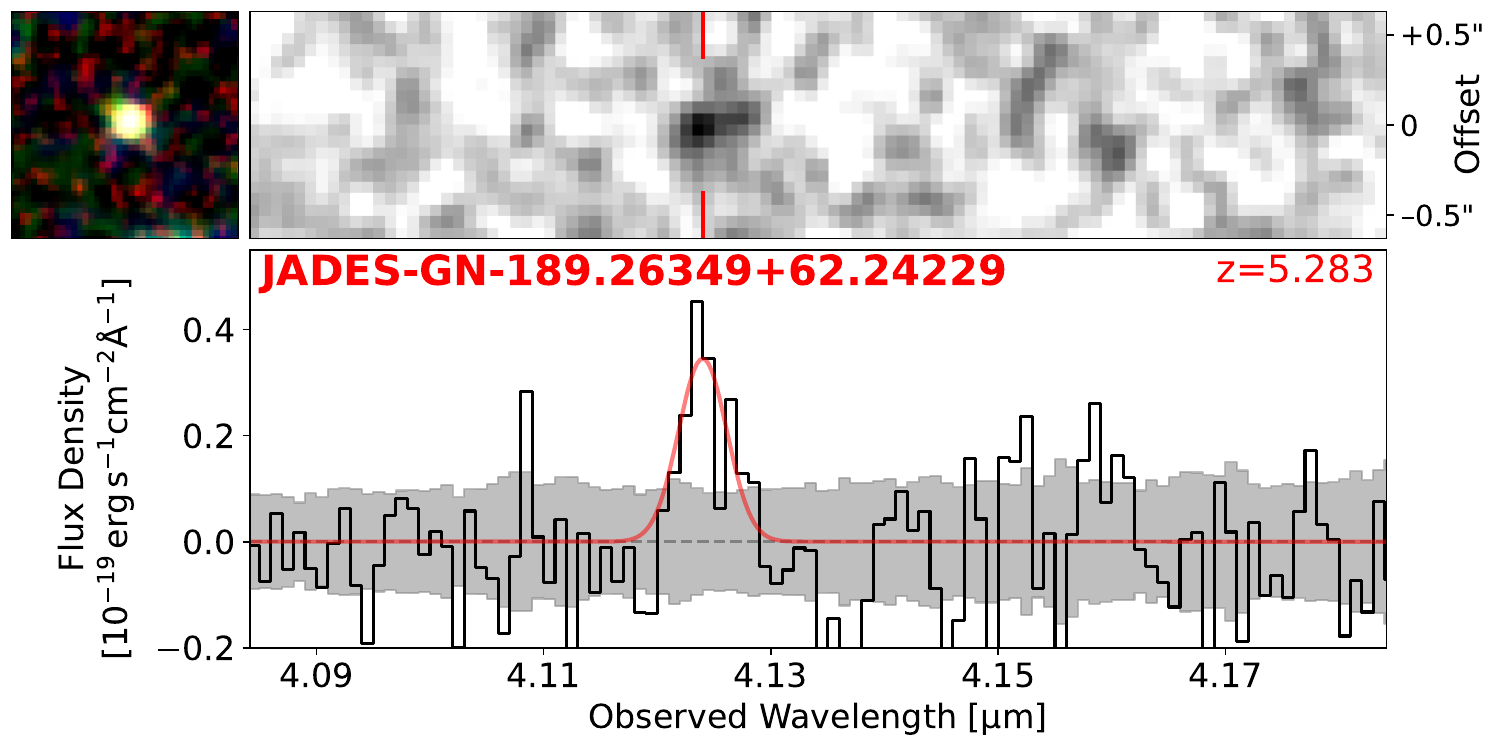}
\includegraphics[width=0.49\linewidth]{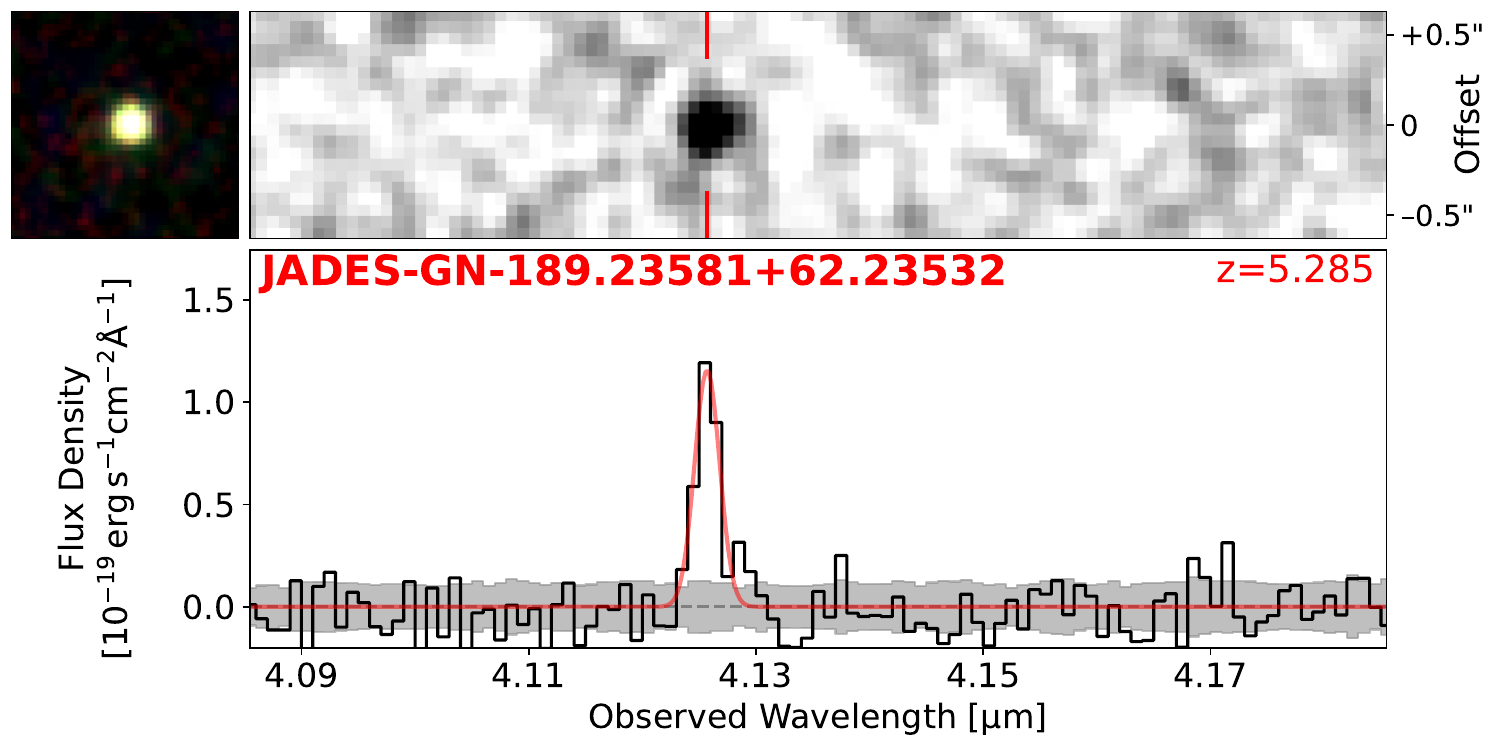}
\caption{Continued.} 
 \end{figure*} 

 \addtocounter{figure}{-1} 
 \begin{figure*}[!ht] 
 \centering
\includegraphics[width=0.49\linewidth]{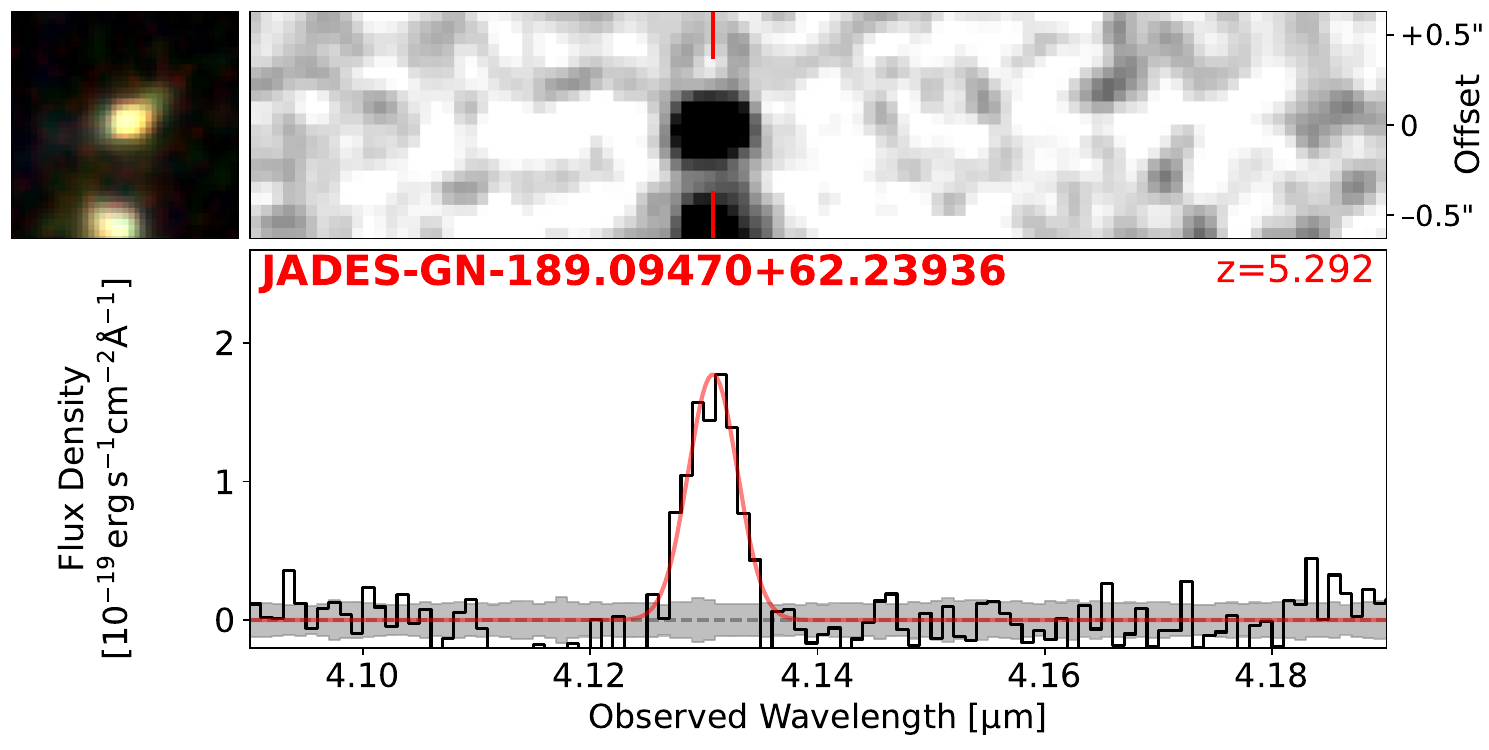}
\includegraphics[width=0.49\linewidth]{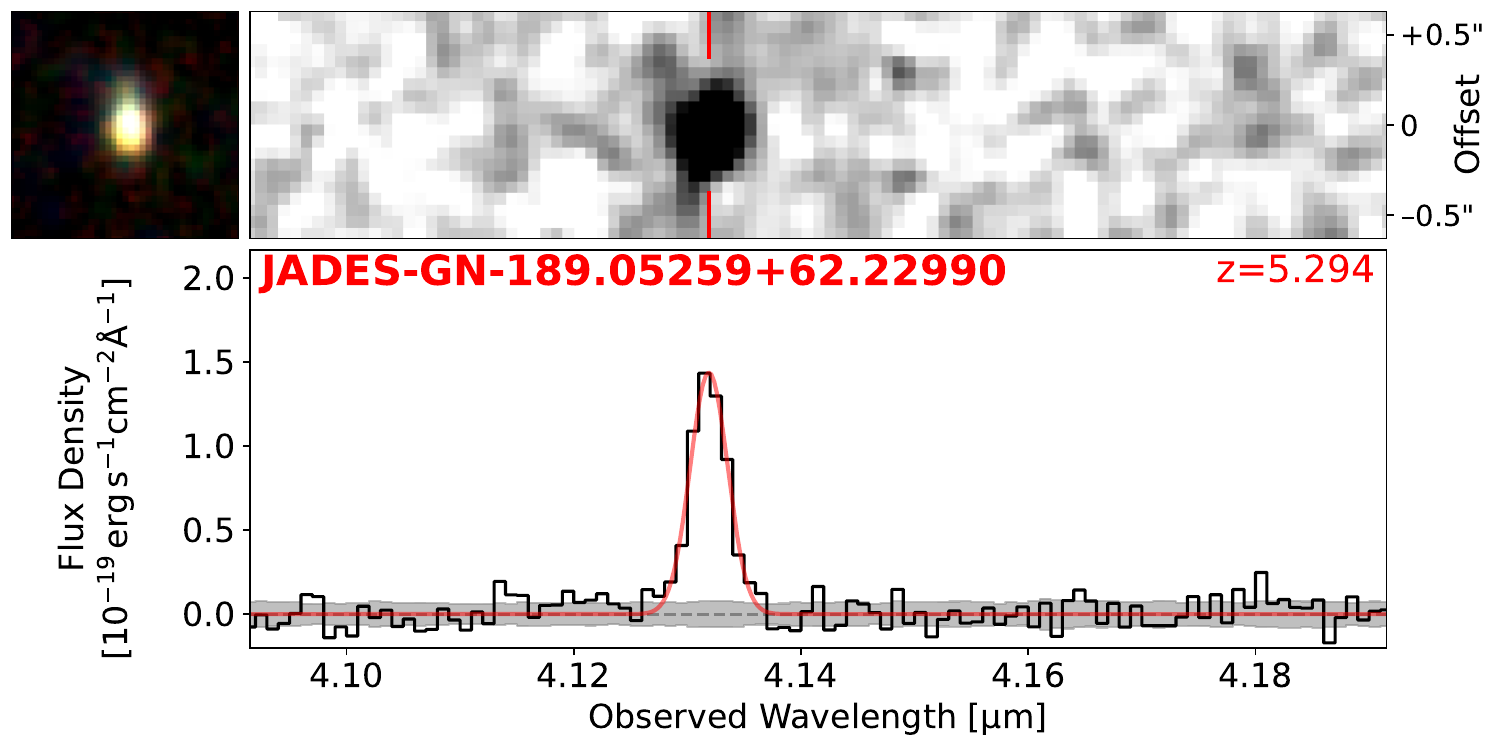}
\includegraphics[width=0.49\linewidth]{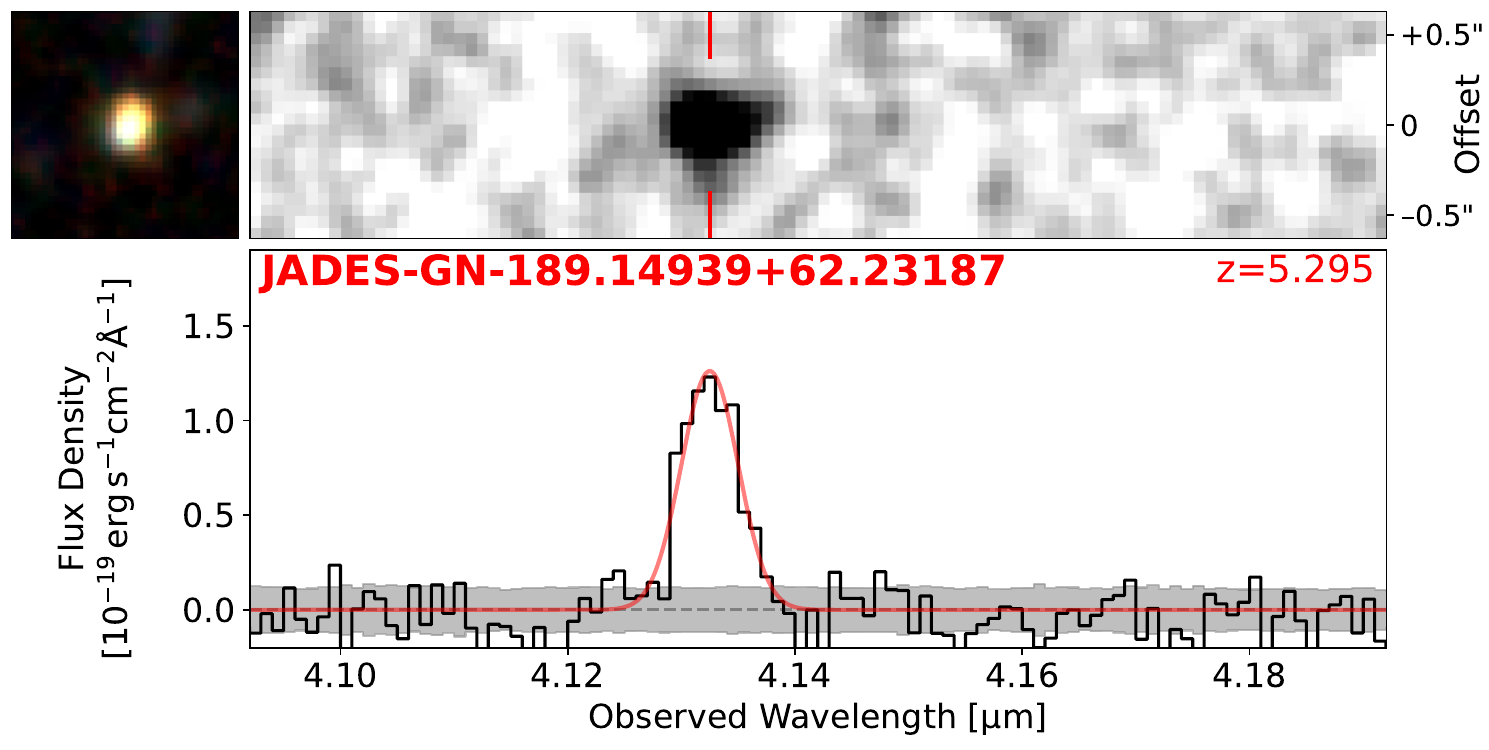}
\includegraphics[width=0.49\linewidth]{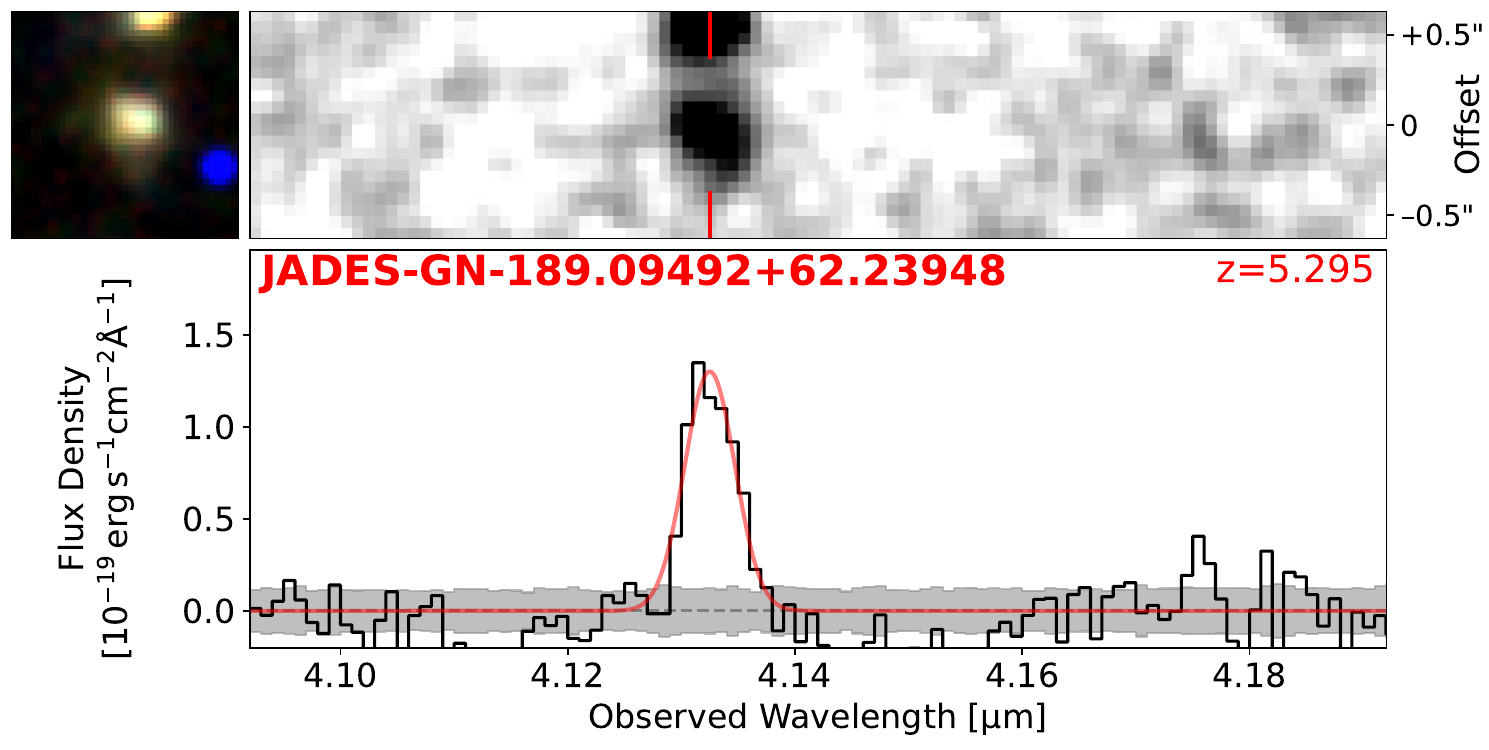}
\includegraphics[width=0.49\linewidth]{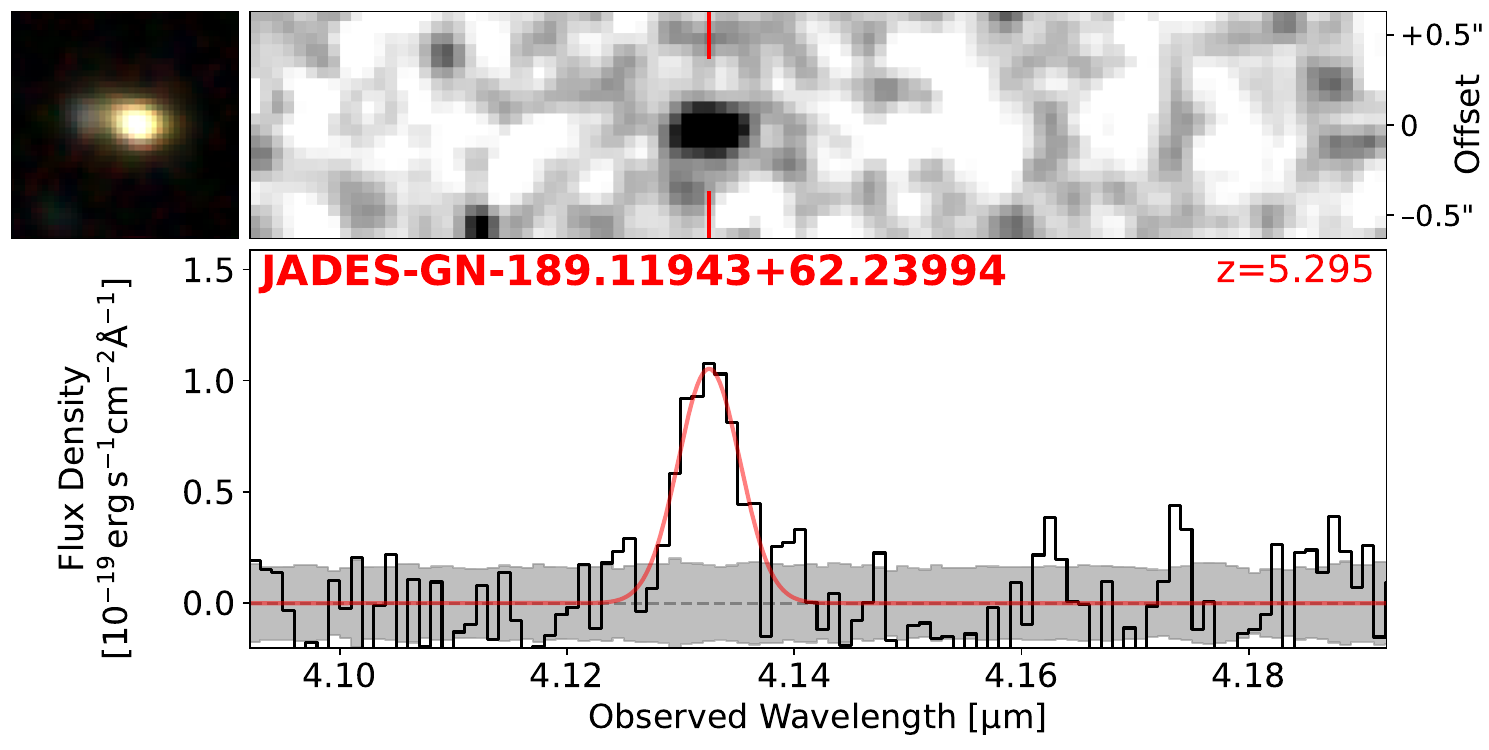}
\includegraphics[width=0.49\linewidth]{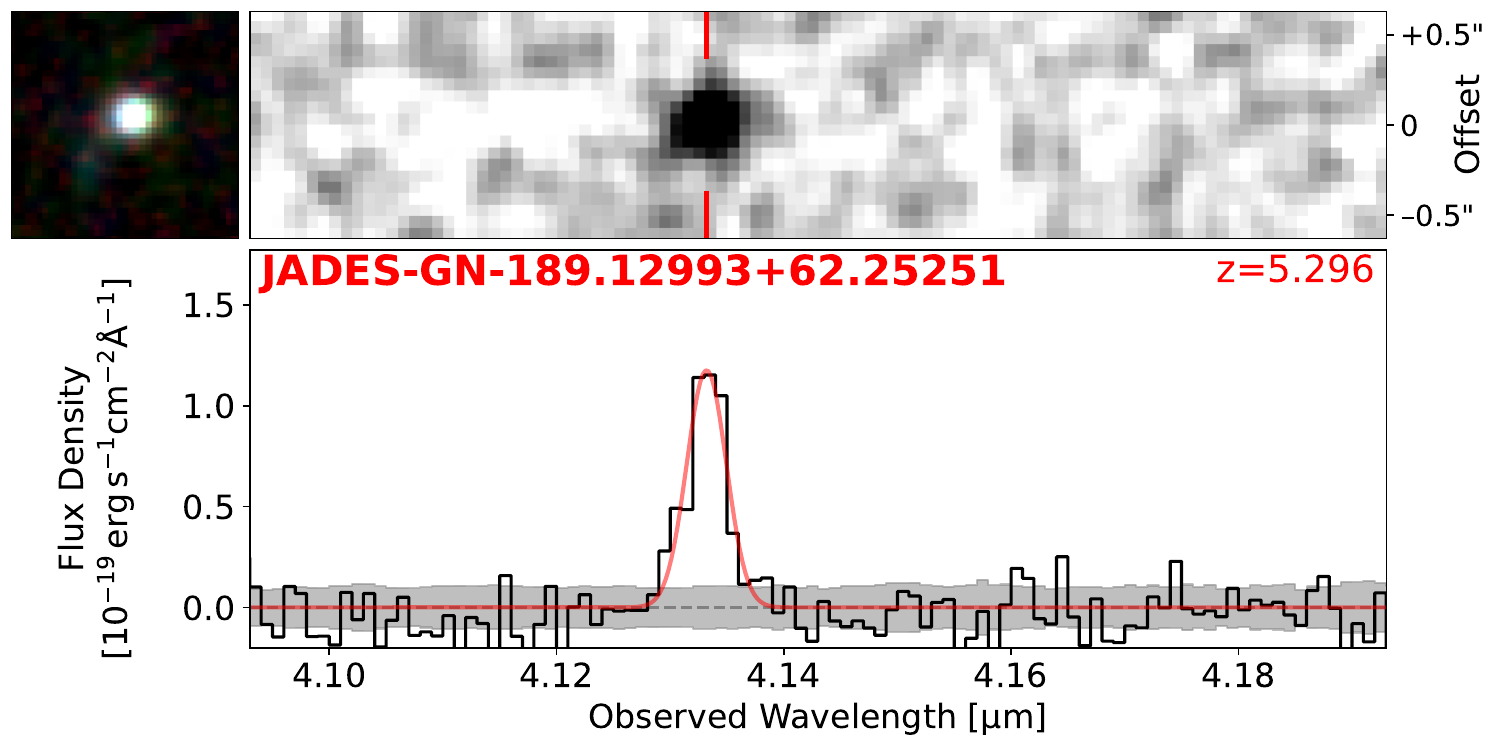}
\includegraphics[width=0.49\linewidth]{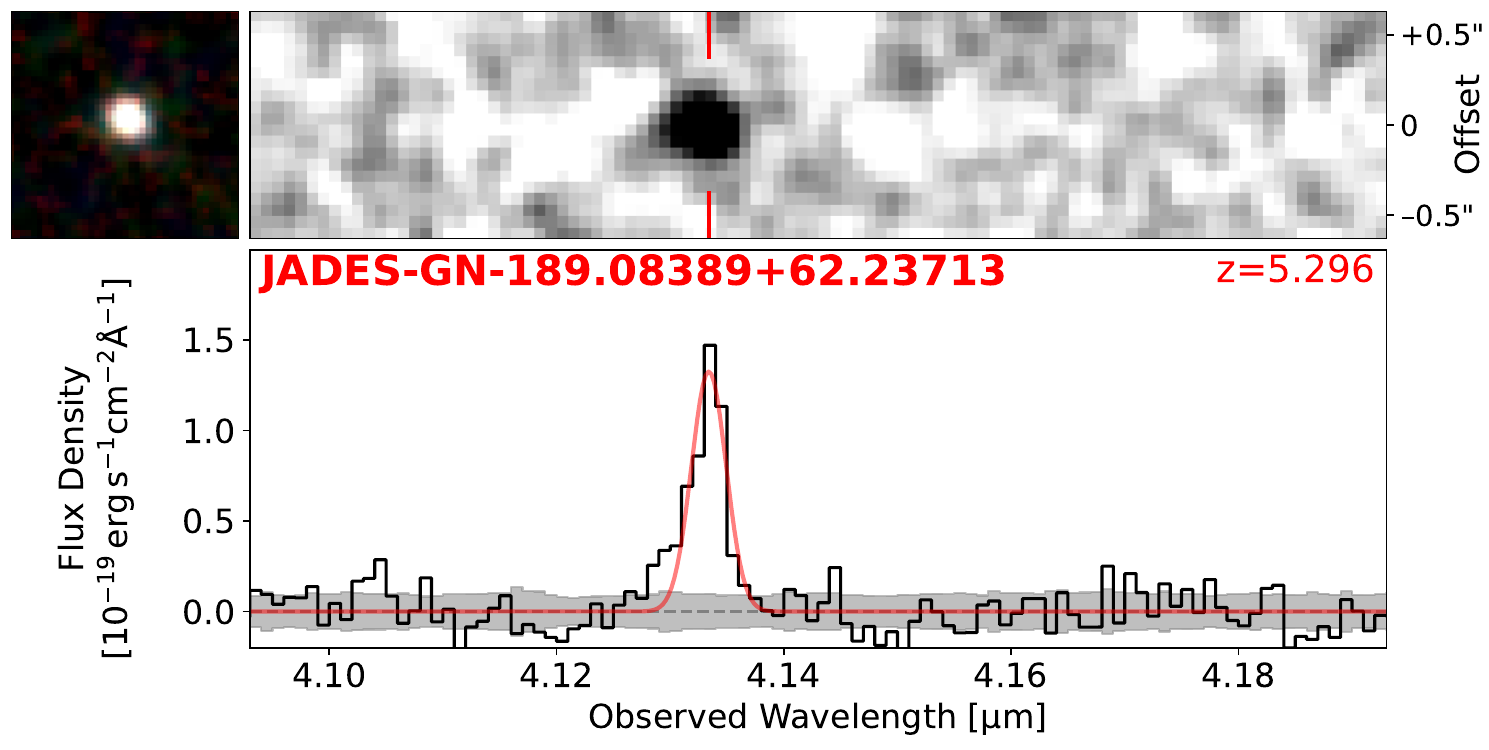}
\includegraphics[width=0.49\linewidth]{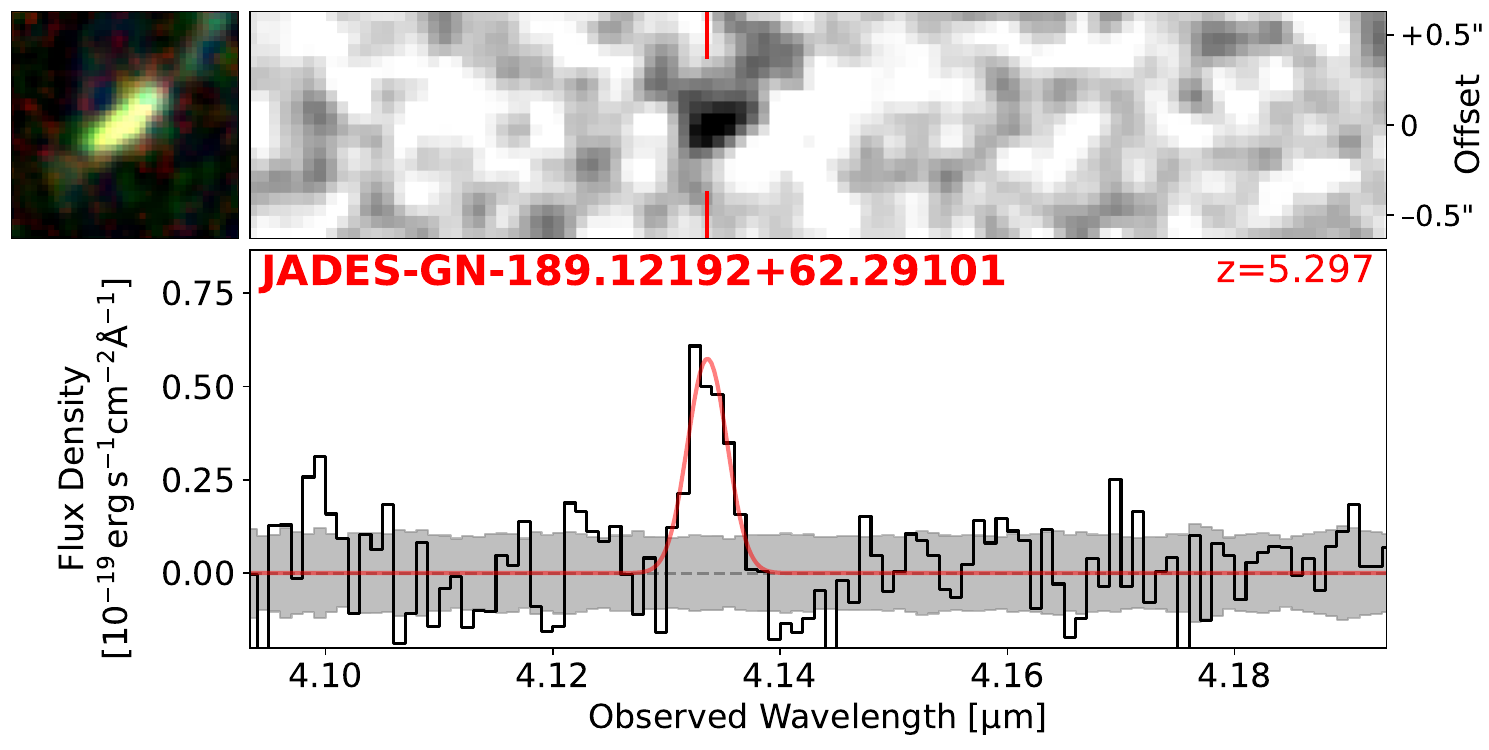}
\includegraphics[width=0.49\linewidth]{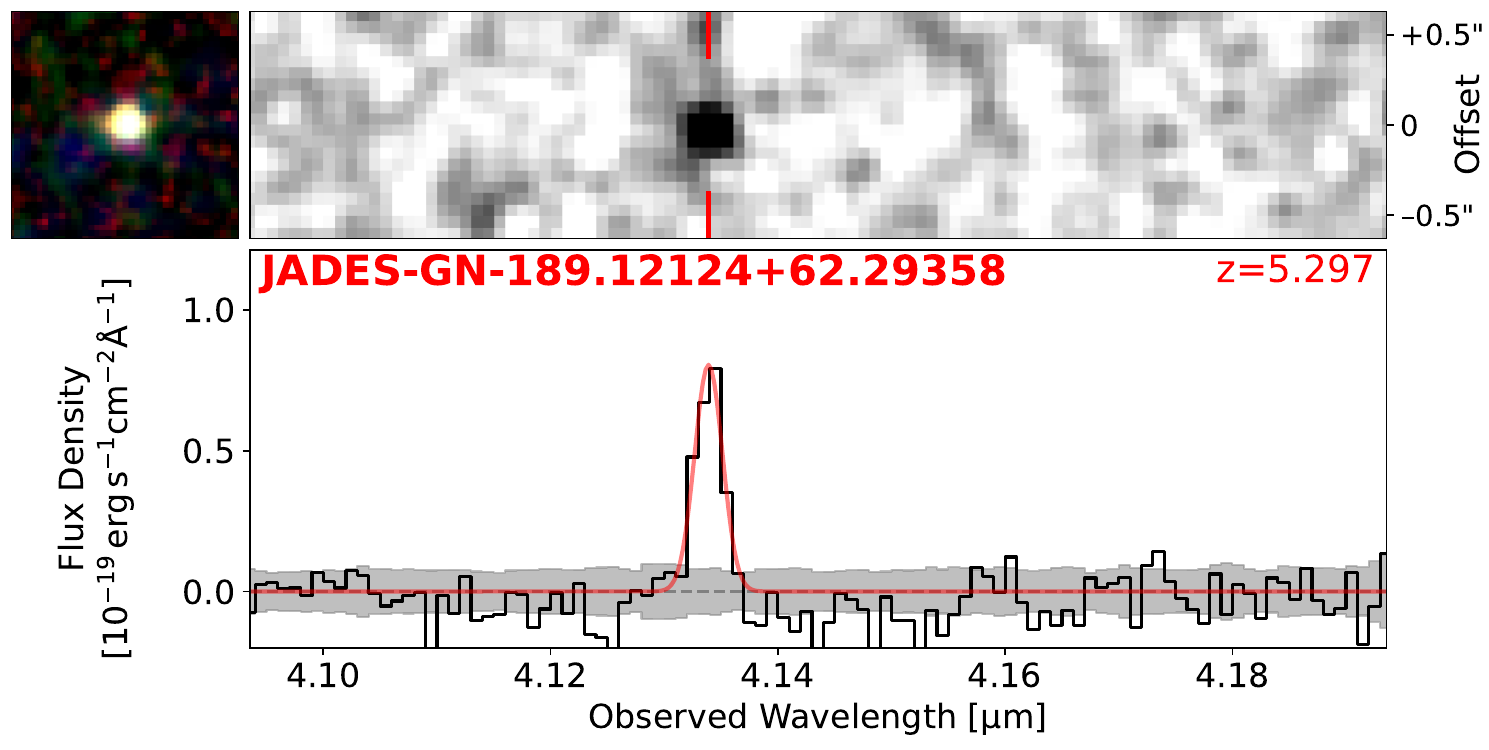}
\includegraphics[width=0.49\linewidth]{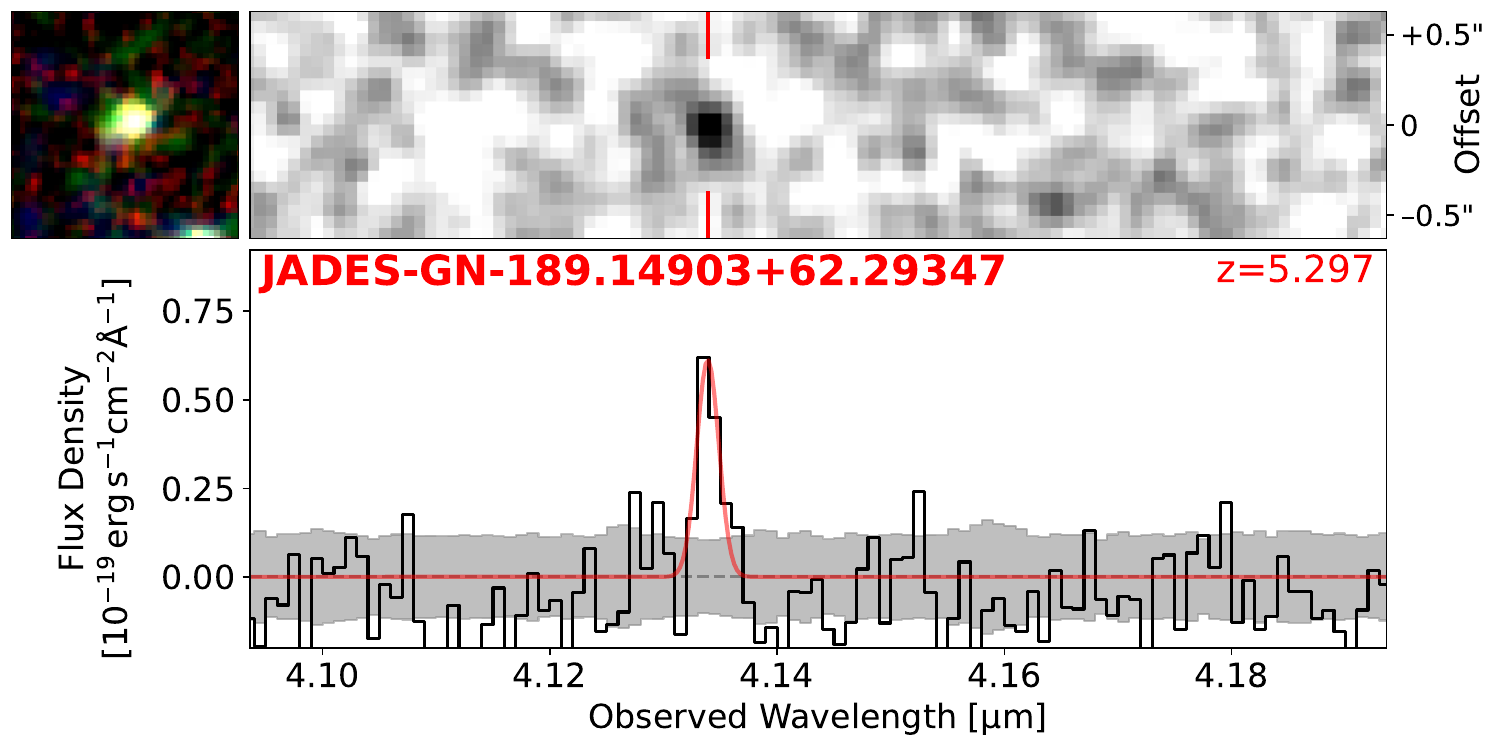}
\caption{Continued.} 
 \end{figure*} 

 \addtocounter{figure}{-1} 
 \begin{figure*}[!ht] 
 \centering
\includegraphics[width=0.49\linewidth]{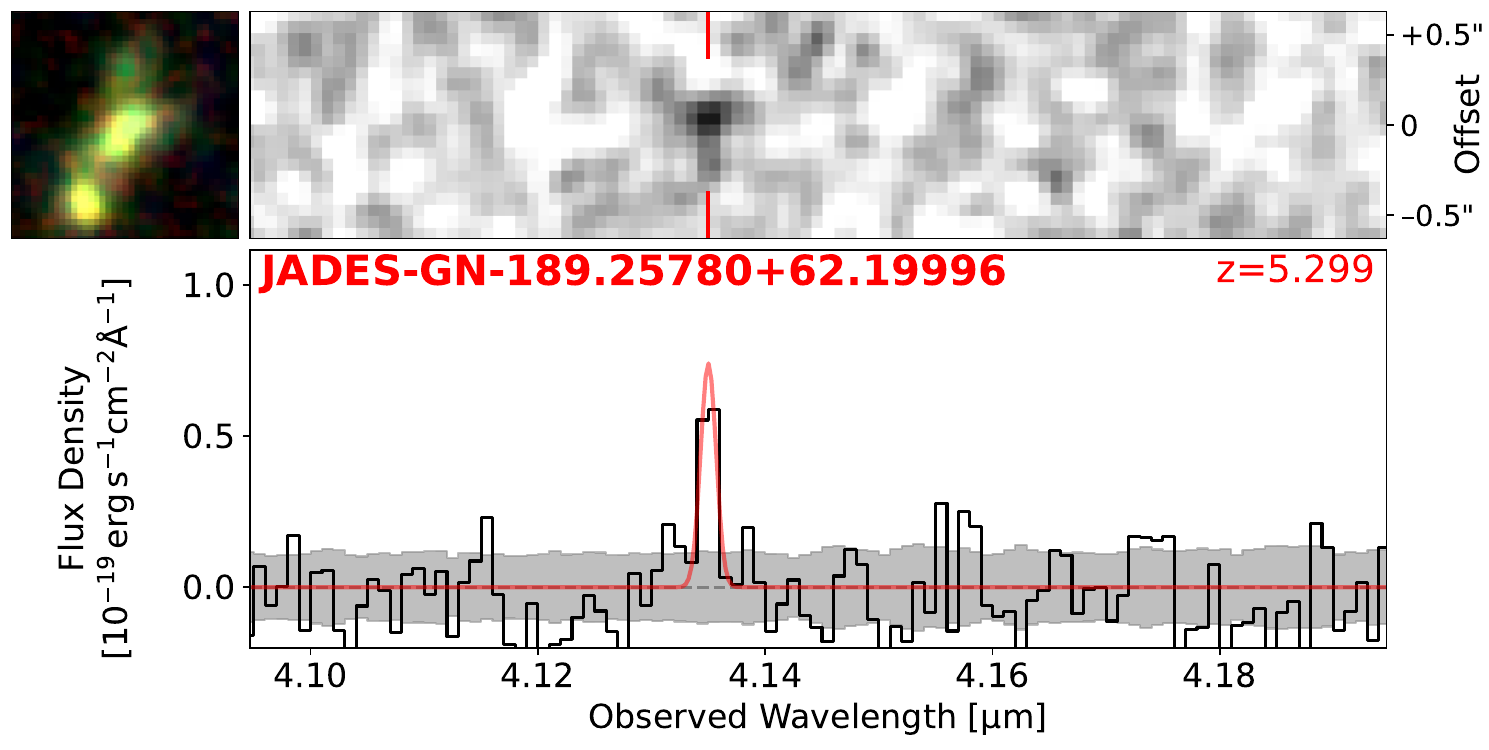}
\includegraphics[width=0.49\linewidth]{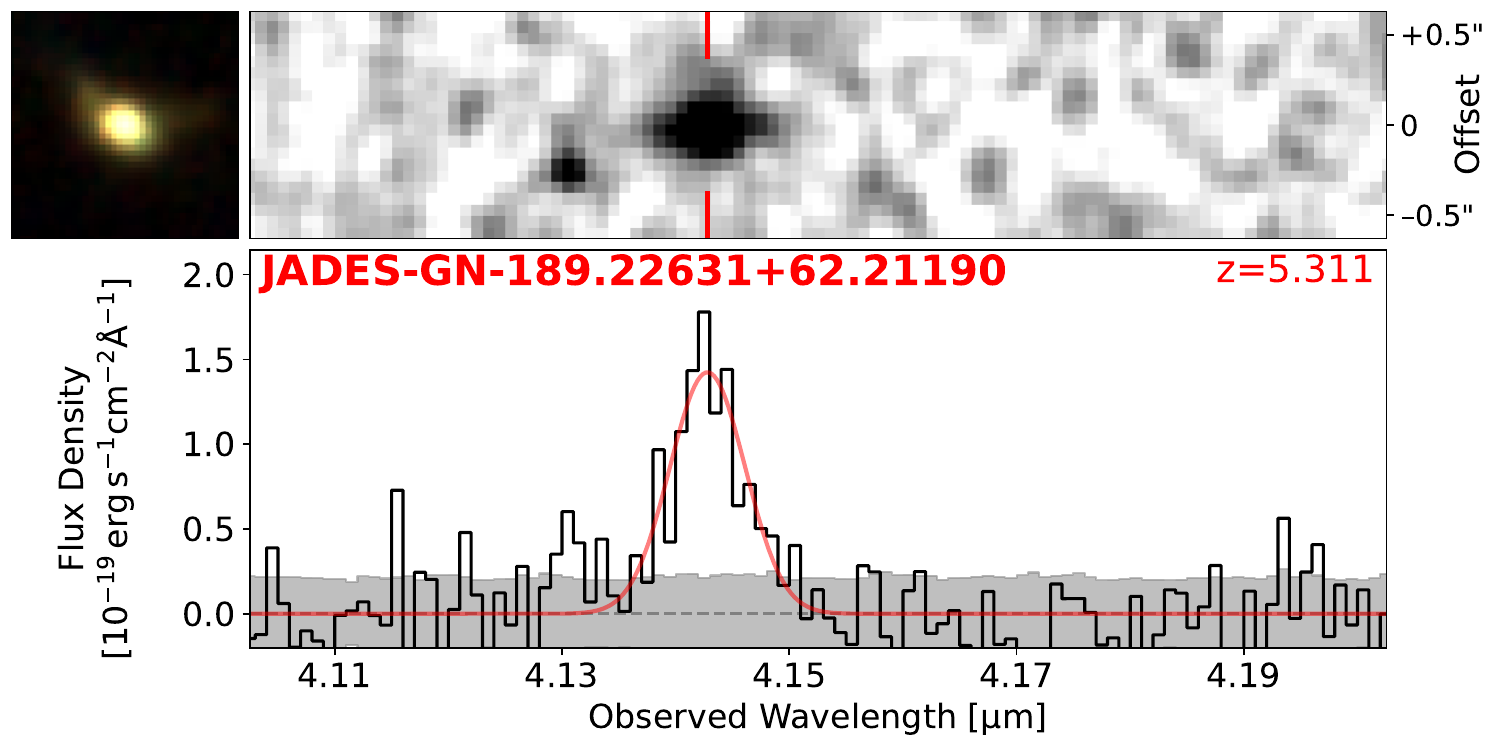}
\includegraphics[width=0.49\linewidth]{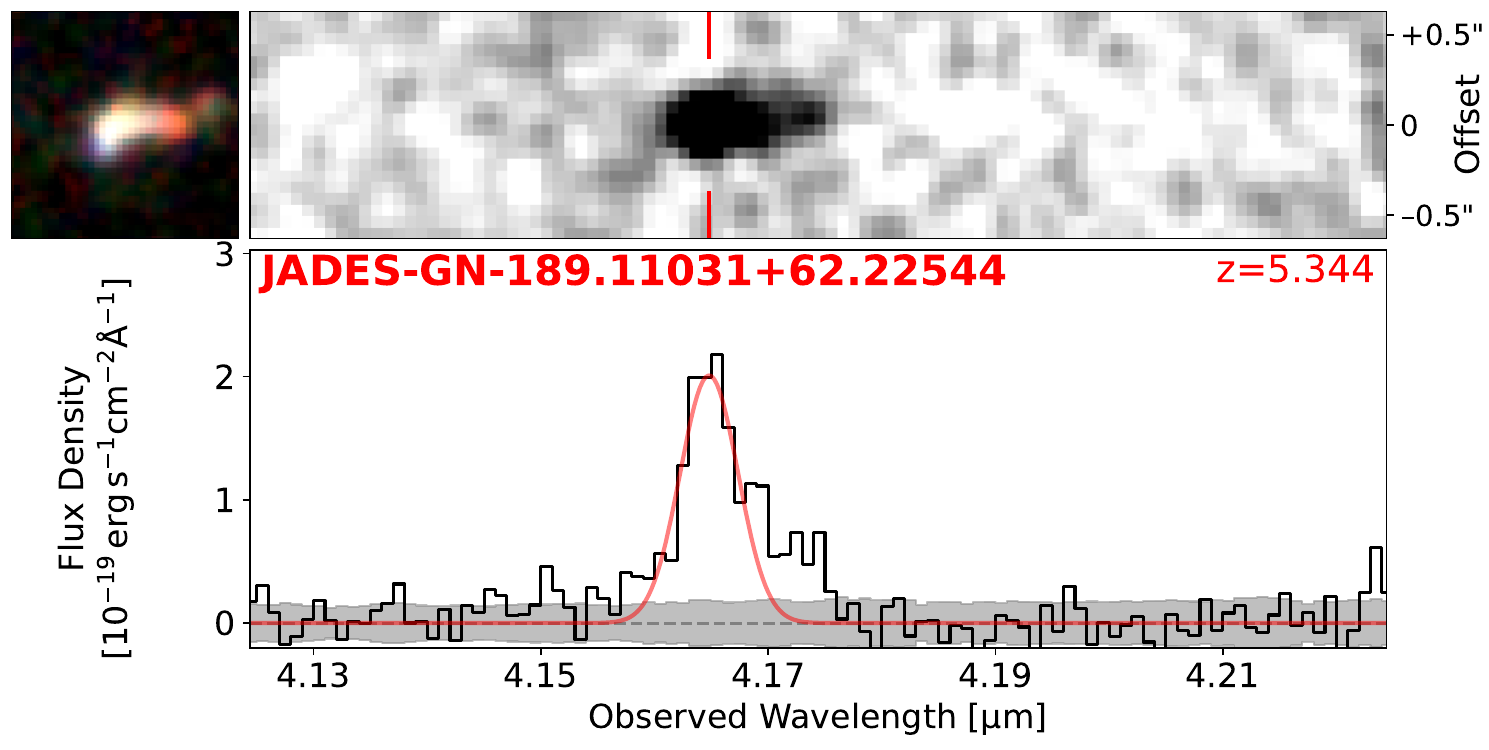}
\includegraphics[width=0.49\linewidth]{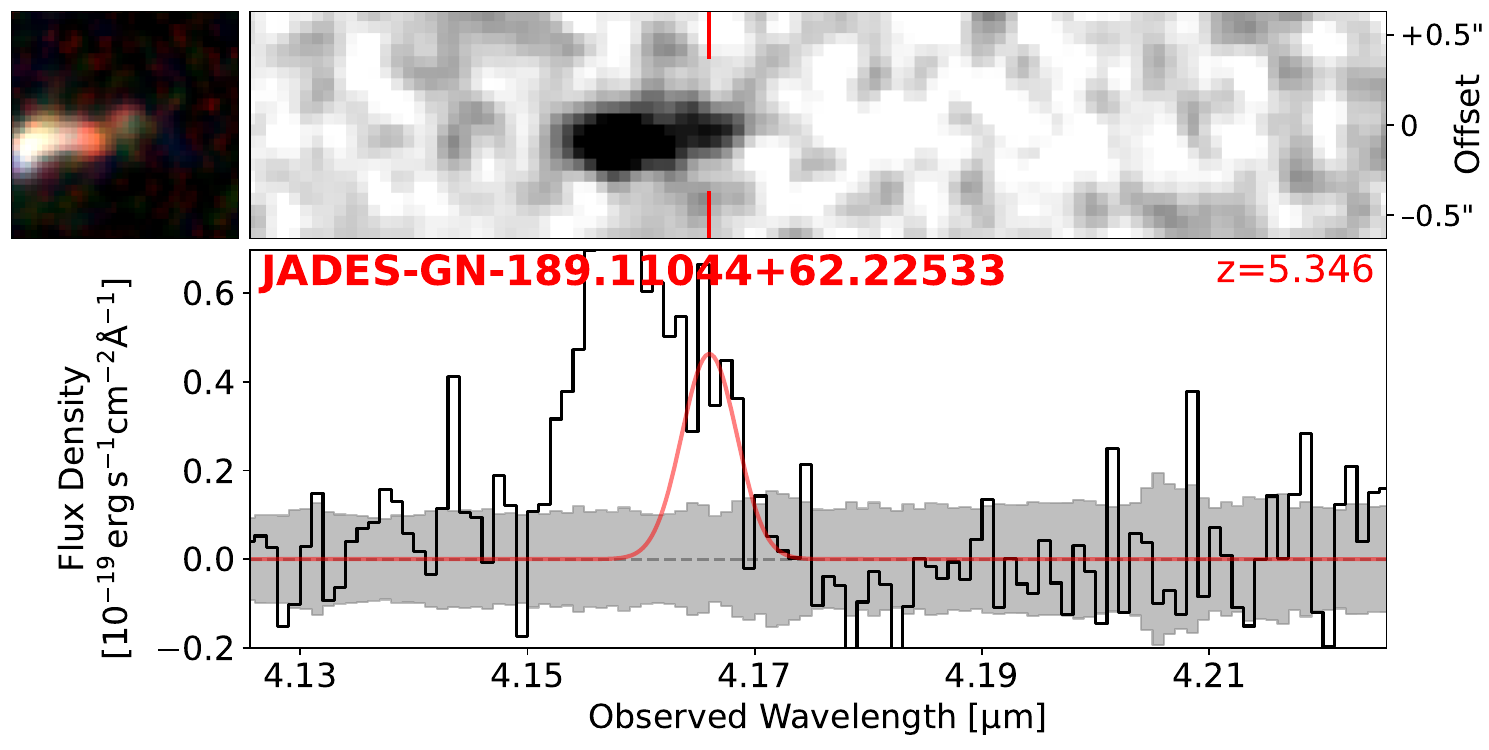}
\includegraphics[width=0.49\linewidth]{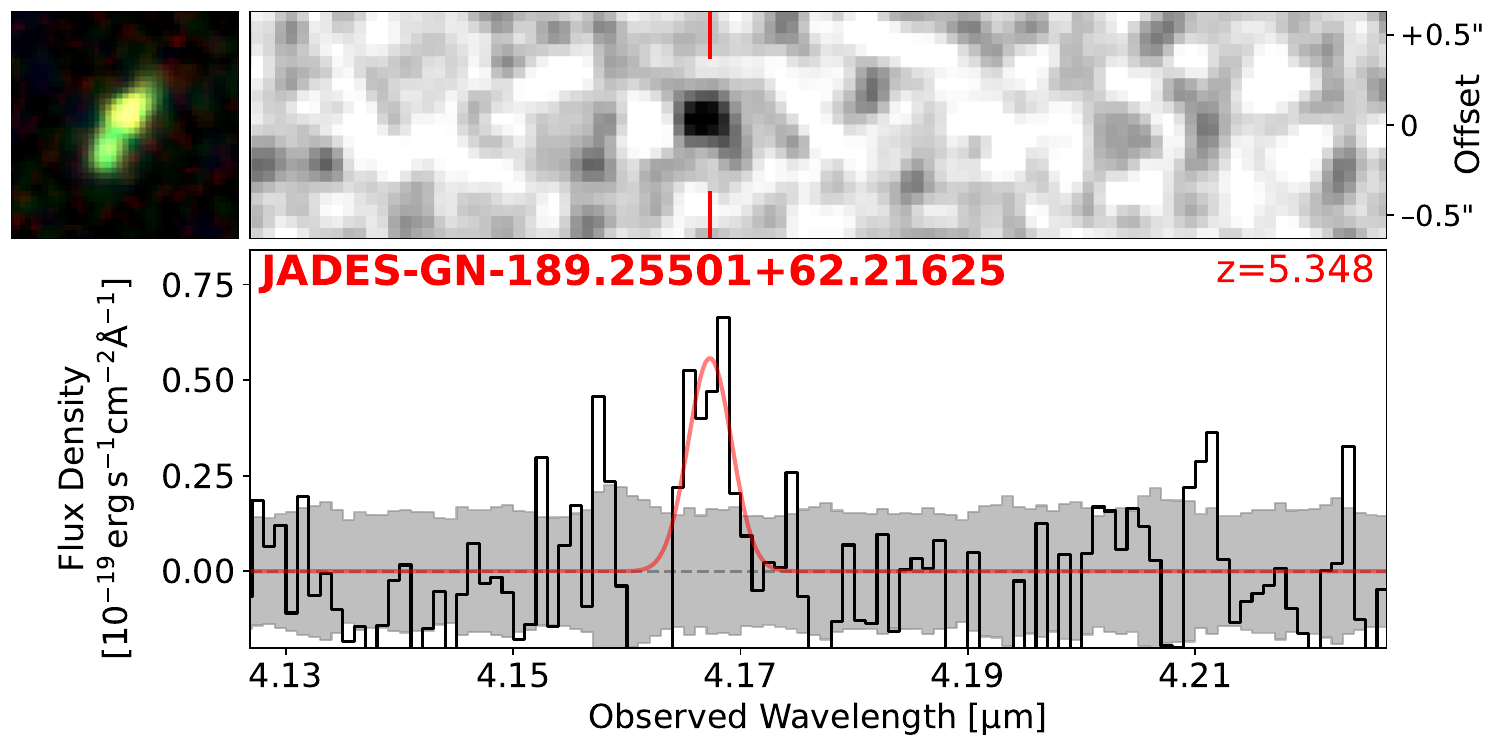}
\includegraphics[width=0.49\linewidth]{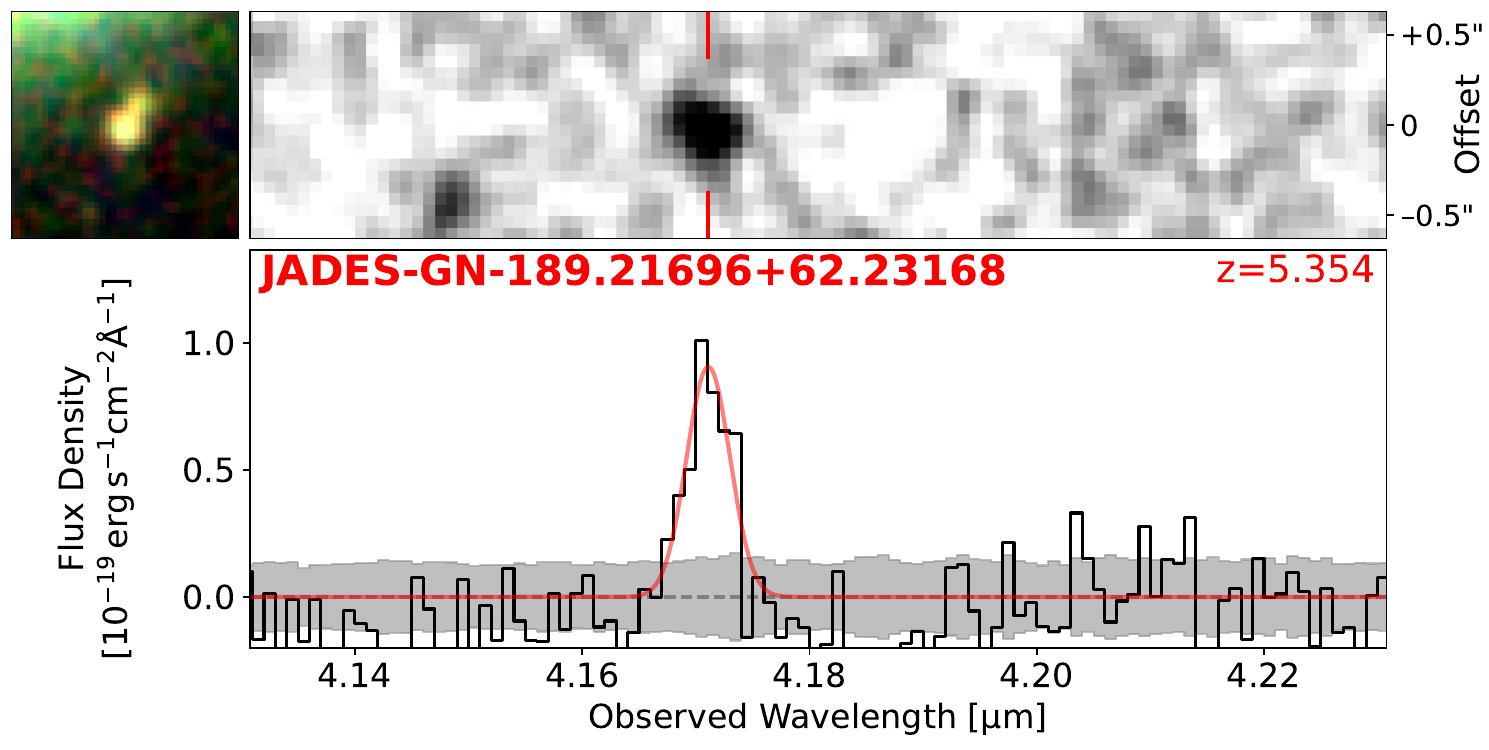}
\includegraphics[width=0.49\linewidth]{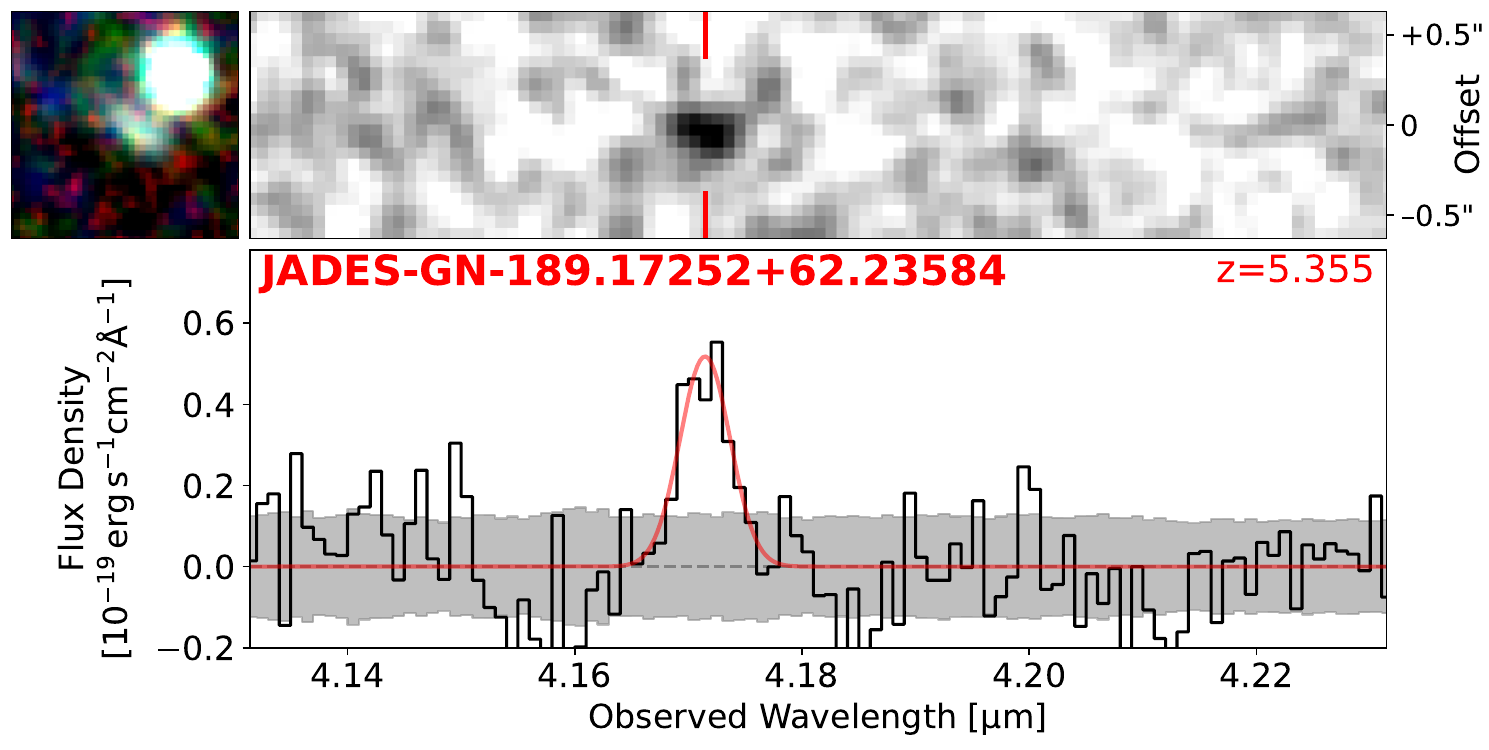}
\includegraphics[width=0.49\linewidth]{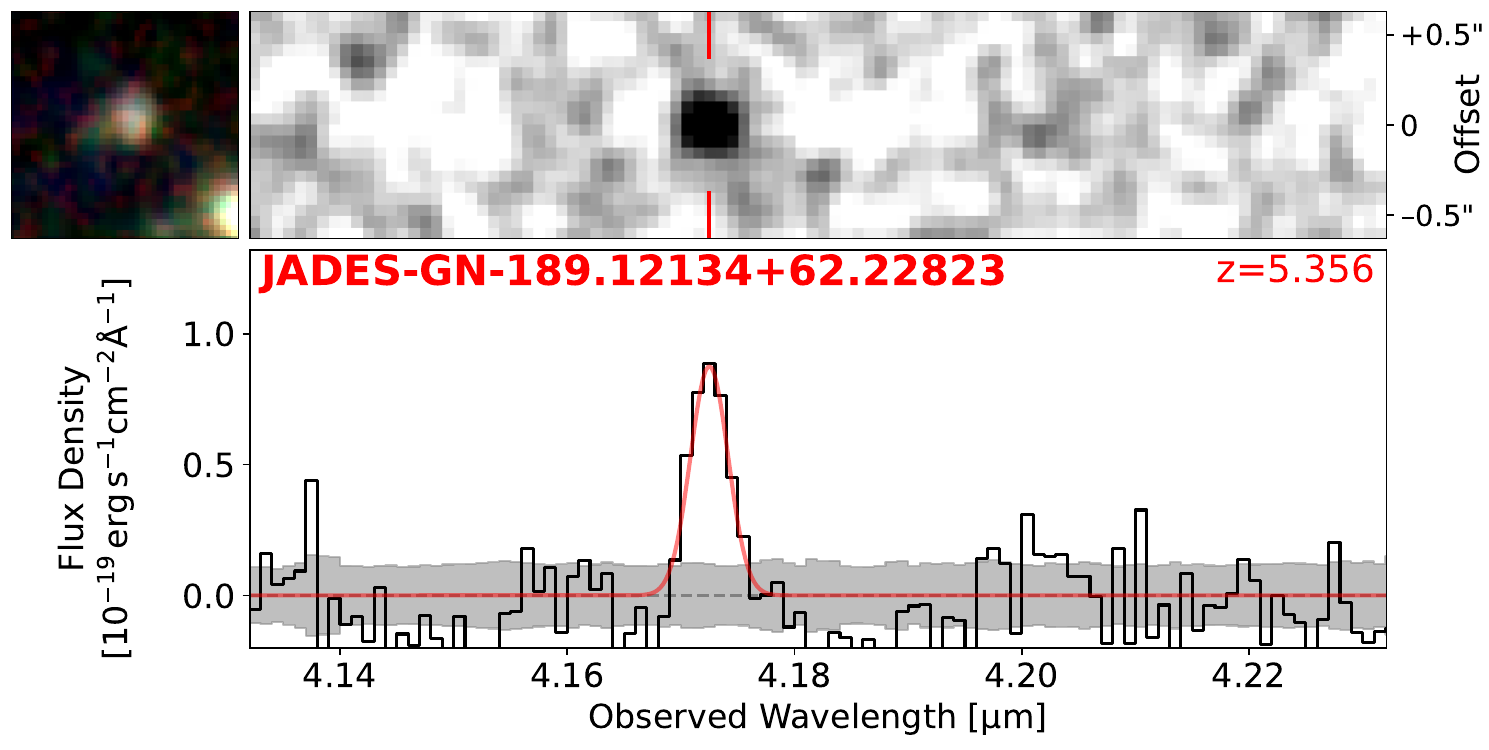}
\includegraphics[width=0.49\linewidth]{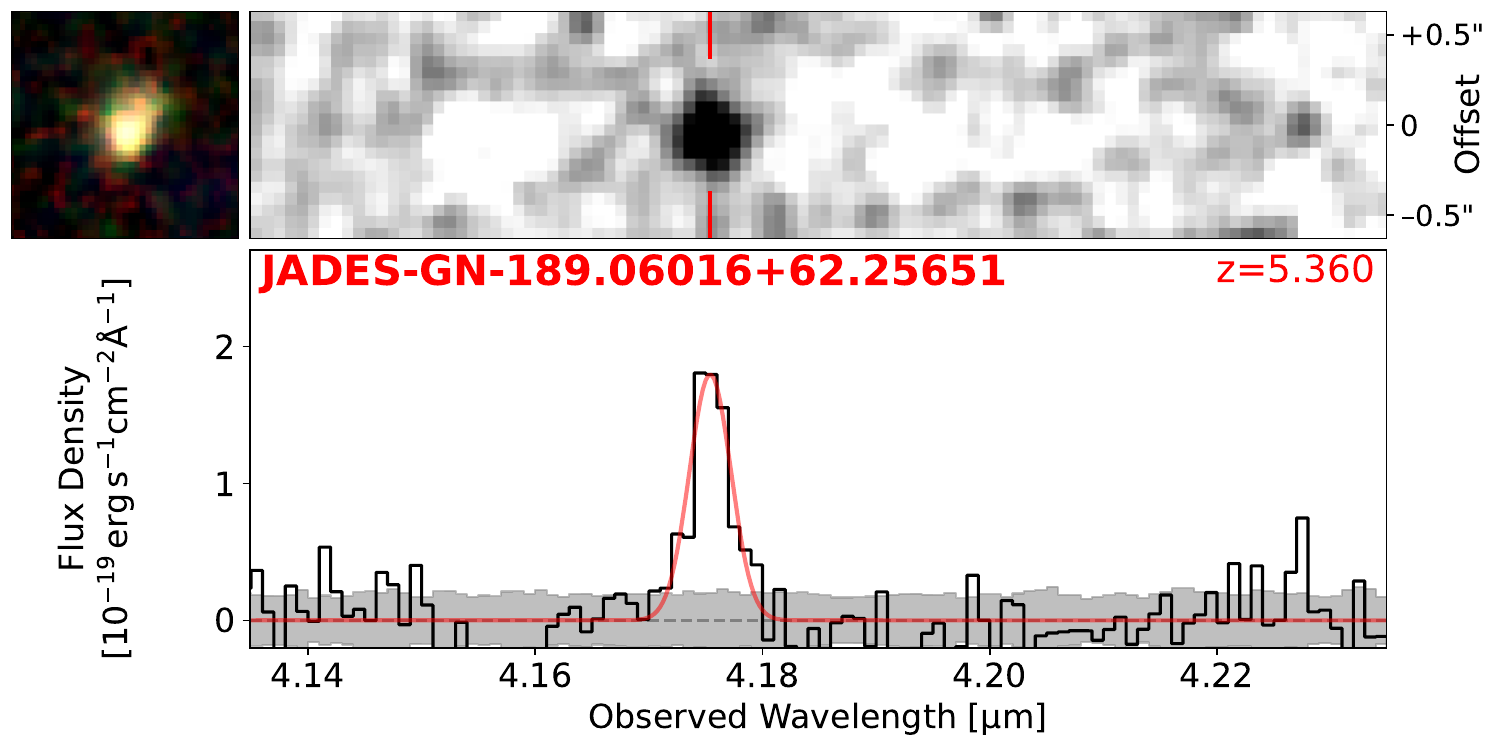}
\includegraphics[width=0.49\linewidth]{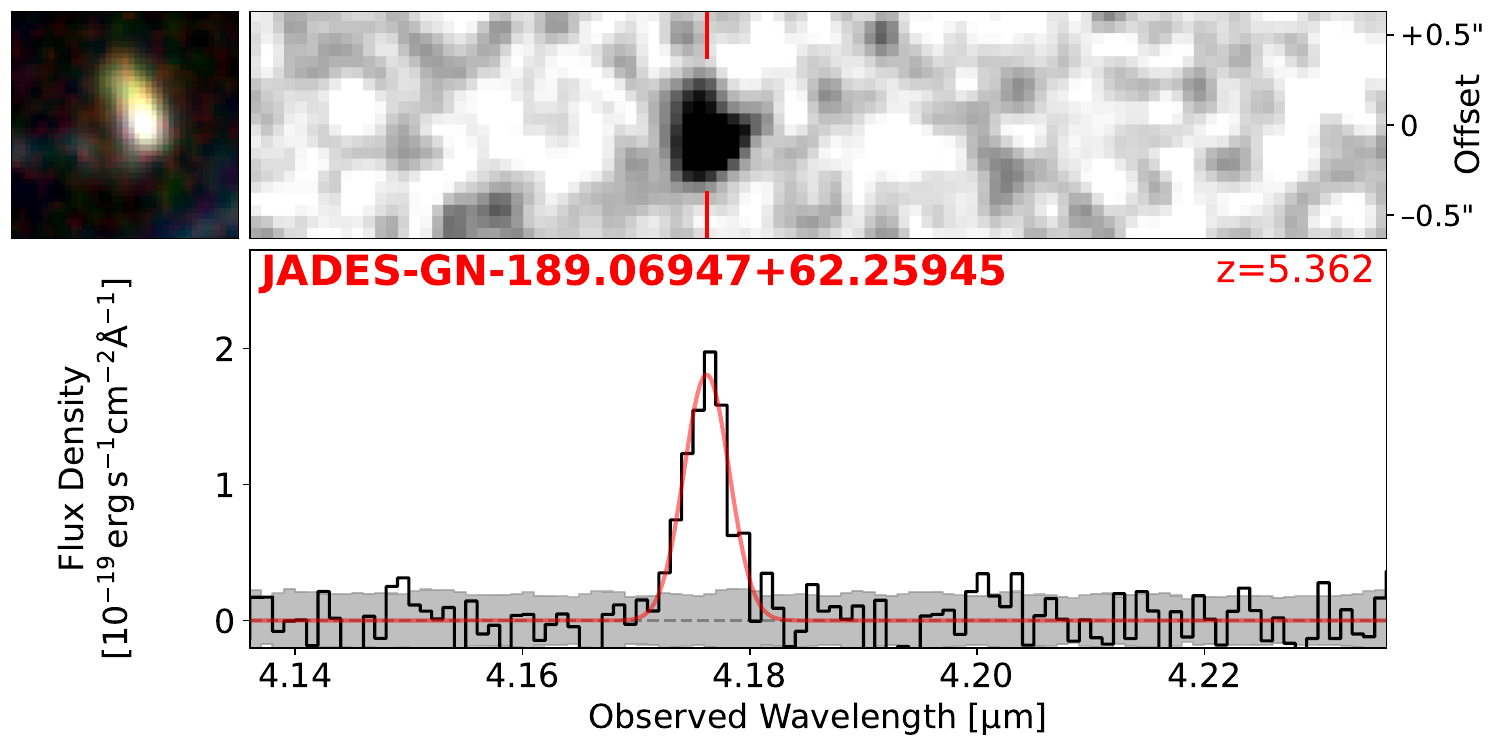}
\caption{Continued.} 
 \end{figure*} 

 \addtocounter{figure}{-1} 
 \begin{figure*}[!ht] 
 \centering
\includegraphics[width=0.49\linewidth]{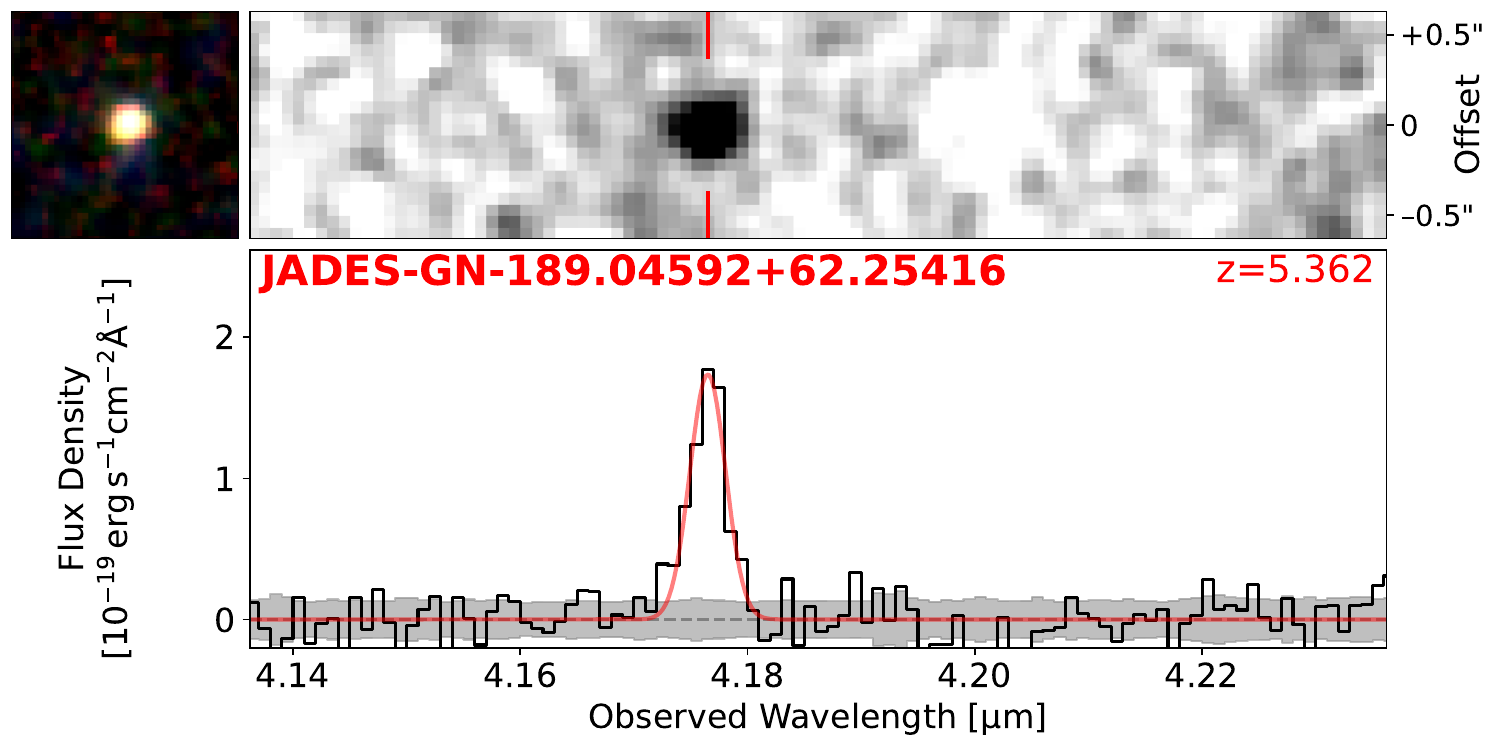}
\includegraphics[width=0.49\linewidth]{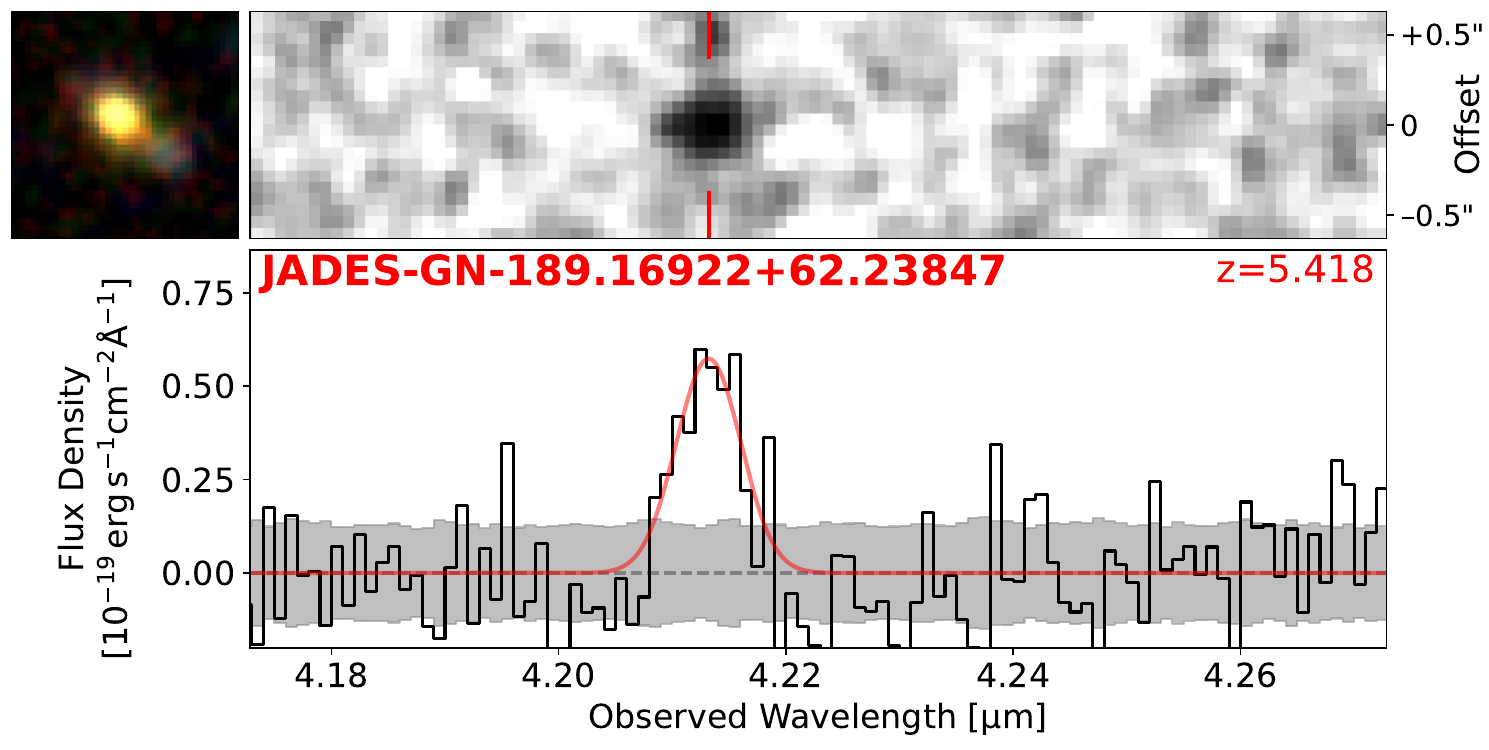}
\includegraphics[width=0.49\linewidth]{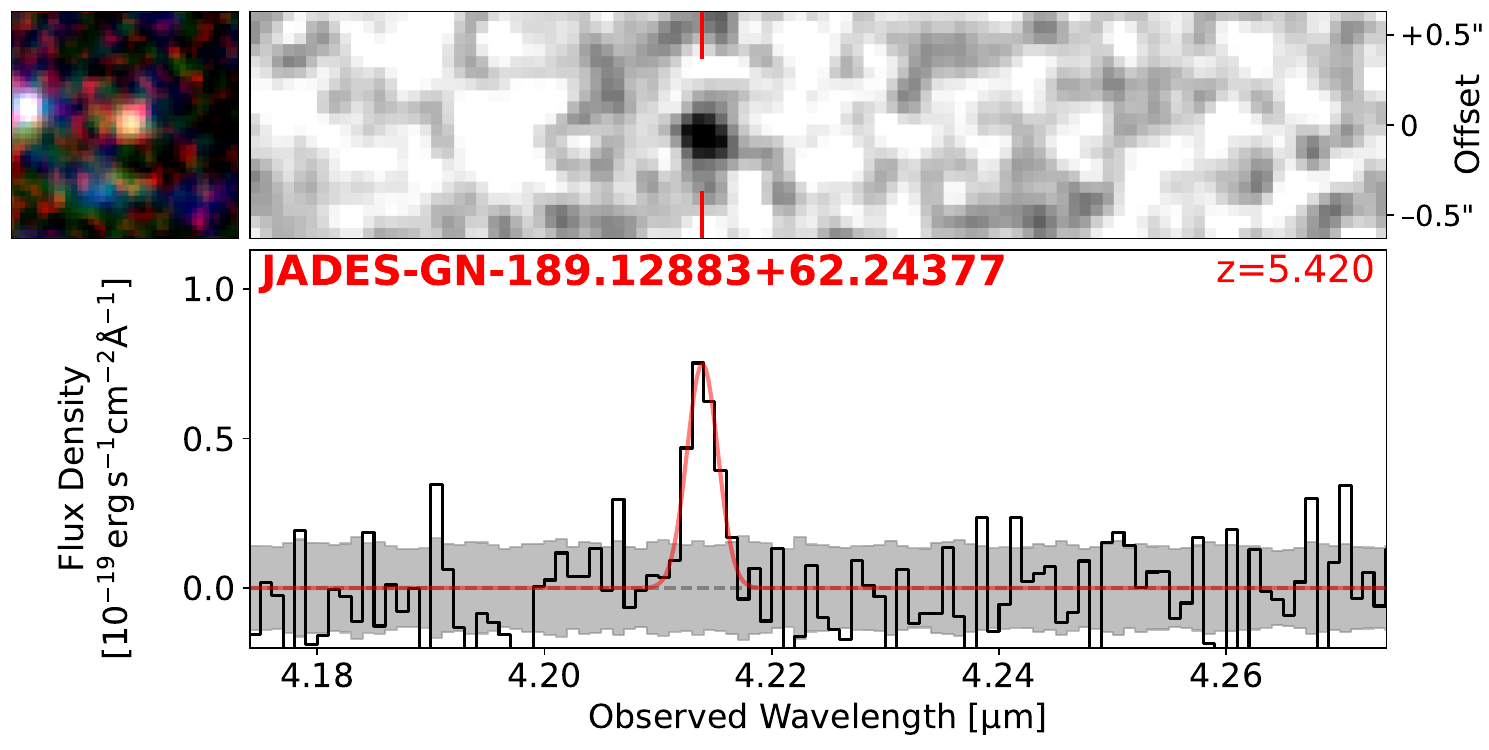}
\includegraphics[width=0.49\linewidth]{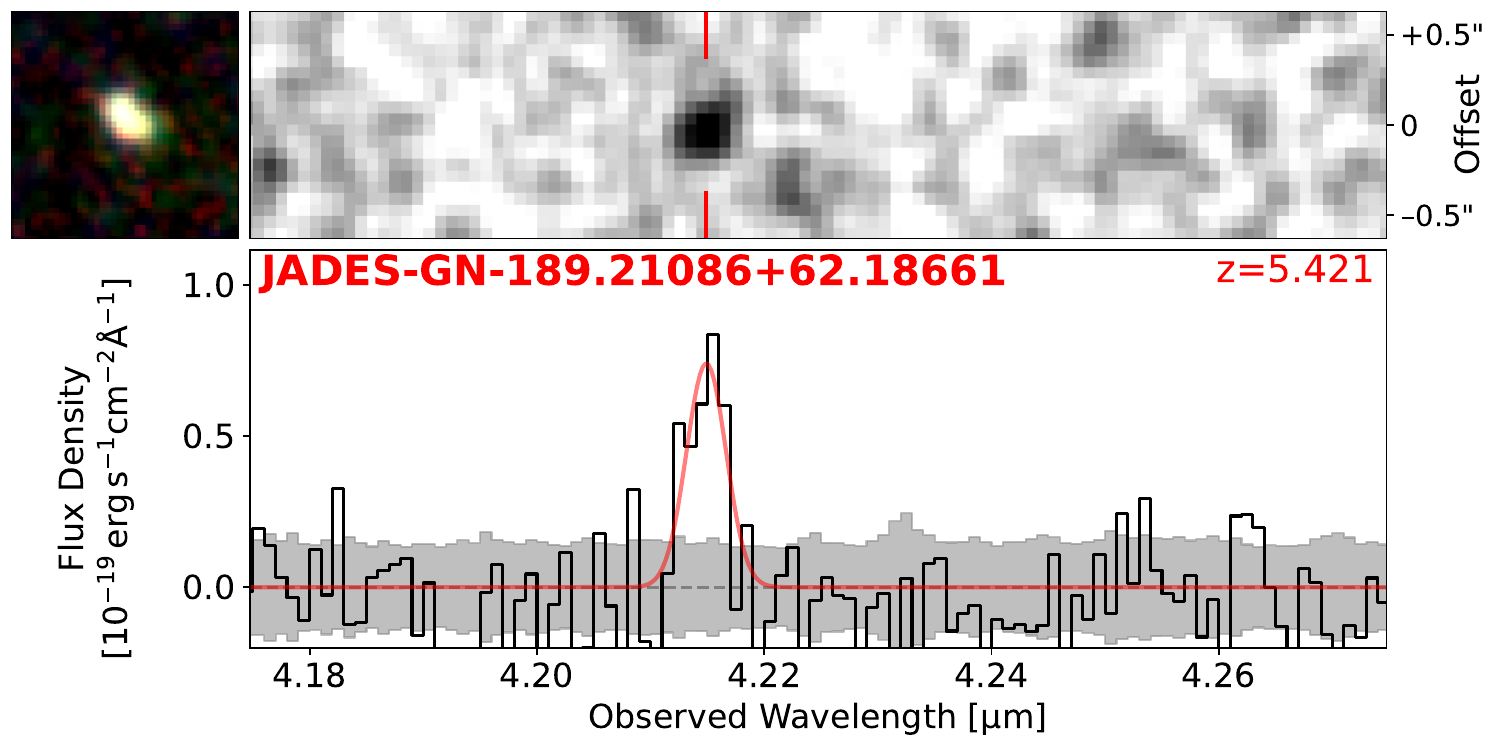}
\includegraphics[width=0.49\linewidth]{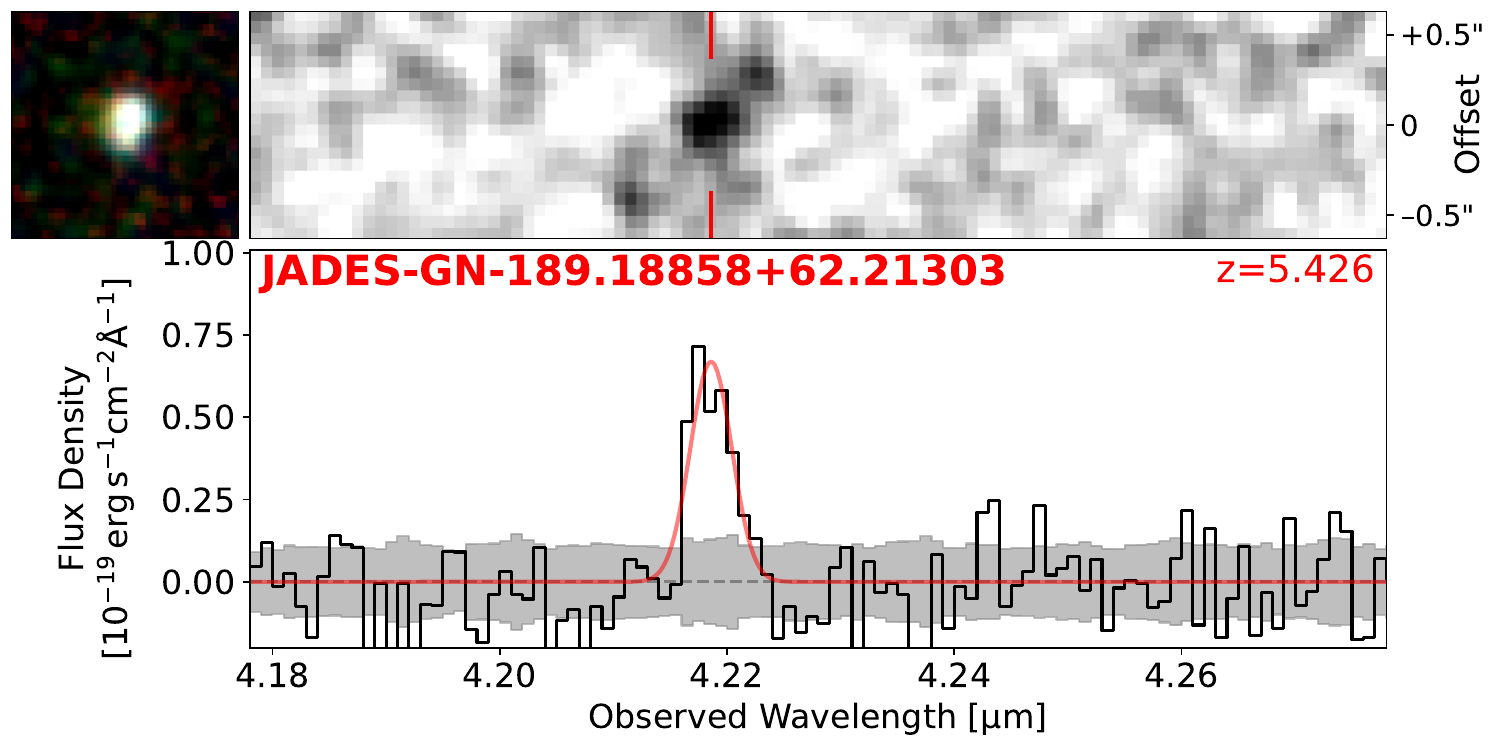}
\includegraphics[width=0.49\linewidth]{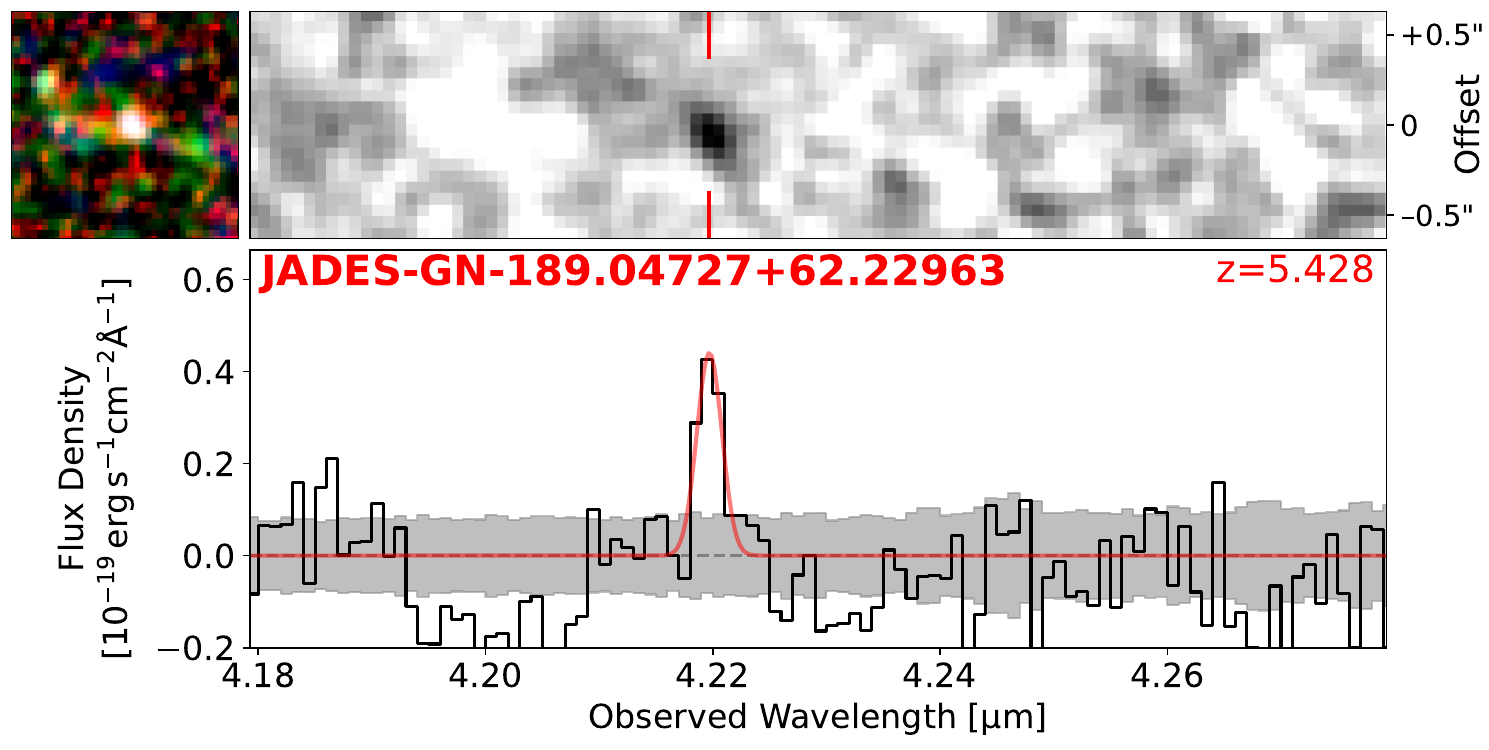}
\includegraphics[width=0.49\linewidth]{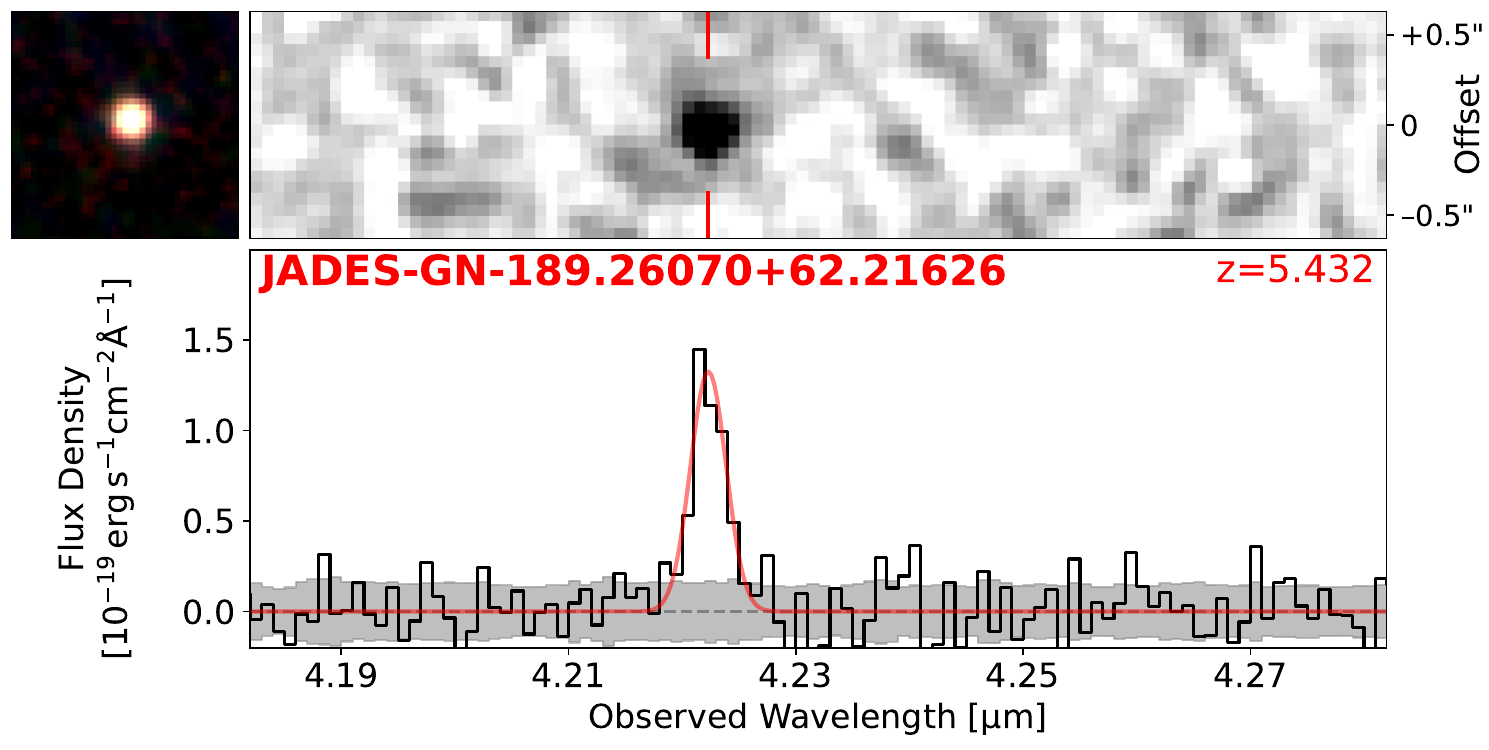}
\includegraphics[width=0.49\linewidth]{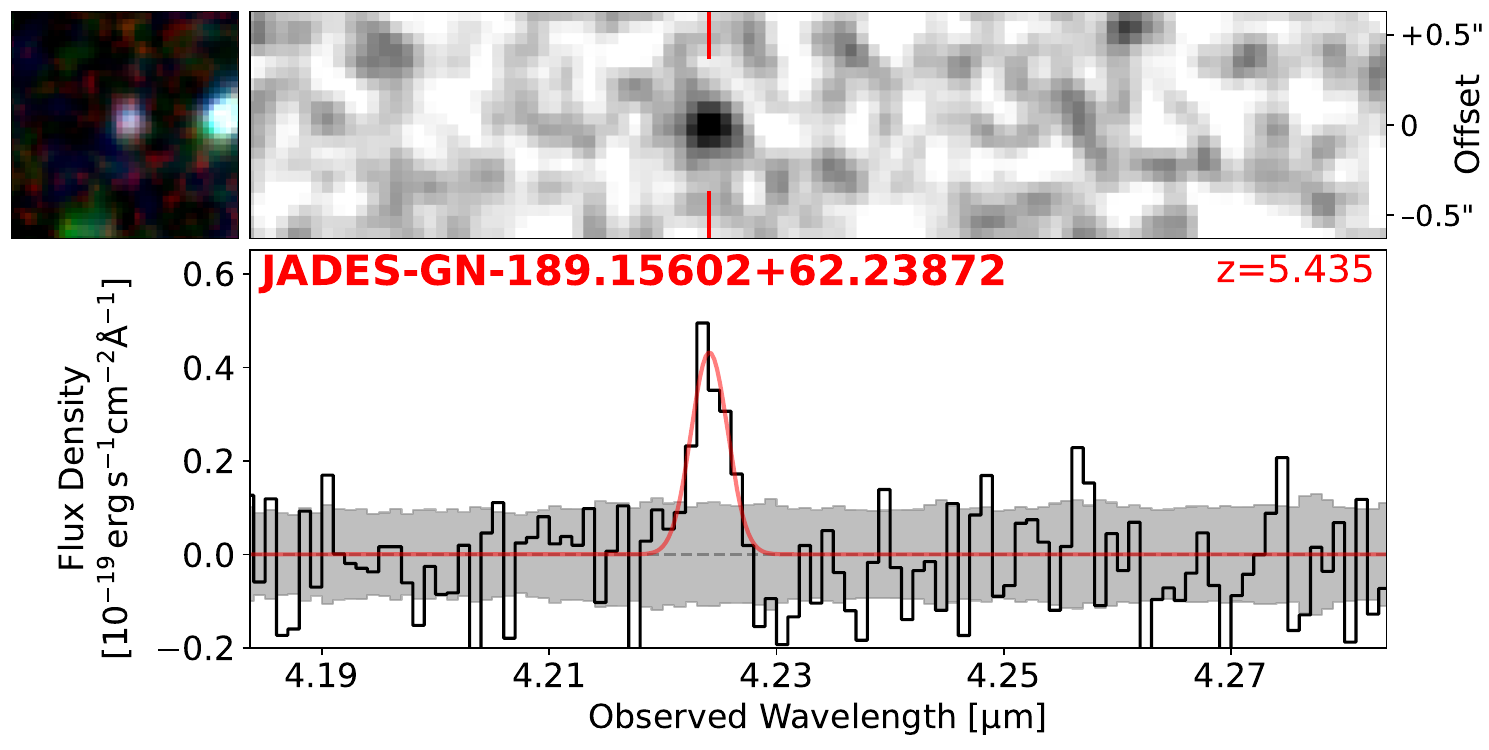}
\includegraphics[width=0.49\linewidth]{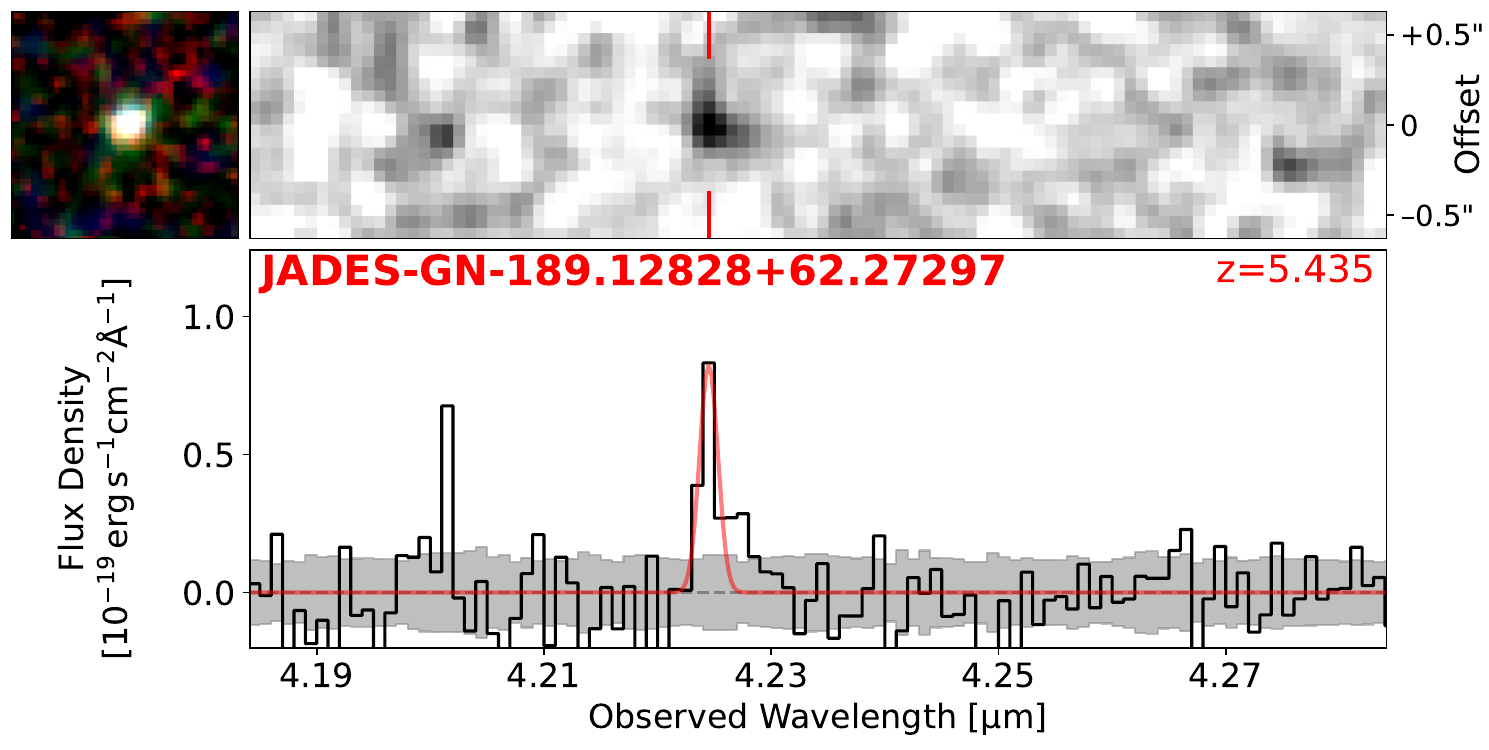}
\includegraphics[width=0.49\linewidth]{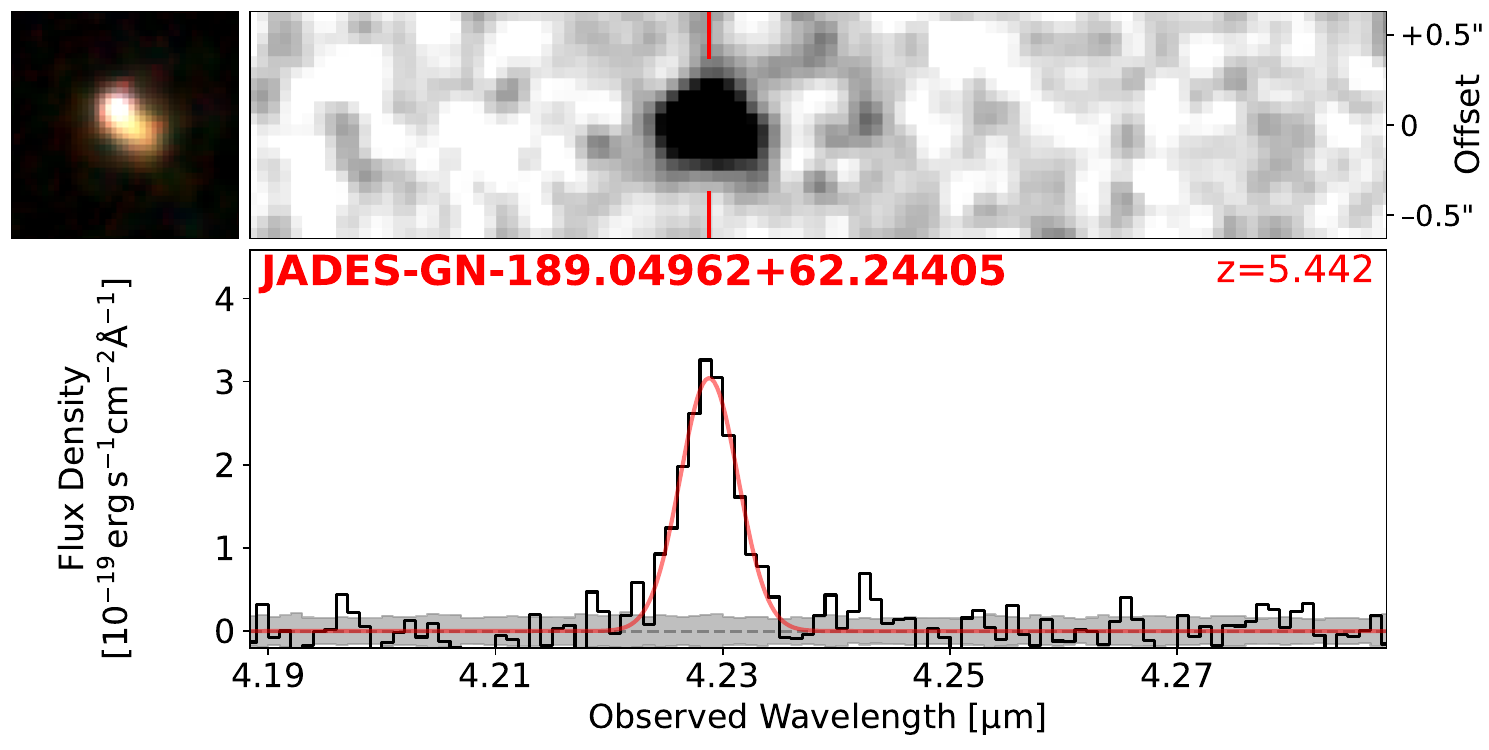}
\caption{Continued.} 
 \end{figure*} 

 \addtocounter{figure}{-1} 
 \begin{figure*}[!ht] 
 \centering
\includegraphics[width=0.49\linewidth]{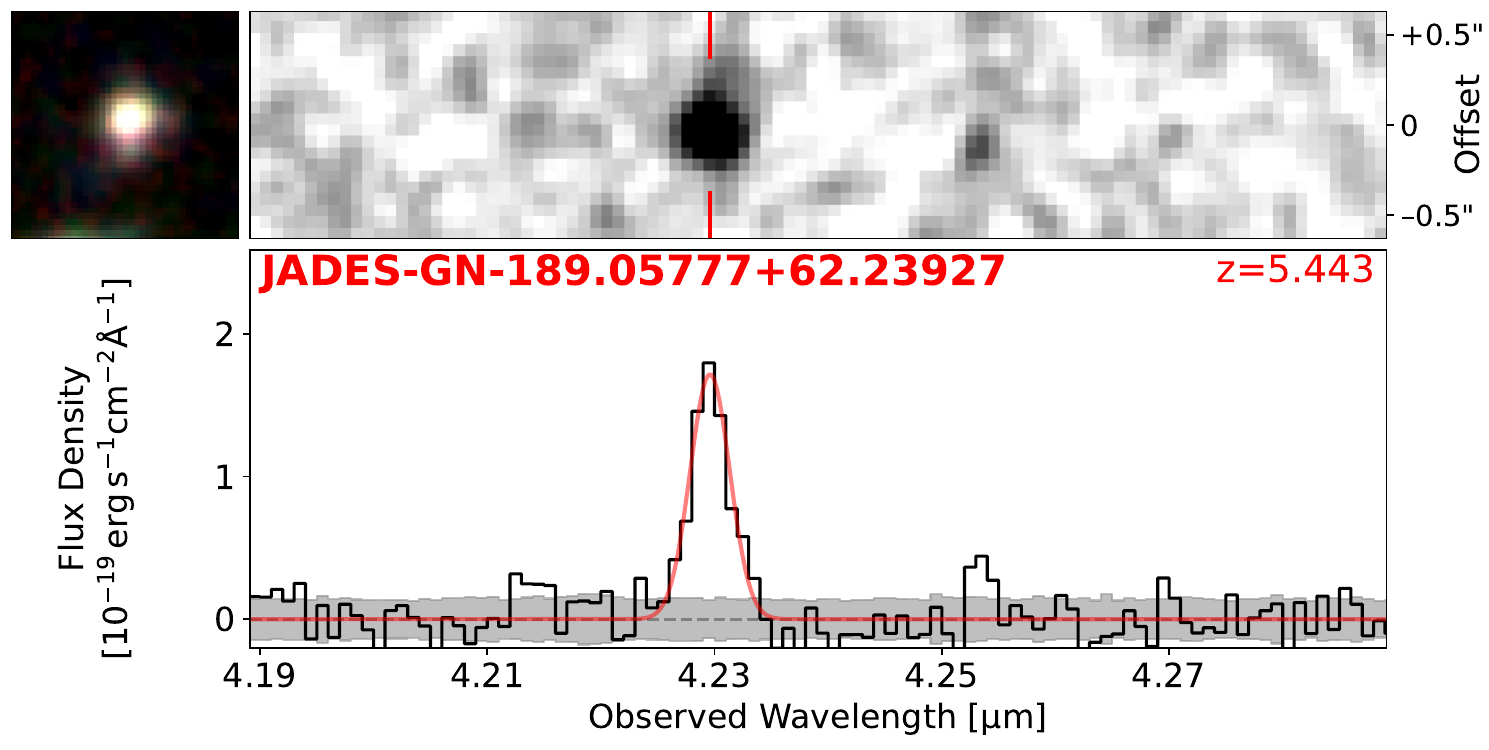}
\includegraphics[width=0.49\linewidth]{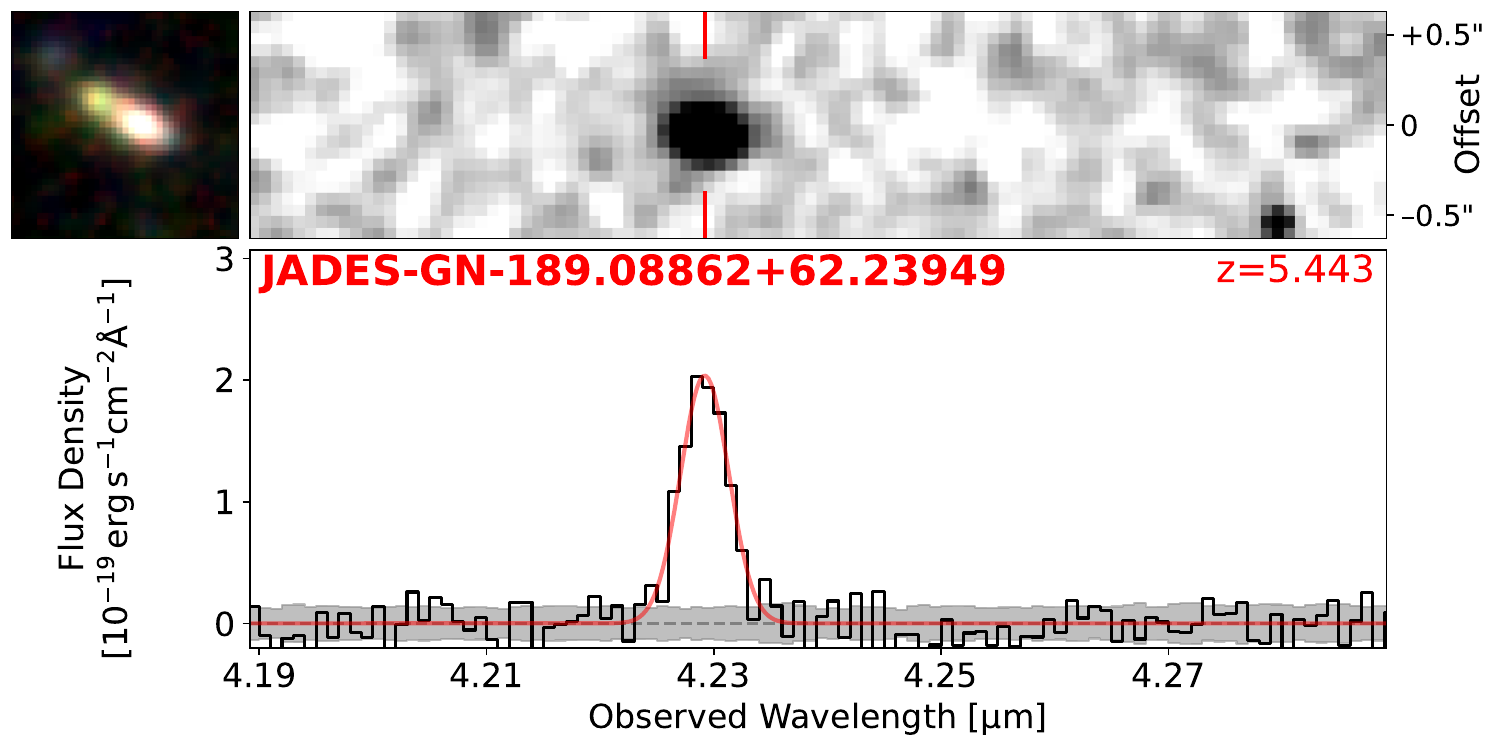}
\includegraphics[width=0.49\linewidth]{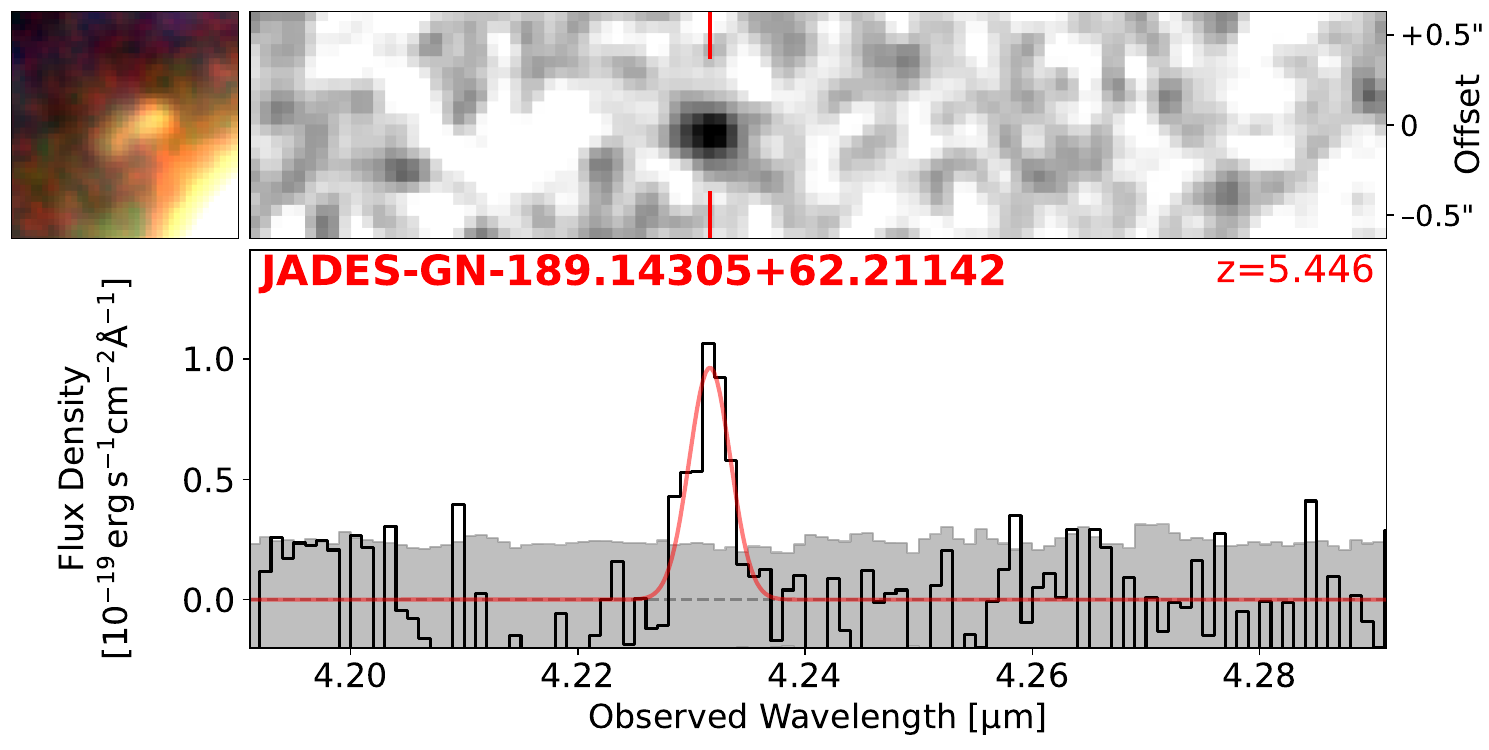}
\includegraphics[width=0.49\linewidth]{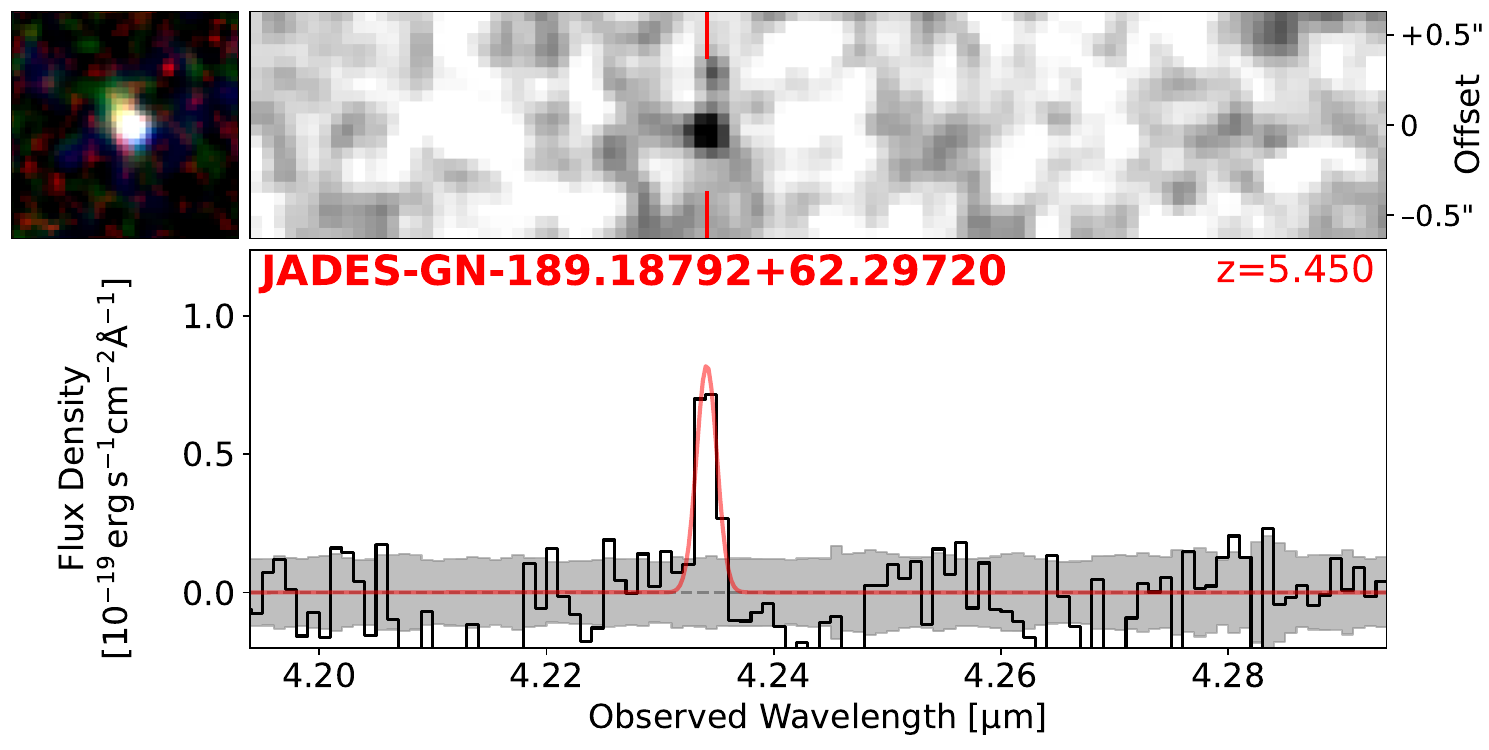}
\includegraphics[width=0.49\linewidth]{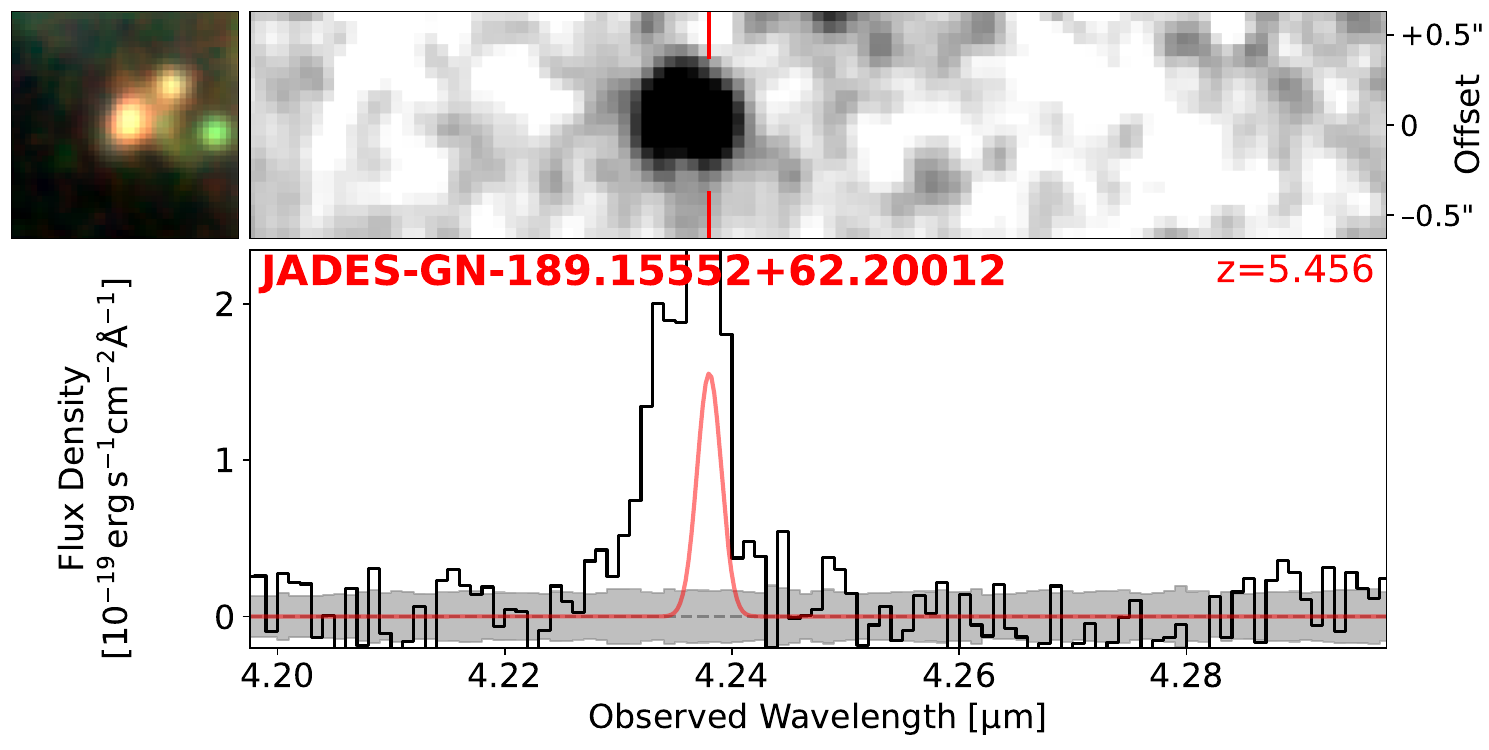}
\includegraphics[width=0.49\linewidth]{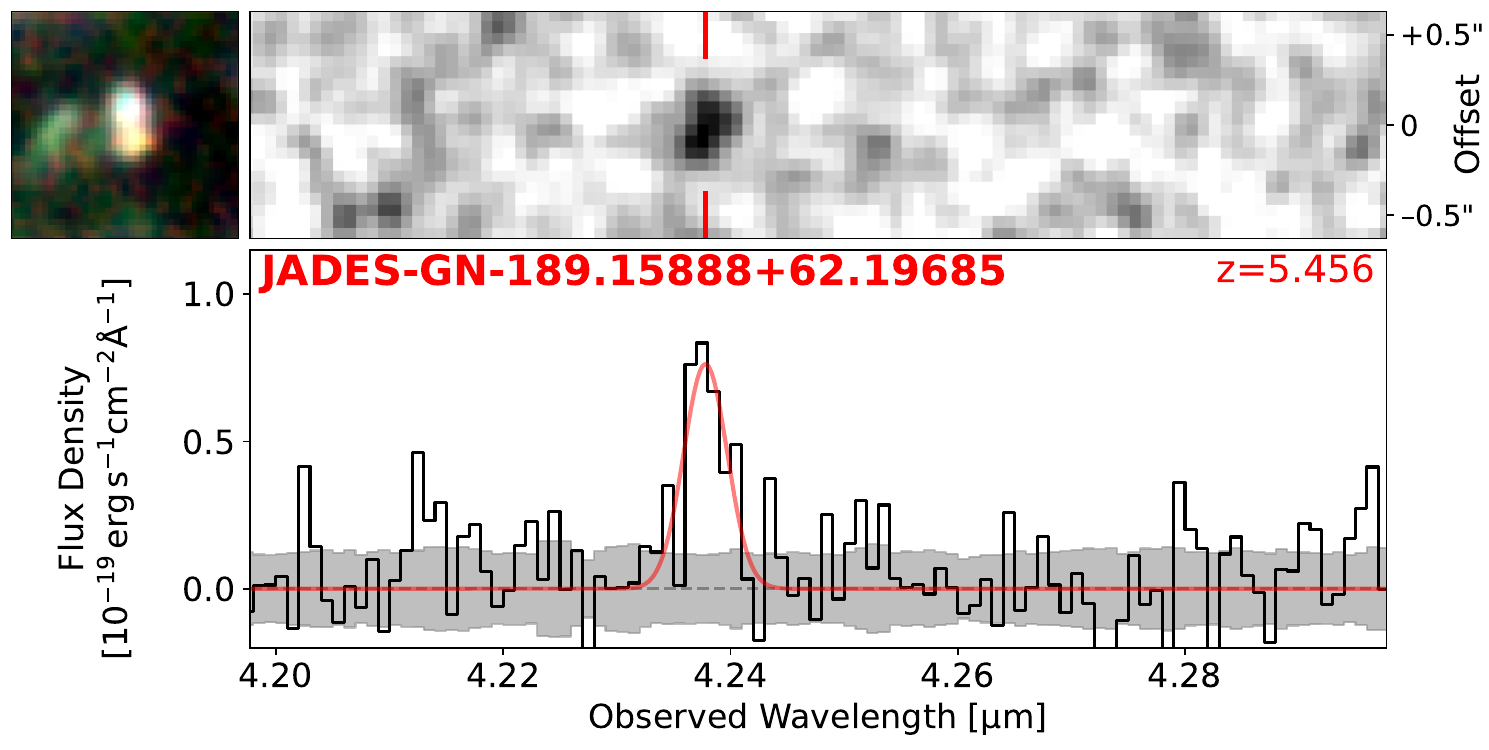}
\includegraphics[width=0.49\linewidth]{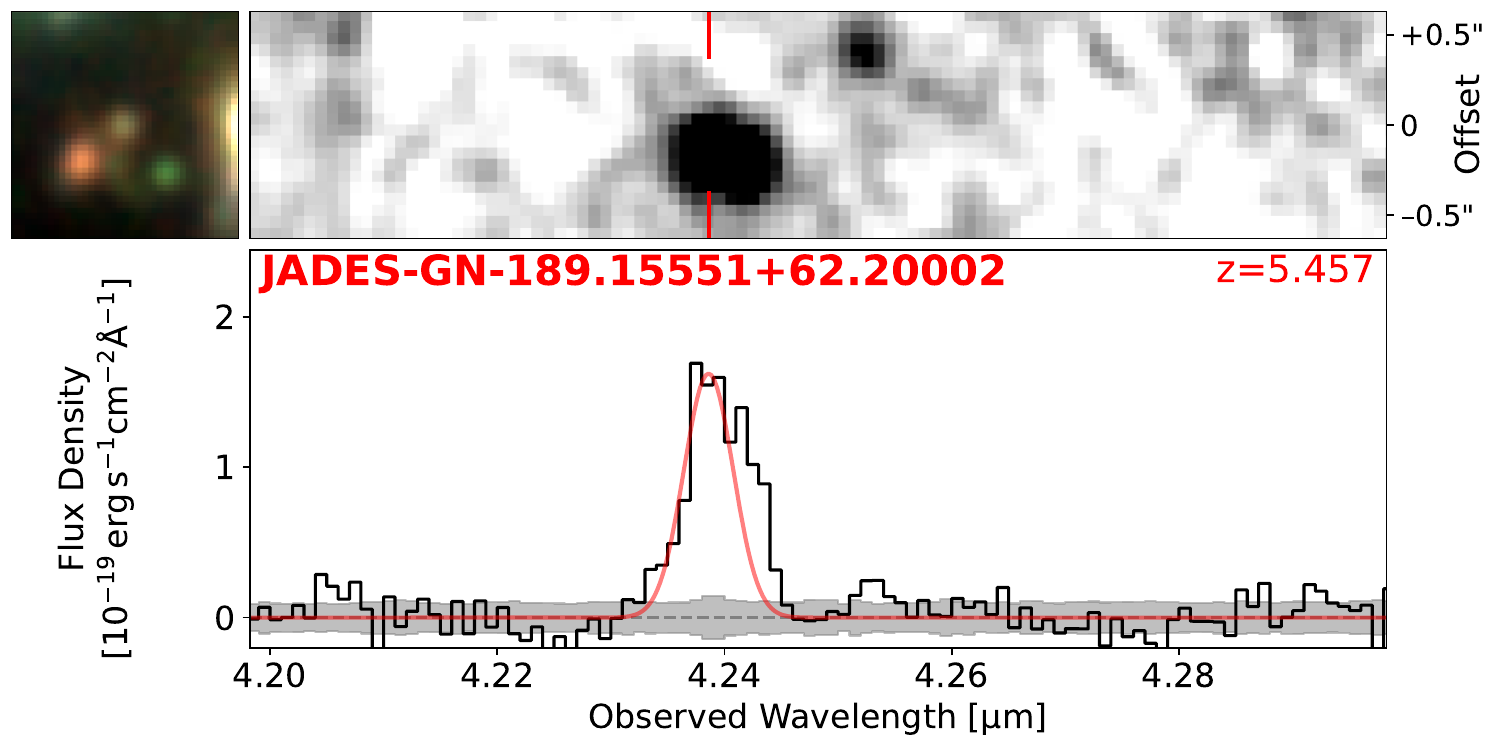}
\includegraphics[width=0.49\linewidth]{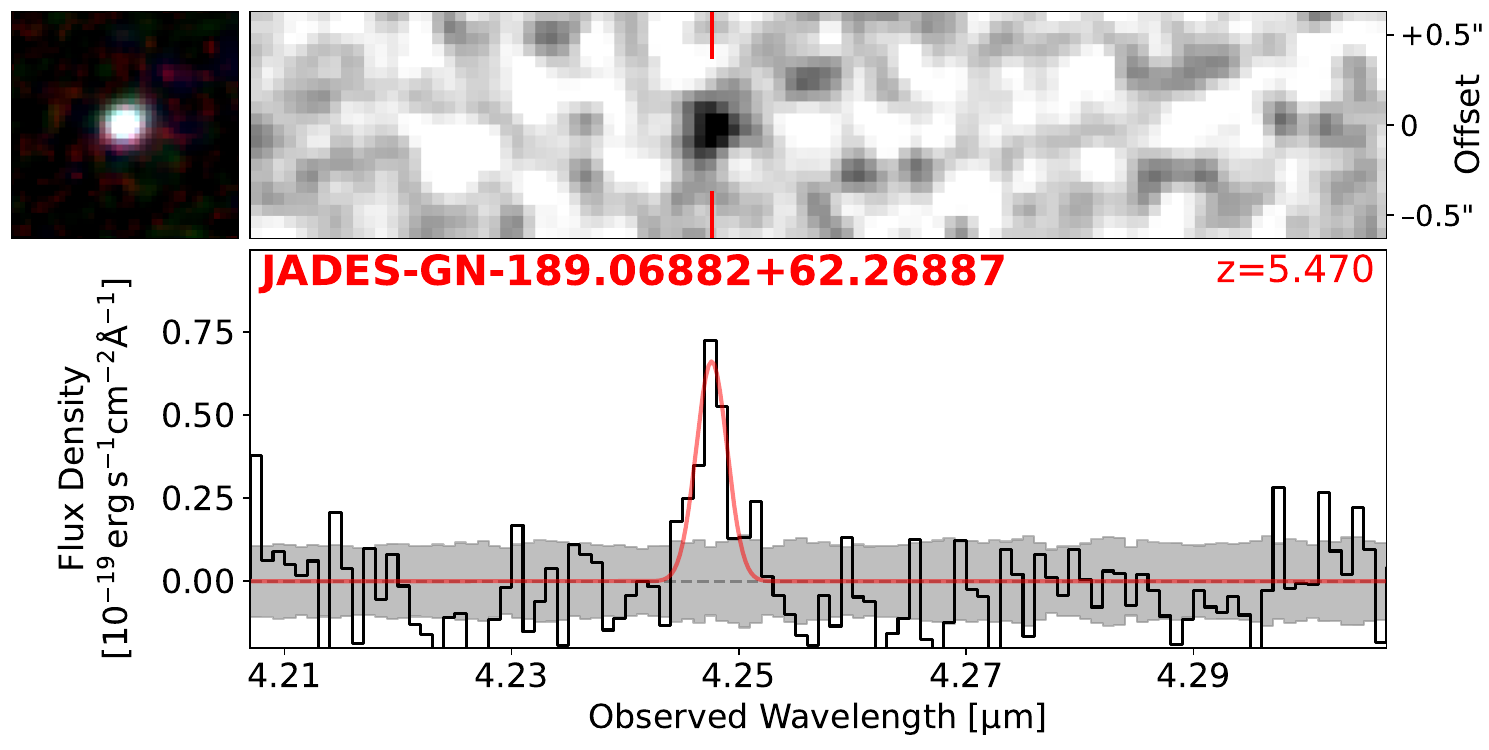}
\includegraphics[width=0.49\linewidth]{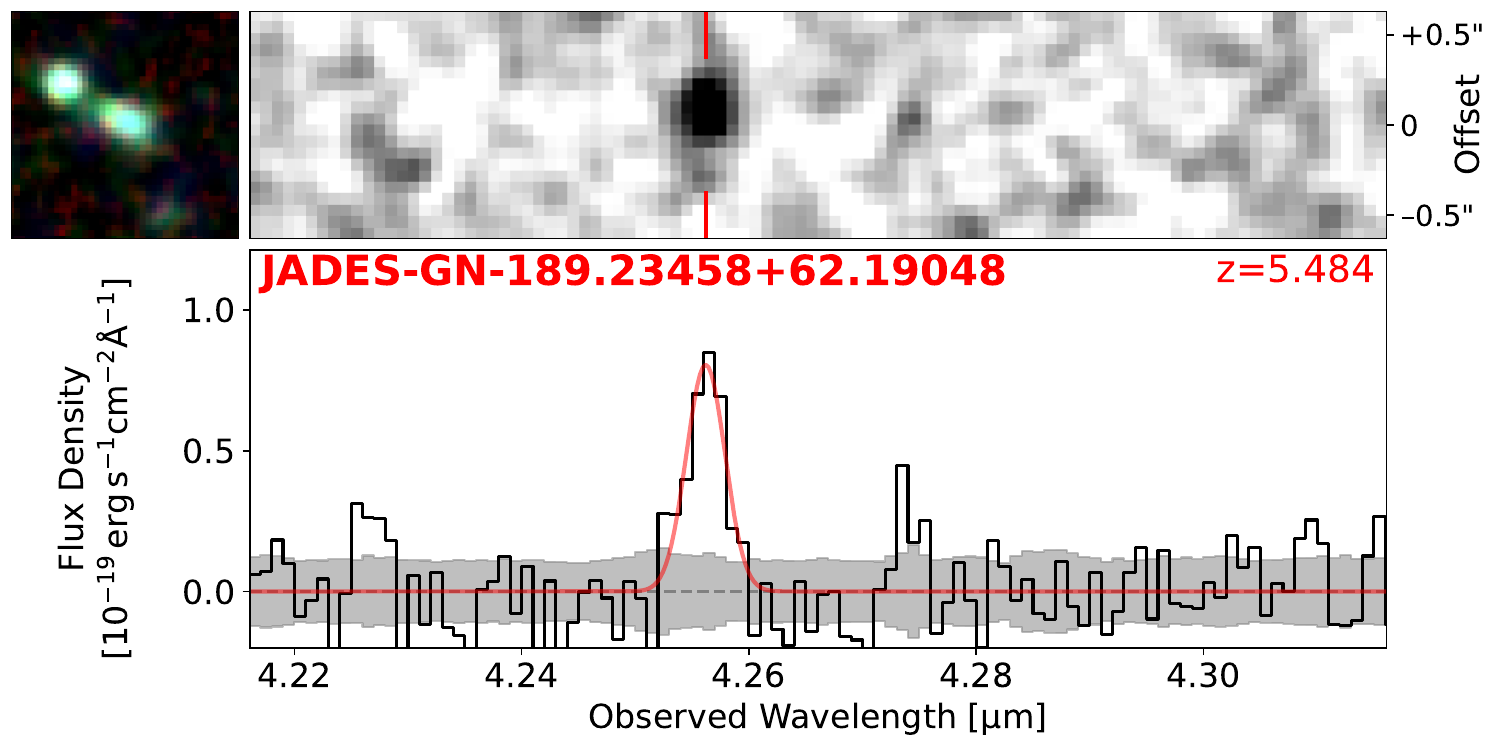}
\includegraphics[width=0.49\linewidth]{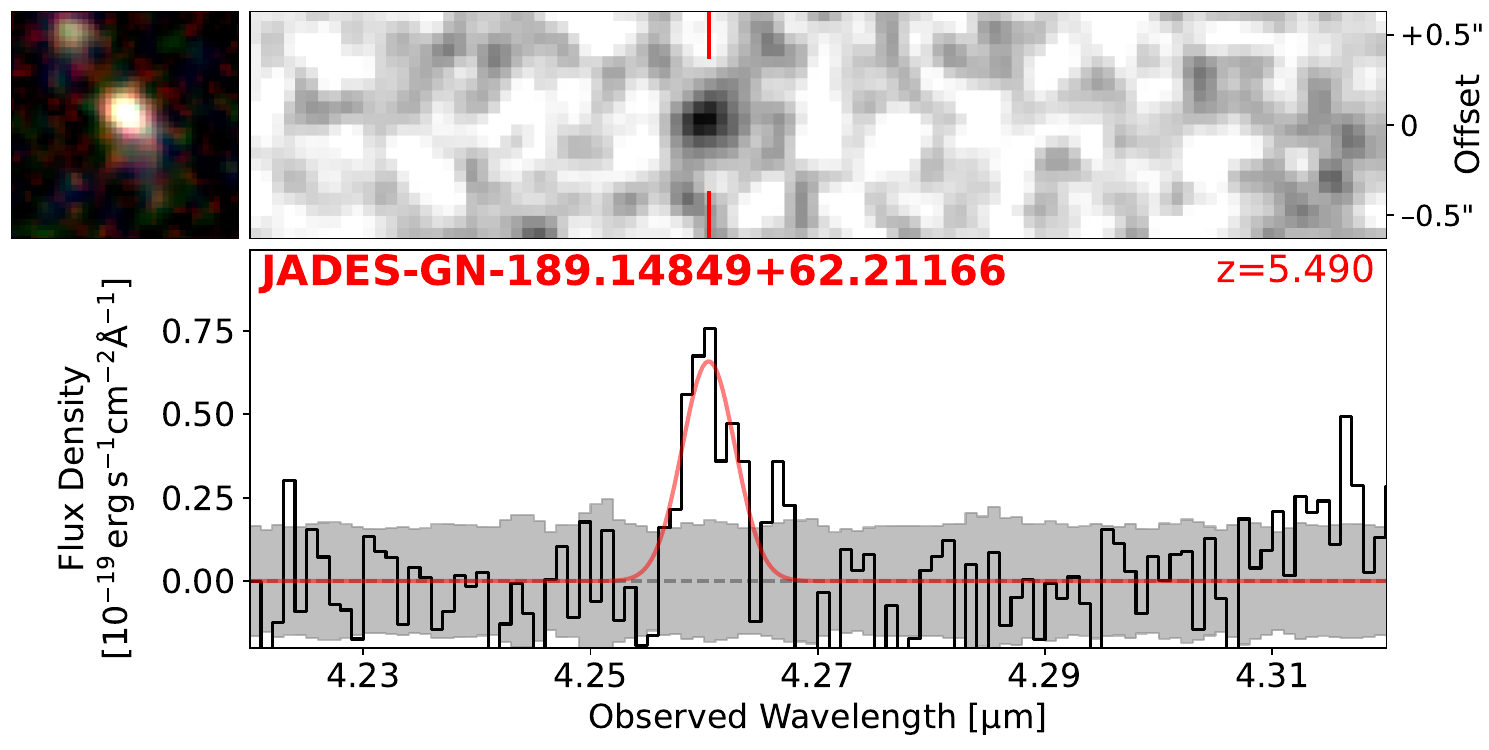}
\caption{Continued.} 
 \end{figure*}


\setcounter{figure}{1}
\renewcommand{\thefigure}{\thesection\arabic{figure}}

\begin{figure*}[!ht]
\centering
\includegraphics[width=0.8\linewidth]{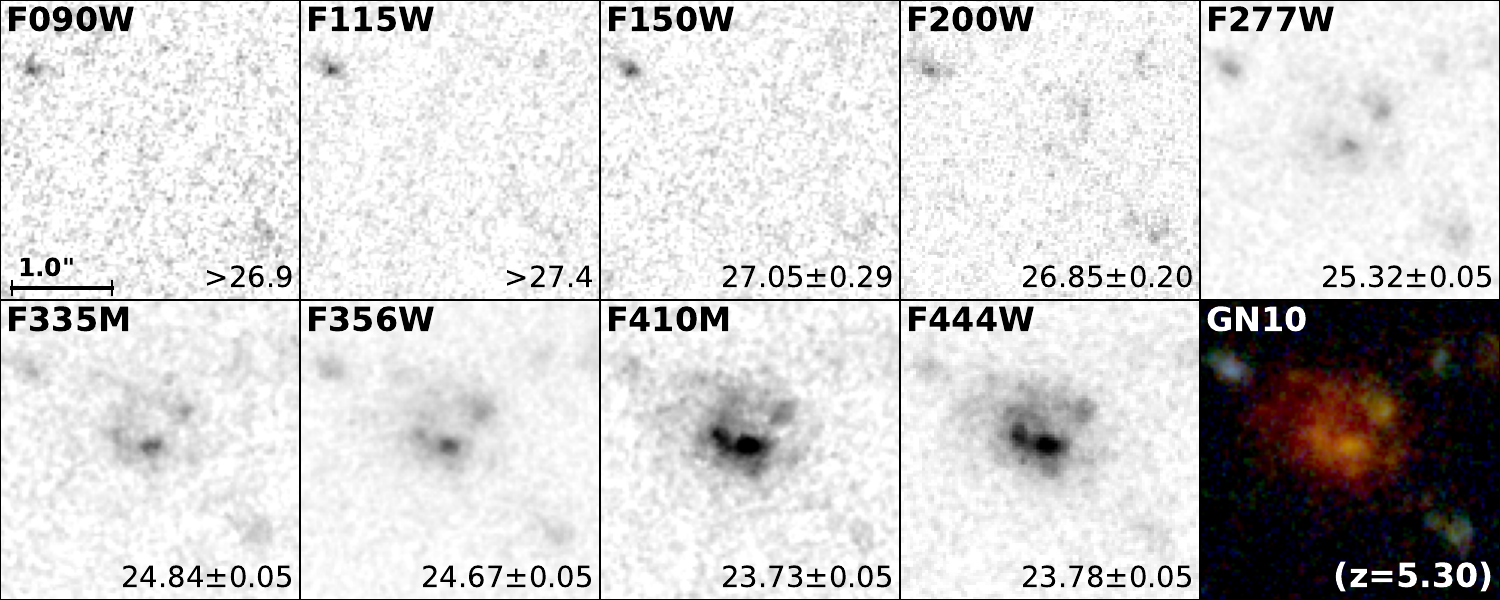}
\includegraphics[width=0.8\linewidth]{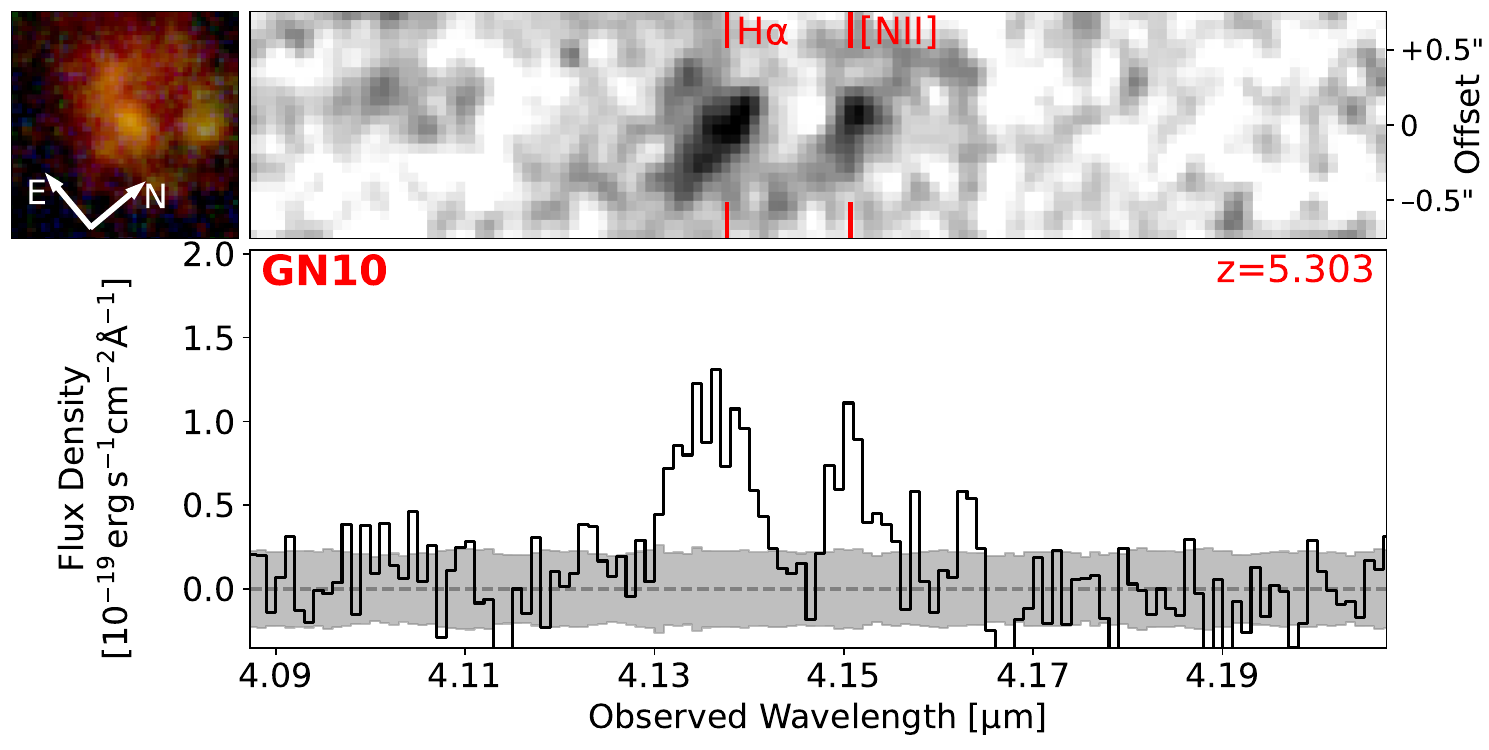}
\caption{\textbf{Top}: NIRCam images of GN10 in the F090W--F444W band. The brightness of GN10 in each band is noted in the lower-right corner of each panel (unit: AB mag). Image sizes are 3\arcsec$\times$3\arcsec\ (north up, east left). 
The last panel shows the true-color NIRCam image.
\textbf{Bottom}: NIRCam grism spectrum of GN10 (Similar to those in Figure~\ref{fig:04_spec} and Figure~\ref{fig:apd_spec}). Images are rotated to align with the dispersion direction. 
\ha\ and \nii\,$\lambda$6583 line can be detected at $z=5.303$, which are indicated by vertical red lines in 2D spectrum.
}
\label{fig:gn10}
\end{figure*}

\begin{figure*}[!hb]
\centering
\includegraphics[width=\linewidth]{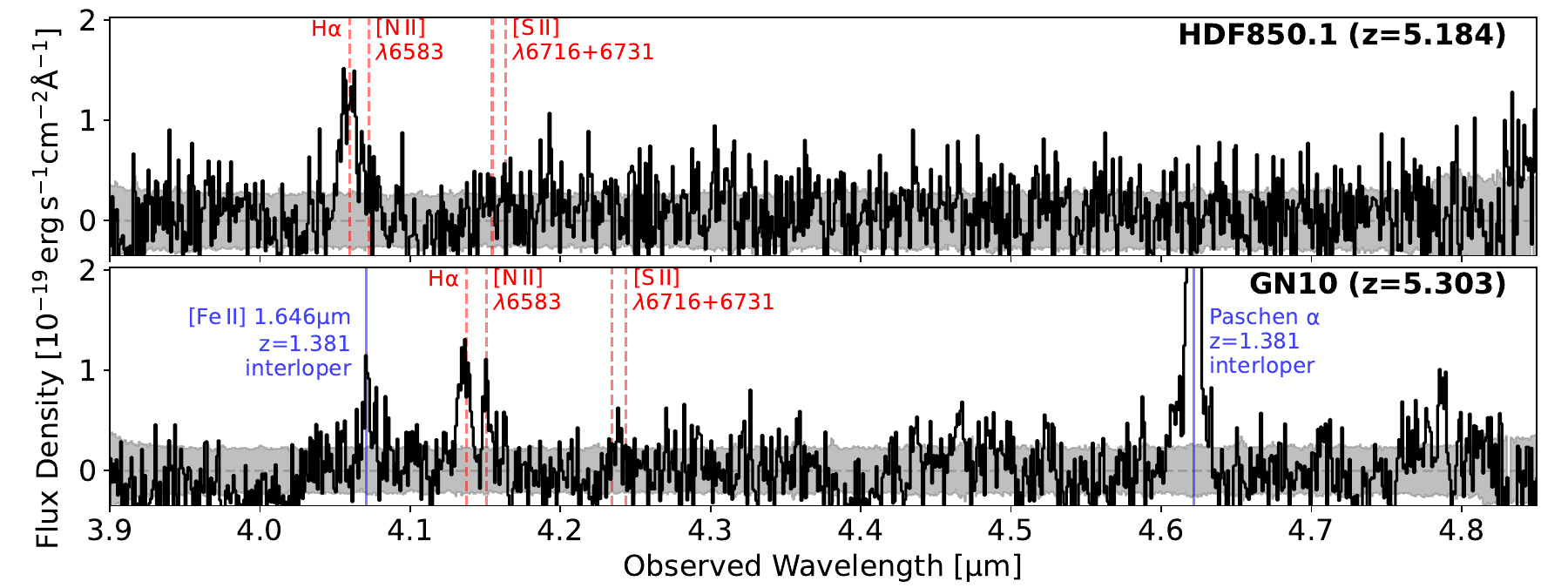}
\caption{Full NIRCam 1D grism spectra of HDF850.1 (top) and GN10 (bottom). 
The expected wavelengths of \ha, \nii\,$\lambda$6583, \sii\,$\lambda$6716 and 6731 lines are labeled with dashed red lines.
The full spectrum of GN10 is contaminated by a $z=1.381$ galaxy (R.A.: 12:36:34.282, Decl.: $+$62\arcdeg14\arcmin00\farcs5; 10\arcsec\ from GN10), whose \feii\,$\lambda$1.646\,\micron\ and Paschen\,$\alpha$ lines are detected (indicated by the solid blue lines).
}
\label{fig:full_spec}
\end{figure*}




\end{document}